\documentclass[12pt,nominion]{robinthesis}


\author{Robin Houston}
\title{Linear Logic without Units}
\begin{thesis}
    \maketitle
    \tableofcontents
    \vfill Word count: 34015
    
	\prefacesection{Linear Logic Without Units: Abstract}

{\Large Robin Houston, University of Manchester}
\vskip \baselineskip \noindent
{\large For the degree of Doctor of Philosophy, 30 September 2007}
\vskip\baselineskip
\begin{singlespace}
	We study categorical models for the unitless fragment of multiplicative linear logic. We find that the appropriate notion of model is a special kind of promonoidal category. Since the theory of promonoidal categories has not been developed very thoroughly, at least in the published literature, we need to develop it here. The most natural way to do this -- and the simplest, once the (substantial) groundwork has been laid -- is to consider promonoidal categories as an instance of the general theory of pseudomonoids in a monoidal bicategory. Accordingly, we describe and explain the notions of monoidal bicategory and pseudomonoid therein.
	
	The higher-di\-mens\-ion\-al nature of monoidal bicategories presents serious notational difficulties, since to use the natural analogue of the commutative diagrams used in ordinary category theory would require the use of three-di\-mens\-ion\-al diagrams. We therefore introduce a novel technical device, which we dub the \emph{calculus of components}, that dramatically simplifies the business of reasoning about a certain class of algebraic structure internal to a monoidal bicategory. When viewed through this simplifying lens, the theory of pseudomonoids turns out to be essentially formally identical to the ordinary theory of monoidal categories -- at least in the absence of permutative structure such as braiding or symmetry. We indicate how the calculus of components may be extended to cover structures that make use of the braiding in a braided monoidal bicategory, and use this to study braided pseudomonoids.
	
	A higher-dimensional analogue of Cayley's theorem is proved, and used to deduce a novel characterisation of the unit of a promonoidal category. This, and the other preceding work, is then used to give two characterisations of the categories that model the unitless fragment of intuitionistic multiplicative linear logic. Finally we consider the non-intuitionistic case, where the second characterisation in particular takes a surprisingly simple form.
\end{singlespace}
    
    \declpage
    \copyrightpage
	\prefacesection{Acknowledgements}

I am indebted to many people for interesting and informative
discussions over the past four years. There follows a list of some
of the people who have
influenced my thinking in varied and often immeasurable ways -- in some cases
through just one or two conversations, and in others through repeated
contact or protracted corespondence. I apologise to
all those whose names I have mistakenly omitted. Thanks, then
(without, as the lawyers say, prejudice to the generality) to
John Baez,
Eugenia Cheng,
Robin Cockett,
Jeff Egger,
Marcelo Fiore,
Nicola Gambino,
Richard Garner,
Martin Hyland,
Achim Jung,
Joachim Kock,
Tom Leinster,
Don MacInnes,
Craig Pastro,
John Power,
Luigi Santocanale,
Peter Selinger,
Harold Simmons,
Paul Taylor,
Todd Trimble,
and
Glynn Winskel.

A particular debt is owed to Dominic Hughes, who set me on the road
that led to this thesis by innocently asking what the definition of
`unitless star-aut\-on\-om\-ous category' should be; and without whose
unceasing encouragement and boundless optimism this work would
never have reached the stage that it has.

I am also deeply indebted to Andrea Schalk, an unfailingly helpful and
supportive supervisor whose seemingly inexhaustible patience I must
have severely tested at times.

My friends and family have helped in more ways than I
could list, and this thesis would not exist without them.

\newpage\thispagestyle{plain}
\vbox to \textheight{%
	\vskip 12em
	\hbox to \textwidth{\hskip 6em To Miranda\hfill}
	\vfill}

\documentclass{robinthesis}

\begin{thesischapter}{Intro}{Introduction}
\section{Linear logic without units}
The starting point of this work is the question: what is a categorical model of the unitless fragment of multiplicative linear logic? The question has at least some intrinsic interest, and we shall see that a proper understanding of the natural answer demands some unexpectedly sophisticated mathematics. Also the category of proof nets \citep{Girard87} is an important part of the proof theory of linear logic, and proof nets do not give a natural interpretation of the units. (Several authors \emph{have} considered extended notions of proof net that include units -- see \citet{TrimbleThesis,BCST,LSUnits,HughesUnits} -- but it must be admitted that these extended proof nets are substantially more complicated, and none succeeds in giving a purely geometric normal form for proofs. As a symptom of this complication, it is an open question whether equivalence of MLL proofs can be decided in polynomial time; whereas Girard's proof nets immediately suggest a polynomial time algorithm to decide proof equivalence in the unitless fragment.)

On a more pragmatic note, linear logic and related systems have a number of applications to computer programming. One example is the linear logic programming developed by \citet{LinearLogicProgramming}. Systems of linear types, and the closely related `uniqueness types', are also increasingly important; the functional programming language Clean \citep{Clean} uses
a system of uniqueness types to facilitate integration of effects such as input and output with purely-functional code.
For many practical purposes, the unit objects (or unit types, in a type system) do not play an important role. What is more, they can create significant complication, illustrated by the remarkable fact that the provability problem for the unit-\emph{only} fragment of multiplicative linear logic is NP-complete \citep{UOLLNP}. So it is reasonable to imagine that the unitless fragment of multiplicative linear logic will prove to be of practical importance.

The answer may appear at first glance to be trivial. After all it is well known, following \citet{SeelyLL}, that the star-autonomous categories of \citet{BarrStac} are the appropriate structures to model multiplicative linear logic, so surely one may simply describe some obvious notion of `unitless star-autonomous category'?
This superficially reasonable idea turns out to be too simple-minded to work. Consider the following proof:
\[\begin{prooftree}\thickness=0.2pt
	\[
		\justifies p \proves p
		\using \mbox{Ax}
	\]
	\quad
	\[
		\[
			\justifies q\proves q
			\using\mbox{Ax}
		\]
		\justifies \proves q\perp\parr q
		\using\mbox{$\bot$R}
	\]
	\justifies
	p \proves p \tn (q\perp\parr q)
	\using\mbox{$\tn$R}
\end{prooftree}\]
Although this proof does not involve any units, it makes essential use of a sequent with an empty left-hand side, and so its interpretation in a star-aut\-on\-om\-ous category \emph{does} necessarily involve the unit object. (A sequent of type $\proves q\perp\parr q$ would be interpreted as an arrow $I\to Q\parr Q\perp$, where $Q$ is the object that interprets the propositional variable $q$.)

So it is clear that the units cannot be dispensed with altogether: we need at least to have some way to interpret sequents that have an empty left- or right-hand side. Having thus isolated the difficulty, we begin by concentrating on the intuitionistic fragment. An intuitionistic sequent always has precisely one formula on the right, so only the left-hand side may be empty. Thus we need some analogue of `arrow $I\to X$', without there actually being a unit object $I$. In other words, we need some functor $\C\to\Set$ (where $\C$ is the category in question) to play the role of the hom functor $\C(I, -)$.
It turns out that structures of this sort have previously been studied, for a very different reason: \citet{DayPro} considered what he at the time called `premonoidal' categories, while studying monoidal structure on presheaf categories. Nowadays these structures are known as \emph{promonoidal} instead,\footnote{Indeed, the term \emph{premonoidal} has since been reused to mean something altogether different \citep{Premonoidal}.} which is the term we shall use.

A promonoidal category is something more general than a monoidal category: instead of having a tensor functor $\tn:\C\times\C\to\C$ and a unit object $I\in\C$, it has a tensor profunctor $P:\C\times\C\rPro\C$ and a unit profunctor $J: 1\rPro\C$. A profunctor $1\rPro\C$ is precisely a functor $\C\to\Set$, which is what we are looking for. So we are interested in particular in the special case of those promonoidal categories whose tensor is an honest functor, but whose unit is a general profunctor.

\section{The multicategory approach}
It is worth briefly contemplating the road not taken (particularly since it looks more attractive at first, but turns out on examination to lead to the same destination). One might take the view that any sequent system -- or at least any sequent system in which the cut rule is admissible, which is to say any sequent system that might reasonably be considered to describe a logic -- may be interpreted in a multicategory. The presence of such rules as the left-tensor and left-unit rules of linear logic allows us to restrict our attention to representable multicategories \citep{RepMulticats}, which are essentially the same as monoidal categories. On this view, in order to model the unitless fragment we should be looking at multicategories in which only \emph{non-empty} sequences of objects admit a representation. Such multicategories are \emph{prima facie} more general than the promonoidal categories considered here, for the following reason. Although a multicategory does have, for example, a natural transformation of the form
\begin{equation}\label{eq-notinv}
	\int^{A} \C(;A) \times \C(A,B;C) \to \C(B;C)
\end{equation}
given by composition, this is not necessarily invertible (as it must be in a promonoidal category). Since there seems to be no intrinsic reason that we should demand invertibility here, the multicategory formulation looks like an improvement over the promonoidal one. But this is an illusion: in the cases that we are considering, there is also an implication connective, so we may suppose that for every object $A$ there is a functor $A \lolli -$ and an isomorphism
\[
	\C(\vec X, A; B) \cong \C(\vec X; A\lolli B)
\]
natural in the sequence $\vec X$ and the object $B$. This does force the transformation (\ref{eq-notinv}) to be invertible, since we now have a sequence of isomorphisms
\[\begin{array}{r@{\;}c@{\;}l}
	\int^{A} \C(;A) \times \C(A,B;C)
		&\cong& \int^{A} \C(;A) \times \C(A;B\lolli C) \\
		&\cong& \C(;B\lolli C) \\
		&\cong& \C(B; C).
\end{array}\]
So we have arrived at the same destination by a different, and arguably more natural, route.

\section{The study of pseudomonoids}
Having established that we are engaged in the study of promonoidal categories, there is the immediate problem that not much has been written about them -- certainly when compared with monoidal categories, which have been very well-studied. It seems clear that most of the known results about monoidal categories have analogies in the promonoidal setting, but it would be unaccountably tedious merely to `translate' huge portions of the monoidal categories literature into the promonoidal setting. Better would be to find a general argument to the effect that 
such a translation is possible.

In fact there is nothing particularly special about promonoidal categories in an abstract sense. They are but one example of the general notion of \emph{pseudomonoid} in a monoidal bicategory, and we expect (and shall prove) that much of what is known about monoidal categories in particular is actually true of pseudomonoids in general, when formulated appropriately. Furthermore, when considering structures internal to a monoidal bicategory there is nothing particularly special about pseudomonoids! The translation procedure can in fact be carried through for a substantial class of structures internal to a monoidal bicategory. So we arrive at a general translation result that has potential applications that go far beyond those we consider here.

The thesis may be approximately divided into three parts. The first part (Chapters~\refchapter{Bicats} and~\refchapter{MonBicats}) consists of background material: we review the basics of bicategories and monoidal bicategories, before going on to define pseudomonoids. The second part (Chapters~\refchapter{Coh}--\refchapter{Psmon}) establishes the `translation' mentioned above. Only in the third part (Chapters~\refchapter{Promon} and~\refchapter{SemiCC}) do we finally return to the original question, and use all this machinery to study the models of unitless linear logic.

\section{Prerequisites}
We assume the basics of linear logic, category theory, and categorical proof theory. As far as linear logic and its categorical interpretation is concerned, these prerequisites are essentially contained in \citet{Girard87} and \citet{SeelyLL}. Of category theory we assume a little more: say the contents of \citet{MacLane}. Some familiarity with the theory of profunctors \citep{LawvereModules,BenabouDist} and promonoidal categories \citep{DayPro} would be useful, but is not strictly assumed.

\section{Other approaches}
Others have recently considered the question of defining categorical models for the unitless fragment of multiplicative linear logic. A preprint of \citet{LSFreeBool} gave a definition that, on examination, appeared
weaker than the one developed in this work.
Correspondence with the authors established that this difference was
not intended, and the final version includes an additional axiom that
makes the definition equivalent to ours.
\citet{ProofNetCats} give a very different-looking definition just for the star-autonomous case, which is nevertheless again equivalent to ours.

\end{thesischapter}

\documentclass{robinthesis}

\newcommand\defeqto{=}

\begin{thesischapter}{Bicats}{Bicategories}
Monoidal and promonoidal categories are both instances of the
general notion of a pseudomonoid in a monoidal bicategory. We
shall need some results about promonoidal categories that are
in fact generally true of pseudomonoids in any monoidal bicategory,
and which we should therefore like to prove as such.

Unfortunately the literature on monoidal bicategories is still fairly sparse.
The relevant definitions may be obtained by specialising to the one-object
case those given for tricategories by \citet{GPS}, and this has been done
explicitly in the unpublished dissertation of \citet{CarmodyThesis}.
Also \citet{MonBicat} and \citet{HDA1} have given explicit definitions for
the important special case of Gray-categories. However there is no explicit,
published account of the general notion.

Even the literature on plain bicategories is rather scanty, and although
the situation is improving \citep[see][for example]{LackCompanion} many
fundamental results have no published proof, and there is still a
substantial gap between what is known to the experts and what has been
written down. Neither is notation
yet standardised. For these reasons, we give in this chapter a rapid
but reasonably thorough account of the bicategory theory that we need,
with the occasional digression.

\section{Bicategories: basic definitions}
\begin{definition} 
	A bicategory $\B$ consists of:
	\begin{itemize}
		\item a set\footnote{Possibly quite a large set: whereas most of ordinary
		category theory can be formalised in a ``one universe'' foundation, it's
		convenient to assume at least two Grothendieck universes for the purposes
		of bicategory theory. (We want to permit e.g.\ the bicategory of large
		categories.) We shall leave these considerations implicit, on the whole.}
		$|\B|$ of objects,
		\item for every pair $A$, $B$ of objects, a hom-category $\B(A,B)$,
			whose objects are called \emph{1-cells} or arrows, and whose morphisms
			are called \emph{2-cells}
		\item for every object $A$, a selected `identity' 1-cell $1_A\in\B(A,A)$,
		\item for every triple $A$,$B$,$C$ of objects, a `horizontal composition' functor
		\[
			\o: \B(B,C)\x\B(A,B) \to \B(A,C),
		\]
		\item for every pair $A$, $B$ of objects, natural isomorphisms
		\[
			\begin{array}{l}
			\l: 1_B\o- \To -: \B(A,B)\to\B(A,B),\\
			\r: -\o 1_A \To -: \B(A,B)\to\B(A,B),
			\end{array}
		\]
		\item for every four objects $A$,$B$,$C$,$D$, a natural isomorphism
		\[
			\a: -\o(-\o-)\To(-\o-)\o-: \B(C,D)\x\B(B,C)\x\B(A,B)\to\B(A,D)
		\]
	\end{itemize}
	subject to two coherence conditions: for all $f\in\B(A,B)$ and $h\in\B(B,C)$,
	the diagram
	\begin{diagram}
		h\o(1_B\o f) &\rTo^{\a_{h,1_B,f}}&(h\o1_B)\o f\\
		&\rdTo[snake=-1ex](1,2)<{h\o \l_f}
			\raise1ex\hbox{$\clr$}%
			\ldTo[snake=1ex](1,2)>{\r_h\o f}\\
		&h\o f,
	\end{diagram}
	commutes in $\B(A,C)$,
	and for all $e\in\B(A,B)$, $f\in\B(B,C)$, $g\in\B(C,D)$, $h\in\B(D,E)$, the diagram
	\begin{diagram}
	  h\o \bigl(g\o (f\o e)\bigr)
	  &\rTo^{\a_{h,g,f\o e}}&(h\o g)\o (f\o e)
	  &\rTo^{\a_{h\o g,f,e}} & \bigl((h\o g)\o f\bigl)\o e
	  \\
	  &\rdTo[snake=-1em](1,2)<{h\o \a_{g,f,e}}
	  &\ca
	  & \ruTo[snake=1em](1,2)>{\a_{h,g,f}\o e}
	  \\
	  & \spleft{h\o\bigl((g\o f)\o e\bigr)}
	  & \rTo_{\a_{h,g\o f,e}}
	  & \spright{\bigl(h\o(g\o f)\bigr)\o e}
	\end{diagram}
	commutes in $\B(A,E)$.
\end{definition}
We shall often omit the subscript on identity 1-cells, writing just $1$
rather than $1_{A}$, when the object can be easily determined from the context.
We also omit the 1-cell subscripts of $\a$, $\l$ and $\r$ from time
to time.
\begin{remark} 
	A bicategory with just one object may be regarded as a monoidal
	category: the definition simply reduces to the usual definition
	of monoidal category, in that case.
	Furthermore, Mac Lane's coherence theorem for monoidal categories
	\citep{MLCoh} equally well applies to bicategories in general:
	the proof goes through essentially unchanged. \Citet{KellyML}
	shows -- again for monoidal categories -- that just two axioms,
	corresponding to our $\ca$ and $\clr$, suffice for coherence; and
	that proof, too, applies equally well to bicategories in general.
	
	This shows that, between any pair of functors built up from identities
	and composition, there is at most one natural transformation built
	from $\a$, $\l$, $\r$ and their inverses.
	Therefore we shall not usually give names to the `structural isomorphisms'
	$\a$, $\l$, $\r$, their inverses, and composites thereof. We shall instead
	use the symbol `$\cong$' as a generic label for a structural isomorphism.
	Since by coherence there is a unique such isomorphism of each type,
	this practice introduces no ambiguity.
\end{remark}
\begin{remark}\label{rem-abstract-coherence} 
	There is another version of the coherence theorem, proven for
	the case of monoidal categories by \citet[section~1]{BTC}, and
	explicitly for bicategories by \citet[chapter~2]{GurskiThesis}%
	\footnote{
		Gurski's account is exceptionally thorough and largely
		self-contained. He does use the bicategorical Yoneda
		lemma without proof, and indeed that proof does not
		appear ever to have been published -- presumably because
		the idea is straightforward, even if the details verge
		on overwhelming. We provide a proof as Prop.~\ref{prop-yoneda}
		below.
	}.
	Anticipating some definitions we have yet to make, this version
	says that the canonical functor from a free bicategory to the
	corresponding free 2-category is a biequivalence. This is a
	powerful result, which gives an honest justification for neglecting
	the structural isomorphisms in many circumstances.
\end{remark}
\begin{remark} 
	We shall often describe 2-cells, and equations between them,
	using pasting diagrams. It's important to be clear about how
	such diagrams are to be interpreted, which we'll explain by
	reference to the following example. Let $\sigma$ and $\tau$
	be 2-cells that fit into the diagram
	\begin{diagram}
		A & \rTo^{f} & B & \rTo^{g} & C \\
		\dTo<h & \Arr\Swarrow\sigma & \dTo>k & \Arr\Swarrow\tau & \dTo>l \\
		X & \rTo_{m} & Y & \rTo_{n} & Z
	\end{diagram}
	Firstly, notice that this diagram does not uniquely define a
	2-cell. Instead it defines a \emph{family} of 2-cells, one
	for each bracketing of the source and target edges, e.g.\ for
	the bracketings $l\o(g\o f)$ and $(h\o m)\o n$,
	the pasting diagram describes the 2-cell
	\[
		l\o(g\o f) \rTo^{\a_{l,g,f}} (l\o g)\o f
			\rTo^{\tau\o f} (n\o k)\o f
			\rTo^{\a^{-1}_{n,k,f}} n\o (k\o f)
			\rTo^{n\o\sigma} (h\o m)\o n.
	\]
	This also demonstrates the second subtlety: the associativity
	$(n\o k)\o f \rTo^{\a^{-1}_{n,k,f}} n\o (k\o f)$ must be implicitly
	inserted between $\tau$ and $\sigma$.
	
	Observe also that an \emph{equation} between pasting diagrams may
	be regarded as a family of equations, one for each bracketing of the
	source and target 1-cells. These equations are all equivalent, in the
	sense that each implies the others, so to prove such a family of
	equations it suffices to prove one of them.
	
	These remarks are intended only to help the reader to understand
	what we mean when we draw a pasting diagram or an equation
	involving them. A rigorous treatment of bicategorical pasting
	is given by \citet{VerityThesis}.
\end{remark}

\begin{remark} 
	As well as pasting diagrams, one can also use \emph{string diagrams}
	\citep{GTC, LowDim} to represent and calculate with 2-cells in a
	bicategory. In a string diagram, an object is represented by a region
	of the plane, a 1-cell by a line, and a 2-cell by a node. For example,
	the 2-cell $\gamma: f \To g: A \to B$ is drawn as
	
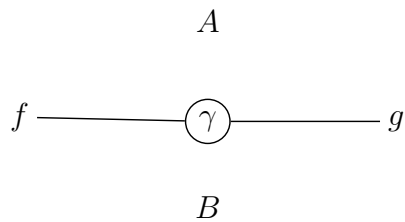
\begin{diagram}
		&& A \\
		\rnode{f}{f} && \circlenode{gamma}{\gamma} && \rnode{g}{g} \\
		&& B
		\ncline[nodesepB=0pt]{-}{f}{gamma}
		\ncline[nodesepA=0pt]{-}{gamma}{g}
	\end{diagram}
	and a 2-cell $\delta: g\o f \To h$ would be drawn as
	
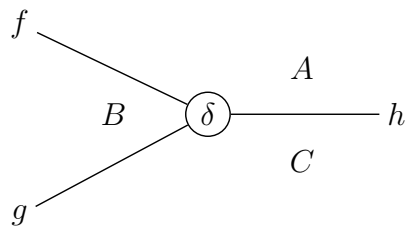
\begin{diagram}[h=1.5em]
		\rnode{f}{f} \\
		&&& A \\
		& B & \circlenode{delta}{\delta} && \rnode{h}{h} \\
		&&& C \\
		\rnode{g}{g}
		\ncline[nodesepB=0pt]{-}{f}{delta}
		\ncline[nodesepB=0pt]{-}{g}{delta}
		\ncline[nodesepA=0pt]{-}{delta}{h}
	\end{diagram}
	for $f: A\to B$, $g:B\to C$ and $h:A\to C$.
	Identity 1-cells are not drawn, so a 2-cell $\zeta: 1_{A}\To f: A\to A$
	is drawn
	
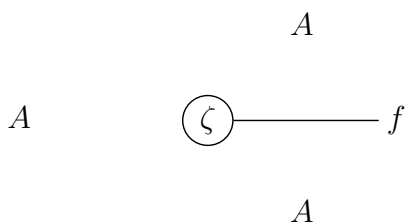
\begin{diagram}
		&& & A \\
		A && \circlenode{zeta}{\zeta} && \rnode{f}{f}\\
		&& & A
		\ncline[nodesepA=0pt]{-}{zeta}{f}
	\end{diagram}
	Composition of 2-cells corresponds to the pasting together of
	string diagrams along the appropriate edge: the orientation we
	have chosen for our diagrams has the unfortunate effect that
	horizontal composition is represented by vertical pasting, and
	vice versa. However it has the psychological advantage that
	the diagrams read from left to right.
	
	The object labels may usually be omitted, since they can be
	inferred from the types of the 1-cells.
	
	It is the case that geometric manipulations of string diagrams
	always correspond to allowable operations. For example, given
	2-cells $\gamma: f\To 1: A\to A$ and $\delta: 1\To g: A\to A$,
	the diagram
	
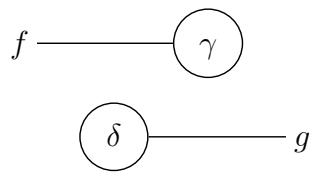
\begin{diagram}
		\rnode{f}{\strut f} & & \circlenode{gamma}{\strut\gamma} \\
		& \circlenode{delta}{\strut\delta} & & \rnode{g}{\strut g}
		\ncline[nodesepB=0pt]{-}{f}{gamma}
		\ncline[nodesepA=0pt]{-}{delta}{g}
	\end{diagram}
	can be deformed to
	
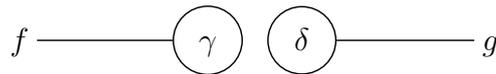
\begin{diagram}
		\rnode{f}{\strut f} & & \circlenode{gamma}{\strut\gamma} & \circlenode{delta}{\strut\delta}
			&& \rnode{g}{\strut g}
		\ncline[nodesepB=0pt]{-}{f}{gamma}
		\ncline[nodesepA=0pt]{-}{delta}{g}
	\end{diagram}
	and then to
	
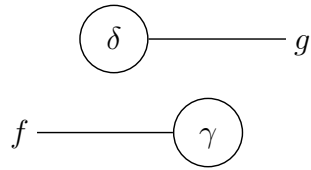
\begin{diagram}
		& \circlenode{delta}{\strut\delta} & & \rnode{g}{\strut g} \\
		\rnode{f}{\strut f} & & \circlenode{gamma}{\strut\gamma}
		\ncline[nodesepB=0pt]{-}{f}{gamma}
		\ncline[nodesepA=0pt]{-}{delta}{g}
	\end{diagram}
	The corresponding sequence of pasting diagrams is
	\vskip4em
	\hbox to \linewidth{
	$
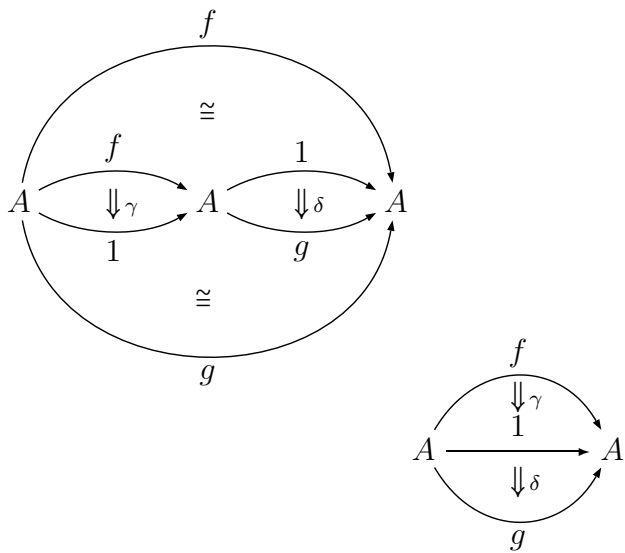
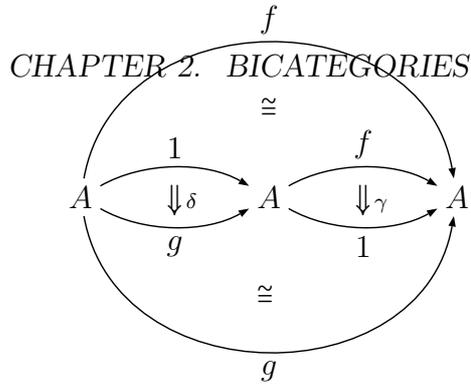
\begin{diagram}
		&&\cong \\
		\rnode{1}{A} & \Arr\Downarrow{\scriptstyle\gamma} & \rnode{2}{A}
			& \Arr\Downarrow{\scriptstyle\delta} & \rnode{3}{A} \\
		&&\cong
		\ncarc[arcangle=30]{->} {1}{2} \Aput{f}
		\ncarc[arcangle=-30]{->}{1}{2} \Bput{1}
		\ncarc[arcangle=30]{->} {2}{3} \Aput{1}
		\ncarc[arcangle=-30]{->}{2}{3} \Bput{g}
		\ncarc[arcangle=80,ncurv=1]{->}{1}{3} \Aput{f}
		\ncarc[arcangle=-80,ncurv=1]{->}{1}{3} \Bput{g}
	\end{diagram}$\hfil}
	\vskip1em
	\begin{diagram}[h=1em]
		&\Arr\Downarrow{\scriptstyle\gamma} \\
		\rnode{l}{A} & \rTo^{1} & \rnode{r}{A} \\
		&\Arr\Downarrow{\scriptstyle\delta}
		\ncarc[arcangle=60,ncurv=1]{->}{l}{r} \Aput{f}
		\ncarc[arcangle=-60,ncurv=1]{->}{l}{r} \Bput{g}
	\end{diagram}
	\hbox to \linewidth{\hfil
	$\begin{diagram}
		&&\cong \\
		\rnode{1}{A} & \Arr\Downarrow{\scriptstyle\delta} & \rnode{2}{A}
			& \Arr\Downarrow{\scriptstyle\gamma} & \rnode{3}{A} \\
		&&\cong
		\ncarc[arcangle=30]{->} {1}{2} \Aput{1}
		\ncarc[arcangle=-30]{->}{1}{2} \Bput{g}
		\ncarc[arcangle=30]{->} {2}{3} \Aput{f}
		\ncarc[arcangle=-30]{->}{2}{3} \Bput{1}
		\ncarc[arcangle=80,ncurv=1]{->}{1}{3} \Aput{f}
		\ncarc[arcangle=-80,ncurv=1]{->}{1}{3} \Bput{g}
	\end{diagram}$}
	\vskip4em\noindent
	which is clearly much harder to follow. This example illustrates
	the way that string diagrams leave the unit constraints $\l$
	and $\r$ implicit, making powerful use of coherence. (Of
	course the string diagram formalism -- in common with pasting
	diagrams -- also leaves implicit the associator $\a$, though that
	is not shown in this particular example. But the real power of string
	diagrams comes from the ease with which they handle the identities.)
\end{remark}

\begin{definition} 
	For any bicategory $\B$ there is a bicategory $\B\op$ obtained from $\B$
	by reversing the direction of the 1-cells, so $\B\op(A,B) = \B(B,A)$; and also a
	bicategory $\B\co$ obtained by reversing the direction of the 2-cells, so
	$\B\co(A,B) = \B(A,B)\op$.
\end{definition}
\begin{definition} 
	Bicategories $\B$ and $\BC$ have a product formed
	in the obvious way: the set of objects is $|\B\x\BC| = |\B|\x|\BC|$,
	the hom-categories are
	$(\B\x\BC)(\langle A,B\rangle,\langle X,Y\rangle) = \B(A,X)\x\BC(B,Y)$, and
	horizontal composition is defined pointwise. This also extends in the obvious
	way to the product of three of more bicategories.
\end{definition}

\begin{definition}\label{def-psfun} 
	Given bicategories $\B$ and $\BC$, a \emph{pseudo-functor}
	$F:\B\to\BC$ consists of:
	\begin{itemize}
		\item for every object $A\in\B$, an object $FA\in\BC$,
		\item for every pair $A$, $B$ of objects of $\B$, a functor
		\[
			F_{A,B}: \B(A,B)\to\B(FA,FB),
		\]
		\item for every $A\in\B$, an invertible 2-cell $F_A: 1_{FA}\To F(1_A): FA\to FA$,
		\item for every $A$ and $B\in\B$, a natural isomorphism with components
		\[
			F_{g,f}: F(g)\o F(f) \To F(g\o f),
		\]
	\end{itemize}
	such that for every $f:A\to B$ in $\B$, the diagrams below commute in the category $\BC(FA,FB)$,
	\[
	\begin{diagram}
		1_{FB}\o F(f) & \rTo^{F_B\o F(f)} & F(1_B)\o F(f)\\
		\dTo<{\l_{F(f)}} &\cFl&\dTo>{F_{1_B, f}}\\
		F(f) & \lTo_{F(\l_f)} & F(1_B\o f)
	\end{diagram}
	\qquad
	\begin{diagram}
		F(f)\o 1_{FA} & \rTo^{F(f)\o F_A} & F(f)\o F(1_A)\\
		\dTo<{\r_{F(f)}} &\cFr&\dTo>{F_{f,1_A}}\\
		F(f) & \lTo_{F(\r_f)} & F(f\o 1_A)
	\end{diagram}
	\]
	and for every $A\rTo^f B\rTo^g C\rTo^h D$ in $\B$, the diagram
	\begin{diagram}
		F(h)\o(F(g)\o F(f)) & \rTo^{\a_{Fh,Fg,Ff}}& (F(h)\o F(g))\o F(f)\\
		\dTo<{F(h)\o F_{g,f}} && \dTo>{F_{h,g}\o F(f)}\\
		F(h)\o F(g\o f) &\cFa& F(h\o g)\o F(f)\\
		\dTo<{F_{h,(g\o f)}} && \dTo>{F_{(h\o g),f}}\\
		F(h\o(g\o f)) & \rTo_{F(\a_{h,g,f})} & F((h\o g)\o f)
	\end{diagram}
	commutes in the category $\BC(FA,FD)$.
	A pseudo-functor is also known as a \emph{homomorphism of bicategories}.
\end{definition}
\begin{remark}
	There is also a notion of \emph{lax functor} between bicategories,
	defined as above, except that the 2-cells that appear in the
	definition need not be invertible. Lax functors have their uses
	-- for example, a lax functor $1\to\B$ is the same as a monad
	in $\B$ -- but we have no need of them.
\end{remark}

\begin{definition}\label{def-psnat} 
	Given pseudo-functors $F$, $G:\B\to\BC$, a \emph{pseudo-natural
	transformation} $\gamma: F\To G$ consists of:
	\begin{itemize}
		\item for every $A\in\B$, a 1-cell $\gamma_A: FA\to GA$,
		\item for every $f:A\to B$ in $\B$, an invertible 2-cell
		\begin{diagram}
			FA &\rTo^{\gamma_A} & GA\\
			\dTo<{Ff} & \Nearrow \gamma_f& \dTo>{Gf}\\
			FB & \rTo_{\gamma_B} & GB
		\end{diagram}
		such that this assignment is natural in $f$. Naturality amounts to
		asking that, for every 2-cell $\tau: f\To g: A\to B$, we have
		\[
		\hskip-1em
		\begin{diagram}[size=4em]
			FA & \rTo^{\gamma_A} & \rnode{GA}{GA}\\
			\dTo<{Ff} & \llap{$\Nearrow \gamma_f$}
				& \begin{array}{c}\Rightarrow\\G(\tau)\end{array}\\
			FB & \rTo_{\gamma_B} & \rnode{GB}{GB}
			\ncarc{->}{GA}{GB}\Aput{Gg}
			\ncarc{<-}{GB}{GA}\Aput{Gf}
		\end{diagram}
		\hskip3em=\hskip3em
		\begin{diagram}[size=4em]
			\rnode{FA}{FA} & \rTo^{\gamma_A} & GA\\
			\begin{array}{c}\Rightarrow\\F(\tau)\end{array}
				& \rlap{$\Nearrow \gamma_g$} & \dTo>{Gg}\\
			\rnode{FB}{FB} & \rTo_{\gamma_B} & GB
			\ncarc{->}{FA}{FB}\Aput{Fg}
			\ncarc{<-}{FB}{FA}\Aput{Ff}
		\end{diagram}
		\]
		\item These data must satisfy the \emph{unit condition}: for every $A\in\B$ we have
		\[
		\hskip-1em
			\begin{diagram}[size=4em]
				\rnode{FA}{FA} & \rTo^{\gamma_A} & GA\\
				\begin{array}{c}\Rightarrow\\F_A\end{array}
					& \hbox to 0pt{$\mkern4mu\Nearrow \gamma_{1_A}$} & \dTo>{G(1_A)}\\
				\rnode{FB}{FA} & \rTo_{\gamma_A} & \rnode{GA}{GA}
				\ncarc{->}{FA}{FB}\Aput{F(1_A)}
				\ncarc{<-}{FB}{FA}\Aput{1_{FA}}
			\end{diagram}
			\hskip3em=\hskip3em
			\begin{diagram}[size=4em]
				FA & \rTo^{\gamma_A} & \rnode{GA}{GA}\\
				\dTo<{1_{FA}} & \llap{$\cong\mkern4mu$}
					& \begin{array}{c}\Rightarrow\\G_A\end{array}\\
				FA & \rTo_{\gamma_A} & \rnode{GB}{GA}
				\ncarc{->}{GA}{GB}\Aput{G(1_A)}
				\ncarc{<-}{GB}{GA}\Aput{1_{GA}}
			\end{diagram}
		\]
		\item and the \emph{composition condition}: for all $A\rTo^f B\rTo^g C$ in $\B$, we have
		\[
		\begin{diagram}[h=2em]
			\rnode{FA}{FA} & \rTo^{Ff}& FB & \rTo^{Fg}& \rnode{FC}{FC}\\
			&&\raise4pt\hbox{$\Downarrow F_{g,f}$}\\
			\dTo<{\gamma_A} &&&& \dTo>{\gamma_C}\\
			&&\Swarrow \gamma_{g\o f}\\
			GA && \rTo_{G(g\o f)} && GC
			\ncarc[arcangle=-60]{->}{FA}{FC}\Bput{F(g\o f)}
		\end{diagram}
		\quad=\quad
		\begin{diagram}[h=2em]
			FA & \rTo^{Ff}& FB & \rTo^{Fg}& FC\\
			&\Swarrow\gamma_f & \dTo>{\gamma_B}&\Swarrow\gamma_g\\
			\dTo<{\gamma_A} && GB && \dTo>{\gamma_C}\\
			&\ruTo^{Gf} & \Downarrow G_{g,f} & \rdTo^{Gg}\\
			GA && \rTo_{G(g\o f)} && GC
		\end{diagram}
		\]
	\end{itemize} 
\end{definition}
\begin{remark}
	Of course there is such a thing as a lax natural transformation,
	where the 2-cells $\gamma_{f}$ need not be invertible. (In
	B{\'e}nabou's original terminology, this would actually be an
	oplax transformation -- his 2-cells point the other way --
	but the direction we use seems to be a more natural and
	useful choice: see \citet[Section~5.7]{LackCompanion} for
	one piece of technical evidence for this assertion.)
	However, it should be noted that the collection of
	lax functors $\B\to\BC$, lax transformations between them,
	and modifications (for which see below) between those does
	\emph{not} form a bicategory, which shows that one must be careful
	what one laxifies. In any case, pseudo-functors and pseudo-natural
	transformations are all that we shall need.
\end{remark}
\begin{definition} 
	Given pseudo-natural transformations $\gamma$, $\delta: F\To G:\B\to\BC$,
	a \emph{modification} $m: \gamma\Tto\delta$ consists of: for every
	$A\in\B$, a 2-cell $m_A: \gamma_A\To\delta_A$ such that for every $f:A\to B$
	in $\B$, we have
	\[
		\begin{diagram}
			\rnode{FA}{FA} & \Uparrow m_A & \rnode{GA}{GA}\\
			\dTo<{Ff} & \raise-4pt\hbox{$\Nearrow\gamma_f$} & \dTo>{Gf}\\
			FB & \rTo_{\gamma_B} & GB
			\ncarc{->}{FA}{GA}\Aput{\delta_A}
			\ncarc{<-}{GA}{FA}\Aput{\gamma_A}
		\end{diagram}
		\qquad=\qquad
		\begin{diagram}
			FA & \rTo^{\delta_A} & GA\\
			\dTo<{Ff} & \raise4pt\hbox{$\Nearrow\delta_f$} & \dTo>{Gf}\\
			\rnode{FB}{FB} & \Uparrow m_B & \rnode{GB}{GB}
			\ncarc{->}{FB}{GB}\Aput{\delta_B}
			\ncarc{<-}{GB}{FB}\Aput{\gamma_B}
		\end{diagram}
	\]
\end{definition}

\begin{remark}
	Recall that, given a commutative diagram in an ordinary category,
	if one of the arrows is invertible, and is replaced by its inverse in such
	a way that the resulting diagram still has a single source and a single
	target object, then this resulting diagram also commutes.
	This fact has a two-dimensional analogue, as follows: a pasting equation may be
	viewed as a polyhedron, by gluing the two pasting diagrams together along their
	(common) boundary. If, in this polyhedron, one of the cells (faces) is
	invertible and is replaced
	by its inverse in such a way that there is still a unique way to decompose the
	resulting polyhedron as a pair of pasting diagrams, then this resulting pasting
	equation also holds.
	We shall sometimes use this implicitly, when it is more convenient to use
	such a variant of some equation.
\end{remark}

The usual 2-categorical notions of adjunction, equivalence, monad, etc.\ may
also be defined in a bicategory in the obvious way: a little of the theory
of adjunctions is developed below. Since it will be useful almost
immediately, we give here the definition of equivalence:
\begin{definition}\label{def-equivalence} 
	An \emph{equivalence} from $A$ to $B$ in a bicategory consists of a pair of 1-cells
	$f: A\to B$ and $g:B\to A$, and a pair of invertible 2-cells $e: 1_A\To g\o f$
	and $e': 1_B\To f\o g$. We also say that the arrow $f$ is an equivalence
	just when there exist $g$, $e$, $e'$ as above.
\end{definition}
In particular, every identity arrow $1_{A}$ is an equivalence, as is any
1-cell isomorphic to an identity. There is also an obvious
notion of isomorphism, but it is not very useful in general. In particular, identity
arrows in a bicategory are generally only equivalences and not isomorphisms,
and an object of a bicategory need not even be isomorphic to itself.

Pseudo-natural transformations compose in the obvious way, as do modifications.
Given any two bicategories $\B$ and $\BC$, there is a \emph{pseudo-functor bicategory}
$\Bicat(\B,\BC)$ whose objects are pseudo-functors $\B\to\BC$,
whose 1-cells are pseudo-natural transformations and whose
2-cells are modifications.\footnote{
	Note that this notation is inconsistent with that of
	\citet{FIB}, which uses $\Bicat(\B,\BC)$ to denote the
	bicategory whose objects are \emph{lax} functors from
	$\B$ to $\BC$.
}
We omit the routine verification that
this is indeed a bicategory, but remark that coherence lifts from
$\BC$. It is significant that if $\BC$ is in fact a 2-category
(i.e.\ its associativity and unit isomorphisms are all identities) then
$\Bicat(\B,\BC)$ is also a 2-category.

A pseudo-functor $F:\B\to\BC$ is said to be a \emph{biequivalence}
if it is a local equivalence and \emph{biessentially surjective} -- i.e.\ every
object of $\BC$ is equivalent to one of the form $FA$. If there is such a
biequivalence then we say that $\B$ is \emph{biequivalent} to $\BC$.
(There is a lot more that could be said about biequivalence, of course\dots)

For any bicategories $\BA$, $\B$, $\BC$, there is a biequivalence
\[
	\Bicat(\BA\x\B, \BC) \simsim \Bicat(\BA, \Bicat(\B,\BC))
\]
defined in the natural way.
%

\section{On the identities}\label{s-identities}
It is a trivial observation that, say, a semigroup may have at
most one unit: if $i$ and $j$ are both units then $ij=i$ -- because
$j$ is a unit -- and also $ij=j$ because $i$ is a unit. So $i=j$.
Thus the existence of a unit is a property of a semigroup, rather
than being real additional structure. In higher dimensions, this
phenomenon persists: for example, let
$A$ be an object of the bicategory $\B$, and let $1^\bullet: A\to A$
be equipped with natural isomorphisms $\l^{\bullet}$ and $\r^{\bullet}$
making $1^{\bullet}$ act as an identity. Then there is an isomorphism
\[
	1^{\bullet} \rTo^{(\r^{\bullet}_{1_{A}})^{-1}} 1_{A}\o 1^{\bullet}
		\rTo^{\l_{1^{\bullet}}} 1_{A},
\]
so identities in a bicategory are unique up to isomorphism, which
is as much as one could reasonably expect in the circumstances.

An important, if partially-submerged, theme of the present work
is that the property-likeness of units has some interesting consequences.
One consequence of this has been studied in detail by
\citet{KockUnits,JoyalKockUnits,KockWeakIdentityArrows};
a different aspect is visible in the present work, especially in
Section~\chref{Promon}{s-promon-unit}, where we
find that a braided promonoidal category has a \emph{canonical} unit,
and identify a simple property that holds just when this unit exists.

In the immediate context, we can illustrate some of the phenomena as
follows: the unit conditions in the
definitions of pseudo-functor and pseudo-natur\-al transformation
are, in a sense, redundant. (This is not the case where lax functors
or lax natural transformations are concerned -- the invertibility of
our 2-cells is essential to the argument.)
Although there is, to my knowledge, no written account available
of the material of this section, in view of its elementary nature
it is reasonable to suppose that it is known to experts in the
field. (The expert whom I asked declined to comment on the question
of how well-known these results are: make of that what you will.)

It will be convenient to introduce some temporary notation,
that we use only in this section. Let $F:\B\to\BC$ be a
pseudo-functor.
Then given a 1-cell $f:FA\to FB$ in $\BC$, let us write $\r^{F}_{f}$
for the composite
\[
	f\o F(1_{A}) \rTo^{f\o F_{A}^{-1}} f\o 1_{FA}
		\rTo^{\r_{f}} f,
\]
and $\l^{F}_{f}$ for
\[
		F(1_{A})\o f \rTo^{F_{A}^{-1}\o f} 1_{FA}\o f
		\rTo^{\l_{f}} f.
\]
Observe that $\r^{F}$ and $\l^{F}$ inherit the coherence of
$\r$ and $\l$, so that in particular the diagrams
\[
	\begin{diagram}[vtriangleheight=3em]
		h\o (k\o F1) && \rTo^{\a} && (h\o k)\o F1 \\
		&\rdTo[snake=-1ex]<{h\o\r^{F}_{k}} && \ldTo[snake=1ex]>{\r^{F}_{h\o k}} \\
		&&h\o k
	\end{diagram}
	and
	\begin{diagram}[vtriangleheight=3em]
		h\o (F1\o k) && \rTo^{\a} && (h\o F1)\o k \\
		& \rdTo[snake=-1ex]<{h\o\l^{F}_{k}} && \ldTo[snake=1ex]>{\r^{F}_{h}\o k} \\
		&& h\o k
	\end{diagram}
\]
commute for all suitably-typed 1-cells $h$ and $k$.

Now we can demonstrate the promised redundancy.
We shall start with the pseudo-natural transformations, since
the situation there is very simple: the unit condition is quite
redundant:
\begin{propn}\label{prop-psnat-redundant}
	Given pseudo-functors $F$, $G:\B\to\BC$ and the data of
	Definition~\ref{def-psnat}, the composition condition
	implies the unit condition.
\end{propn}
\begin{proof}
	Let $\gamma$ be given as in Definition~\ref{def-psnat},
	and suppose it to satisfy the composition condition.
	Now, for every $f: A\to B$ in $\B$, we have the following
	diagram of 1-cells and 2-cells (with associativities left
	implicit):
	
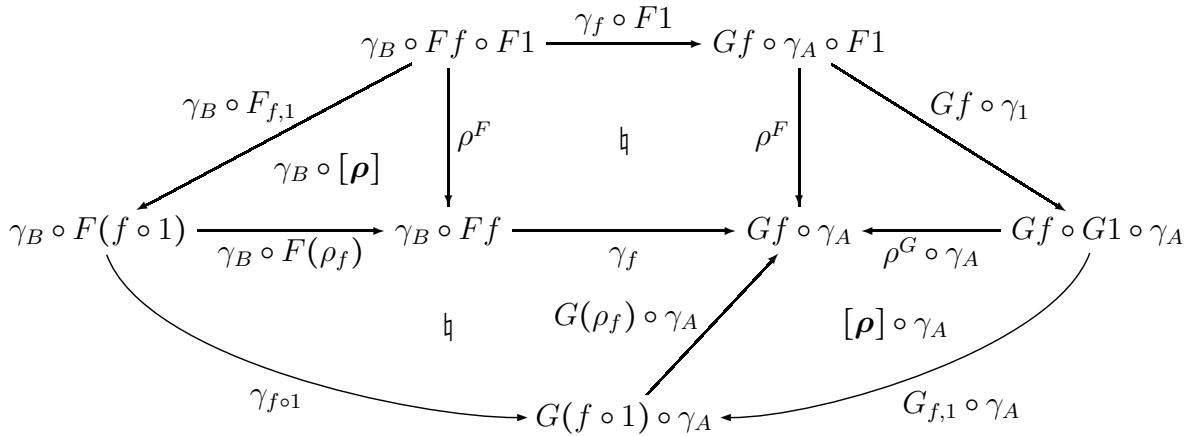
\begin{diagram}
		&&\gamma_{B}\o Ff\o F1 & \rTo^{\gamma_{f}\o F1}
			& Gf\o \gamma_{A}\o F1 \\
		&\ldTo^{\gamma_{B}\o F_{f,1}} \raise-1em\rlap{$\gamma_{B}\o\cFr$}
			& \dTo>{\r^{F}} & \natural & \dTo<{\r^{F}}
			& \rdTo^{Gf\o\gamma_{1}} \\
		\rnode{l}{\gamma_{B}\o F(f\o 1)} & \rTo_{\gamma_{B}\o F(\r_{f})}
			& \gamma_{B}\o Ff & \rTo_{\gamma_{f}}
			& Gf\o\gamma_{A} & \lTo_{\r^{G}\o\gamma_{A}}
			& \rnode{r}{Gf\o G1\o\gamma_{A}} \\
		& & \natural
			& \ruTo[snake=-1em](1,2)<{G(\r_{f})\o\gamma_{A}}
			& & \llap{$\cFr\o\gamma_{A}$} \\
		& & & \rnode{b}{G(f\o 1)\o\gamma_{A}}
		\nccurve[angleA=290,angleB=180,ncurv=.5]{->}{l}{b} \Bput{\gamma_{f\o 1}}
		\nccurve[angleA=250,angleB=0,ncurv=.5]{->}{r}{b} \Aput{G_{f,1}\o\gamma_{A}}
	\end{diagram}
	The marked cells commute for the reasons shown, and the outside
	commutes by the composition condition. Since $\gamma_{f}\o F1$ is
	invertible, it follows that the unlabelled triangle commutes. By
	the observation above, about coherence of $r^{F}$ and $\l^{F}$,
	this triangle is equivalent to
	
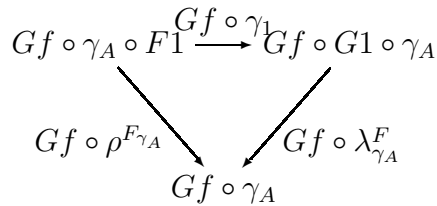
\begin{diagram}[htrianglewidth=4em,tight]
		Gf\o\gamma_{A}\o F1 & \rTo^{Gf\o \gamma_{1}} & Gf \o G1 \o \gamma_{A} \\
		&\rdTo(1,2)<{Gf\o\r^{F_{\gamma_{A}}}} \ldTo(1,2)>{Gf\o\l^{F}_{\gamma_{A}}}\\
		&Gf\o\gamma_{A}
	\end{diagram}
	so, by letting $f=1$, we conclude that
	
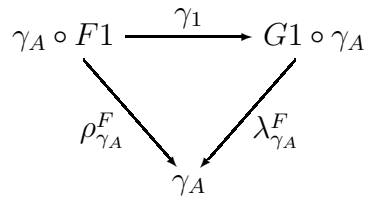
\begin{diagram}[htrianglewidth=4em,tight]
		\gamma_{A}\o F1 & \rTo^{\gamma_{1}} & G1 \o \gamma_{A} \\
		&\rdTo(1,2)<{\r^{F}_{\gamma_{A}}} \ldTo(1,2)>{\l^{F}_{\gamma_{A}}}\\
		&\gamma_{A}
	\end{diagram}
	commutes, which is equivalent to the unit condition.
\end{proof}
For pseudo-functors, the situation is a little more subtle.
The following notion will be useful, both here
and later in Chapter~\refchapter{Cayley}.
\begin{definition}\label{def-fully-faithful}
	A 1-cell $f: A\to B$ is \emph{(representably) fully-faithful}
	if, for every object $X$, the functor
	$\B(X, f): \B(X, A)\to \B(X, B)$ is fully faithful.
	Concretely, this means that, for every pair of arrows
	$h$, $k: X\to A$, every 2-cell
	
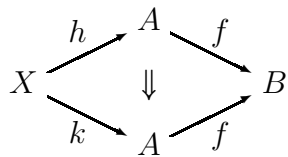
\begin{diagram}[w=2em,h=1em,tight]
		&& A \\
		&\ruTo^{h} && \rdTo^{f} \\
		X && \Downarrow && B \\
		&\rdTo_{k} && \ruTo_{f} \\
		&& A
	\end{diagram}
	is equal to
	\begin{diagram}
		\\
		\rnode{X}{X} &\Downarrow\gamma& \rnode{A}{A} & \rTo^{f} & B \\
		\ncarc[arcangle=50,ncurv=1]{->}{X}{A}  \Aput{h}
		\ncarc[arcangle=-50,ncurv=1]{->}{X}{A} \Bput{k}
	\end{diagram}
	for some unique $\gamma$.
	
	Dually, $f$ is \emph{co-fully-faithful} if, for every object $X$,
	the functor $\B(f, X): \B(B, X)\to \B(A, X)$ is fully faithful.
\end{definition}
\begin{remark} 
	Some remarks on the definition:
	\begin{itemize}
	\item It's easy to check that the fully-faithful 1-cells in
		$\Cat$ are precisely the full and faithful functors. On
		the other hand, the analogous property does not generally
		hold for enriched categories: the $\V$-fully-faithful
		functors do not always coincide with those 1-cells of
		$\V$-$\Cat$ that are representably fully-faithful.
	\item The co-fully-faithful 1-cells of $\Cat$ are characterised
	by \citet{LaxEpis}, who call them `lax epimorphisms'.
	\item Every equivalence is both fully-faithful and
	co-fully-faithful.
\end{itemize}
\end{remark}
\begin{lemma}\label{lemma-psnat-unit-only-one}
	The conditions $\cFl$ and $\cFr$ of Definition~\ref{def-psfun}
	are redundant, in the sense that each implies the other.
\end{lemma}
\begin{proof}
	Let $F:\B\to\BC$ be a pseudo-functor, and let
	\[
		A \rTo^{f} B \rTo^{g} C
	\]
	be 1-cells in $\B$. Consider the following diagram
	in $\BC(FA, FC)$:
	\begin{diagram}[w=4em,h=2em,tight,hug]
		Fg\o(F1\o Ff) && \rTo^{\a} && (Fg\o F1)\o Ff \\
		& \rdTo_{Fg\o\l^{F}} && \ldTo_{\r^{F}\o Ff} \\
		\dTo<{Fg\o F_{1,f}} &\hskip-2em \raise -1ex\hbox{$Fg\o\cFl$}& Fg\o Ff
			& \hskip2em \raise -1ex\hbox{$\cFr\o Ff$} & \dTo>{F_{g,1}\o Ff} \\
		& \ruTo_{Fg\o F(\l)} && \luTo_{F(\r)\o Ff} \\
		Fg\o F(1\o f) && \dTo>{F_{g,f}} && F(g\o 1)\o Ff \\
		& \natural && \natural \\
		\dTo<{F_{g,1\o f}} && F(g\o f) && \dTo>{F_{g\o 1, f}} \\
		&\ruTo^{F(g\o\l)} && \luTo^{F(\r\o f)} \\
		F(g\o(1\o f)) && \rTo_{F(\a)} && F((g\o 1)\o f)
	\end{diagram}
	The upper and lower triangles commute, as does the outside edge.
	Since all the 2-cells in the diagram are invertible, if $\cFl$
	holds then so does $\cFr\o Ff$. Taking $f=1$ and using the fact
	that $F1$ is co-fully-faithful, we conclude that $\cFr$ holds.
	In the other direction, if $\cFr$ holds then so does $Fg\o\cFl$,
	whence taking $g=1$ and using the fact that $F1$ is fully-faithful,
	we conclude that $\cFl$ holds.
\end{proof}
\begin{lemma}\label{lemma-psnat-unit-uq}
	The invertible 2-cells $F_{A}: 1_{FA}\to F(1_{A})$
	of Definition~\ref{def-psfun} are uniquely determined
	by the other data.
\end{lemma}
\begin{proof}
	Let $F: \B\to\BC$ be a pseudo-functor. Condition~$\cFl$
	implies that, for every object $A\in\B$, the 2-cell $F_{A}\o F(1_{A})$
	is equal to
	\[
		1_{FA}\o F(1_{A}) \rTo^{\l_{F(f)}} F(1_{A})
			\rTo^{F(\lambda_{1_{A}}^{-1})} F(1_{A}\o 1_{A})
			\rTo^{F_{1_{A},1_{A}}^{-1}} F(1_{A})\o F(1_{A})
	\]
	(just by taking $f=1_{A}$). Since $F(1_{A})$ is fully-faithful,
	this equation uniquely determines $F_{A}$.
\end{proof}
\begin{propn}
	Let $F: \B\to\BC$ be as in the definition of
	pseudo-functor, though without the 2-cells $F_{A}$.
	These data may be augmented to give a pseudo-functor $F$,
	in a unique way, if and only if $F(1_{A})$ is both
	fully-faithful and co-fully-faithful for each object $A\in\B$,
	if and only if $F(1_{A})$ is \emph{either}
	fully-faithful \emph{or} co-fully-faithful for each object $A\in\B$.
\end{propn}
\begin{proof}
	If we have invertible 2-cells $F_{A}$, then each $F(1_{A})$ is
	isomorphic to the identity, hence an equivalence, so in particular
	is fully-faithful and co-fully-faithful. For the converse,
	take an object $A\in\B$, and suppose that $F(1_{A})$ is co-fully-%
	faithful. Let $F_{A}: 1_{FA}\To F(1_{A})$ be the unique
	2-cell for which $F_{A}\o F(1_{A})$ is equal to the composite
	\[
		1_{FA}\o F(1_{A}) \rTo^{\l} F(1_{A})
			\rTo^{F(\l^{-1})} F(1_{A}\o 1_{A})
			\rTo^{F_{1_{A},1_{A}}^{-1}} F(1_{A})\o F(1_{A}).
	\]
	Since this $F_{A}$ is invertible, $F(1_{A})$ is isomorphic to
	the identity, hence also fully-faithful.
	For any $f: A\to B$,
	we have the following diagram in $\B(FA, FB)$ (with associativities
	left implicit):
	\begin{diagram}
		Ff\o 1\o F1 & \rTo^{Ff\o\l} & \rnode{FfF1}{Ff\o F1}
			& \rTo^{Ff\o F(\l^{-1})} & Ff\o F(1\o 1)
			& \rTo^{Ff\o F_{1,1}^{-1}} & Ff\o F1\o F1 \\
		&& && \dTo<{F_{f,f\o 1}} &\cFa& \dTo>{F_{f,1}\o F1} \\
		&& &\raise 1em\hbox{$\natural$}& F(f\o 1\o 1) & \rTo_{F_{f\o 1,1}^{-1}}
			& F(f\o 1)\o F1 \\
		&& && \dTo<{\begin{array}{r}F(f\o\l)\\=F(\r\o 1)\end{array}}
			&\natural& \dTo>{F(\r)\o F1} \\
		&& && \rnode{Ff1}{F(f\o 1)} & \rTo_{F_{f,1}^{-1}} & Ff\o F1
		\nccurve[angleA=270,angleB=180,ncurv=1]{->}{FfF1}{Ff1} \Bput{F_{f,1}}
	\end{diagram}
	The top row is equal to $Ff\o F_{A}\o F1$ by definition, hence
	this diagram is equivalent to
	\begin{diagram}
		Ff\o 1\o F1 & \rTo^{Ff\o F_{A}\o F1} & Ff\o F1\o F1 \\
		\dTo<{Ff\o\l_{F1}} && \dTo>{F_{f,1}\o F1} \\
		Ff\o F1 & \lTo_{F(\r_{f})\o F1} & F(f\o 1)\o F1
	\end{diagram}
	and since $F1$ is co-fully-faithful, it follows that $\cFl$ holds.
	By Lemma~\ref{lemma-psnat-unit-only-one}, it follows that condition
	$\cFr$ holds too.
	
	If we start instead with the assumption that $F(1_{A})$ is
	fully-faithful, the dual argument applies.
	Finally, the uniqueness is a consequence of Lemma~\ref{lemma-psnat-unit-uq}.
\end{proof}

\section{The bicategorical Yoneda lemma}
The Yoneda lemma for bicategories was first stated by \citet{FIB},
in a long paper that states many basic results without proof. Since
the proof, even if it is in some sense routine, is rather intricate,
we give a detailed account here.
\begin{definition} 
	Let $A$ be an object of the bicategory $\B$. The \emph{covariant
	representable pseudo-functor determined by $A$},
	\[
		\B(A,-): \B\to\Cat,
	\]
	is defined as follows. On an object $X\in\B$, the category $\B(A,X)$ is
	just the the hom-category of the same name. The action
	\[\B(A,-)_{X,Y}: \B(X,Y)\to[\B(A,X),\B(A,Y)]\] is defined to be the
	currying of the composition functor \[\o:\B(X,Y)\x\B(A,X)\to\B(A,Y).\]
	
	For an object $X\in\B$, the unit isomorphism $\B(A,-)_X: 1_{\B(A,X)}\To1_X\o-$
	is defined to be $\l^{-1}$, and for a composable pair
	\[
		X \rTo^f Y \rTo^g Z
	\]
	in $\B$, the isomorphism $\B(A,-)_{g,f}: \B(A,g)\o\B(A,f)\To\B(A,g\o f)$
	is defined to be $\a_{g,f,-}$. I.e.\ for $x\in\B(A,X)$, the component
	$(\B(A,-)_{g,f})_x: g\o(f\o x)\to (g\o f)\o x$ is just $\a_{g,f,x}$.
	
	By duality there is also a \emph{contravariant representable pseudo-functor},
	$\B(-,A) \defeqto \B\op(A,-) : \B\op\to\Cat$.
\end{definition}
\begin{propn}\label{prop-yoneda}
	For any bicategory $\B$, pseudo-functor $F: \B\to\Cat$
	and object $A\in\B$, there is an equivalence of categories
	\[
		\psi: FA \simeq \Bicat(\B,\Cat)(\B(A,-), F).
	\]
\end{propn}
\begin{proof}
	Fix $F$ and $A$. We shall define a functor
	\[
		\psi: FA \to \Bicat(\B,\Cat)(\B(A,-), F),
	\]
	and show that it is an equivalence.
	(This gets rather dizzying, so take a deep breath.)
	First we shall define $\psi$ on objects: for any $a\in FA$, we need a
	pseudo-natural transformation $\psi(a): \B(A,-)\To F$. Thus for every
	object $X\in\B$, we must give a functor
	\[
		\psi(a)_X: \B(A,X)\to FX.
	\]
	This functor is defined as follows. On objects: for $f\in\B(A,X)$, let
	$\psi(a)_X(f) \defeqto F(f)(a)$. On morphisms: for $\beta: f\To g: A\to X$,
	let $\psi(a)_X(\beta) \defeqto F(\beta)_a$. (Note that $F(\beta)$ is a natural
	transformation $F(f)\To F(g)$, whose component $F(\beta)_a$ is
	therefore indeed a map $F(f)(a)\to F(g)(a)$.)
	
	Also, for every 1-cell $k: X\to Y$ in $\B$, we must give a natural
	isomorphism
	\[
		\psi(a)_k: \psi(a)_Y\cdot\B(A,k) \To F(k)\cdot\psi(a)_X.
	\]
	For $x\in\B(A,X)$, we define the component
		\[(\psi(a)_k)_x: F(k\o x)(a)\to F(k)(F(x)(a))\]
	to be $(F_{k,x}^{-1})_a$.
	The naturality of $F_{k,x}$ ensures that $\psi(a)_k$ is natural.
	
	This completes the definition of $\psi$ on objects, though we must
	check that $\psi(a)$ is indeed a pseudo-natural transformation.
	That follows from the pseudo-functoriality of $F$, as the reader
	may verify: for $\psi(a)$ to satisfy the unit condition is precisely
	for $F$ to satisfy condition~$\cFl$, and for $\psi(a)$ to satisfy the composition
	condition is precisely for $F$ to satisfy~$\cFa$.
	
	Next we define $\psi$ on morphisms. For each morphism $h:a\to b$ in $FA$,
	we must define a modification
	\(
		\psi(h): \psi(a) \Tto \psi(b),
	\)
	so for each $X\in\B$ we need a natural transformation
	\(
		\psi(h)_X: \psi(a)_X \To \psi(b)_X,
	\)
	which means that for every $f:A\to X$ in $\B$ we require a map
	\(
		(\psi(h)_X)_f: F(f)(a) \to F(f)(b)
	\)
	in the category $FX$. So we define $(\psi(h)_X)_f$ to be the map $F(f)(h)$.
	It's easy to check that this makes $\psi(h)_X$ into a natural transformation.
	To complete the definition of $\psi$, we must
	confirm that $\psi(h)$ is indeed a modification. This amounts to
	checking that, for every $f: A\to X$ and $k:X\to Y$ in $\B$, and every
	$h:a\to b$ in $FA$, the square
	\begin{diagram}
		F(kf)(a) & \rTo^{(F_{k,f}^{-1})_a} & (Fk \cdot Ff)(a)\\
		\dTo<{F(kf)(h)} && \dTo>{(Fk\cdot Ff)(h)}\\
		F(kf)(b) & \rTo_{(F_{k,f}^{-1})_b} & (Fk \cdot Ff)(b)
	\end{diagram}
	commutes, which is of course precisely the naturality of $F_{k,f}^{-1}$.
	
	We have defined $\psi$, and need to check that it is indeed a
	functor. Consider the modification $\psi(1_{a}): \psi(a)\Tto\psi(a)$:
	for $f:A\to X$ we have $(\psi(h)_{X})_{f} = F(f)(1_{a})$, and
	since $F(f)$ is a functor, this is equal to $1_{F(f)(a)}$ as
	required.
	For composition, a similar argument applies: given arrows
	$h:a\to b$ and $j:b\to c$ in $FA$, we have
	$(\psi(jh)_{X})_{f} = F(f)(jh)$; and since $F(f)$ is a functor,
	this is equal to $F(f)(j)\cdot F(f)(h)$ as required.
	
	It remains to show that $\psi$ is an equivalence. We begin by exhibiting
	a local inverse, showing that $\psi$ is full and faithful. Fix objects $a$
	and $b\in FA$. We shall define a function $\psi_{a,b}^{-1}$ from the set
	of modifications $\psi(a)\Tto\psi(b)$ to the set $FA(a,b)$, and show that
	it is inverse to $\psi_{a,b}$. The definition is as follows. For a modification
	$\mu: \psi(a)\Tto\psi(b)$, let $\psi_{a,b}^{-1}(\mu)$ be the composite
	\[
		a = 1_{FA}(a) \rTo^{(F_A)_a} F(1_A)(a)
			\rTo^{(\mu_A)_{1_A}} F(1_A)(b) \rTo^{(F_A^{-1})_b} 1_{FA}(b) = b.
	\]
	It is easy to check that, for any $h: a\to b$, we have $\psi_{a,b}^{-1}(\psi(h))= h$:
	indeed it is immediate from the definition of $\psi$, and the naturality of $F_A$.
	The other direction is more interesting. Fix some $\mu: \psi(a)\Tto\psi(b)$, and
	take $X\in\B$ and $f:A\to X$. We wish to show that 
	$(\psi(\psi_{a,b}^{-1}(\mu))_X)_f$ is equal to $(\mu_X)_f$. Consider the diagram
	\begin{diagram}[h=2em,w=4em]
		&&(Ff.1_{FA})(a)\\
		&\ruTo^= && \rdTo^{Ff((F_A)_a)}\\
		F(f)(a) && \cFr && (Ff.F1_A)(a)\\
		&\rdTo^{F(\r_f^{-1})(a)} && \ruTo^{(F_{f,1_A})_a}\\
		&&F(f.1_A)(a)\\
		\dTo<{(\mu_X)_f} &\natural_{\mu_X}& \dTo>{(\mu_X)_{f.1_A}} &\qquad\S_\mu& \dTo>{(Ff)((\mu_A)_{1_A})}\\
		&&F(f.1_A)(b)\\
		&\ldTo^{F(\r_f)(b)} && \luTo^{(F_{f,1_A})_b}\\
		F(f)(b) && \cFr && (Ff.F1_A)(b)\\
		&\luTo_= && \ldTo_{Ff((F_A^{-1})_b)}\\
		&&(Ff.1_{FA})(b)
	\end{diagram}
	whose regions commute for the reasons marked: $\natural_{\mu_X}$ means that
	the square commutes because $\mu_X$ is natural, and $\S_\mu$ means that the
	square commutes because $\mu$ is a modification.
	
	The composite around the upper, right, and bottom edges is equal to
	$(\psi(\psi_{a,b}^{-1}(\mu))_X)_f$ by definition, which is therefore equal
	to $(\mu_X)_f$ as required. Thus $\psi$ is indeed full and faithful. It remains
	only to show that $\psi$ is essentially surjective on objects. Consider an
	arbitrary pseudo-natural transformation
	\(
		\gamma: \B(A,-)\To F.
	\)
	We intend to show that $\psi(\gamma_A(1_A))$ is isomorphic to $\gamma$.
	For any $X\in\B$ and $f:A\to X$, we have an invertible 2-cell
	\begin{diagram}
		\B(A,A) & \rTo^{\gamma_A} & FA \\
		\dTo<{\B(A,f)} & \Nearrow\gamma_f & \dTo>{Ff}\\
		\B(A,X) & \rTo_{\gamma_X} & FX
	\end{diagram}
	thus an isomorphism
	\[
		\gamma_X(f) \rTo^{\gamma_X(\r_f^{-1})} \gamma_X(f.1_A)
			\rTo^{(\gamma_f)_{1_A}} F(f)(\gamma_A(1_A)) = \psi(\gamma_A(1_A))_X(f).
	\]
	This isomorphism is natural in $f$, since $\r_f$ and $\gamma_f$ both are.
	So, for every $X\in\B$ we have defined a natural isomorphism
	$\gamma_X \To \psi(\gamma_A(1_A))_X$. Finally it remains to check
	that this collection constitutes a modification. Take $k:X\to Y$: we need to
	check the commutativity of the diagram
	\begin{equation}\label{diag-ymod}
	\begin{diagram}
		\gamma_Y(kf) & \rTo^{(\gamma_k)_f} & F(k)(\gamma_X(f))\\
		\dTo<{\gamma_Y(\r_{kf}^{-1})} && \dTo>{Fk(\gamma_X(\r_f^{-1}))}\\
		\gamma_Y((kf)1) && Fk(\gamma_X(f1))\\
		\dTo<{(\gamma_{kf})_1} && \dTo>{Fk((\gamma_f)_1)}\\
		F(kf)(\gamma_A(1)) & \rTo_{(F_{k,f}^{-1})_{\gamma_A(1)}} & (Fk\cdot Ff)(\gamma_A(1))
	\end{diagram}
	\end{equation}
	
	Since $\gamma$ is pseudo-natural, we know that
	\[\hskip-1cm
	\begin{diagram}
		\B(A,A) & \rTo^{\gamma_A} & \rnode{FA}{FA}\\
		\dTo<{\B(A,f)} & \Nearrow\gamma_f & \dTo>{Ff}\\
		\B(A,X) & \rTo_{\gamma_X} & FX\\
		\dTo<{\B(A,k)} & \Nearrow\gamma_k & \dTo>{Fk}\\
		\B(A,Y) & \rTo_{\gamma_Y} & \rnode{FY}{FY}
		\nccurve{->}{FA}{FY}\Aput{F(kf)}
	\end{diagram}
	\hskip4em=
	\begin{diagram}
		&& \rnode{BAA}{\B(A,A)} & \rTo^{\gamma_A} & FA \\
		&\ldTo^{\B(A,f)}\\
		\B(A,X) & \hbox to1em{$\begin{array}c\To\\[-4pt]\a_{k,f,-}\end{array}$\hss}
			&\dTo[snake=1em]>{\B(A,kf)} && \dTo>{F(kf)}\\
		&\rdTo_{\B(A,k)} && \raise1em\hbox{$\Nearrow\gamma_{kf}$} \\
		&& \rnode{BAY}{\B(A,Y)} & \rTo_{\gamma_Y} & FY
	\end{diagram}
	\]
	hence in particular that the diagram
	\refstepcounter{equation}
	\begin{diagram}[eqno=\textup(\theequation\textup)]\label{diag-yon1}
		\gamma_Y(k(f1)) & \rTo^{\gamma_k)_{f1}} & Fk(\gamma_X(f1))\\
		&&\dTo>{Fk((\gamma_f)_1)}\\
		\dTo<{\gamma_Y(k)} && (Fk\cdot Ff)(\gamma_A(1))\\
		&&\dTo>{(F_{k,f})_{\gamma_A(1)}}\\
		\gamma_Y((kf)1) & \rTo_{(\gamma_{kf})_1} & F(kf)(\gamma_A(1))
	\end{diagram}
	commutes. Now we have
	\begin{diagram}
		\gamma_Y(kf) && \rTo^{(\gamma_k)_f} && F(k)(\gamma_X(f))\\
		\dTo<{\gamma_Y(\r_{kf}^{-1})} &\rdTo^{\gamma_Y(k\r_f^{-1})}&&& \dTo>{Fk(\gamma_X(\r_f^{-1}))}\\
		\gamma_Y((kf)1) &\rTo_{\gamma_Y(\a_{k,f,1}^{-1})}&\gamma_Y(k(f1))
			&\rTo_{(\gamma_k)_{f1}}& Fk(\gamma_X(f1))\\
		\dTo<{(\gamma_{kf})_1} &&&& \dTo>{Fk((\gamma_f)_1)}\\
		F(kf)(\gamma_A(1)) && \rTo_{(F_{k,f}^{-1})_{\gamma_A(1)}} && (Fk\cdot Ff)(\gamma_A(1))
	\end{diagram}
	where the triangle commutes by coherence, the upper-right quadrilateral by
	naturality of $\gamma_k$ and the lower region by \pref{diag-yon1}.
	Thus diagram \pref{diag-ymod} does indeed commute, and we are done.
\end{proof}
\begin{definition}
	For any bicategory $\B$, the \emph{bicategorical Yoneda embedding}
	\[
		Y: \B\op\to\Bicat(\B,\Cat)
	\]
	is defined as follows.
	On objects $A\in\B$, we define $YA \defeqto \B(A,-)$; and
	on hom-categories $\B(A,B)$ we define the component
	\[
		Y_{B,A} : \B\op(A,B) = \B(B,A) \to \Bicat(\B,\Cat)(\B(A,-), \B(B,-))
	\]
	to be $\psi^{\B(B,-)}_A$.
\end{definition}
\begin{corollary}
	The Yoneda embedding is locally an equivalence.
\end{corollary}
\begin{proof}
	Immediate from the definition, by Prop.~\ref{prop-yoneda}.
\end{proof}
\begin{remark}
	Since $\Cat$ is a 2-category, so is $\Bicat(\B\op,\Cat)$.
	And since the Yoneda embedding $Y$ is locally an equivalence, any
	bicategory $\B$ is biequivalent to the full sub-bicategory of
	$\Bicat(\B\op,\Cat)$ determined by the objects $YA$ for $A\in\B$, which
	is of course still a 2-category. Thus any bicategory is biequivalent
	to a 2-category. This is a simple coherence result, which can serve
	as a stepping-stone to more sophisticated coherence theorems
	\citep[Chapter~2]{GurskiThesis}.
\end{remark}

\section{Adjunctions}
\begin{definition} 
	An \emph{adjunction} in a bicategory $\B$ consists of 1-cells
	$f: A\to B$ and $g: B\to A$, and 2-cells $\eta: 1_A\To gf$ and
	$\e: fg\To 1_B$ with the property that
	\[
	\begin{diagram}
		A & \rTo^{f} & B \\
		\dTo<{1_{A}}>{\raise1em\rlap{$\mkern10mu
			\begin{array}c\Rightarrow\\[-4pt]\eta\end{array}$}}
			& \ldTo[snake=1em]_{g}
			& \dTo>{1_{B}}<{\raise-1.5em\llap{$
				\begin{array}c\Rightarrow\\[-4pt]\e\end{array}\mkern10mu$}}\\
		A & \rTo_{f} & B
	\end{diagram}
	\quad=\quad
	\begin{diagram}
		A & \rTo^{f} & B \\
		\dTo<{1_{A}} & \cong & \dTo>{1_{B}}\\
		A & \rTo_{f} & B
	\end{diagram}
	\]
	and
	\[
	\begin{diagram}
		B & \rTo^{g} & A \\
		\dTo<{1_{B}}>{\raise1em\rlap{$\mkern10mu
			\begin{array}c\Leftarrow\\[-4pt]\e\end{array}$}}
			& \ldTo[snake=1em]_{f}
			& \dTo>{1_{A}}<{\raise-1.5em\llap{$
				\begin{array}c\Leftarrow\\[-4pt]\eta\end{array}\mkern10mu$}}\\
		B & \rTo_{g} & A
	\end{diagram}
	\quad=\quad
	\begin{diagram}
		B & \rTo^{g} & A \\
		\dTo<{1_{B}} & \cong & \dTo>{1_{A}}\\
		B & \rTo_{g} & A
	\end{diagram}
	\]
	We write $f\dashv g$ to indicate that there is such an adjunction, and say that
	$f$ is \emph{left adjoint} to $g$, and $g$ is \emph{right adjoint} to $f$. We
	also write $f\dashv g : A \to B$ to mean that $f\dashv g$ for $f: A\to B$
	and $g: B\to A$.
\end{definition}
\begin{remark} 
	There is an \emph{identity adjunction} on every object $A$,
	which is just $1_{A} \dashv 1_{A}$, with the unit and counit
	being structural isomorphisms.
	
	Also, adjunctions may be composed: given adjunctions
	$f\dashv g: A \to B$ and $f'\dashv g': B\to C$, there
	is a composite adjunction $f'\o f \dashv g\o g'$ with
	unit
	
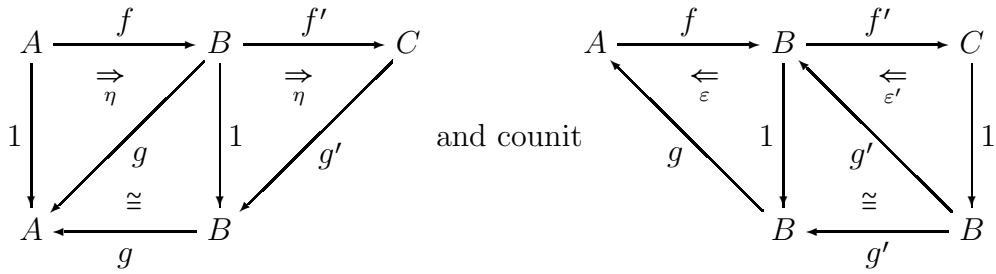
\[
	\begin{diagram}
		A & \rTo^{f} & B & \rTo^{f'} & C \\
		\dTo<1 & \raise 2em\llap{$\Right_{\eta}$}
			\ldTo_{g}
			\raise -2em\rlap{$\cong$}
			& \dTo>1
			& \raise 2em\llap{$\Right_{\eta}$}
			\ldTo_{g'} \\
			A & \lTo_{g} & B
	\end{diagram}
	\mbox{and counit}
	\begin{diagram}
		A & \rTo^{f} & B & \rTo^{f'} & C \\
		& \luTo_{g} \raise 2em \rlap{$\Left_{\e}$} & \dTo<1
			& \luTo_{g'} \raise2em\rlap{$\Left_{\e'}$}
			\raise-2em\llap{$\cong$}
			& \dTo>1 \\
		&& B & \lTo_{g'} & B
	\end{diagram}
	\]
	Indeed, every bicategory $\B$ has a \emph{bicategory of adjunctions},
	whose objects are the objects of $\B$ and whose 1-cells are adjunctions.
\end{remark}
\begin{remark}\label{rem-adj-functor} 
	Pseudofunctors preserve adjunctions, in the following sense.
	If $F:\B\to\BC$ is a pseudo-functor, and
	$f\dashv g: A\to B$ is an adjunction in $\B$,
	then there is an adjunction $Ff \dashv Fg$ in $\BC$
	with the following unit and counit:
	
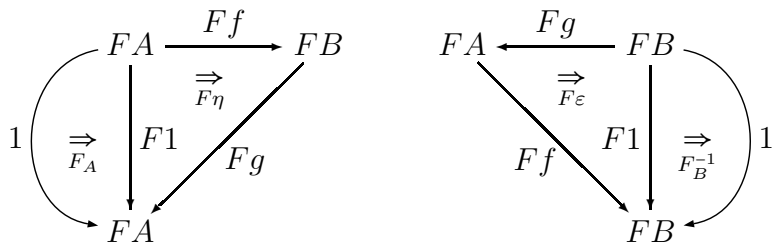
\[
		\begin{diagram}
			&\rnode{t}{FA} & \rTo^{Ff} & FB \\
			\Right_{F_{A}}\hskip-3em&\dTo>{F1} & \ldTo_{Fg}
				\raise 2em\llap{$\Right_{F\eta}$} \\
			&\rnode{b}{FA}
			\ncarc[arcangle=-80,ncurv=1]{->}tb\Bput{1}
		\end{diagram}
		\hskip3em
		\begin{diagram}
			FA & \lTo^{Fg} & \rnode{t}{FB} \\
			& \rdTo_{Ff}
				 \raise 2em\rlap{$\Right_{F\e}$}
				&\dTo<{F1} & \hskip-3em\Right_{F_{B}^{-1}}\\
			&&\rnode{b}{FB}
			\ncarc[arcangle=80,ncurv=1]{->}tb\Aput{1}
		\end{diagram}
	\]
\end{remark}
\subsection{Mates}
The theory of mates \citep[\S2]{ks1} is a useful tool
for dealing with adjunctions in a bicategory. We shall give
a brief overview here.
\begin{definition}\label{def-mate}
	Let $f \dashv g: A \to B$ and $f' \dashv g': A' \to B'$.
	Given a 2-cell
	
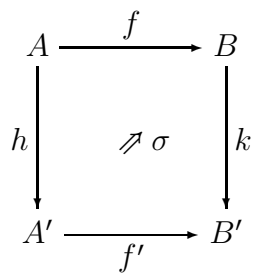
\begin{diagram} 
		A & \rTo^{f} & B \\
		\dTo<{h} & \Arr\Nearrow\sigma & \dTo>{k} \\
		A' & \rTo_{f'} & B'
	\end{diagram}
	we may form a 2-cell
	\begin{diagram} 
		A & \lTo^{g} & B \\
		\dTo<{h} & \Arr\Searrow\tau & \dTo>{k} \\
		A' & \lTo_{g'} & B'
	\end{diagram}
	as the pasting
	\begin{diagram} 
	\rnode{A}{A} & \lTo^{g} & \rnode{B}{B}\\
	\dTo<{h}&\rdTo_{f}^{\raise4pt\hbox{$\begin{array}c\To\\[-4pt]\e\end{array}$}}
		& \dTo>1\\
	A'&\begin{array}c\To\\[-4pt]\sigma\end{array}&B\\
	\dTo<1
		& \rdTo^{f'}_{\raise-4pt\hbox{$\begin{array}c\To\\[-4pt]\eta'\end{array}$}}
		& \dTo>{k}\\
	\rnode{A'}{A'} & \lTo_{g'} & \rnode{B'}{B'.}
	\nccurve[angleA=210,angleB=140]{->}{A}{A'}\Bput{h}\Aput{\ \ \ \cong}
	\nccurve[angleA=-30,angleB=30]{->}{B}{B'}\Aput{k}\Bput{\cong\ \ \ }
	\end{diagram}
	We say that $\tau$ is the \emph{right mate} of $\sigma$.
	Conversely, given a 2-cell $\tau$ as above, we may form its
	\emph{left mate} as
	\begin{diagram} 
	\rnode{A}{A} & \rTo^{f} & \rnode{B}{B}\\
	\dTo<{1}&\ldTo_{g}^{\raise4pt\hbox{$\begin{array}c\To\\[-4pt]\eta\end{array}$}}
		& \dTo>k\\
	A&\begin{array}c\To\\[-4pt]\tau\end{array}&B'\\
	\dTo<h
		& \ldTo^{g'}_{\raise-4pt\hbox{$\begin{array}c\To\\[-4pt]\e'\end{array}$}}
		& \dTo>{1}\\
	\rnode{A'}{A'} & \rTo_{f'} & \rnode{B'}{B'}
	\nccurve[angleA=210,angleB=140]{->}{A}{A'}\Bput{h}\Aput{\ \ \ \cong}
	\nccurve[angleA=-30,angleB=30]{->}{B}{B'}\Aput{k}\Bput{\cong\ \ \ }
	\end{diagram}
\end{definition}
\begin{propn}\label{prop-mate-inv}
	The `left mate' and `right mate' operations are mutually
	inverse, i.e.\ $\sigma$ is the left mate of $\tau$ if and
	only if $\tau$ is the right mate of $\sigma$.
	In this case we say that \emph{$\sigma$ and $\tau$ are
	mates} (with respect to the adjunctions $f\dashv g$ and $f'\dashv g'$).
\end{propn}
\begin{proof}
	This follows easily from the definition of adjunction and
	the coherence of structural isomorphisms.
\end{proof}
Matehood can be characterised in terms of either the units or
counits of the adjunctions.
\begin{propn} 
	The 2-cells $\sigma$ and $\tau$ are mates if and only if
	\begin{equation}\label{eq-mate-eta}
	\begin{diagram}
		\rnode{A}{A}\\
		\dTo<{h}&\rdTo^{f} \\
		A'&\begin{array}c\To\\[-4pt]\sigma\end{array}&B\\
		\dTo<1
			& \rdTo^{f'}_{\raise-4pt\hbox{$\begin{array}c\To\\[-4pt]\eta'\end{array}$}}
			& \dTo>{k}\\
		\rnode{A'}{A'} & \lTo_{g'} & B'
		\nccurve[angleA=220,angleB=140]{->}{A}{A'}\Bput{h}\Aput{\ \ \ \cong}
	\end{diagram}
	\qquad=\qquad
	\begin{diagram}
		& \rnode{A}{A} & \rTo^{f} & \rnode{B}{B}\\
		& \dTo<{1}&\ldTo_{g}^{\raise4pt\hbox{$\begin{array}c\To\\[-4pt]\eta\end{array}$}}
		     & \dTo>k\\
		& A &\begin{array}c\To\\[-4pt]\tau\end{array}&B'\\
		& \dTo<h & \ldTo_{g'} \\
		& \rnode{A'}{A'}
		\nccurve[angleA=220,angleB=140]{->}{A}{A'}\Bput{h}\Aput{\ \ \ \cong}
	\end{diagram}
	\end{equation}
	if and only if
	\begin{equation}\label{eq-mate-eps}
	\begin{diagram}
	A & \lTo^{g} & \rnode{B}{B} &\\
	\dTo<{h}&\rdTo_{f}^{\raise4pt\hbox{$\begin{array}c\To\\[-4pt]\e\end{array}$}}
		& \dTo>1 &\\
	A'&\begin{array}c\To\\[-4pt]\sigma\end{array}&B& \\
	& \rdTo_{f} & \dTo>{k} & \\
	& & \rnode{B'}{B'} &
	\nccurve[angleA=-30,angleB=30]{->}{B}{B'}\Aput{k}\Bput{\cong\ \ \ }
	\end{diagram}
	\qquad=\qquad
	\begin{diagram}
	& & \rnode{B}{B}\\
	& \ldTo^{g} & \dTo>k\\
	A & \begin{array}c\To\\[-4pt]\tau\end{array}&B'\\
	\dTo<h
		& \ldTo^{g'}_{\raise-4pt\hbox{$\begin{array}c\To\\[-4pt]\e'\end{array}$}}
		& \dTo>{1}\\
	A' & \rTo_{f'} & \rnode{B'}{B'}
	\nccurve[angleA=-30,angleB=30]{->}{B}{B'}\Aput{k}\Bput{\cong\ \ \ }
	\end{diagram}
	\hskip4em
	\end{equation}
\end{propn}
\begin{proof}
	By definition we have
	\[
	\begin{diagram} 
		A & \rTo^{f} & B \\
		\dTo<{h} & \Arr\Nearrow\sigma & \dTo>{k} \\
		A' & \rTo_{f'} & B'
	\end{diagram}
	\qquad=\qquad
	\begin{diagram} 
	\rnode{A}{A} & \rTo^{f} & \rnode{B}{B}\\
	\dTo<{1}&\ldTo_{g}^{\raise4pt\hbox{$\begin{array}c\To\\[-4pt]\eta\end{array}$}}
		& \dTo>k\\
	A&\begin{array}c\To\\[-4pt]\tau\end{array}&B'\\
	\dTo<h
		& \ldTo^{g'}_{\raise-4pt\hbox{$\begin{array}c\To\\[-4pt]\e'\end{array}$}}
		& \dTo>{1}\\
	\rnode{A'}{A'} & \rTo_{f'} & \rnode{B'}{B'}
	\nccurve[angleA=210,angleB=140]{->}{A}{A'}\Bput{h}\Aput{\ \ \ \cong}
	\nccurve[angleA=-30,angleB=30]{->}{B}{B'}\Aput{k}\Bput{\cong\ \ \ }
	\end{diagram}
	\]
	Onto both sides, we paste the 2-cell
	
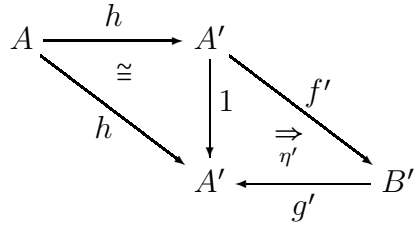
\begin{diagram}[h=2em]
		A & \rTo^{h} & A' \\
		&\rdTo_{h} \raise 1em\rlap{$\cong$} & \dTo>1
			& \rdTo^{f'} \raise-1em\llap{$\Right_{\eta'}$} \\
		&& A' & \lTo_{g'} & B'
	\end{diagram}
	along the edge $A \rTo^{h} A' \rTo^{f'}B'$,
	then use coherence (on the right) to deduce \pref{eq-mate-eta}.
	
	Similarly we can deduce \pref{eq-mate-eps} by taking the
	equation displayed above and pasting the 2-cell
	
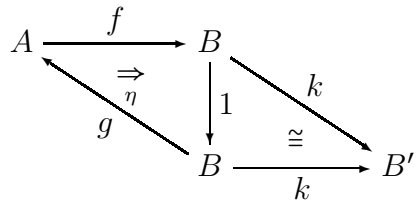
\begin{diagram}[h=2em]
		A & \rTo^{f} & B \\
		&\luTo_{g} \raise 1em\rlap{$\Right_{\eta}$} & \dTo>1
			& \rdTo^{k} \raise-1em\llap{$\cong$} \\
		&& B & \rTo_{k} & B'
	\end{diagram}
	onto both sides along the edge $A\rTo^{f} B\rTo^{k}B'$.
\end{proof}
Next we list some useful elementary properties of mates.
\begin{propn}\label{prop-mates}
	Mates have the following properties:
	\begin{enumerate}
		\item Mating is natural in $h$ and $k$, i.e.\ if $\sigma$
			and $\tau$ are mates then so are
			
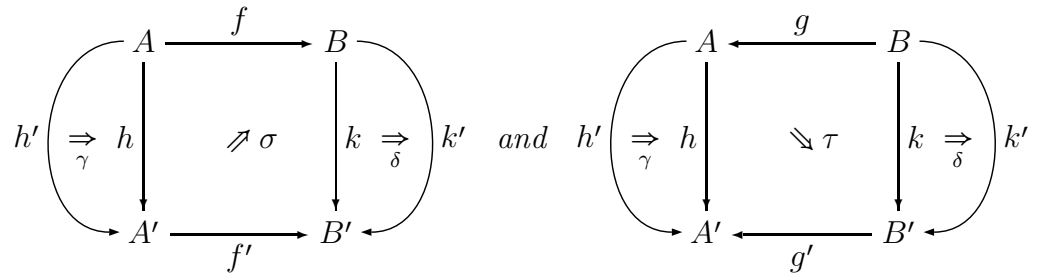
\[
			\begin{diagram} 
				&\rnode{A}{A} & \rTo^{f} & \rnode{B}{B} \\
				\Right_{\gamma}\hskip-2.2em
					& \dTo<{h}
					& \Arr\Nearrow\sigma & \dTo>{k}
					& \hskip-2.3em\Right_{\delta} \\
				&\rnode{A'}{A'} & \rTo_{f'} & \rnode{B'}{B'}
				\ncarc[arcangle=-90,ncurv=1]{->}{A}{A'}\Bput{h'}
				\ncarc[arcangle=90,ncurv=1]{->}{B}{B'}\Aput{k'}
			\end{diagram}
			\qquad\mbox{and}\qquad
			\begin{diagram}
				&\rnode{A}{A} & \lTo^{g} & \rnode{B}{B} \\
				\Right_{\gamma}\hskip-2.2em
					& \dTo<{h}
					& \Arr\Searrow\tau & \dTo>{k}
					& \hskip-2.3em\Right_{\delta} \\
				&\rnode{A'}{A'} & \lTo_{g'} & \rnode{B'}{B'}
				\ncarc[arcangle=-90,ncurv=1]{->}{A}{A'}\Bput{h'}
				\ncarc[arcangle=90,ncurv=1]{->}{B}{B'}\Aput{k'}
			\end{diagram}
			\]
			for all appropriately-typed 2-cells $\gamma$ and $\delta$.
			\label{mate-natural}
		\item Mating preserves horizontal pasting:
			if $\sigma_{1}$ and $\tau_{1}$ are mates with respect to the
			adjunctions $f_{1} \dashv g_{1}$ and $f'_{1} \dashv g'_{1}$,
			and $\sigma_{2}$ and $\tau_{2}$ are mates with respect to
			$f_2 \dashv g_2$ and $f_2' \dashv g_2'$, then
			\begin{diagram} 
				A & \rTo^{f_{1}} & B & \rTo^{f_2} & C\\
				\dTo<{h} & \Arr\Nearrow{\sigma_{1}} & \dTo>{k}
				 	& \Arr\Nearrow{\sigma_2} & \dTo>n\\
				A' & \rTo_{f_{1}'} & B' & \rTo_{f_{2}'} & C'
			\end{diagram}
			and
			\begin{diagram} 
				A & \lTo^{g_{1}} & B & \lTo^{g_2} & C\\
				\dTo<{h} & \Arr\Searrow{\tau_{1}} & \dTo>{k}
				 	& \Arr\Searrow{\tau_2} & \dTo>n\\
				A' & \lTo_{g_{1}'} & B' & \lTo_{g_{2}'} & C'
			\end{diagram}
			are mates, with respect to the adjunctions
			$f_{2}\o f_{1} \dashv g_{1}\o g_{2}$
			and $f_{2}'\o f_{1}' \dashv g_{1}'\o g_{2}'$.
			\label{mate-horiz}
		\item Mating preserves vertical pasting: if $\sigma$ and $\tau$
			are mates with respect to $f\dashv g$ and $f'\dashv g'$,
			and $\sigma'$ and $\tau'$
			are mates with respect to $f'\dashv g'$ and $f''\dashv g''$,
			then
			\[
			\begin{diagram} 
				A & \rTo^{f} & B \\
				\dTo<{h} & \Arr\Nearrow\sigma & \dTo>{k} \\
				A' & \rTo_{f'} & B' \\
				\dTo<{h'} &  \Arr\Nearrow{\sigma'} & \dTo>{k'} \\
				A'' & \rTo_{f''} & B''
			\end{diagram}
			\qquad\mbox{and}\qquad
			\begin{diagram} 
				A & \lTo^{g} & B \\
				\dTo<{h} & \Arr\Searrow\tau & \dTo>{k} \\
				A' & \lTo_{g'} & B' \\
				\dTo<{h'} & \Arr\Searrow{\tau'} & \dTo>{k'} \\
				A'' & \lTo_{g''} & B''
			\end{diagram}
			\]
			are mates with respect to $f\dashv g$ and $f''\dashv g''$.
			\label{mate-vert}
		\item For every adjunction $f\dashv g: A\to B$, the
			structural isomorphisms
			\[
				\begin{diagram}
					A & \rTo^{f} & B \\
					\dTo<1 & \cong & \dTo>1 \\
					A & \rTo_{f} & B
				\end{diagram}
				\mbox{\qquad and\qquad}
				\begin{diagram}
					A & \lTo^{g} & B \\
					\dTo<1 & \cong & \dTo>1 \\
					A & \lTo_{g} & B
				\end{diagram}
			\]
			are mates.
			\label{mate-structural}
	\end{enumerate}
\end{propn}
We omit the (routine) verification of this proposition,
which requires nothing more than the definitions of mate
and adjunction, and the coherence of the unit isomorphisms.
The next proposition is essential for some of our applications of
mates in later chapters.
\begin{propn} 
	Let $\gamma: F \To G: \B \to \BC$ be a pseudo-natural transformation,
	and let $f\dashv g: A\to B$ be an adjunction in $\B$. Then
	\[
		\begin{diagram}
			FA & \rTo^{Ff} & FB \\
			\dTo<{\gamma_{A}} & \Arr\Nearrow{\gamma_{f}^{-1}}
				& \dTo>{\gamma_{B}} \\
			GA & \rTo_{Gf} & GB
		\end{diagram}
		\qquad\mbox{and}\qquad
		\begin{diagram}
			FA & \lTo^{Fg} & FB \\
			\dTo<{\gamma_{A}} & \Arr\Searrow{\gamma_{g}}
				& \dTo>{\gamma_{B}} \\
			GA & \lTo_{Gg} & GB
		\end{diagram}
	\]
	are mates with respect to the adjunctions $Ff\dashv Fg$ and $Gf\dashv Gg$.
\end{propn}
\begin{proof}
	Consider the 2-cell
	\begin{diagram}[h=2em]
		\rnode{t}{FA} & \rTo^{\gamma_{A}} & GA \\
		& \rdTo^{Ff} & \Arr\Uparrow{\gamma_{f}} & \rdTo^{Gf} \\
		\dTo<{F1} & \llap{$\Right_{F\eta}$} & FB
			& \rTo^{\gamma_{B}} & GB \\
		&\ldTo_{Fg} & \Arr\Nearrow{\gamma_{g}} & \ldTo_{Gg} \\
		\rnode{b}{FA} & \rTo_{\gamma_{A}} & GA
		\ncarc[arcangle=-80,ncurv=1]{->}tb\Bput{1}
	\end{diagram}
	Since $\gamma$ is pseudo-natural, by the naturality condition
	in Definition~\ref{def-psnat} this is equal to
	\begin{diagram}[h=1.5em]
		\rnode{t}{FA} & \rTo^{\gamma_{A}} & GA \\
		& & & \rdTo^{Gf} \\
		\dTo>{F1} & \Arr\Nearrow{\gamma_{1_{A}}} & \dTo>{G1}
			& \llap{$\Right_{G\eta}$} & GB \\
		& & & \ldTo_{Gg} \\
		\rnode{b}{FA} & \rTo_{\gamma_{A}} & GA
		\ncarc[arcangle=-80,ncurv=1]{->}tb\Bput{1}\Aput{\ \ \Right_{F_{A}}}
	\end{diagram}
	which, by the unit condition, is equal to
	\begin{diagram}[h=1.5em,w=3em]
		FA & \rTo^{\gamma_{A}} & \rnode{t}{GA} \\
		& & & \rdTo^{Gf} \\
		\dTo<{1} & \llap{$\cong$\quad} & \dTo>G1
			& \llap{$\Right_{G\eta}$} & GB \\
		& & & \ldTo_{Gg} \\
		FA & \rTo_{\gamma_{A}} & \rnode{b}{GA}
		\ncarc[arcangle=-60]{->}tb\Bput{1}\Aput{\ \Right_{G_{A}}}
	\end{diagram}
	If we paste $\gamma_{f}^{-1}$ onto the top of this equation,
	and $\r_{\gamma_{A}}^{-1}$ onto the bottom, then we get
	\[
		\begin{diagram}
		& \rnode{t}{FA} \\
		\rlap{\quad$\Right_{F_{A}}$} & \dTo>{F1}
			& \llap{$\Right_{F\eta}$}\rdTo(2,1)^{Ff} & FB \\
		& \rnode{m}{FA} & \ldTo(2,1)_{Fg} & \dTo>{\gamma_{B}} \\
		\raise 2em\hbox{$\cong$} & \dTo<{\gamma_{A}}
			& \raise 1.5em\hbox{$\Arr\Swarrow{\gamma_{g}}$} & GB \\
		& \rnode{b}{GA} & \ldTo(2,1)_{Gg}
		\ncarc[arcangle=-80,ncurv=1]{->}tm\Bput{1}
		\ncarc[arcangle=-90,ncurv=1]{->}tb\Bput{\gamma_{A}}
		\end{diagram}
		\quad=\hskip5em
		\begin{diagram}
		& \rnode{t}{FA} \\
		\raise0em\hbox{$\cong$}& \dTo<{\gamma_{A}} & \rdTo(2,1)^{Ff} & FB \\
		& \rnode{m}{GA} & \raise 2em\hbox{$\Arr\Nearrow{\gamma_{f}^{-1}}$}
			& \dTo>{\gamma_{B}}\\
		\rlap{\quad$\Right_{G_{A}}$} & \dTo>{G1}
			& \llap{$\Right_{G\eta}$}\rdTo(2,1)^{Gf} & FB \\
		& \rnode{b}{GA} & \ldTo(2,1)_{Fg}
		\ncarc[arcangle=-80,ncurv=1]{->}mb\Bput{1}
		\ncarc[arcangle=-90,ncurv=1]{->}tb\Bput{\gamma_{A}}
		\end{diagram}
	\]
	Thus, by equation~\pref{eq-mate-eta} and Remark~\ref{rem-adj-functor},
	$\gamma_{f}^{-1}$ and $\gamma_{g}$ are indeed mates.
\end{proof}
A special case of mating that is sometimes useful occurs when the
vertical 1-cells $h$ and $k$ are identities. In this case we may
omit them entirely, and speak of the left mate, or right mate, of
a 2-cell $f'\To f$, for adjunctions $f\dashv g$, $f'\dashv g': A\to B$.
It should be clear what is meant by this: for
example, to take the right mate of a 2-cell $\gamma: f'\To f$, one
forms the composite
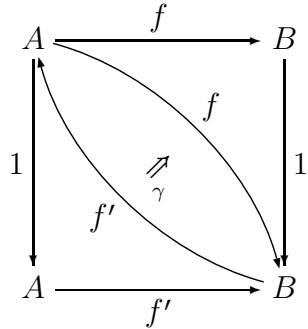
\begin{diagram}[s=4em]
	\rnode{A}{A} & \rTo^{f} & B \\
	\dTo<1 & \mathop\Nearrow\limits_{\gamma} & \dTo>1 \\
	A & \rTo_{f'} & \rnode{B}{B}
	\ncarc[arcangle=30]{->}{A}{B} \Aput{f}
	\ncarc[arcangle=30]{->}{B}{A} \Aput{f'}
\end{diagram}
(where the triangular cells contain unit isomorphisms), and takes its
right mate, say
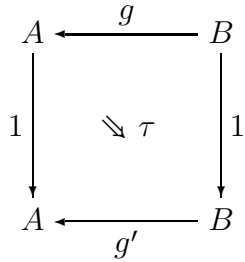
\begin{diagram}
	A & \lTo^{g} & B \\
	\dTo<1 & \Searrow\tau & \dTo>1 \\
	A & \lTo_{g'} & B
\end{diagram}
and then forms the composite
\vskip2em
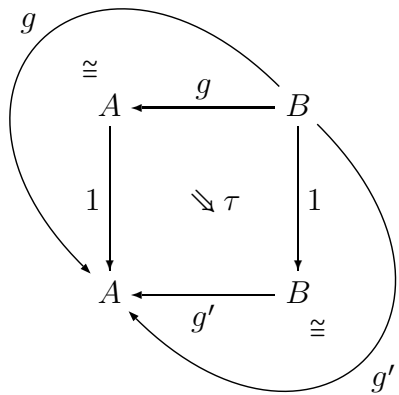
\begin{diagram}
	A & \lTo^{g} & \rnode{B}{B} \\
	\dTo<1 & \Arr\Searrow\tau & \dTo>1 \\
	\rnode{A}{A} & \lTo_{g'} & B
	\nccurve[angleA=135,angleB=135,ncurv=2]{->}{B}{A}
		\Bput{g} \Aput{\raise -1em\rlap{\quad$\cong$}}
	\nccurve[angleA=-45,angleB=-45,ncurv=2]{->}{B}{A}
		\Aput{g'} \Bput{\raise 0.5em\llap{$\cong$\quad}}
\end{diagram}
\vskip3em
In general it is necessary to distinguish between left mate and right mate
in this situation, because one cannot tell from the context which is intended.
However, see Prop.~\ref{prop-adjeq-mate-dual} below for a case where they coincide.

\subsection{Adjunctions and mates in terms of string diagrams}
\renewcommand\graphicscale{1}
Adjunctions and mates have a particularly elegant string diagram
representation. Let $f\dashv g: A\to B$ be an adjunction with unit
$\eta$ and counit $\e$. Then the unit and counit are drawn as
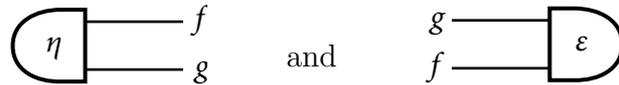
\[
	\cdiag{d-adj/eta}
	\qquad\mbox{and}\qquad
	\cdiag{d-adj/eps}
\]
The adjunction axioms say precisely that
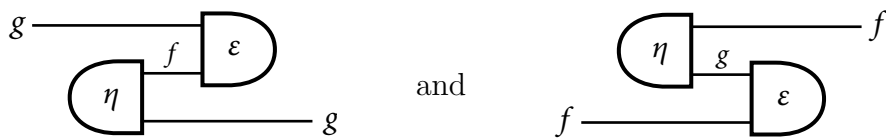
\[\cdiag{d-adj/snake1} \qquad\mbox{and}\qquad \cdiag{d-adj/snake2}\]
are both identities. Given a 2-cell $\sigma$:
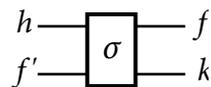
\[\cdiag{d-adj/sigma}\]
its right mate is
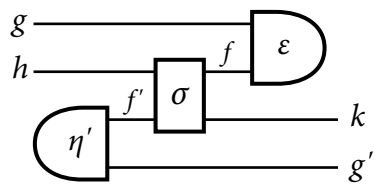
\[\cdiag{d-adj/right-mate-of-sigma}\]
and given a 2-cell $\tau$:
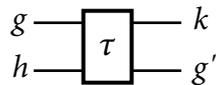
\[\cdiag{d-adj/tau}\]
its left mate is
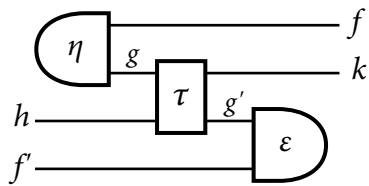
\[\diag{d-adj/left-mate-of-tau}\]

\subsection{Adjoint pseudo-natural transformations}
Now we turn our attention to adjoint pseudo-natural transformations,
i.e.\ adjunctions in a pseudo-functor bicategory.
\begin{propn}\label{prop-adj-1}
	Let there be given an adjoint pair of pseudo-natural
	transformations
	\[
		\phi \dashv \gamma: F \To G : \B\to\BC
	\]
	with unit $\eta: 1\Tto\gamma\phi$
	and counit $\e: \phi\gamma \Tto 1$.
	Then
	\begin{enumerate}
		\item For every object $A\in\B$, there is an adjunction
		\[
			\phi_{A} \dashv \gamma_{A}
		\]
		with unit $\eta_{A}$ and counit $\e_{A}$.
		\item For every 1-cell $f: A\to B$ in $\B$, the
		2-cells
		\[
			\begin{diagram}
				FA & \rTo^{\phi_{A}} & GA \\
				\dTo<{F(f)} & \Arr\Nearrow{\phi_{f}} & \dTo>{G(f)} \\
				FB & \rTo_{\phi_{B}} & GB
			\end{diagram}
			\mbox{\quad and\quad}
			\begin{diagram}
				FA & \lTo^{\gamma_{A}} & GA \\
				\dTo<{F(f)} & \Arr\Searrow{\gamma_{f}^{-1}} & \dTo>{G(f)} \\
				FB & \lTo_{\gamma_{B}} & GB
			\end{diagram}
		\]
		are mates with respect to the adjunctions $\phi_{A}\dashv\gamma_{A}$
		and $\phi_{B}\dashv\gamma_{B}$.
	\end{enumerate}
\end{propn}
\begin{proof}
	The first part is immediate by definition, so let's consider the second.
	Since $\eta$ is a modification, we know that for every $f:A\to B$,
	\[
	\begin{diagram}
		FA & \rTo^{\phi_A} & GA & \rTo^{\gamma_A} & FA\\
		\dTo<{Ff} & \Nearrow\phi_f & \dTo<{Gf} &\Nearrow\gamma_f & \dTo>{Ff}\\
		\rnode{1}{FB} & \rTo_{\phi_B} & GB & \rTo_{\gamma_B} & \rnode{2}{FB}\\
		\nccurve[angleA=-50,angleB=-130]{->}12\Bput1\Aput{\raise6pt\hbox{$\Uparrow\eta_B$}}
	\end{diagram}
	=
	\begin{diagram}[h=2em]
		&&GA\\
		&\ruTo^{\phi_A} & \Uparrow\eta_A & \rdTo^{\gamma_A}\\
		FA && \rTo_1 && FA\\
		\dTo<{Ff} && \cong && \dTo>{Ff}\\
		FB && \rTo_1 && FB
	\end{diagram}
	\]
	Onto both sides of the equation, we paste the 2-cells
	$\gamma_{f}^{-1}$ and $\l^{-1}: Ff\to 1\o Ff$, yielding
	the equation
	\[
	\begin{diagram}
		\rnode{FA}{FA} & \rTo^{\phi_A} & GA\\
		\dTo<{Ff} & \Nearrow\phi_f & \dTo>{Gf}\\
		FB & \rTo^{\phi_B} & GB\\
		\dTo<1>{\raise6pt\hbox{$\ \begin{array}c\To\\[-4pt]\eta_B\end{array}$}} & \ldTo_{\gamma_B}\\
		\rnode{FB}{FB}
		\nccurve[angleA=210,angleB=140]{->}{FA}{FB}\Bput{Ff}\Aput{\ \ \ \cong}
	\end{diagram}
	\quad=\hskip4em
	\begin{diagram}
		\rnode{FA}{FA}\\
		\dTo<1>{\raise-6pt\hbox{$\ \begin{array}c\To\\[-4pt]\eta_A\end{array}$}} & \rdTo^{\phi_A}\\
		FA & \lTo_{\gamma_A} & GA\\
		\dTo<{Ff} & \Searrow\gamma_f^{-1}&\dTo>{Gf}\\
		\rnode{FB}{FB} & \lTo_{\gamma_B} & GB
		\nccurve[angleA=210,angleB=140]{->}{FA}{FB}\Bput{Ff}\Aput{\ \ \ \cong}
	\end{diagram}
	\]
	which, by \pref{eq-mate-eta}, is what we require.
\end{proof}

\begin{propn}\label{prop-adj-2}
	Let there be given a pseudo-natural transformation
	\[
		\phi: F\To G: \B \to \BC.
	\]
	To give a right adjoint $\gamma: G\To F$ for $\phi$ is to give
	\begin{itemize}
		\item for each $A\in\B$, a 1-cell $\gamma_{A}: GA\to FA$ and
			an adjunction $\phi_{A}\dashv \gamma_{A}$,			
		\item such that for every $f:A\to B$ in $\B$, the mate of $\phi_{f}$
			with respect to the adjunctions $\phi_{A}\dashv \gamma_{A}$
			and $\phi_{B}\dashv \gamma_{B}$ is invertible.
	\end{itemize}
\end{propn}
\begin{proof}
	We have already shown (Prop.~\ref{prop-adj-1}) that every
	right-adjoint pseudo-natural transformation $\gamma$ has these
	properties, so suppose that we have a collection of
	adjunctions $\phi_{A}\dashv \gamma_{A}$ as in the statement,
	each with unit $\eta_{A}$ and counit $\e_{A}$, say.
	For each 1-cell $f: A\to B$ in $\B$, define the 2-cell
	\begin{diagram}
		GA & \rTo^{\gamma_{A}} & FA \\
		\dTo<{Gf} & \Arr\Nearrow{\gamma_{f}} & \dTo>{Ff} \\
		GB & \rTo_{\gamma_{B}} & FB
	\end{diagram}
	to be the inverse of the mate of $\phi_{f}$.
	We shall show that these data constitute a pseudo-natural
	transformation $\gamma$, checking the naturality and composition
	conditions
	of Definition~\ref{def-psnat}. This is essentially a
	matter of writing down the corresponding conditions for
	$\phi$ and taking mates.
	\begin{itemize}
	\item The naturality condition for $\phi$ states that
		\[
		\hskip-1em
		\begin{diagram}[size=4em]
			FA & \rTo^{\phi_A} & \rnode{GA}{GA}\\
			\dTo<{Ff} & \llap{$\Nearrow \phi_f$}
				& \begin{array}{c}\Rightarrow\\G(\tau)\end{array}\\
			FB & \rTo_{\phi_B} & \rnode{GB}{GB}
			\ncarc{->}{GA}{GB}\Aput{Gg}
			\ncarc{<-}{GB}{GA}\Aput{Gf}
		\end{diagram}
		\hskip3em=\hskip3em
		\begin{diagram}[size=4em]
			\rnode{FA}{FA} & \rTo^{\phi_A} & GA\\
			\begin{array}{c}\Rightarrow\\F(\tau)\end{array}
				& \rlap{$\Nearrow \phi_g$} & \dTo>{Gg}\\
			\rnode{FB}{FB} & \rTo_{\phi_B} & GB
			\ncarc{->}{FA}{FB}\Aput{Fg}
			\ncarc{<-}{FB}{FA}\Aput{Ff}
		\end{diagram}
		\]
		Taking mates of both sides, by Prop.~\ref{prop-mates}(\ref{mate-natural})
		we have
		\[
		\hskip-1em
		\begin{diagram}[size=4em]
			FA & \lTo^{\gamma_A} & \rnode{GA}{GA}\\
			\dTo<{Ff} & \llap{$\Nearrow \gamma_f^{-1}$}
				& \begin{array}{c}\Rightarrow\\G(\tau)\end{array}\\
			FB & \lTo_{\gamma_B} & \rnode{GB}{GB}
			\ncarc{->}{GA}{GB}\Aput{Gg}
			\ncarc{<-}{GB}{GA}\Aput{Gf}
		\end{diagram}
		\hskip3em=\hskip3em
		\begin{diagram}[size=4em]
			\rnode{FA}{FA} & \lTo^{\gamma_A} & GA\\
			\begin{array}{c}\Rightarrow\\F(\tau)\end{array}
				& \rlap{$\Nearrow \gamma_g^{-1}$} & \dTo>{Gg}\\
			\rnode{FB}{FB} & \lTo_{\gamma_B} & GB
			\ncarc{->}{FA}{FB}\Aput{Fg}
			\ncarc{<-}{FB}{FA}\Aput{Ff}
		\end{diagram}
		\]
		which may be rearranged into the naturality condition for $\gamma$.
	\item The composition condition for $\phi$ states that
		\[\hskip-1cm
		\begin{diagram}[h=2em]
			\rnode{FA}{FA} & \rTo^{Ff}& FB & \rTo^{Fg}& \rnode{FC}{FC}\\
			&&\raise4pt\hbox{$\Downarrow F_{g,f}$}\\
			\dTo<{\phi_A} &&&& \dTo>{\phi_C}\\
			&&\Swarrow \phi_{g\o f}\\
			GA && \rTo_{G(g\o f)} && GC
			\ncarc[arcangle=-60]{->}{FA}{FC}\Bput{F(g\o f)}
		\end{diagram}
		\quad=\quad
		\begin{diagram}[h=2em]
			FA & \rTo^{Ff}& FB & \rTo^{Fg}& FC\\
			&\Swarrow\phi_f & \dTo>{\phi_B}&\Swarrow\phi_g\\
			\dTo<{\phi_A} && GB && \dTo>{\phi_C}\\
			&\ruTo^{Gf} & \Downarrow G_{g,f} & \rdTo^{Gg}\\
			GA && \rTo_{G(g\o f)} && GC
		\end{diagram}
		\]
		Taking mates, and using Prop.~\ref{prop-mates}(\ref{mate-natural}, \ref{mate-vert}), gives
		\[\hskip-1cm
		\begin{diagram}[h=2em]
			\rnode{FA}{FA} & \rTo^{Ff}& FB & \rTo^{Fg}& \rnode{FC}{FC}\\
			&&\raise4pt\hbox{$\Downarrow F_{g,f}$}\\
			\uTo<{\gamma_A} &&&& \uTo>{\gamma_C}\\
			&&\Searrow \gamma_{g\o f}^{-1}\\
			GA && \rTo_{G(g\o f)} && GC
			\ncarc[arcangle=-60]{->}{FA}{FC}\Bput{F(g\o f)}
		\end{diagram}
		\quad=\quad
		\begin{diagram}[h=2em]
			FA & \rTo^{Ff}& FB & \rTo^{Fg}& FC\\
			&\Searrow\gamma_f^{-1} & \dTo>{\gamma_B}&\Searrow\gamma_g^{-1}\\
			\uTo<{\gamma_A} && GB && \uTo>{\gamma_C}\\
			&\ruTo^{Gf} & \Downarrow G_{g,f} & \rdTo^{Gg}\\
			GA && \rTo_{G(g\o f)} && GC
		\end{diagram}
		\]
		which may be rearranged into the composition condition for $\gamma$.
	\end{itemize}
	It remains to show that each of the collections $\eta_{A}$
	and $\e_{A}$ constitutes a modification. By equation
	\pref{eq-mate-eta}, we know that
	\[
	\begin{diagram}
		\rnode{A}{FA}\\
		\dTo<{Ff}&\rdTo^{\phi_{A}} \\
		FB&\begin{array}c\To\\[-4pt]\phi_{f}\end{array} & GA \\
		\dTo<1
			& \rdTo^{\phi_{B}}_{\raise-4pt\hbox{$\begin{array}c\To\\[-4pt]\eta_B\end{array}$}}
			& \dTo>{Gf}\\
		\rnode{A'}{FB} & \lTo_{\gamma_{B}} & GB
		\nccurve[angleA=220,angleB=140]{->}{A}{A'}\Bput{Ff}\Aput{\ \ \ \cong}
	\end{diagram}
	\qquad=\qquad
	\begin{diagram}
		& \rnode{A}{FA} & \rTo^{\phi_{A}} & \rnode{B}{GA}\\
		& \dTo<{1}&\ldTo_{\gamma_{A}}^{\raise4pt\hbox{$\begin{array}c\To\\[-4pt]\eta_{A}\end{array}$}}
		     & \dTo>Gf\\
		& FA &\begin{array}c\To\\[-4pt]\gamma_{f}^{-1}\end{array} & GB \\
		& \dTo<{Ff} & \ldTo_{\gamma_{B}} \\
		& \rnode{A'}{FB}
		\nccurve[angleA=220,angleB=140]{->}{A}{A'}\Bput{Ff}\Aput{\ \ \ \cong}
	\end{diagram}
	\]
	This can be rearranged to give
	\[
		\begin{diagram}
			FB & \rTo^{Ff} & FA \\
			&\rdTo^{\phi_{B}} & \Arr\Swarrow{\phi_{f}} & \rdTo^{\phi_{A}} \\
			\dTo<1 & \Right_{\eta_{B}} & GB & \rTo_{Gf} & GA \\
			& \ldTo_{\gamma_{A}} & \Arr\Searrow{\gamma_{f}} & \ldTo_{\gamma_{A}} \\
			FB & \lTo_{Ff} & FA
		\end{diagram}
		\quad=\quad
		\begin{diagram}
			FB & \lTo^{Ff} & FA \\
			&&&\rdTo^{\phi_A} \\
			\dTo<1 & \cong & \dTo<1 & \Right_{\eta_{A}} & GA \\
			&&&\ldTo_{\gamma_{A}} \\
			FB & \lTo_{Ff} & FA
		\end{diagram}
	\]
	which shows that $\eta$ is a modification.
	Similarly, we may use equation \pref{eq-mate-eps} to show that
	$\e$ is a modification.
\end{proof}

In the bicategory $\Cat$, adjunctions $F \dashv G: \C\to\D$ are
characterised by the existence of a natural isomorphism
\[
	\D(FA, B) \cong \C(A, GB),
\]
natural in $A$ and $B$. It is interesting to observe that a similar
characterisation exists for adjunctions in an \emph{arbitrary} bicategory,
as the next proposition shows.
\begin{propn} 
	Let there be given 1-cells $f:A\to B$ and $g:B\to A$ in a bicategory $\B$.
	To give an adjunction $f\dashv g$ is to give, for every $A\lTo^a X\rTo^b B$,
	an isomorphism
	\[
		\phi_{a,b}: \B(X,B)(f\o a, b) \cong \B(X,A)(a, g\o b),
	\]
	natural in the sense that:
	\begin{itemize}
	\item	for every $\sigma: a\To a'$, $\tau: b\To b'$,
		and $\zeta: f\o a' \To b$, we have
		\[
		\begin{diagram}[w=2em]
			&&\rnode{X}{X}\\
			&\ldTo[snake=1em]_{a'}
				& \raise-1em\hbox{$\begin{array}c\To\\[-4pt]\phi_{a',b}(\zeta)\end{array}$}
				& \rdTo[snake=-1em]_b\\
			\rnode{A}{A} && \lTo_{g} && \rnode{B}{B}
			\ncarc[arcangle=-50,ncurv=1]{->}XA\Bput{a}
				\Aput{\Searrow\sigma}
			\ncarc[arcangle=50,ncurv=1]{->}XB\Aput{b'}
				\Bput{\Nearrow\tau}
		\end{diagram}
		\qquad=\qquad
		\begin{diagram}[w=2em]
			&&\rnode{X}{X}\\
			&& \begin{array}c\To\\\phi_{a,b'}(\tau\cdot\zeta\cdot(f\o\sigma))\end{array}
				&\\
			\rnode{A}{A} && \lTo_{g} && \rnode{B}{B}
			\ncarc[arcangle=-30]{->}XA\Bput{a}
			\ncarc[arcangle=30]{->}XB\Aput{b'}
		\end{diagram}
	\]
	\item for every $k:Y\to X$ and $\zeta:fa\To b$, we have
	\[
		\begin{diagram}[w=2em]
			&&Y\\
			&&\dTo>k\\
			&&X\\
			&\ldTo^{a} & \begin{array}c\To\\[-4pt]\phi_{a,b}(\zeta)\end{array} & \rdTo^b\\
			A && \lTo_{g} && B
		\end{diagram}
		\quad=\hskip3em
		\begin{diagram}[w=2em]
			&&Y\\
			&\ldTo^k&&\rdTo^k\\
			X&&&&X\\
			\dTo<{a} && \begin{array}c\To\\[-4pt]\phi_{ak,bk}(\zeta\o k)\end{array} && \dTo>b\\
			A && \lTo_{g} && B
		\end{diagram}
	\]
	\end{itemize}
\end{propn}
\begin{proof}
	By Yoneda, we know that to give an adjunction $f\dashv g$ is to give
	an adjunction $\B(-,f)\dashv\B(-,g)$. By Prop.~\ref{prop-adj-2}, we know that
	to give such an adjunction is to give, for every $X\in\B$, an adjunction
	$\B(X,f)\dashv\B(X,g)$, collectively subject to condition \pref{eq-mate-eta}.
	This is just an ordinary adjunction, which can therefore be given as a
	natural isomorphism
	\[
		\B(X,B)(\B(X,f)(a), b) \cong \B(X,A)(a, \B(X,g)(b))
	\]
	natural in $a\in\B(X,A)$ and $b\in\B(X,B)$, i.e.\ a natural isomorphism
	\[
		\B(X,B)(f\o a, b) \cong \B(X,A)(a, g\o b).
	\]
	This corresponds to the data in the statement of this Proposition,
	subject to our first naturality condition. We shall write the unit
	of this adjunction as \[\eta_a: a\To g\o(f\o a)\] for $a:X\to A$.
	
	It remains to show that the second naturality condition is satisfied just
	when \pref{eq-mate-eta} is. Writing down the concrete interpretation
	of~\pref{eq-mate-eta} in our setting, we find that it holds
	when, for every $a:X\to A$ and $k:Y\to X$, the diagram
	\begin{diagram}
		a\o k & \rTo^{\eta_{a\o k}} & g\o (f\o (a\o k))\\
		\dTo<{\eta_a\o k} && \dTo>{g\o\a_{f,a,k}}\\
		(g\o (f\o a))\o k & \rTo_{\a_{g,f\o a,k}} & g\o((f\o a)\o k)
	\end{diagram}
	commutes in $\B(Y,B)$.
	In pictures, this says that
	
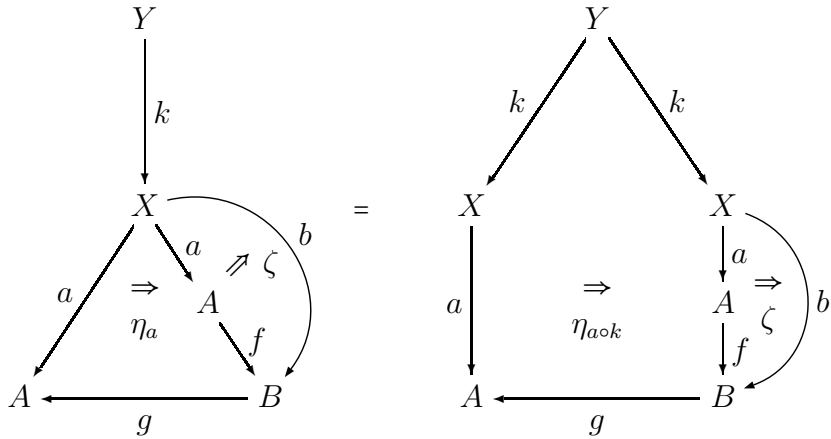
\[
		\begin{diagram}[w=1em,h=1.5em]
			&& &&Y\\ \\
			&& &&\dTo>k\\ \\
			&& &&\rnode{X}{X}\\
			&& &\ldTo(4,4)^{a} && \rdTo^a\\
			&& &&\begin{array}c\To\\[-4pt]\eta_a\end{array} &&A\\
			&&&&&&&\rdTo^f\\
			A && && \lTo_{g}&&  && \rnode{B}{B}
			\ncarc[arcangle=70,ncurv=1]{->}XB\Aput{b}\Bput{\Nearrow\zeta}
		\end{diagram}
		\qquad=\qquad
		\begin{diagram}[w=2em,h=1.5em]
			&&Y\\
			&\ldTo(2,4)^k&&\rdTo(2,4)^k\\
			\\
			\\
			X&&&&\rnode{X}{X}\\
			&&&& \dTo>a\\
			\dTo<{a} && \begin{array}c\To\\[-4pt]\eta_{a\o k}\end{array} && A\\
			&&&&\dTo>{f}\\
			A && \lTo_{g} && \rnode{B}{B}
			\ncarc[arcangle=70,ncurv=1]{->}XB\Aput{b}\Bput{\begin{array}c\To\\[-4pt]\zeta\end{array}}
		\end{diagram}
	\]
	And since $\eta_a = \phi_{a,fa}(1_{fa})$ by definition,
	this is equivalent to the particular case of our second naturality condition
	with $b=f\o a$ and $\zeta = 1_{f\o a}$.
	
	But this special case
	implies the general case, by the first naturality condition: for we have
	\[\begin{array}{rl@{\qquad}c@{\qquad}r}
		&
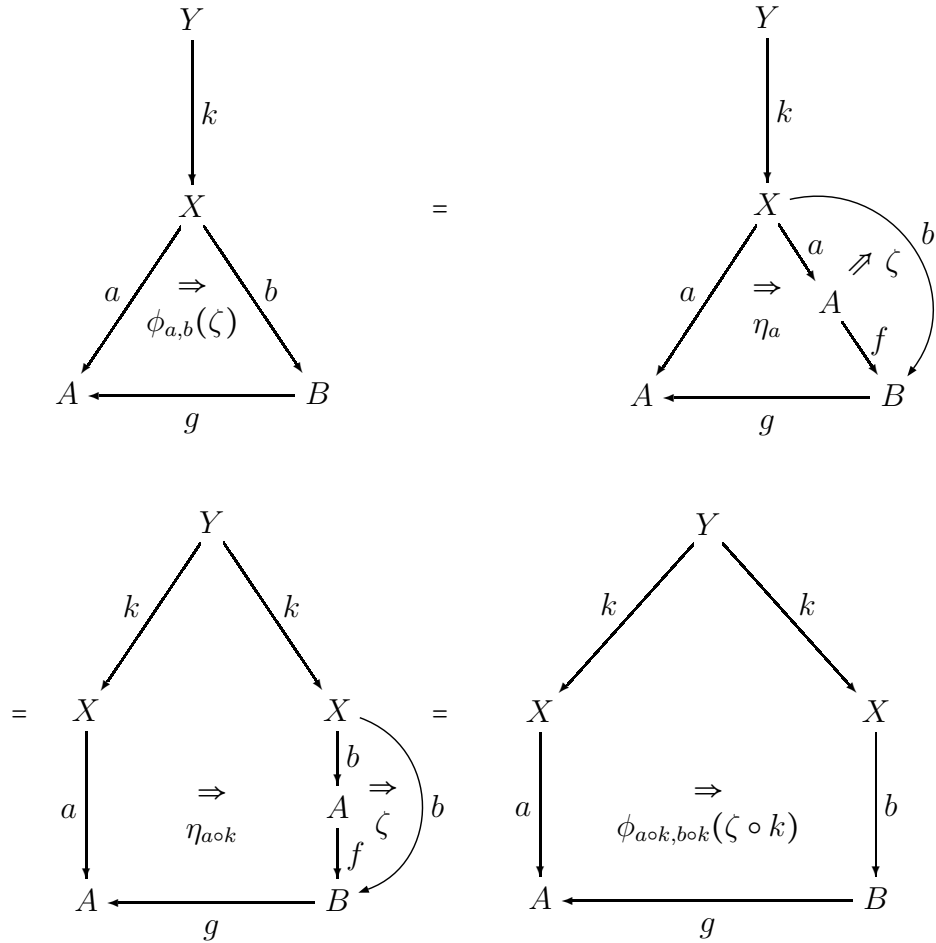
\begin{diagram}[w=2em]
			&&Y\\
			&&\dTo>k\\
			&&X\\
			&\ldTo^{a} & \begin{array}c\To\\[-4pt]\phi_{a,b}(\zeta)\end{array} & \rdTo^b\\
			A && \lTo_{g} && B
		\end{diagram}
		&=&
		\begin{diagram}[w=1em,h=1.5em]
			&& &&Y\\ \\
			&& &&\dTo>k\\ \\
			&& &&\rnode{X}{X}\\
			&& &\ldTo(4,4)^{a} && \rdTo^a\\
			&& &&\begin{array}c\To\\[-4pt]\eta_a\end{array} &&A\\
			&&&&&&&\rdTo^f\\
			A && && \lTo_{g}&&  && \rnode{B}{B}
			\ncarc[arcangle=70,ncurv=1]{->}XB\Aput{b}\Bput{\Nearrow\zeta}
		\end{diagram}
		\\[9em]
		=&
		\begin{diagram}[w=2em,h=1.5em]
			&&Y\\
			&\ldTo(2,4)^k&&\rdTo(2,4)^k\\
			\\
			\\
			X&&&&\rnode{X}{X}\\
			&&&& \dTo>b\\
			\dTo<{a} && \begin{array}c\To\\[-4pt]\eta_{a\o k}\end{array} && A\\
			&&&&\dTo>{f}\\
			A && \lTo_{g} && \rnode{B}{B}
			\ncarc[arcangle=70,ncurv=1]{->}XB\Aput{b}\Bput{\begin{array}c\To\\[-4pt]\zeta\end{array}}
		\end{diagram}
		&=&
		\begin{diagram}[w=2em]
			&&Y\\
			&\ldTo^k&&\rdTo^k\\
			X&&&&X\\
			\dTo<{a} && \begin{array}c\To\\[-4pt]\phi_{a\o k,b\o k}(\zeta\o k)\end{array} && \dTo>b\\
			A && \lTo_{g} && B
		\end{diagram}
	\end{array}\]
	as required.
\end{proof}

\begin{remark} 
	In this section we have considered adjunctions \emph{in} a bicategory.
	It is also possible to consider \emph{pseudo-adjunctions between bicategories}:
	to exhibit $G:\BC\to\B$ as right pseudo-adjoint to $F:\B\to\BC$ is to give
	an equivalence $\BC(FA,X)\simeq\B(A, GX)$ pseudo-natural in $A$ and $X$.
	We do not pursue pseudo-adjunctions further here.
\end{remark}

\section{On equivalence}\label{s-equiv}
Recall the definition of equivalence (Definition\ref{def-equivalence}).
\begin{definition} 
	An \emph{adjoint equivalence} is an adjunction whose unit and
	counit are invertible.
\end{definition}
\begin{remark}
	If $f \dashv g$ is an adjoint equivalence with unit $\eta$ and counit $\e$,
	then $g\dashv f$ is an adjoint equivalence with unit $\e^{-1}$
	and counit $\eta^{-1}$.
\end{remark}
In later chapters, particularly Chapter~\refchapter{Psmon}, we will often consider
mates with respect to adjoint equivalences. Such mating has some special
properties which are crucial for our applications.
\begin{lemma}\label{lemma-adjeq-mate}
	If we have mates
	\[
	\begin{diagram} 
		A & \rTo^{f} & B \\
		\dTo<{h} & \Arr\Nearrow\sigma & \dTo>{k} \\
		A' & \rTo_{f'} & B'
	\end{diagram}
	\qquad\mbox{and}\qquad
	\begin{diagram} 
		A & \lTo^{g} & B \\
		\dTo<{h} & \Arr\Searrow\tau & \dTo>{k} \\
		A' & \lTo_{g'} & B'
	\end{diagram}
	\]
	with respect to adjoint \emph{equivalences} $f\dashv g$
	and $f'\dashv g'$, then $\sigma$ is invertible if and only
	if $\tau$ is.
\end{lemma}
\begin{proof}
	Immediate from the definition of mate.
\end{proof}
\begin{lemma}\label{lemma-adjeq-twisted}
	Given adjoint equivalences $f\dashv g: A\to B$
	and $f'\dashv g': A'\to B'$,
	and an invertible 2-cell
	\begin{diagram} 
		A & \rTo^{f} & B \\
		\dTo<{h} & \Arr\Nearrow\sigma & \dTo>{k} \\
		A' & \rTo_{f'} & B',
	\end{diagram}
	the inverse of the right mate of $\sigma$ is equal to the
	left mate of its inverse.
\end{lemma}
\begin{proof}
	The right mate of $\sigma$ is
	\begin{diagram} 
	\rnode{A}{A} & \lTo^{g} & \rnode{B}{B}\\
	\dTo<{h}&\rdTo_{f}^{\raise4pt\hbox{$\begin{array}c\To\\[-4pt]\e\end{array}$}}
		& \dTo>1\\
	A'&\begin{array}c\To\\[-4pt]\sigma\end{array}&B\\
	\dTo<1
		& \rdTo^{f'}_{\raise-4pt\hbox{$\begin{array}c\To\\[-4pt]\eta'\end{array}$}}
		& \dTo>{k}\\
	\rnode{A'}{A'} & \lTo_{g'} & \rnode{B'}{B',}
	\nccurve[angleA=210,angleB=140]{->}{A}{A'}\Bput{h}\Aput{\ \ \ \cong}
	\nccurve[angleA=-30,angleB=30]{->}{B}{B'}\Aput{k}\Bput{\cong\ \ \ }
	\end{diagram}
	whose inverse is
	\begin{diagram} 
	\rnode{A}{A} & \lTo^{g} & \rnode{B}{B}\\
	\dTo<{h}&\rdTo_{f}^{\raise4pt\hbox{$\begin{array}c\Leftarrow\\[-4pt]\e^{-1}\end{array}$}}
		& \dTo>1\\
	A'&\begin{array}c\Leftarrow\\[-4pt]\sigma^{-1}\end{array}&B\\
	\dTo<1
		& \rdTo^{f'}_{\raise-4pt\hbox{$\begin{array}c\Leftarrow\\[-4pt]\eta'^{-1}\end{array}$}}
		& \dTo>{k}\\
	\rnode{A'}{A'} & \lTo_{g'} & \rnode{B'}{B',}
	\nccurve[angleA=210,angleB=140]{->}{A}{A'}\Bput{h}\Aput{\ \ \ \cong}
	\nccurve[angleA=-30,angleB=30]{->}{B}{B'}\Aput{k}\Bput{\cong\ \ \ }
	\end{diagram}
	which is the left mate of $\sigma^{-1}$.
\end{proof}
\begin{propn}\label{prop-adjeq-mate-dual}
	Given adjoint equivalences $f\dashv g: A\to B$
	and $f'\dashv g': A\to B$,
	and an invertible 2-cell
	\vskip0pt
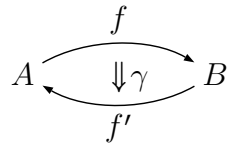
	\begin{diagram}
		\rnode{A}{A} & \Arr\Downarrow\gamma & \rnode{B}{B}
		\ncarc[arcangle=30]{->}{A}{B} \Aput{f}
		\ncarc[arcangle=30]{->}{B}{A} \Aput{f'}
	\end{diagram}
	\vskip2em
	the left mate of $\gamma$ is equal to its right mate.
\end{propn}
\begin{proof}
	The proof is surprisingly intricate, and rather difficult to
	follow unless string diagrams are used. (In fact, this proof is
	the reason that we have introduced string diagrams into this
	chapter.) In string diagram terms, what we have to prove is that
	\[
		\cdiag{d-adj/gamma-up} \qquad=\qquad \cdiag{d-adj/gamma-down}
	\]
	The proof is as follows:
	\[ \cdiag{d-adj/gamma-up} \]
	\[ =\quad \cdiag{d-adj/gamma-pf-1} \]
	\[ =\quad \cdiag{d-adj/gamma-pf-2} \]
	\[ =\quad \cdiag{d-adj/gamma-pf-3} \]
	\[ =\quad \cdiag{d-adj/gamma-pf-4} \]
	\[ =\quad \cdiag{d-adj/gamma-pf-5} \]
	\[ =\quad \cdiag{d-adj/gamma-down} \]
\end{proof}
\begin{propn}\label{prop-adjeq} 
	If there is an equivalence $(f,g,e,e')$ from $A$ to $B$, then there
	is an adjoint equivalence $f\dashv g$ with unit $e$.
\end{propn}
\begin{proof}
	It is well known\footnote{And easy to prove using
	the ordinary Yoneda lemma.} that this is true in $\Cat$.
	We shall use Yoneda to
	infer that it is therefore true in an arbitrary bicategory. Let there be
	given an equivalence $(f,g,e,e')$ from $A$ to $B$. This induces
	an equivalence from $\B(-,A)$ to $\B(-,B)$ in $\Bicat(\B\op,\Cat)$.
	Thus for every $X\in\B$ we have an adjoint equivalence in $\Cat$
	from $\B(X,A)$ to $\B(X,B)$, with unit $\B(X,e)$. By Prop.~\ref{prop-adj-2}
	and Lemma~\ref{lemma-adjeq-mate}
	this induces an adjoint equivalence in $\Bicat(\B\op,\Cat)$, and Yoneda
	therefore yields the desired adjoint equivalence in $\B$.
\end{proof}
\begin{remark} 
	It is possible to give a more elementary proof of the preceding
	Proposition, by directly constructing a counit for the adjoint
	equivalence. This parallels the situation that one frequently
	encounters in ordinary category theory, where there is a choice
	between a concise, perspicuous Yoneda proof and an obscure but
	elementary equational one.
\end{remark}
\begin{remark} 
	An adjunction in $\B$ is an adjunction in ${\B\co}\op$, with the
	unit and counit reversed. Thus in the situation of Prop.~\ref{prop-adjeq}
	there is also a (generally different) adjoint equivalence with counit $e'$.
\end{remark}
Just as an ordinary natural transformation is invertible just when all its
components are, so a modification is invertible just when all its components
are. Furthermore a pseudo-natural transformation is an equivalence
in $\Bicat(\B,\BC)$ just when all its components are equivalences in
their respective hom-categories. This latter fact, though unsurprising,
is not altogether trivial to prove -- though we have done the hard work
already.
\begin{propn}\label{prop-pneq}
	Let there be given a pseudo-natural transformation
	\[
		\gamma: F\To G: \B\to\BC.
	\]
	This $\gamma$ is an equivalence in $\Bicat(\B,\BC)$ just
	when for every $A\in\B$ the 1-cell $\gamma_A: FA\to GA$
	is an equivalence.
\end{propn}
\begin{proof}
	Suppose that for every $A$, the component $\gamma_A: FA\to GA$
	is an equivalence. By Prop.~\ref{prop-adjeq} we may suppose that
	there is an arrow $\delta_A: $ and an adjoint equivalence
	$\gamma_A\dashv\delta_A$. Now the claim follows from
	Lemma~\ref{lemma-adjeq-mate} and Prop.~\ref{prop-adj-2}.
\end{proof}

\section{Normal pseudo-functors}
Here we prove a useful coherence-type result about pseudo-functors.
It is certainly well-known, but I am not aware of a published proof.%
\begin{definition} 
	A pseudo-functor $F:\B\to\BC$ is \emph{normal} if
	$F_A: 1_{FA}\to F(1_A)$ is an identity map for every $A\in\B$.
\end{definition}
\begin{lemma}\label{l-normal}
	Let $F:\B\to\BC$ be a pseudo-functor, and let there be given,
	for all $A$, $B\in\B$, a functor $G_{A,B}: \B(A,B)\to\BC(FA,FB)$
	and a natural isomorphism $\phi_{A,B}: F_{A,B}\To G_{A,B}$.
	
	Then $G$ may be extended to a pseudo-functor that coincides
	with $F$ on objects, such that $\phi$ becomes a pseudo-natural
	equivalence between $F$ and $G$.
\end{lemma}
\begin{proof}
	Define $G$ like $F$ on objects, and let its action on the hom-category
	$\B(A,B)$ be the functor $G_{A,B}$. For an object $A$, let $G_A$ be the composite
	\[
		1_{GA} = 1_{FA} \rTo^{F_A} F(1_A) \rTo^{(\phi_{A,A})_{1_A}} G(1_A),
	\]
	and for a composable pair
	\(
		A\rTo^f B\rTo^g C
	\)
	let $G_{g,f}$ be the composite
	\[
		G(g)\o G(f) \rTo^{(\phi_{B,C}^{-1})_g\o(\phi_{A,B}^{-1})_f} F(g)\o F(f)
			\rTo^{F_{g,f}} F(g\o f) \rTo^{(\phi_{A,C})_{g\o f}} G(g\o f).
	\]
	It is necessary to check that $G$ is indeed a pseudo-functor. For example,
	for condition $\cFl$ take an arrow $f:A\to B$. We have the diagram
	\begin{diagram}[w=6em]
		1\o Gf & \rTo^{F_B\o Gf} & F1\o Gf & \rTo^{\phi_1\o Gf} & G1\o Gf\\
		&\rdTo(0,6)<{\l_{Gf}}\rdTo(1,2)>{1\o\phi_f^{-1}} &&\rdTo(1,2)<{F1\o \phi_f^{-1}}
			\ldTo(1,2)^{\phi_1^{-1}\o\phi_f^{-1}}\ruTo(0,6)<{G_{1,f}}\\
		& 1\o Ff & \rTo_{F_B\o Ff} & F1\o Ff\\
		\rlap{\qquad$\natural_\l$}&\dTo<{\l_{Ff}} &\cFl& \dTo<{F_{1,f}}\\
		&Ff & \rTo_{F(\l_f)} & F(1\o f)\\
		\ldTo(1,2)>{\phi_f} &&\natural_\phi&& \rdTo(1,2)_{\phi_{1\o f}}\\
		Gf && \rTo_{G(\l_f)} && G(1\o f)
	\end{diagram}
	where we have omitted the object subscripts of $\phi$.
	The regions commute for the marked reasons, or functoriality of composition,
	except for the rightmost region which commutes by definition of $G_{1,f}$.
	Thus the outside commutes, and $G$ satisfies condition $\cFl$.
	The other conditions may be checked similarly.
	
	We make $\phi$ into a pseudo-natural transformation by defining $\phi_A$
	to be the identity at $FA=GA$, for every object $A$. For an arrow $f:A\to B$,
	$\phi_f$ is defined to be the pasting
	\begin{diagram}[size=5em]
		\rnode{FA}{FA} & \rTo^{\phi_A = 1} & GA\\
		\dTo<{Ff} &\hskip4pt\Nearrow (\phi_{A,B})_f& \dTo>{Gf}\\
		FA & \rTo_{\phi_B = 1} & \rnode{GB}{GB}
		\ncarc{->}{FA}{GB}\Aput{Gf}
		\ncarc{<-}{GB}{FA}\Aput{Ff}
	\end{diagram}
	It is easy to check that this constitutes a pseudo-natural transformation,
	and its components are identities (hence equivalences), so by
	Prop.~\ref{prop-pneq} it is a pseudo-natural equivalence, as required.
\end{proof}

\begin{propn}\label{prop-normal}
	Every pseudo-functor $F$ is equivalent to a normal pseudo-functor
	that agrees with $F$ on objects, as well as on non-identity 1-cells and
	the 2-cells between them.
\end{propn}
\begin{proof}
	Let there be given a pseudofunctor $F:\B\to\BC$. We shall
	construct an equivalent normal pseudofunctor $G$. By Lemma~\ref{l-normal}
	it suffices to do so for each hom-category separately. The definition
	of $G_{A,B}$ and $\phi_{A,B}$ is by cases, as follows.
	\begin{itemize}
		\item Given objects $A\neq B$, let $G_{A,B} \defeqto F_{A,B}$
			and let $\phi_{A,B}$ be the identity.
		\item For each object $A$, let $G_{A,A}(1) \defeqto 1$, and let
			$G_{A,A}(f) \defeqto F_{A,A}(f)$ for $f \neq 1$.
		\item Given a 2-cell $\beta:f\To g: A\to A$ with $f\neq1\neq g$,
			let $G_{A,A}(\beta) \defeqto \beta$.
		\item Given a 2-cell $\beta:1\To f: A\to A$ with $f\neq 1$,
			let $G_{A,A}(\beta) \defeqto F(\beta)\cdot F_A$.
		\item Given a 2-cell $\gamma:f\To 1: A\to A$ with $f\neq 1$,
			let $G_{A,A}(\gamma) \defeqto F_A^{-1}\cdot F(\gamma)$.
		\item Given a 2-cell $\delta: 1\To 1: A\to A$,
			let $G_{A,A}(\delta) \defeqto F_A^{-1}\cdot F(\delta)\cdot F_A$.
		\item Let $(\phi_{A,A})_1: F(1_A)\to G(1_A) = 1_{GA} = 1_{FA}$ be $F_A^{-1}$,
		\item and let $(\phi_{A,A})_f \defeqto 1$.
	\end{itemize}
	It is straightforward to check (four cases) that this
	makes $\phi_{A,A}$ a natural transformation.
	
	Now we may extend $G$ to a pseudo-functor using Lemma~\ref{l-normal},
	in which $G_A$ is defined to be $(\phi_{A,A})_1\cdot F_A$. Since
	$(\phi_{A,A})_1 = F_A^{-1}$ by definition, $G_A$ is the identity and
	so $G$ is normal, as required.
\end{proof}

\end{thesischapter}
\documentclass{robinthesis}

\begin{thesischapter}{MonBicats}{Monoidal Bicategories}
\begin{definition}\label{def-monbicat}
	A monoidal bicategory $\B$ is a bicategory equipped with
	a unit object $\I\in\B$, a pseudo-functor
	\[
		\tn: \B\x\B \to \B,
	\]
	pseudo-natural equivalences $a$, $l$ and $r$ with components
	\[\begin{array}l
		a_{A,B,C}: A\tn (B\tn C) \to (A\tn B)\tn C,\\
		l_A: \I\tn A\to A,\\
		r_A: A\tn \I\to A,
	\end{array}\]
	and invertible modifications $\pi$, $\mu$, $L$ and $R$ with components
	\begin{diagram}
	  A\tensor \bigl(B\tensor (C\tensor D)\bigr)
	  &\rTo^{a_{A,B,C\tn D}}&(A\tensor B)\tensor (C\tensor D)
	  &\rTo^{a_{A\tn B,C,D}} & \bigl((A\tensor B)\tensor C\bigl)\tensor D
	  \\
	  &\rdTo[snake=-1em](1,2)<{A\tn a_{B,C,D}}
	  &\Downarrow\pi_{A,B,C,D}
	  & \ruTo[snake=1em](1,2)>{a_{A,B,C}\tn D}
	  \\
	  & \spleft{A\tensor\big((B\tensor C)\tensor D\big)}
	  & \rTo_{a_{A,B\tn C,D}}
	  & \spright{\bigl(A\tensor(B\tensor C)\bigr)\tensor D}
	\end{diagram}
	\begin{diagram}
		A\tn(\I\tn C) &\rTo^{a_{A,\I,C}}&(A\tn \I)\tn C\\
		&\rdTo[snake=-1ex](1,2)<{A\tn l_C}
			\raise1ex\hbox{$\begin{array}c\Rightarrow\\[-5pt]\mu_{A,C}\end{array}$}%
			\ldTo[snake=1ex](1,2)>{r_A\tn C}\\
		&A\tn C
	\end{diagram}
\[
	\begin{diagram}
		\I\tn(B\tn C) &\rTo^{a_{\I,B,C}}&(\I\tn B)\tn C\\
		&\rdTo[snake=-1ex](1,2)<{l_{B\tn C}}
			\raise1ex\hbox{$\begin{array}c\Rightarrow\\[-5pt]L_{B,C}\end{array}$}%
			\ldTo[snake=1ex](1,2)>{l_B\tn C}\\
		&B\tn C
	\end{diagram}
	\hskip 3em
	\begin{diagram}
		A\tn(B\tn \I) &\rTo^{a_{A,B,\I}}&(A\tn B)\tn \I\\
		&\rdTo[snake=-1ex](1,2)<{A\tn r_B}
			\raise1ex\hbox{$\begin{array}c\Rightarrow\\[-5pt]R_{A,B}\end{array}$}%
			\ldTo[snake=1ex](1,2)>{r_{A\tn B}}\\
		&A\tn B
	\end{diagram}
\]
such that for all $A$, $B$, $C$, $D$ and $E$ in $\B$, the condition
shown in Fig.~\ref{fig-4coc} holds,
\begin{sidewaysfigure}
\ \hskip4cm\scalebox{0.75}{\begin{diagram}
	  &&
	  (A\tn B)\tensor \bigl(C\tensor (D\tensor E)\bigr)
	  &\rTo[rightshortfall=4.5em]^{a}&\spboth{((A\tn B)\tensor C)\tensor (D\tensor E)}
	  &\rTo[leftshortfall=4.5em]^{a} & \bigl(((A\tn B)\tensor C)\tensor D\bigl)\tensor E
	  \\
	  &\ruTo^a&
	  &\rdTo[snake=-1em](1,2)<{(A\tn B)\tn a}
	  &\spboth{\Downarrow\pi_{(A\tn B),C,D,E}}
	  & \ruTo[snake=1em](1,2)>{a\tn E}
	  && \spboth{((A\tn(B\tn C))\tn D)\tn E} \luTo[snake=1.5em](1,1)^{(a\tn D)\tn E}
	  \\
	  A\tn(B\tn(C\tn(D\tn E)))
	  && \Downarrow a^{-1}
	  & {(A\tn B)\tensor\big((C\tensor D)\tensor E\big)}
	  & \rTo_{a}
	  &{\bigl((A\tn B)\tensor(C\tensor D)\bigr)\tensor E}
	  &\begin{array}c\To\\[-4pt]\pi_{A,B,C,D}\tn E\end{array}
	  & \uTo>{a\tn E}
	  \\
	  & \rdTo_{A\tn(B\tn a)}
	  & \ruTo(1,2)^a
	  &&\spboth{\Searrow\pi_{A,B,(C\tn D),E}}
	  &&\luTo(1,2)>{a\tn E}
	  &\spboth{(A\tn((B\tn C)\tn D))\tn E}
	  \\
	  &&{A\tn(B\tn((C\tn D)\tn E))}
	  & \rTo[rightshortfall=4.5em]_{A\tn a} & \spboth{A\tn((B\tn(C\tn D))\tn E)}
	  & \rTo[leftshortfall=4.5em]_a & (A\tn(B\tn(C\tn D)))\tn E
	   \ruTo[snake=1.5em](1,1)_{(A\tn a)\tn E}
\end{diagram}}
\\ must be equal to\\
\ \hskip4cm\scalebox{0.75}{\begin{diagram}
	&(A\tn B)\tn(C\tn(D\tn E))
	& \rTo^a & ((A\tn B)\tn C)\tn(D\tn E)
	& \rTo^{a} & (((A\tn B)\tn C)\tn D)\tn E
	\\
	\ruTo(1,3)^a &&\raise-2em\spleft{\Searrow\pi_{A,B,C,(D\tn E)}}
	& \uTo<{a\tn(D\tn E)} & \Searrow a
	& \uTo>{(a\tn D)\tn E}
	\\
	&&& (A\tn(B\tn C))\tn(D\tn E) & \rTo^a
	& ((A\tn(B\tn C))\tn D)\tn E
	\\
	A\tn(B\tn(C\tn(D\tn E)))
	& \rTo[rightshortfall=4.5em]^{A\tn a}
	& \spboth{A\tn((B\tn C)\tn(D\tn E))} \ruTo[snake=-1.5em](1,1)^a
	&& \Searrow\pi_{A,(B\tn C),D,E} & \uTo>{a\tn E}
	\\
	&\rdTo(1,3)_{A\tn(B\tn a)}
	&& A\tn(((B\tn C)\tn D)\tn E) \rdTo[snake=-1.5em](1,1)_{A\tn a}
	& \rTo_a & (A\tn((B\tn C)\tn D))\tn E
	\\
	&&\raise2em\spleft{\Downarrow A\tn\pi_{B,C,D,E}}
	&\uTo>{A\tn(a\tn E)} & \Searrow a & \uTo>{(A\tn a)\tn E}
	\\
	& A\tn(B\tn((C\tn D)\tn E)) & \rTo_{A\tn a} & A\tn((B\tn(C\tn D)\tn E)
	& \rTo_a & (A\tn(B\tn(C\tn D)))\tn E
\end{diagram}}
\caption{The associativity axiom used in the definition of monoidal bicategory (sometimes
	called the \textit{non-abelian 4-cocycle condition}).
}\label{fig-4coc}
\end{sidewaysfigure}
and for all $A$, $B$ and $C$, the following two conditions hold:
\begin{diagram}
	&&A\tn(B\tn C) & \rTo^{a_{A,B,C}} & (A\tn B)\tn C
	\\
	&\ruTo^{A\tn l_{B\tn C}}_{\Searrow \mu_{A,B\tn C}}
	&\uTo>{r_A\tn (B\tn C)} && \uTo>{(r_A\tn B)\tn C}<{\Searrow a_{r_A,B,C}\hskip2em}
	\\
	A\tensor \bigl(\I\tensor (B\tensor C)\bigr)
	 &\rTo^{a_{A,\I,B\tn C}}&(A\tensor \I)\tensor (B\tensor C)
	&\rTo^{a_{A\tn \I,B,C}} & \bigl((A\tensor \I)\tensor B\bigl)\tensor C
	\\
	&\rdTo[snake=-1em](1,2)<{A\tn a_{\I,B,C}}
	&\Downarrow\pi_{A,\I,B,C}
	& \ruTo[snake=1em](1,2)>{a_{A,\I,B}\tn C}
	\\
	& \spleft{A\tensor\big((\I\tensor B)\tensor C\big)}
	& \rTo_{a_{A,\I\tn B,C}}
	& \spright{\bigl(A\tensor(\I\tensor B)\bigr)\tensor C}
\end{diagram}
is equal to
\begin{diagram}
	&&A\tn(B\tn C) & \rTo^{a_{A,B,C}} & (A\tn B)\tn C
	\\
	&\ruTo^{A\tn l_{B\tn C}} &&&& \luTo^{(r_A\tn B)\tn C}
	\\
	A\tn(\I\tn(B\tn C)) & \spright{\Searrow A\tn L_{B,C}} & \uTo[snake=1.5em]>{A\tn(l_B\tn C)}
	& \Searrow a_{A,l_B,C} & \uTo[snake=-1.5em]<{(A\tn l_B)\tn C}
	& \spleft{\begin{array}c\To\\[-4pt]\mu_{A,B}\tn C\end{array}}
	& ((A\tn \I)\tn B)\tn C
	\\
	&\rdTo_{A\tn a_{\I,B,C}}
	&&&& \ruTo_{a_{A,\I,B}\tn C}
	\\
	&&A\tn((\I\tn B)\tn C) & \rTo_{a_{A,\I\tn B,C}} & (A\tn(\I\tn B))\tn C
\end{diagram}
and
\scalebox{0.75}{\begin{diagram}
	A\tn(B\tn C) & \rTo^{a_{A,B,C}} & (A\tn B)\tn C
	\\
	\uTo<{A\tn (B\tn l_C)}>{\hskip3em\Searrow a_{A,B,l_C}}
	&& \uTo<{(A\tn B)\tn l_C}>{\quad\begin{array}c\To\\[-4pt]\mu_{(A\tn B),C}\end{array}}
	& \luTo^{r_{A\tn B}\tn C}
	\\
	A\tn (B\tn (\I\tn C))
	&\rTo^{a_{A,B,\I\tn C}}
	& (A\tn B)\tn (\I\tn C)
	&\rTo_{a_{(A\tn B),\I,C}}& ((A\tn B)\tn \I)\tn C
	\\
	&\rdTo[snake=-1em](1,2)<{A\tn a_{B,\I,C}}
	&\Downarrow\pi_{A,\I,B,C}
	& \ruTo[snake=1em](1,2)>{a_{A,\I,B}\tn C}
	\\
	& \spleft{A\tn\big((B\tn \I)\tn C\big)}
	& \rTo_{a_{A,B\tn \I,C}}
	& \spright{\bigl(A\tn(B\tn \I)\bigr)\tn C}
\end{diagram}}
is equal to
\scalebox{0.75}{\begin{diagram}
	&&A\tn(B\tn C) & \rTo^{a_{A,B,C}} & (A\tn B)\tn C
	\\
	&\ruTo^{A\tn (B\tn l_C)} &&&& \luTo^{r_{A\tn B}\tn C}
	\\
	A\tn(B\tn(\I\tn C)) & \spright{\Searrow A\tn \mu_{B,C}} & \uTo[snake=1.5em]>{A\tn(r_B\tn C)}
	& \Searrow a_{A,r_B,C} & \uTo[snake=-1.5em]<{(A\tn r_B)\tn C}
	& \spleft{\begin{array}c\To\\[-4pt]R_{A,B}\tn C\end{array}}
	& ((A\tn B)\tn \I)\tn C
	\\
	&\rdTo_{A\tn a_{B,\I,C}}
	&&&& \ruTo_{a_{A,B,\I}\tn C}
	\\
	&&A\tn((B\tn \I)\tn C) & \rTo_{a_{A,B\tn \I,C}} & (A\tn(B\tn \I))\tn C
\end{diagram}}
\end{definition}
This is not \emph{quite} the most general possible definition, since we have
merely specified an object $\I$ rather than a pseudo-functor $1\to\B$. But since
every pseudo-functor is equivalent to a normal one, there is no essential loss of
generality.

When we have occasion to refer explicitly to the equivalence-inverse of
$a$, $l$ or $r$, we shall denote it as $a'$, $l'$ or $r'$. Furthermore,
we shall assume where necessary that we have \emph{adjoint} equivalences
$a\dashv a'$, $l\dashv l'$ etc.

When working in
a monoidal bicategory, we extend our convention of not explicitly naming
structural isomorphisms to the isomorphisms representing the pseudo-functoriality
of tensor. Instead we mark them with the symbol $\sim$. Note that there
will usually be some implicit structural isomorphisms too:
for example, given 1-cells $f:A\to B$ and $g: C\to D$, the diagram
\begin{diagram}
	A\tn B & \rTo^{A\tn g} & A\tn D\\
	\dTo<{f\tn B} & \sim & \dTo>{f\tn D}\\
	C\tn B & \rTo_{C\tn g} & C\tn D
\end{diagram}
indicates the 2-cell
\[
	(f\tn D)\o(A\tn g) \rTo^{\tn_{(f,1),(1,g)}} (f\o 1)\tn(1\o g) \rTo^{\r_f\tn\l_g}f.g
\hskip8em\]
\[\hskip8em
		\rTo^{\l_f^{-1}\tn\r_g^{-1}} (1.f)\tn(g.1) \rTo^{\tn_{(1,f),(g,1)}^{-1}} (C\tn g)\o(f\tn B).  
\]

\begin{remark}\label{rem-defining-L} 
	Notice that the first unit equation -- since all its cells are invertible -- allows
	$A\tn L_{B,C}$ to be expressed in terms of $\pi$, $\mu$ and $a$. In particular
	$\I\tn L_{B,C}$ may be so expressed
	which, since the pseudofunctor $\I\tn-$ is equivalent, via $l$, to the identity,
	allows $L_{B,C}$ to be expressed in terms of $\pi$, $\mu$, $a$ and the implicit
	data that make $l$ an equivalence. Furthermore it is not hard to see that the
	components $L_{A,B}$ thus defined constitute a modification. Therefore the
	modification $L$ may be defined in terms of the other data. It is perhaps
	tempting to conclude that $L$ is redundant, but there
	is a subtlety: the modification $L$ so defined does not necessarily -- at least as far
	as I can tell -- satisfy the necessary equation. Thus we \emph{could} suppress $L$
	(and $R$ too, since its definition is symmetrical to that of $L$) from our
	definition, but only at the expense of introducing a new and complicated
	equation involving the other data.
\end{remark}

\section{Monoidal pseudo-functors and transformations}
\begin{definition} 
	A \emph{monoidal pseudofunctor} $F: \B\to\BC$,
	between monoidal bicategories $\B$ and $\BC$, consists of a pseudofunctor
	equipped with:
	\begin{itemize}
	\item a 1-cell $F^\tn_\I: \I\to F\I$,
	\item a pseudo-natural transformation $F^\tn$ with components
	\[
		F^\tn_{A,B}: FA\tn FB \to F(A\tn B)
	\]
	\item an invertible modification $F^a$ with components
	\begin{diagram}
		FA\tn(FB\tn FC) & \rTo^{a_{FA,FB,FC}} & (FA\tn FB)\tn FC\\
		\dTo<{FA\tn F^\tn_{B,C}} && \dTo>{F^\tn_{A,B}\tn FC}\\
		FA\tn F(B\tn C) &\Nearrow F^a_{A,B,C}& F(A\tn B)\tn FC\\
		\dTo<{F^\tn_{A,B\tn C}} && \dTo>{F^\tn_{A\tn B,C}}\\
		F(A\tn(B\tn C)) & \rTo_{F(a_{A,B,C})} & F((A\tn B)\tn C)
	\end{diagram}
	\end{itemize}
	\item invertible modifications $F^l$ and $F^r$ with components
	\[
	\begin{diagram}
		\I\tn FA & \rTo^{F^\tn_\I\tn FA} & F\I \tn FA\\
		\dTo<{l_{FA}} & \begin{array}c\Rightarrow\\F^l_{A}\end{array} & \dTo>{F^\tn_{\I,A}}\\
		FA & \lTo_{F(l_{A})} & F(\I\tn A)
	\end{diagram}
	\qquad
	\begin{diagram}
		FA\tn \I & \rTo^{FA\tn F^\tn_\I} & FA \tn F\I\\
		\dTo<{r_{FA}} & \begin{array}c\Rightarrow\\F^r_{A}\end{array} & \dTo>{F^\tn_{A,\I}}\\
		FA & \lTo_{F(r_{A})} & F(A\tn \I)
	\end{diagram}
	\]
	\item satisfying the equation shown in Figs.~\ref{fig-monpsf-1}
		and~\ref{fig-monpsf-2}, and
	\begin{sidewaysfigure}
		\scalebox{0.75}{\begin{diagram}
			FA\tn(FB\tn(FC\tn FD))
			& \rTo^{FA\tn(FB\tn F^\tn_{C,D})} & FA\tn(FB\tn F(C\tn D))
			& \rTo^{FA\tn F^\tn_{B,C\tn D}} & FA\tn F(B\tn(C\tn D))
			& \rTo^{F^\tn_{A,B\tn(C\tn D)}} & \rnode{t}{F(A\tn(B\tn(C\tn D)))}
			\\
			\dTo<{FA\tn a_{FB,FC,FD}}
			&& \Swarrow FA\tn F^a_{B,C,D}
			&& \dTo<{FA\tn F(a_{B,C,D})}
			& \Swarrow (F^\tn_{A,a_{B,C,D}})^{-1}
			& \dTo>{F(A\tn a_{B,C,D})}
			\\
			FA\tn((FB\tn FC)\tn FD)
			& \rTo^{FA\tn(F^\tn_{B,C}\tn FD)} & FA\tn(F(B\tn C)\tn FD)
			& \rTo^{FA\tn F^\tn_{B\tn C, D}} & FA\tn F((B\tn C)\tn D)
			& \rTo^{F^\tn_{A,(B\tn C)\tn D}} & F(A\tn((B\tn C)\tn D)
			\\
			\dTo<{a_{FA,FB\tn FC, FD}}
			& {\Swarrow \a_{FA,F^\tn_{B,C}, FD}}
			& \dTo>{a_{FA,F(B\tn C), FD}}
			&& \Swarrow F^a_{A,B\tn C,D}
			&& \dTo<{F(a_{A,B\tn C,D})}
			& \spleft{\begin{array}c\Leftarrow\\ F(\pi_{A,B,C,D})\end{array}}
			& \spright{\rnode{r}{F((A\tn B)\tn(C\tn D))}}
			\\
			(FA\tn(FB\tn FC))\tn FD
			& \rTo_{(FA\tn F^\tn_{B,C})\tn FD} & (FA\tn F(B\tn C))\tn FD
			& \rTo_{F^\tn_{A,B\tn C}\tn FD} & F(A\tn(B\tn C))\tn FD
			& \rTo_{F^\tn_{A\tn(B\tn C), D}} & F((A\tn(B\tn C))\tn D)
			\\
			\dTo<{a_{FA,FB,FC}\tn FD}
			&& \Swarrow F^a_{A,B,C}\tn FD
			&& \dTo<{F(a_{A,B,C})\tn FD}
			& \Swarrow (F^\tn_{a_{A,B,C},D})^{-1}
			& \dTo>{F(a_{A,B,C}\tn D)}
			\\
			((FA\tn FB)\tn FC)\tn FD
			& \rTo_{(F^\tn_{A,B}\tn FC)\tn FD} & (F(A\tn B)\tn FC)\tn FD
			& \rTo_{F^\tn_{A\tn B,C}\tn FD} & F((A\tn B)\tn C)\tn FD
			& \rTo_{F^\tn_{(A\tn B)\tn C, D}} & \rnode{b}{F(((A\tn B)\tn C)\tn D)}
			\nccurve[angleA=0,angleB=90]{->}tr\Aput{F(a_{A,B,C\tn D})}
			\nccurve[angleA=270,angleB=0]{->}rb\Aput{F(a_{A,B,C\tn D})}
		\end{diagram}}
		\caption{Left-hand side of an equation used in the definition of monoidal pseudo-functor:
		for all $A$, $B$, $C$, $D\in\B$, this pasting must be equal to the one shown
		in Fig.~\ref{fig-monpsf-2}.}\label{fig-monpsf-1}
	\end{sidewaysfigure}
	\begin{sidewaysfigure}
		\scalebox{0.75}{%
		\begin{diagram}
			&&
			\rnode{1}{FA\tn(FB\tn(FC\tn FD))}
			& \rTo^{FA\tn(FB\tn F^\tn_{C,D})} & FA\tn(FB\tn F(C\tn D))
			& \rTo^{FA\tn F^\tn_{B,C\tn D}} & FA\tn F(B\tn(C\tn D))
			& \rTo^{F^\tn_{A,B\tn(C\tn D)}} & \rnode{t}{F(A\tn(B\tn(C\tn D)))}
			\\
			&&
			& \Swarrow a_{FA,FB,F(C\tn D)}
			& \dTo>{a_{FA,FB,F(C\tn D)}}
			\\
			\rnode{2}{FA\tn((FB\tn FC)\tn FD)}
			&& \dTo>{a_{FA,FB,FC\tn FD}}
			&& (FA\tn FB)\tn F(C\tn D)
			&& \Swarrow F^a_{A,B,C\tn D}
			&& \dTo>{F(a_{A,B,C\tn D})}
			\\
			&&
			& \ruTo^{(FA\tn FB)\tn F^\tn_{C,D}}
			&& \rdTo^{F^\tn_{A,B}\tn F(C\tn D)}
			\\
			\dTo<{a_{FA,FB\tn FC, FD}}
			& \spboth{\begin{array}c\Leftarrow\\\pi_{FA,FB,FC,FD}\end{array}}
			& \spright{(FA\tn FB)\tn(FC\tn FD)}
			&& \cong
			&& F(A\tn B) \tn F(C\tn D)
			& \rTo^{F^\tn_{A\tn B,C\tn D}}
			& F((A\tn B)\tn(C\tn D))
			\\
			&&
			& \rdTo_{F^\tn_{A,B}\tn(FC\tn FD)}
			&& \ruTo_{F(A\tn B)\tn F^\tn_{C,D}}
			\\
			\rnode{3}{(FA\tn(FB\tn FC))\tn FD}
			&& \dTo>{a_{FA\tn FB, FC, FD}}
			&& F(A\tn B)\tn(FC\tn FD)
			&& \Swarrow F^a_{A\tn B,C,D}
			&& \dTo>{F(a_{A\tn B,C,D})}
			\\
			&&
			& \Swarrow a_{F^\tn_{A,B},FC,FD}
			& \dTo>{a_{F(A\tn B),FC,FD}}
			\\
			&&
			\rnode{4}{((FA\tn FB)\tn FC)\tn FD}
			& \rTo_{(F^\tn_{A,B}\tn FC)\tn FD} & (F(A\tn B)\tn FC)\tn FD
			& \rTo_{F^\tn_{A\tn B,C}\tn FD} & F((A\tn B)\tn C)\tn FD
			& \rTo_{F^\tn_{(A\tn B)\tn C, D}} & \rnode{b}{F(((A\tn B)\tn C)\tn D)}
			\nccurve[angleA=180,angleB=90]{->}12\Aput{FA\tn a_{FB,FC,FD}}
			\nccurve[angleA=270,angleB=180]{->}34\Aput{a_{FA,FB,FC}\tn FD}
		\end{diagram}}
		\caption{Right-hand side of the equation: this pasting must be equal to the one
		shown in Fig.~\ref{fig-monpsf-1}}\label{fig-monpsf-2}
	\end{sidewaysfigure}
	also such that for all $A$, $B\in\B$ the pasting
	\[\scalebox{0.75}{\begin{diagram}
		&&
		\rnode{1}{FA\tn(\I\tn FB)}
		& \rTo^{a_{FA,\I,FB}} & (FA\tn \I)\tn FB
		\\
		&&
		\dTo<{FA\tn(F^\tn_\I\tn FB)}
		& \Nearrow a_{FA,F^\tn_\I,FB}
		& \dTo>{(FA\tn F^\tn_\I)\tn FB}
		\\
		&&
		FA\tn(F\I\tn FB)
		& \rTo_{a_{FA,F\I,FB}} & (FA\tn F\I)\tn FB
		& \rTo^{F^\tn_{A,\I}\tn FB} & F(A\tn \I)\tn FB
		\\
		&\begin{array}c\Rightarrow\\FA\tn F^l_B\end{array}
		& \dTo>{FA\tn F^\tn_{\I,B}}
		&& \Nearrow F^a_{A,\I,B}
		&& \dTo>{F^\tn_{A\tn \I, B}}
		\\
		&&
		FA\tn F(\I\tn B)
		& \rTo^{F^\tn_{A,\I,B}} & F(A\tn(\I\tn B))
		& \rTo^{F(a_{A,\I,B})} & F((A\tn \I)\tn B)
		\\
		&&\dTo<{FA\tn FB}
		& \Nearrow F^\tn_{A,l_B}
		& \dTo<{F(A\tn l_B)}>{\begin{array}c\To\\[-4pt]F(\mu_{A,B})\end{array}}
		& \ldTo_{F(r_A\tn B)}
		\\
		&&
		\rnode{2}{FA\tn FB}
		& \rTo_{F^\tn_{A,B}} & F(A\tn B)
		\nccurve[angle=180,ncurv=1]{->}12\Bput{FA\tn l_{FB}}
	\end{diagram}}\]
	is equal to
	\[\scalebox{0.75}{\begin{diagram}
		FA\tn(\I\tn FB)
		& \rTo^{a_{FA,\I,FB}} & (FA\tn \I)\tn FB
		& \rTo^{(FA\tn F^\tn_\I)\tn FB} & (FA\tn F\I)\tn FB
		\\
		& \raise1ex\hbox{$\begin{array}c\Rightarrow\\[-5pt]\mu_{FA,FB}\end{array}$}
		\rdTo(1,2)_{FA\tn l_{FB}}
		\ldTo(1,2)>{r_{FA}\tn FB}
		&& \begin{array}c\To\\ F^r_A\tn FB\end{array}
		& \dTo>{F^\tn_{A,\I}\tn FB}
		\\
		& FA\tn FB
		& \lTo_{F(r_A)\tn FB}
		&& F(A\tn \I)\tn FB
		\\
		& \dTo<{F^\tn_{A,B}}
		&& \Searrow F^\tn_{r_A,B}
		& \dTo>{F^\tn_{A\tn \I, B}}
		\\
		& F(A\tn B)
		&& \lTo_{F(r_A\tn B)}
		& F((A\tn \I)\tn B) 
	\end{diagram}}\]
\end{definition}
\begin{definition}
	A \emph{strong} monoidal pseudofunctor is a monoidal pseudofunctor $F$
	for which the 1-cell $F^{\tn}_{\I}$ is an equivalence, and the pseudo-natural
	transformation $F^{\tn}$ is a pseudo-natural equivalence.
\end{definition}
\begin{remark}
	One might try to define a \emph{strict} map of monoidal bicategories
	to be a monoidal strict functor $F$ such that $F^{\tn}_{I}$ and
	$F_{\tn}$ are identities. However it's easy to see that, with this
	definition, the composite of two strict maps is not necessarily strict
	(because the composite of two identities is not necessarily an identity).
	On the other hand, there certainly \emph{is} a category whose objects
	are monoidal bicategories and whose arrows preserve all the structure
	on the nose. Moreover, these strict maps are in one-one correspondence
	with monoidal strict functors defined as above; but their composition
	is subtly different.
	
	The moral of the story is that we cannot, in general, expect to be
	able to define strict maps merely as special pseudo-maps. That
	this is possible for bicategories is merely a low-dimensional
	quirk.
	
	In a similar way, composition of monoidal pseudofunctors is not
	associative on the nose; again, the existence of a category (as
	opposed to a bi- or tricategory) of bicategories and pseudofunctors
	is a low-dimensional phenomenon that cannot be expected to persist
	at higher dimensions.
\end{remark}
\begin{definition} 
	A \emph{monoidal pseudo-natural transformation} $\gamma: F\To G: \B\to\BC$
	between monoidal pseudo-functors $F$ and $G$ is a pseudo-natural transformation
	equipped with an invertible 2-cell
	\[
		\gamma^\tn_\I: \gamma_\I\o F^\tn_\I \To G^\tn_\I
	\]
	and an invertible modification with components
	\begin{diagram}
		FA\tn FB & \rTo^{F^\tn_{A,B}} & F(A\tn B)
		\\
		\dTo<{\gamma_A \tn \gamma_B} & \Swarrow\gamma^\tn_{A,B}
			& \dTo>{\gamma_{A\tn B}}
		\\
		GA\tn GB & \rTo_{G^\tn_{A,B}} & G(A\tn B) 
	\end{diagram}
	such that for all $A$, $B$, $C\in\B$, the pasting
	\[\scalebox{0.75}{\begin{diagram}
		&
		FA \tn(FB\tn FC)
		& \rTo^{FA\tn F^\tn_{B,C}} & FA\tn F(B\tn C)
		& \rTo^{F^\tn_{A,B\tn C}} & F(A\tn(B\tn C))
		\\
		&
		\dTo<{a_{FA,FB,FC}}
		&& \Swarrow F^a_{A,B,C}
		&& \dTo>{F(a_{A,B,C})}
		\\
		&
		\rnode{FAFBFC}{(FA\tn FB)\tn FC}
		& \rTo^{F^\tn_{A,B}\tn FC} & \rnode{FABFC}{F(A\tn B)\tn FC}
		& \rTo^{F^\tn_{A\tn B,C}} & F((A\tn B)\tn C)
		\\
		&
		\dTo<{(FA\tn FB)\tn\gamma_C}
		&\hbox to 0pt{\hss$\cong$\qquad}
		& \dTo<{F(A\tn B)\tn\gamma_C}>{\enskip\cong}
		\\
		&
		(FA\tn FB)\tn GC
		& \rTo_{F^\tn_{A,B}\tn GC} & F(A\tn B)\tn GC
		& \raise-1em\spright{\Swarrow \gamma^\tn_{A\tn B,C}}
		& \dTo>{\gamma_{A\tn B,C}}
		\\
		&
		\dTo<{(\gamma_A\tn\gamma_B)\tn GC}
		& \raise-.5em\spleft{\Swarrow\gamma^\tn_{A,B}\tn GC}
		& \dTo[snake=.5em]<{\gamma_{A\tn B}\tn GC}
		\\
		&
		\rnode{GAGBGC}{(GA\tn GB)\tn GC}
		& \rTo_{G^\tn_{A,B}\tn GC} & \rnode{GABGC}{G(A\tn B)\tn GC}
		& \rTo_{G^\tn_{A\tn B, C}} & G((A\tn B)\tn C)
		\nccurve[angle=180,ncurv=1]{->}{FAFBFC}{GAGBGC}
			\Bput{\gamma_A\tn(\gamma_B\tn\gamma_C)}
		\nccurve[angleA=-45,angleB=45,ncurv=.9]{->}{FABFC}{GABGC}
			\aput(.4){\gamma_{A\tn B}\tn\gamma_C}
	\end{diagram}}\]
	is equal to
	\[\scalebox{0.75}{\begin{diagram}
		&&
		\rnode{1}{FA\tn(FB\tn FC)}
		& \rTo^{FA\tn F^\tn_{B,C}} & \rnode{4}{FA\tn F(B\tn C)}
		& \rTo^{F^\tn_{A,B\tn C}} & \spright{F(A\tn(B\tn C))}
		\\
		&&
		\dTo[snake=.5em]>{FA\tn(\gamma_B\tn\gamma_C)}
		& \raise-1em\spboth{\Swarrow FA\tn\gamma^\tn_{B,C}}
		& \dTo<{FA\tn\gamma_{B\tn C}}>{\enskip\cong}
		&&& \rdTo(2,3)^{F(a_{A,B,C})}
		\\
		&&
		FA\tn(GB\tn GC)
		& \rTo_{FA\tn G^\tn_{B,C}} & FA\tn G(B\tn C)
		&& \dTo[snake=-1em]<{\gamma_{A\tn(B\tn C)}}
		\\
		&&
		\dTo[snake=-1em]>{\gamma_A\tn(GB\tn GC)}
		&\cong& \dTo[snake=1em]<{\gamma_A\tn G(B\tn C)}
		& \spboth{\Swarrow \gamma^\tn_{A,B\tn C}}
		&& \begin{array}c\Leftarrow\\[-4pt]\gamma_{\alpha_{A,B,C}}\end{array}
		& F((A\tn B)\tn C)
		\\
		&&
		\rnode{2}{GA\tn(GB\tn GC)}
		& \rTo_{GA\tn G^\tn_{B,C}} & \rnode{5}{GA\tn G(B\tn C)}
		& \rTo_{G^\tn_{A,B\tn C}} & \spright{G(A\tn(B\tn C))}
		& \ldTo(2,3)_{\gamma_{(A\tn B)\tn C}}
		\\
		& \Swarrow a_{\gamma_A,\gamma_B,\gamma_C}
		& \dTo>{a_{GA,GB,GC}}
		&& \Swarrow G^a_{A,B,C}
		&& \dTo<{G(a_{A,B,C})}
		\\
		\rnode{3}{(FA\tn FB)\tn FC} & \rTo_{(\gamma_A\tn\gamma_B)\tn\gamma_C}
		& (GA\tn GB)\tn GC
		& \rTo_{G^\tn_{A,B}\tn GC} & G(A\tn B)\tn GC
		& \rTo_{G^\tn_{A\tn B,C}} & \spright{G((A\tn B)\tn C)}
		\nccurve[angleA=-170,angleB=170]{->}12
			\bput{90}{\gamma_A\tn(\gamma_B\tn\gamma_C)}
			\Aput{\quad\cong}
		\nccurve[angleA=180,angleB=90]{->}13
			\aput(.7){a_{FA,FB,FC}}
		\nccurve[angleA=-45,angleB=45,ncurv=.9]{->}45
			\aput(.4){\gamma_A\tn\gamma_{B\tn C}}
	\end{diagram}}\]
	and for all $A\in\B$, the pasting
	\begin{diagram}
		&& \rnode{1}{FA\tn \I}
		& \rTo^{FA\tn F^\tn_\I} & \rnode{FAFI}{FA\tn F\I}
		\\
		&& \dTo<{r_{FA}} & \begin{array}c\To\\F^r_A\end{array}
		& \dTo>{F^\tn_{A,\I}}
		\\
		\rnode{2}{GA\tn \I}
		& \begin{array}c\To\\[-4pt]r_{\gamma_A}\end{array}
		& FA & \lTo_{F(r_A)} & F(A\tn \I)
		& \begin{array}c\To\\[-4pt]\gamma^\tn_{A,\I}\end{array}
		& \rnode{GAGI}{GA\tn G\I}
		\\
		&& \dTo<{\gamma_A}
		& \Nearrow\gamma_{r_A} & \dTo>{\gamma_{A\tn \I}}
		\\
		&& \rnode{3}{GA} & \lTo_{G(r_A)} & \rnode{GAI}{G(A\tn \I)}
		\nccurve[angleA=180,angleB=90]{->}12
			\Bput{\gamma_A\tn \I}
		\nccurve[angleA=-90,angleB=180]{->}23
			\Bput{r_{GA}}
		\nccurve[angleA=0,angleB=90]{->}{FAFI}{GAGI}
			\Aput{\gamma_A\tn\gamma_\I}
		\nccurve[angleA=-90,angleB=0]{->}{GAGI}{GAI}
			\Aput{G^\tn_{A,\I}}
	\end{diagram}
	is equal to
	\begin{diagram}
		FA\tn \I & \rTo^{FA\tn F^\tn_\I} & FA\tn F\I
		\\
		\dTo<{\gamma_A\tn \I}
		& \cong & \dTo<{\gamma_A\tn F\I}>{\quad\cong}
		& \rdTo^{\gamma_A\tn\gamma_\I}
		\\
		\rnode{GAI}{GA\tn \I}
		& \rTo_{GA\tn F^\tn_\I} & GA\tn F\I
		& \rTo_{GA\tn\gamma_\I} & \rnode{GAGI}{GA\tn G\I}
		\\
		&& \raise1.5em\spboth{\Uparrow GA\tn\gamma^\tn_\I}
		\\
		\dTo[snake=1.5em]<{r_{GA}}
		&& \begin{array}c\To\\[-4pt]G^r_A\end{array}
		&& \dTo[snake=1.5em]>{G^\tn_{A,\I}}
		\\
		GA && \lTo_{G(r_A)} && G(A\tn \I)
		\ncarc[arcangle=-35]{->}{GAI}{GAGI}\Bput{GA\tn G^\tn_\I}
	\end{diagram}
	and
	\begin{diagram}
		&& \rnode{1}{\I\tn FA}
		& \rTo^{F^\tn_\I\tn FA} & \rnode{FIFA}{F\I\tn FA}
		\\
		&& \dTo<{l_{FA}} & \begin{array}c\To\\F^l_A\end{array}
		& \dTo>{F^\tn_{\I,A}}
		\\
		\rnode{2}{\I\tn GA}
		& \begin{array}c\To\\[-4pt]l_{\gamma_A}\end{array}
		& FA & \lTo_{F(l_A)} & F(\I\tn A)
		& \begin{array}c\To\\[-4pt]\gamma^\tn_{\I,A}\end{array}
		& \rnode{GIGA}{G\I\tn GA}
		\\
		&& \dTo<{\gamma_A}
		& \Nearrow\gamma_{l_A} & \dTo>{\gamma_{\I\tn A}}
		\\
		&& \rnode{3}{GA} & \lTo_{G(l_A)} & \rnode{GIA}{G(\I\tn A)}
		\nccurve[angleA=180,angleB=90]{->}12
			\Bput{\I\tn \gamma_A}
		\nccurve[angleA=-90,angleB=180]{->}23
			\Bput{l_{GA}}
		\nccurve[angleA=0,angleB=90]{->}{FIFA}{GIGA}
			\Aput{\gamma_\I\tn\gamma_A}
		\nccurve[angleA=-90,angleB=0]{->}{GIGA}{GIA}
			\Aput{G^\tn_{\I,A}}
	\end{diagram}
	is equal to
	\begin{diagram}
		\I\tn FA & \rTo^{F^\tn_\I\tn FA} & F\I\tn FA
		\\
		\dTo<{\I\tn\gamma_A}
		& \cong & \dTo<{F\I\tn\gamma_A}>{\quad\cong}
		& \rdTo^{\gamma_I\tn\gamma_A}
		\\
		\rnode{IGA}{\I\tn GA}
		& \rTo_{F^\tn_\I\tn GA} & F\I\tn GA
		& \rTo_{\gamma_\I\tn GA} & \rnode{GIGA}{G\I\tn GA}
		\\
		&& \raise1.5em\spboth{\Uparrow\gamma^\tn_\I\tn GA}
		\\
		\dTo[snake=1.5em]<{l_{GA}}
		&& \begin{array}c\To\\[-4pt]G^l_A\end{array}
		&& \dTo[snake=1.5em]>{G^\tn_{\I,A}}
		\\
		GA && \lTo_{G(l_A)} && G(\I\tn A)
		\ncarc[arcangle=-35]{->}{IGA}{GIGA}\Bput{G^\tn_\I\tn GA}
	\end{diagram}
\end{definition}
\begin{definition} 
	A \emph{monoidal modification} $m:\gamma\Tto\delta: F\To G$, between monoidal
	pseudo-natural transformations $F$ and $G$, is a modification with the property that
	\[
		\begin{diagram}[h=2em]
			&& \rnode{FI}{F\I}
			\\
			& \ruTo^{F^\tn_\I}
			\\
			\I & \Swarrow\delta^\tn_\I & \dTo<{\delta_\I}
			\\
			& \rdTo_{G^\tn_\I}
			\\
			&& \rnode{GI}{G\I}
			\nccurve[angle=0]{->}{FI}{GI}\Aput{\gamma_\I}
			\Bput{\begin{array}c\Leftarrow\\[-4pt]m_\I\end{array}}
		\end{diagram}
		\hskip 5em=\quad
		\gamma^\tn_\I
	\]
	and
	\[
	\begin{diagram}
		& \rnode{FAFB}{FA\tn FB} & \rTo^{F^\tn_{A,B}} & F(A\tn B)
		\\
		\spright{\mkern-12mu\begin{array}c\Leftarrow\\[-4pt]m_A\tn m_B\end{array}}
		& \dTo[snake=-1em]>{\gamma_A\tn\gamma_B}
		& \raise1em\spboth{\Swarrow\gamma^\tn_{A,B}}
		& \dTo>{\gamma_{A\tn B}}
		\\
		& \rnode{GAGB}{GA\tn GB}
		& \rTo_{G^\tn_{A,B}}
		& G(A\tn B)
		\nccurve[angle=180,ncurv=1]{->}{FAFB}{GAGB}
			\Bput{\delta_A\tn\delta_B}
	\end{diagram}
	\quad=\quad
	\begin{diagram}
		FA\tn FB & \rTo^{F^\tn_{A,B}} & \rnode{FAB}{F(A\tn B)}
		\\
		\dTo<{\delta_A\tn\delta_B}
		& \raise-1em\spboth{\Swarrow\delta^\tn_{A,B}}
		& \dTo[snake=1em]<{\delta_{A\tn B}}
		& \spleft{\begin{array}c\Leftarrow\\[-4pt]m_{A\tn B}\end{array}}
		\\
		\rnode{GAGB}{GA\tn GB}
		& \rTo_{G^\tn_{A,B}}
		& \rnode{GAB}{G(A\tn B)}
		\nccurve[angle=0,ncurv=1]{->}{FAB}{GAB}
			\Aput{\gamma_{A\tn B}}
	\end{diagram}
	\]
\end{definition}

\section{On coherence}\label{s-coherence}
This section aims to give a concise overview of the available coherence
results for monoidal bicategories. In a later chapter (\refchapter{Coh}), we shall
prove some further coherence results, which are then used to develop
a framework that allows us to reason about pseudomonoids without becoming
bogged down in coherence conditions. First, we survey
the coherence results that do (and do not) exist in the literature.
The existing coherence theorems apply to tricategories, of which monoidal
bicategories are a special case: a tricategory with one object is essentially
a monoidal bicategory, in just the same way that a bicategory with one
object is essentially a monoidal category. We shall state the results
as they apply to monoidal bicategories, since that is the situation of
interest here.

The original coherence theorem \citep{GPS} implies that every monoidal bicategory
is monoidally biequivalent to a Gray monoid:
\begin{definition}
A \emph{Gray monoid} is a monoidal
bicategory $\B$ in which:
\begin{itemize}
\item the underlying bicategory $\B$ is a 2-category,
\item given composable pairs $f$, $g$ and $h$, $k$ of 1-cells,
	if either $f$ or $k$ is an identity then the structural 2-cell
	\[
		(f\tn h)\o(g\tn k) \To (f\o g)\tn(h\o k)
	\]
	is an identity,
\item the structural equivalences $a$, $l$ and $r$ are identities.
\end{itemize}
\end{definition}
The second condition here is the most mysterious. It means that
for every object $A\in\B$, the pseudofunctors $A\tn-$ and $-\tn A$
are 2-functors, and that furthermore the tensor product $f\tn g$ of
1-cells $f:A\to C$ and $g:B\to D$ is equal to the composite
\[
	A\tn B \rTo^{f\tn B} C\tn B \rTo^{C\tn g} C\tn D.
\]
When working in a Gray monoid, it will often be convenient to decompose
tensor products of 1-cells in this way. Then the only structural
2-cells are of the form
\begin{diagram}
	A\tn B & \rTo^{f\tn B} & C\tn B \\
	\dTo<{A\tn g} & \sim & \dTo>{C\tn g} \\
	A\tn D & \rTo_{f\tn D} & C\tn D,
\end{diagram}
or composites thereof. We shall label them merely with the symbol
$\sim$ (as above), since there is no possible ambiguity.

This line of work is extended by \citet{GurskiThesis}, who shows that
every \emph{free} monoidal bicategory
is monoidally biequivalent to a free Gray monoid on the same generators.
In particular, Gurski shows that the \emph{canonical} monoidal pseudofunctor
from the free monoidal bicategory to the free Gray monoid is a monoidal
biequivalence.
For our purposes, this theorem is not as useful as it sounds. For example,
we should like to say that the free monoidal bicategory on a pseudomonoid
object is canonically monoidally biequivalent to the free Gray monoid on
a pseudomonoid object. I imagine this is true\footnote{
	Let me be bold: I conjecture that the free monoidal bicategory on
	a pseudomonoid object is canonically monoidally biequivalent to the free
	\emph{strict 2-monoidal 2-category} on a pseudomonoid object, where
	`2-monoidal' means that the tensor product is given by a 2-functor
	rather than a general pseudofunctor.
}, but it does not follow from
Gurski's theorem. The reason is that the theorem applies only to free
constructions of a particular sort, in which the 1-cell generators are of
the form $X\to Y$, where $X$ and $Y$ are 0-cell generators. A pseudomonoid
can not be described in this form, since it requires a `tensor' 1-cell
$P:\C\tn\C\to\C$.

Gurski also provides a strictification (Grayification?) construction,
showing explicitly how to construct, for any monoidal bicategory $\B$,
a Gray monoid $\Gr(\B)$ together with explicit monoidal biequivalences
$e: \Gr(\B)\to\B$ and $f: \B\to\Gr(\B)$. Furthermore, for any strong
monoidal pseudofunctor $F:\B\to\BC$, Gurski constructs a strict functor
$\Gr F: \Gr(\B)\to\Gr(\BC)$ such that the diagrams
\[
	\begin{diagram}
		\Gr(\B) & \rTo^{\Gr F} & \Gr(\BC) \\
		\dTo<{e} && \dTo>{e} \\
		\B & \rTo_{F} & \BC
	\end{diagram}
	\mbox{\quad and\quad}
	\begin{diagram}
		\B & \rTo^{F} & \BC \\
		\dTo<{f} && \dTo>{f} \\
		\Gr(\B) & \rTo_{\Gr F} & \Gr(\BC)
	\end{diagram}
\]
commute up to monoidal equivalence. We make essential use of this
construction in Chapter~\refchapter{Language}.

\section{Braided monoidal bicategories}\label{s-braiding}
The history of attempts to define the concept of braided monoidal bicategory highlights
the difficulties inherent in even finding the correct definition. The
first definition was given by \citet{KV}, with some errors and omissions.
The errors and some omissions were pointed out by \citet{CarmodyThesis}
and \citet{HDA1} -- presumably independently, since neither cites the
other -- though \citeauthor{HDA1} go further than \citeauthor{CarmodyThesis}
and give their own definition of braiding for a Gray monoid. This
definition includes an additional axiom (our third axiom below), which they
attribute to \citet{Breen-ator}.
\Citet{HDA1} do not give any axioms relating the unit object to the
braiding. This is mentioned in their Section~5(1) as an issue that
remains to be resolved. Subsequently \citet{GeneralizedCenters}
noticed that this omission causes an error in \citeauthor{HDA1}'s
`center' construction, and showed how it could be fixed by adding
six axioms for the unit. \Citet{MonBicat} also gave a definition of
braiding with just one axiom for the unit, and despite the superficial
differences this axiomatisation is equivalent to
\citeauthor{GeneralizedCenters}'s. (It turns out that four of Crans's
six unit axioms are redundant.)

These definitions both take the unit to be strict -- in terms of
our definition below, they take the 1-cells $s_{A,\I}$ and $s_{\I,A}$
and the 2-cells $U_{A|\I}$ and $U_{\I|A}$ to be \emph{identities}.
The justification for such a restriction is presumably the reasonable
expectation that a coherence theorem for tetracategories would
show every braided monoidal bicategory to be suitably equivalent
to one with such a strict unit. Since an expectation, however
reasonable, is not a proof, we have opted to eschew the modest
simplification that such a restriction brings.
Thus the definition below is conceivably the first that
includes all the structure known to be necessary at its natural level
of strictness, though we do not claim that any new insight was needed
to formulate it. (It is not unimaginable that further axioms should yet
be found wanting, though the results of Chapter~\refchapter{Coh} provide strong
evidence that these axioms do suffice.)
\begin{definition}\label{def-braiding}
	A \defn{braiding} for a monoidal bicategory $\B$
	consists of a pseudo-natural equivalence $s$ with 1-cell components
	\[
		s_{A,B}: A\tn B \to B\tn A,
	\]
	invertible modifications $S_{-|-,-}$ and $S_{-,-|-}$ with components
	\begin{diagram}
		 A\tn (B\tn C) &\rTo^{a} & (A\tn B)\tn C
		   & \rTo^{s_{A\tn B,C}} & C\tn(A\tn B) \\
		 \dTo<{A\tn s_{B,C}} && \Arr\Swarrow{S_{A,B|C}} && \dTo>{a}\\
		 A\tn(C\tn B) & \rTo_{a} & (A\tn C)\tn B
		   & \rTo_{s_{A,C}\tn B} & (C\tn A)\tn B,
	\end{diagram}
	\begin{diagram}
		 (A\tn B)\tn C &\rTo^{a'} & A\tn(B\tn C)
		   & \rTo^{s_{A,B\tn C}} & (B\tn C)\tn A \\
		 \dTo<{s_{A,B}\tn C} && \Arr\Nearrow{S_{A|B,C}} && \dTo>{a'}\\
		 (B\tn A)\tn C & \rTo_{a'} & B\tn(A\tn C)
		   & \rTo_{B\tn s_{A,C}} & B\tn(C\tn A),
	\end{diagram}
	and invertible modifications $U_{\I|-}$ and $U_{-|\I}$ with components
	\[
	\begin{diagram}[h=2em]
		\I\tn A & \rTo^{s_{\I,A}} & A\tn \I \\
		&\raise 1em\hbox{$\mathop\Leftarrow\limits_{U_{\I|A}}$} \rdTo(1,2)_{l_A} \ldTo(1,2)_{r_A}\\
		&A
	\end{diagram}
	\hskip 3em
	\begin{diagram}[h=2em]
		A\tn \I & \rTo^{s_{A,\I}} & \I\tn A \\
		&\raise 1em\hbox{$\mathop\Rightarrow\limits_{U_{A|\I}}$} \rdTo(1,2)_{r_A} \ldTo(1,2)_{l_A}\\
		&A,
	\end{diagram}
	\]
	subject to various axioms. We have chosen to state these axioms in the
	Gray monoid setting, since the general versions are horribly unwieldy.
	This can be justified by the observation that the data
	above may be transported across a monoidal biequivalence: given a
	monoidal biequivalence (preferably a monoidal pseudo-adjoint biequivalence)
	\[
		\B \pile{\rTo^{F}\\\lTo_{G}} \BC
	\]
	between monoidal bicategories, braiding data for $\B$, say, induces
	braiding data for $\BC$ in a canonical way. For example, the 1-cell
	$s_{A,B}: A\tn B \to B\tn A$ in $\BC$ is defined as
	\[
		A\tn B \approx GFA\tn GFB \approx G(FA\tn FB)
			\rTo_{G(s_{FA,FB})} G(FB\tn FA) \approx GFB\tn GFA \approx B\tn A.
	\]
	It is easy, though tedious, to check that all the braiding
	data may be transported in this way. Given this, and the axioms below,
	we may define a braided monoidal bicategory to be a monoidal bicategory
	$\B$ equipped with braiding data, such that these data satisfy the axioms
	when transported to $\Gr(\B)$. This is unsatisfyingly indirect, and
	complicates the task of verifying these axioms of any particular monoidal
	bicategory, but seems preferable to the alternative. One should
	also verify that, given two Gray monoids $\B$ and $\BC$ and a monoidal
	biequivalence as above, the braiding data for $\B$ satisfy the axioms
	if and only if the transported data satisfy the axioms in $\BC$. I confess
	I have not explicitly done this, though it seems unlikely to be
	false.
	
	In a more pragmatic (if less precise) vein, one could regard these axioms as being mere
	abbreviations of the general versions. In practice it is clear how to write down
	the equivalent axioms for a general monoidal bicategory, by inserting structural 2-cells
	where necessary.
	
	The axioms follow. There are four axioms
	that relate the $S$ modifications to each other:
	\[
	\hskip-1em
	\begin{diagram}[h=4em,labelstyle=\scriptstyle]
		A\tn B\tn C\tn D & \rTo^{s_{A\tn B\tn C,D}} & D\tn A\tn B\tn C \\
		\dTo<{A\tn B\tn s_{C,D}} & \rdTo(2,2)[hug]_{A\tn s_{B\tn C,D}}
			\raise 1.5em\hbox to 0pt{$\Downarrow\scriptstyle S_{A,B\tn C|D}$\hss}
			\raise-2em\hbox to 0pt{\hss$\Swarrow\scriptstyle A\tn S_{B,C|D}$}
			& \uTo>{s_{A,D}\tn B\tn C} \\
		A\tn B\tn D\tn C &\rTo_{A\tn s_{B,D}\tn C} & A\tn D\tn B\tn C
	\end{diagram}
	=
	\begin{diagram}[h=4em,labelstyle=\scriptstyle]
		A\tn B\tn C\tn D & \rTo^{s_{A\tn B\tn C,D}} & D\tn A\tn B\tn C \\
		\dTo<{A\tn B\tn s_{C,D}} & \ruTo(2,2)[hug]_{s_{A\tn B,D}\tn C}
			\raise 1.5em\hbox to 0pt{\hss$\Downarrow\scriptstyle S_{A\tn B,C|D}$}
			\raise-2em\hbox to 0pt{$\Searrow\scriptstyle S_{A,B|D}\tn C$\hss}
			& \uTo>{s_{A,D}\tn B\tn C} \\
		A\tn B\tn D\tn C &\rTo_{A\tn s_{B,D}\tn C} & A\tn D\tn B\tn C
	\end{diagram}
	\]
	\[
	\hskip-1em
	\begin{diagram}[h=4em,labelstyle=\scriptstyle]
		A\tn B\tn C\tn D & \rTo^{s_{A\,B\tn C\tn D}} & B\tn C\tn D\tn A \\
		\dTo<{s_{A,B}\tn C\tn D} & \rdTo(2,2)[hug]_{s_{A,B\tn C}\tn D}
			\raise 1.5em\hbox to 0pt{$\Uparrow\scriptstyle S_{A|B\tn C,D}$\hss}
			\raise-2em\hbox to 0pt{\hss$\Nearrow\scriptstyle S_{A|B,C}\tn D$}
			& \uTo>{B\tn C\tn s_{A,D}} \\
		B\tn A\tn C\tn D &\rTo_{B\tn s_{A,C}\tn D} & B\tn C\tn A\tn D
	\end{diagram}
	=
	\begin{diagram}[h=4em,labelstyle=\scriptstyle]
		A\tn B\tn C\tn D & \rTo^{s_{A\,B\tn C\tn D}} & B\tn C\tn D\tn A \\
		\dTo<{s_{A,B}\tn C\tn D} & \ruTo(2,2)[hug]_{B\tn s_{A,C\tn D}}
			\raise 1.5em\hbox to 0pt{\hss$\Uparrow\scriptstyle S_{A|B,C\tn D}$}
			\raise-2em\hbox to 0pt{$\Nwarrow\scriptstyle B\tn S_{A|C,D}$\hss}
			& \uTo>{B\tn C\tn s_{A,D}} \\
		B\tn A\tn C\tn D &\rTo_{B\tn s_{A,C}\tn D} & B\tn C\tn A\tn D
	\end{diagram}
	\]
	\[\hskip-1em\begin{array}{l}
	\hskip-2em
	\begin{diagram}[hug,labelstyle=\scriptstyle,w=4em,tight]
		A\tn B\tn C\tn D &&& \rTo^{\displaystyle A\tn s_{B,C\tn D}}
			&&& A\tn C\tn D\tn B \\
		&\rdTo^{A\tn s_{B,C}\tn D} && \Uparrow{\scriptstyle A\tn S_{B|C,D}}
			&& \ruTo(4,2)^{A\tn C\tn s_{B,D}} \ldTo(2,2)_{s_{A,C}\tn B\tn D} \\
		\dTo<{\displaystyle s_{A\tn B,C}\tn D}
			& \hbox to 0pt{\hss$\mathop\Rightarrow\limits_{S_{A,B|C}\tn D}$}
			& A\tn C\tn B\tn D & \sim & C\tn A\tn D\tn B
			& \hbox to 0pt{$\mathop\Rightarrow\limits_{S_{A|C,D}\tn B}$\hss}
			& \dTo>{\displaystyle s_{A,C\tn D}\tn B} \\
		&\ldTo_{s_{A,C}\tn B\tn D} && \ruTo(4,2)_{C\tn A\tn s_{B,D}}
			& \Uparrow{\scriptstyle C \tn S_{A,B|D}}
			& \rdTo^{C \tn s_{A,D}\tn B} \\
		C\tn A\tn B\tn D &&& \rTo_{C \tn s_{A\tn B,D}} &&& C\tn D\tn A\tn B
	\end{diagram}
	\\[8em]
	\multicolumn 1r{
	= \qquad \begin{diagram}[h=4em]
		A\tn B\tn C\tn D & \rTo^{A\tn s_{B,C\tn D}} & A\tn C\tn D\tn B \\
		\dTo<{s_{A\tn B,C}\tn D}
			& \rdTo(2,2)[hug]_{s_{A\tn B,C\tn D}}
			\raise 1.5em\hbox to 0pt{$\Nearrow\scriptstyle S_{A,B|C\tn D}$\hss}
			\raise-2em\hbox to 0pt{\hss$\Nearrow\scriptstyle S_{A\tn B|C,D}$}
			& \dTo>{s_{A,C\tn D}\tn B} \\
			C\tn A\tn B\tn D & \rTo_{C \tn s_{A\tn B,D}} & C\tn D\tn A\tn B
	\end{diagram}
	}
	\end{array}\]
	\[
		\begin{diagram}[hug,s=2.5em,tight]
			A\tn B\tn C && \rTo^{s_{A\tn B,C}} && C\tn A\tn B \\
			& \rdTo_{A\tn s_{B,C}}
				&\raise 1em\hbox{$\Arr\Downarrow{\scriptstyle S_{A,B|C}}$}
				& \ruTo_{s_{A,C}\tn B} \\
			\dTo<{s_{A,B\tn C}} && A\tn C\tn B & \Arr\Swarrow{\scriptstyle S_{A|C,B}}
				& \dTo>{C\tn s_{A,B}} \\
			&\Arr\Swarrow{s_{A,s_{B,C}}} && \rdTo_{s_{A,C\tn B}} \\
			B\tn C\tn A&&\rTo_{s_{B,C}\tn A} && C\tn B\tn A
		\end{diagram}
		\quad=\quad
		\begin{diagram}[hug,s=2.5em,tight]
			A\tn B\tn C && \rTo^{s_{A\tn B,C}} && C\tn A\tn B \\
			& \rdTo_{s_{A,B}\tn C} &&\Swarrow{s_{s_{A,B},C}^{-1}} \\
			\dTo<{s_{A,B\tn C}}
				&\hbox to 0pt{\hss$\mathop\Leftarrow\limits_{S_{A|B,C}}$}
				& B\tn A\tn C && \dTo>{C\tn s_{A,B}} \\
			&\ldTo^{B\tn s_{A,C}}
				&\mathop\Leftarrow\limits_{S_{B,A|C}}
				& \rdTo_{s_{B\tn A,C}} \\
			B\tn C\tn A&&\rTo_{s_{B,C}\tn A} && C\tn B\tn A
		\end{diagram}
	\]
	and two axioms that relate the $U$ and $S$ modifications:
	the 2-cells pictured below should each be equal to the identity on $s_{A,B}$:
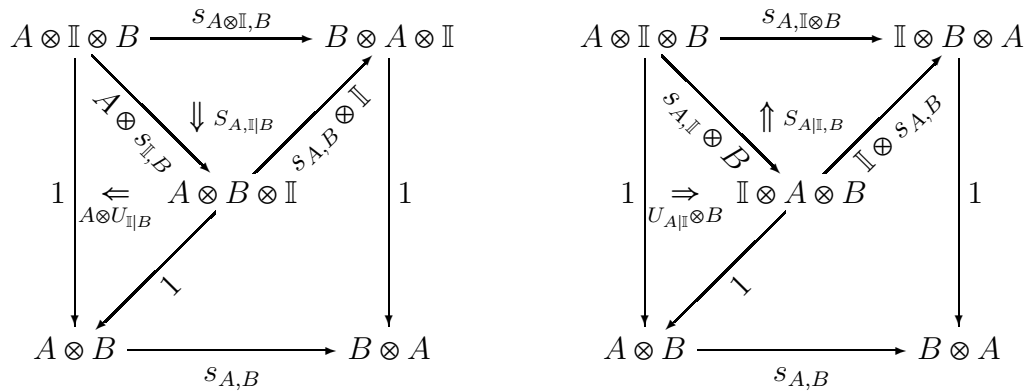
	\[\begin{array}{c@{\hskip4em}c}
		\begin{diagram}[hug,s=2.5em]
			A\tn \I\tn B && \rTo^{s_{A\tn \I,B}} && B\tn A\tn \I \\
			&\rdTo_{A\tn s_{\I,B}} & \Downarrow{\scriptstyle S_{A,\I|B}}
				& \ruTo_{s_{A,B}\tn \I} \\
			\dTo<1 & \hbox to 0pt{\hss$\mathop\Leftarrow\limits_{A\tn U_{\I|B}}$}
				& A\tn B\tn \I && \dTo>1 \\
			&\ldTo_{1} \\
			A\tn B && \rTo_{s_{A,B}} && B\tn A
		\end{diagram}
		&
		\begin{diagram}[hug,s=2.5em]
			A\tn \I\tn B && \rTo^{s_{A,\I\tn B}} && \I\tn B\tn A \\
			&\rdTo_{s_{A,\I}\tn B} & \Uparrow{\scriptstyle S_{A|\I,B}}
				& \ruTo_{\I\tn s_{A,B}} \\
			\dTo<1 & \hbox to 0pt{\hss$\mathop\Rightarrow\limits_{U_{A|\I}\tn B}$}
				& \I\tn A\tn B && \dTo>1 \\
			&\ldTo_{1} \\
			A\tn B && \rTo_{s_{A,B}} && B\tn A
		\end{diagram}
	\end{array}\]
\end{definition}
\begin{definition}
	A \emph{braided monoidal bicategory} is a monoidal bicategory
	equipped with a braiding.
\end{definition}

\subsection{The duality of the definition}
Since $s$ is a pseudo-natural equivalence, it has an adjoint
equivalence-inverse $s'$. We shall assume that such an $s'$
has been chosen: by definition it has 1-cell components
\[
	s'_{A,B}: B\tn A \to A\tn B,
\]
and by Prop.~\chref{Bicats}{prop-adj-1} each 2-cell component $s'_{f,g}$
is the inverse of the mate of $s_{f,g}$. Now we may define
a pseudo-natural equivalence $s^*$ with 1-cell components
$s^*_{A,B} = s'_{A,B}$ and 2-cell components $s^{*}_{f,g} = s'_{g,f}$.
The duality is expressed by the following:
\begin{propn}
	The pseudo-natural transformation $s^{*}$ can be made into
	a braiding in the following way.
	\begin{itemize}
		\item $S^{*}_{A|B,C}$ is defined to be the mate of $S_{B,C|A}$
			with respect to the adjoint equivalences
			$(s_{A,C}\tn B)\o a\o(A\tn s_{B,C})$
			and $a\o(s_{A\tn B,C})\o a$,
		\item $S^{*}_{A,B|C}$ is defined to be the mate of $S_{C|A,B}$
			with respect to the adjoint equivalences
			$a'\o(s_{A,B\tn C})\o a'$ and $(B\tn s_{A,C})\o a'\o (s_{A,B}\tn C)$;
			Note that, by Prop.~\chref{Bicats}{prop-adjeq-mate-dual}, the
			left and right mates are equal in this case and the previous one.
		\item $U^{*}_{\I|A}$ is defined to be the right mate of $U_{A|\I}$
			with respect to the adjoint equivalences $s_{A,\I}$ and $1_{A}$,
		\item $U^{*}_{A|\I}$ is defined to be the left mate of $U_{\I|A}$
			with respect to the adjoint equivalences $s_{\I,A}$ and $1_{A}$.
	\end{itemize}
\end{propn}
\begin{proof}
	The first two axioms for $S$ are duals of each other: taking mates in
	one of them yields the other for $S^{*}$. The third axiom is self-dual,
	in that taking mates gives the corresponding equation for $S^{*}$.
	The fourth axiom is also self-dual, using the fact that the mate of
	$s_{A,s_{B,C}}$ is the inverse of ${s^*_{s^*_{C,B},A}}$. Finally, the
	two unit axioms are duals of each other.
\end{proof}

This symmetry can sometimes spare us a certain amount of repetition,
in Section~\chref{Psmon}{s-braided-facts}, for example.

\subsection{The unit axioms}
The first thing to say about the unit axioms is that we have here
another instance of the phenomenon discussed in Remark~\ref{rem-defining-L}:
the unit axioms show that the 2-cells $U_{\I|A}$ and $U_{A,\I}$ are
definable in terms of the other data. However, the $U$ cells defined
in this way do not themselves necessarily satisfy the unit axioms!
So these cells are redundant data in a sense, but if we were to
eliminate them, we should instead have to impose rather unnatural-looking
axioms involving $S_{A,\I|B}$ and $S_{A|\I,B}$.

One could write down several other natural conditions on the unit
cells. These turn out to be derivable from the two conditions we
have. In particular:
\begin{propn}\label{prop-braiding-unit-1}
	In any braided Gray monoid, the 2-cells
	\[
		\begin{diagram}[hug,s=4em]
			\I\tn A\tn B & \rTo^{\I\tn s_{A,B}} & \I\tn B\tn A \\
			\dTo<{s_{\I\tn A,B}} & \ldTo_{s_{\I,B}\tn A}
				\raise 2em\hbox to 0pt{\hss$\mathop\Rightarrow\limits_{S_{\I,A|B}}$}
				\raise-2em\hbox to 0pt{$\mathop\Rightarrow\limits_{U_{\I|B}\tn A}$\hss}
				& \dTo>{1}\\
			B\tn \I\tn A & \rTo_{1} & B\tn A
		\end{diagram}
		\qquad\mbox{and}\qquad
		\begin{diagram}[hug,s=4em]
			A\tn B\tn \I & \rTo^{s_{A,B}\tn \I} & B\tn A\tn \I \\
			\dTo<{s_{A,B\tn \I}} & \ldTo_{B\tn s_{A,\I}}
				\raise 2em\hbox to 0pt{\hss$\mathop\Leftarrow\limits_{S_{A|B,\I}}$}
				\raise-2em\hbox to 0pt{$\mathop\Leftarrow\limits_{B\tn U_{A|\I}}$\hss}
				& \dTo>{1}\\
			B\tn \I\tn A & \rTo_{1} & B\tn A
		\end{diagram}
	\]
	are both identities.
\end{propn}
\begin{proof}
	By duality, it suffices to prove one of the two: we'll prove
	the second one.
	If we set $B=\I$ in our first axiom for $S$, we obtain
	\[
	\hskip-1em
	\begin{diagram}[h=4em,labelstyle=\scriptstyle]
		A\tn \I\tn C\tn D & \rTo^{s_{A\tn \I\tn C,D}} & D\tn A\tn \I\tn C \\
		\dTo<{A\tn \I\tn s_{C,D}} & \rdTo(2,2)[hug]_{A\tn s_{\I\tn C,D}}
			\raise 1.5em\hbox to 0pt{$\Downarrow\scriptstyle S_{A,\I\tn C|D}$\hss}
			\raise-2em\hbox to 0pt{\hss$\Swarrow\scriptstyle A\tn S_{\I,C|D}$}
			& \uTo>{s_{A,D}\tn \I\tn C} \\
		A\tn \I\tn D\tn C &\rTo_{A\tn s_{\I,D}\tn C} & A\tn D\tn \I\tn C
	\end{diagram}
	=
	\begin{diagram}[h=4em,labelstyle=\scriptstyle]
		A\tn \I\tn C\tn D & \rTo^{s_{A\tn \I\tn C,D}} & D\tn A\tn \I\tn C \\
		\dTo<{A\tn \I\tn s_{C,D}} & \ruTo(2,2)[hug]_{s_{A\tn \I,D}\tn C}
			\raise 1.5em\hbox to 0pt{\hss$\Downarrow\scriptstyle S_{A\tn \I,C|D}$}
			\raise-2em\hbox to 0pt{$\Searrow\scriptstyle S_{A,\I|D}\tn C$\hss}
			& \uTo>{s_{A,D}\tn \I\tn C} \\
		A\tn \I\tn D\tn C &\rTo_{A\tn s_{\I,D}\tn C} & A\tn D\tn \I\tn C
	\end{diagram}
	\]
	Cancelling the invertible 2-cell $S_{A,\I\tn C|D}=S_{A\tn\I,C|D}$, we have
	\[
	\hskip-1em
	\begin{diagram}[w=5em,labelstyle=\scriptstyle,tight]
		A\tn \I\tn C\tn D & & D\tn A\tn \I\tn C \\
		\dTo<{A\tn \I\tn s_{C,D}} & \rdTo(2,2)[hug]^{A\tn s_{\I\tn C,D}}
			\raise-2em\hbox to 0pt{\hss$\Swarrow\scriptstyle A\tn S_{\I,C|D}$}
			& \uTo>{s_{A,D}\tn \I\tn C} \\
		A\tn \I\tn D\tn C &\rTo_{A\tn s_{\I,D}\tn C} & A\tn D\tn \I\tn C
	\end{diagram}
	=
	\begin{diagram}[w=5em,labelstyle=\scriptstyle,tight]
		A\tn \I\tn C\tn D && D\tn A\tn \I\tn C \\
		\dTo<{A\tn \I\tn s_{C,D}} & \ruTo(2,2)[hug]^{s_{A\tn \I,D}\tn C}
			\raise-2em\hbox to 0pt{$\Searrow\scriptstyle S_{A,\I|D}\tn C$\hss}
			& \uTo>{s_{A,D}\tn \I\tn C} \\
		A\tn \I\tn D\tn C &\rTo_{A\tn s_{\I,D}\tn C} & A\tn D\tn \I\tn C
	\end{diagram}
	\]
	Thus
	\begin{diagram}[w=5em,labelstyle=\scriptstyle,tight]
		A\tn \I\tn C\tn D & & D\tn A\tn \I\tn C \\
		\dTo<{A\tn \I\tn s_{C,D}} & \rdTo(2,2)[hug]^{A\tn s_{\I\tn C,D}}
			\raise-1em\hbox to 0pt{\hss$\Swarrow\scriptstyle A\tn S_{\I,C|D}$}
			& \uTo[snake=1em]<{s_{A,D}\tn \I\tn C}
			&\rdTo^{1}
			\\
		A\tn \I\tn D\tn C &\rTo_{A\tn s_{\I,D}\tn C} & A\tn D\tn \I\tn C
		& & D\tn A\tn C \\
		& \rdTo_{1} \raise1em\rlap{$\Swarrow\scriptstyle  A\tn U_{\I|D}\tn C$}
			& \dTo>1 & \ruTo_{s_{A,D}\tn C} \\
		& & A\tn D\tn C
	\end{diagram}
	is equal to
	\begin{diagram}[w=5em,labelstyle=\scriptstyle,tight]
		A\tn \I\tn C\tn D && D\tn A\tn \I\tn C \\
		\dTo<{A\tn \I\tn s_{C,D}} & \ruTo(2,2)[hug]^{s_{A\tn \I,D}\tn C}
			\raise-1.5em\hbox to 0pt{$\Searrow\scriptstyle S_{A,\I|D}\tn C$\hss}
			& \uTo<{s_{A,D}\tn \I\tn C}
			&\rdTo^{1}
			\\
		A\tn \I\tn D\tn C &\rTo_{A\tn s_{\I,D}\tn C} & A\tn D\tn \I\tn C
		& & D\tn A\tn C \\
		& \rdTo_{1} \raise1em\rlap{$\Swarrow\scriptstyle  A\tn U_{\I|D}\tn C$}
			& \dTo>1 & \ruTo_{s_{A,D}\tn C} \\
		& & A\tn D\tn C
	\end{diagram}
	which, by our first unit axiom, is the identity.
	Cancelling the equivalence $s_{A,D}\tn C$, we find that
	\begin{diagram}[h=3.5em,labelstyle=\scriptstyle]
		A\tn \I\tn C\tn D \\
		\dTo<{A\tn \I\tn s_{C,D}} & \rdTo(2,2)[hug]^{A\tn s_{\I\tn C,D}}
			\raise-1.5em\hbox to 0pt{\hss$\Swarrow\scriptstyle A\tn S_{\I,C|D}$}
			\\
		A\tn \I\tn D\tn C &\rTo_{A\tn s_{\I,D}\tn C} & A\tn D\tn \I\tn C \\
		& \rdTo_{1} \raise1em\rlap{$\Swarrow\scriptstyle A\tn U_{\I|D}\tn C$}
			& \dTo>1 \\
		& & A\tn D\tn C
	\end{diagram}
	is the identity, and setting $A=\I$ yields the claim.
\end{proof}

\begin{propn}\label{prop-braiding-unit-2}
	In any braided Gray monoid, the equations
		\[
		\begin{diagram}[hug,s=4em]
			A\tn B\tn \I & \rTo^{A\tn s_{B,\I}} & A\tn \I\tn B \\
			\dTo<1 & \rdTo(2,2)^{s_{A\tn B,\I}}
				\raise 2em\hbox to 0pt{$\Arr\Nearrow{\scriptstyle S_{A,B|\I}}$\hss}
				\raise-1em\hbox to 0pt{\hss$\mathop\Rightarrow\limits_{U_{A\tn B|\I}}$}
				& \dTo>{s_{A,\I}\tn B} \\
			A\tn B & \lTo_{1} & \I\tn A\tn B
		\end{diagram}
		\hskip3em=\hskip-1em
		\begin{diagram}[hug,s=4em]
			A\tn B\tn \I & \rTo^{A\tn s_{B,\I}} & A\tn \I\tn B \\
			\dTo<1 & \ldTo(2,2)^{1}
				\raise-1em\hbox to 0pt{$\Arr\Searrow{\scriptstyle U_{A|\I}\tn B}$\hss}
				\raise 2em\hbox to 0pt{\hss$\mathop\Rightarrow\limits_{A\tn U_{B|\I}}$}
				& \dTo>{s_{A,\I}\tn B} \\
			A\tn B & \lTo_{1} & \I\tn A\tn B
		\end{diagram}
	\]
	\[
		\begin{diagram}[hug,s=4em]
			\I\tn A\tn B & \rTo^{s_{\I,A}\tn B} & A\tn \I\tn B \\
			\dTo<1 & \rdTo(2,2)^{s_{A\tn B,\I}}
				\raise 2em\hbox to 0pt{$\Arr\Swarrow{\scriptstyle S_{\I|A,B}}$\hss}
				\raise-1em\hbox to 0pt{\hss$\mathop\Leftarrow\limits_{U_{\I|A\tn B}}$}
				& \dTo>{A\tn s_{\I,B}} \\
			A\tn B & \lTo_{1} & A\tn B\tn \I
		\end{diagram}
		\hskip3em=\hskip-1em
		\begin{diagram}[hug,s=4em]
			\I\tn A\tn B & \rTo^{s_{\I,A}\tn B} & A\tn \I\tn B \\
			\dTo<1 & \ldTo(2,2)^{1}
				\raise-1em\hbox to 0pt{$\Arr\Nwarrow{\scriptstyle A\tn U_{\I|B}}$\hss}
				\raise 2em\hbox to 0pt{\hss$\mathop\Leftarrow\limits_{U_{\I|A}\tn B}$}
				& \dTo>{A\tn s_{\I,B}} \\
			A\tn B & \lTo_{1} & A\tn B\tn \I
		\end{diagram}
	\]
	hold.
\end{propn}
\begin{proof}[Proof sketch]
	This is proved in a similar way to the preceding proposition, though the
	argument is lengthier. Start with the third axiom, setting $C=\I$.
\end{proof}	
\end{thesischapter}
    
\documentclass{robinthesis}
\tikzstyle{every picture}+=[remember picture, >=latex]

\begin{thesischapter}{Coh}{Some Coherence Results}
In this chapter, we prove some new coherence results for
Gray monoids. These results pave the way for the following
chapter, which describes a simple but apparently novel
technique for stating definitions and performing calculations
in a monoidal bicategory.

At the time of writing, the sum of human knowledge about
coherence for monoidal bicategories -- and, more generally, for
tricategories -- is contained in the PhD dissertation of
\citet{GurskiThesis}, which builds on the pioneering work
of \citet{GPS}. Nothing has yet been written about coherence
for \emph{braided} monoidal bicategories. It is gradually
becoming clear that tricategories, and other such higher-dimensional
structures, enjoy more coherence than they are generally credited with.
One manifestation of this is that many diagrams of 2-cells commute
in any Gray monoid. To be precise, we shall prove:
\begin{thm}\label{thm-coh}
	In the free Gray monoid generated by a multigraph,
	every diagram of 2-cells commutes.
\end{thm}
This strengthens Gurski's Theorem~10.2.2, which (specialised to
monoidal bicategories) addresses the free Gray monoid generated
by a mere graph, rather than a multigraph. It is apparently possible
to strengthen it still further, but the statement here is
adequate for our present purposes.

Our second result encompasses the braided case:
\begin{thm}\label{thm-coh-braiding}
	In the free braided Gray monoid generated by
	a multigraph, every diagram of 2-cells whose
	source and target are `positive' 1-cells commutes.
\end{thm}
Here `positive' means that the 1-cell is built without using
$s'$, the equivalence inverse of $s$. This restriction is made
because it significantly simplifies the proof, yet remains sufficient
for our applications in this work. The result does appear in fact
to hold for all 2-cells, and we conjecture that a more complex
application of the techniques of this chapter suffices to prove
the general version.

In both cases, the method of proof is essentially that used
by \citet{MacLane} in his coherence theorem for monoidal categories,
using term rewriting. This syntactic technique seems
to permit a finer analysis of coherence than the
powerful but blunt semantic methods employed by \citet{GPS}
and refined by \citet{GurskiThesis}.

\section{Non-braided Case}
A multigraph consists of the data for a multicategory, but
without the identity arrows or the composition operation.
More formally,
	a multigraph $G$ consists of:
	\begin{itemize}
		\item a set $G_{0}$ of objects,
		\item for each finite `source' sequence $A_{1}, \dots, A_{n}$
			of objects and `target' object $B$, a set $G(A_{1},\dots,A_{n}; B)$
			of multiarrows.
	\end{itemize}
A morphism of multigraphs
$f: G\to H$ consists of an object function $f_{0}: G_{0}\to H_{0}$ 
together with a family of functions
\[
	f_{A_{1},\dots,A_{n};B}: G(A_{1},\dots,A_{n};B) \to H(f_{0}(A_{1}),\dots,f_{0}(A_{n});f_{0}(B)).
\]
The category $\cat{GrayMon}_\mathrm{str}$ of Gray monoids and \emph{strictly} structure-preserving
maps has an obvious forgetful functor to the category of multigraphs, and clearly this forgetful functor can be furnished with a left adjoint $F$ by the usual syntactic construction. Naively, then, the objects,
1-cells and 2-cells of the Gray category $F(G)$ are formal expressions built from the objects and
multiarrows of $G$, quotiented by the smallest equivalence relation that makes the resulting
structure into a Gray monoid, i.e. the equivalence relation generated by the axioms that define
Gray monoid. But of course this description can be simplified, as follows.

Since the objects of a Gray monoid form a monoid under tensor, an object of $F(G)$ can be represented
by a finite sequence $\tuple{X_{1},\dots,X_{n}}$ of objects of $G$;
the tensor of two objects is their concatenation as sequences, and
the unit object is represented by the empty sequence.
Turning next to the 1-cells, notice that since the tensor of $f: A\to B$ and $g: C\to D$
is equal to $(B\tn g)(f\tn C)$, the tensor of 1-cells can be expressed in terms of
the tensor of a 1-cell with an object, and composition. Therefore we need
not regard the tensor of two 1-cells as a primitive operation, provided that we can
take the tensor of a 1-cell with an object.
Furthermore, every 1-cell is equal to a finite composite of \emph{multiarrow 1-cells},
where a multiarrow 1-cell is obtained by tensoring a multiarrow of $G$ with objects, on
either side. More formally, a multiarrow 1-cell
\[
	\langle X_{1},\dots,X_{m}\rangle \to \langle Y_{1},\dots,Y_{n}\rangle
\]
consists of a natural number $1\leq j\leq n$ and a multiarrow
\[
	f \in G(X_{j},\dots,X_{j+m-n}; Y_{j})
\]
such that $\langle X_{1}, \dots, X_{j-1}\rangle = \langle Y_{1}, \dots, Y_{j-1}\rangle$
and $\langle X_{j+m-n+1}, \dots, X_{m}\rangle = \langle Y_{j+1},\dots, Y_{n}\rangle$.
This might be pictured as follows:
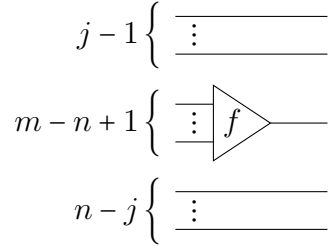
\[\begin{tikzpicture}
	\newcommand\wires[1]{
		\begin{scope}
			\pgftransformscale{0.25}
			\node[left] at (0,0) {
				$#1
				\left\{\hbox{\vrule height 8pt depth 8pt width 0pt}\right.$
			} ;
			\draw (0,1) -- +(8,0) ;
			\node at (1,0) {\vdots} ;
			\draw (0,-1) -- +(8,0) ;
		\end{scope}
	}
	\newcommand\multiarrow[2]{
		\begin{scope}
			\pgftransformscale{0.25}
			\node[left] at (0,0) {
				$#1
				\left\{\hbox{\vrule height 8pt depth 8pt width 0pt}\right.$
			} ;
			\draw (0,1) -- +(2,0) ;
			\node at (1,0) {\vdots} ;
			\draw (0,-1) -- +(2,0) ;
			\draw (2,-2) -- ++(0,4) -- ++(3,-2) -- cycle ;
				\node at (3,0) {$#2$} ;
			\draw (5,0) -- ++(3,0) ;
		\end{scope}
	}
	\matrix {
		\wires{j-1} \\ \multiarrow{m-n+1}{f} \\ \wires{n-j} \\
	} ;
\end{tikzpicture}\]
Similarly, the tensor of 2-cells can be expressed in terms of
the tensor of a 1-cell with an object, and horizontal composition.
In turn, horizontal composition can be expressed in terms of whiskering
and vertical composition. Therefore our 2-cells may be built from
\emph{interchange cells}, where an interchange cell is either
a \emph{positive interchange cell} $b_{\vec X,f,\vec Y,g,\vec Z}$ like this:
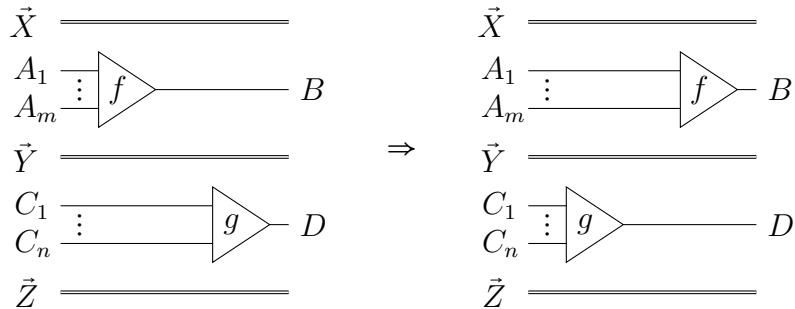
\[
	\newcommand\tube[1]{
		\draw[double] node[left] {$#1\phantom{\scriptstyle1}$} (0,0) -- +(12,0) ;
	}
	\newcommand\multiarrow[5][2]{
		\draw (0, 1) node[left] {$#2_{1}$} -- +(#1,0) ;
		\node at (1,0) {\vdots} ;
		\draw (0,-1) node[left] {$#2_{\rlap{$\scriptstyle #3$}\phantom{1}}$} -- +(#1,0) ;
		\draw (#1,-2) -- ++(0,4) -- ++(3,-2) -- cycle ;
			\node at (1+#1,0) {$#4$} ;
		\draw (3+#1,0) -- (12,0) node[right] {$#5$} ;
	}
	\tikzstyle{every picture}+=[baseline]
	\tikzstyle{every cell}	 +=[scale=0.25]
	\begin{tikzpicture}
	\matrix {
		\tube{\vec X} \\
		\multiarrow{A}{m}{f}{B} \\
		\tube{\vec Y} \\
		\multiarrow[8]{C}{n}{g}{D} \\
		\tube{\vec Z} \\
	} ;
	\end{tikzpicture}
	\quad \To \quad
	\begin{tikzpicture}
	\matrix {
		\tube{\vec X} \\
		\multiarrow[8]{A}{m}{f}{B} \\
		\tube{\vec Y} \\
		\multiarrow{C}{n}{g}{D} \\
		\tube{\vec Z} \\
	} ;
	\end{tikzpicture}
\]
or a \emph{negative interchange cell} $b_{\vec X,f,\vec Y,g,\vec Z}^{-1}$ in
the other direction. In case it is not clear from the picture, a positive
interchange cell is a 2-cell
\[
  b_{\vec X,f,\vec Y,g,\vec Z}:
	(\vec X\tn B\tn\vec Y\tn g\tn\vec Z)
	\cdot
	(\vec X\tn f\tn\vec Y\tn\vec C\tn\vec Z)
	\To
	(\vec X\tn f\tn\vec Y\tn D\tn\vec Z),
	\cdot
	(\vec X\tn \vec A\tn\vec Y\tn g\tn\vec Z)
\]
where $f \in G(\vec A;B)$ and $g\in G(\vec C;D)$.
In context, we shall omit some of the subscripts and write just $b_{f,g}$.
Now a \emph{basic 2-cell} is the result of whiskering an interchange cell on both
sides, by arbitrary 1-cells, and every 2-cell is a finite vertical composite
of basic 2-cells.

To summarise, a 1-cell $f$ may be canonically represented as a finite sequence of multiarrow cells.
There is a basic 2-cell $f\To g$ whenever $g$ can be obtained from $f$ by interchanging
two adjacent (but non-interfering) multiarrows. (In particular, $f$ and $g$ contain the
same number of multiarrow cells.)

As explained above, we plan to use Mac Lane's technique for proving coherence.
So the next step is to define a rewriting system on 1-cells,
where each rewrite rule corresponds to a basic 2-cell. We shall
show that this rewriting system is strongly normalising and locally confluent,
hence that every 1-cell is isomorphic to one in normal form. Now, local confluence
means that, for every 1-cell $f$, if we have rewriting steps $\gamma: f \To f_{1}$
and $\delta: f\To f_{2}$ then there are sequences of rewrites $\gamma^{*}: f_{1}\To g$
and $\delta^{*}: f_{2}\To g$ for some 1-cell $g$. The final requirement is to show
that the corresponding diagram of 2-cells commutes:
\begin{diagram}
	f & \rTo^{\gamma} & f_{1} \\
	\dTo<\delta && \dTo>{\gamma^{*}} \\
	f_{2} & \rTo_{\delta^{*}} & g.
\end{diagram}
It will be convenient, of course, to prove this at the same time as we establish
local confluence. This extended confluence property does not seem to have a standard
name: we shall call it \emph{local coherent confluence}. Notice that, in a strongly
normalising system, local coherent confluence implies \emph{(global) coherent confluence}:
i.e. given any \emph{sequences} of rewrites $\gamma: f \To f_{1}$
and $\delta: f\To f_{2}$, there are sequences of rewrites $\gamma^{*}: f_{1}\To g$
and $\delta^{*}: f_{2}\To g$ for some 1-cell $g$ such that the corresponding diagram
of 2-cells commutes.

In this, the non-braided case, the required rewriting system is very simple:
we treat every 1-cell as a composite of multiarrow 1-cells, and a rewriting step
simply corresponds to a positive interchange cell applied to a pair of
consecutive multiarrow 1-cells.
\begin{lemma}
	This rewriting system is strongly normalising, i.e.\ every reduction
	path is finite.
\end{lemma}
\begin{proof}
	We shall define a weighting function that assigns a natural number to each 1-cell,
	so that each rewriting step strictly decreases the weight. The appropriate weighting function
	here is the \emph{total prefix weight}, which is defined as follows. Recall that
	a 1-cell is represented as a sequence $f_{1}, \dots, f_{n}$ of multiarrow 1-cells. If we
	imagine the 1-cell drawn as a diagram like those above, then each output wire (at the
	far right of the diagram, according to our convention above) is attached to a distinct
	tree of multiarrows. Define the \emph{weight} of an output wire to be the number of multiarrows
	in the tree. So for each multiarrow, the weight of its output wire is $1+$ the sum of
	the weights of its input wires.
	
	A multiarrow 1-cell consists of a single multiarrow with a number of bare wires (objects)
	above and below it. Let us refer to the bare wires above as the `prefix' of the multiarrow 1-cell.
	In the context of a 1-cell $f_{1}, \dots, f_{n}$, define the \emph{prefix weight} of a
	constituent multiarrow 1-cell $f_{i}$ to be the sum of the weights of each wire in the prefix.
	For example, in the diagram below each wire is annotated with its weight, and each
	multiarrow is annotated with its prefix weight.
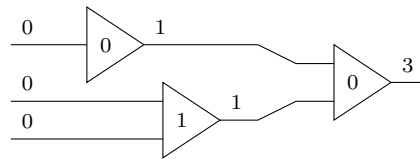
\[
	\def\triangle(#1,#2)#3 ; {
		\draw (#1,#2-2) -- ++(0,4) -- ++(3,-2) -- cycle ;
			\node at (1+#1,#2) {#3} ;
	}
	\tikzstyle{every picture}+=[baseline,scale=0.25]
	\begin{tikzpicture}
		\draw (0,3)	 node[above right] {$\scriptstyle 0$} -- +(4,0) ;
			\triangle (4,3) {$\scriptstyle 0$} ;
			\draw (7,3) node[above right] {$\scriptstyle 1$} -- +(6,0) ;
		\draw (0,0)	 node[above right] {$\scriptstyle 0$} -- +(8,0) ;
		\draw (0,-2) node[above right] {$\scriptstyle 0$} -- +(8,0) ;
			\triangle (8,-1) {$\scriptstyle 1$} ;
			\draw (11,-1) node[above right] {$\scriptstyle 1$} -- +(2,0) ;
		\draw (13,3)  -- ++(2,-1) -- +(2,0) ;
		\draw (13,-1) -- ++(2,1)  -- +(2,0) ;
		\triangle (17,1) {$\scriptstyle 0$} ;
		\draw (20,1) node[above right] {$\scriptstyle 3$} -- +(2,0) ;
	\end{tikzpicture}
\]
	The total prefix weight of a 1-cell is simply the sum of the prefix weights of the
	constituent multiarrow cells, so the example above has a total prefix weight of $1$.
	Clearly the target of a positive interchange cell must have a strictly smaller total
	prefix weight than the source.
\end{proof}
It remains to show confluence. Let $\vec f$ be a 1-cell, and suppose that we
have two different interchanges $\gamma$ and $\delta$ applied to $\vec f$. If
$\gamma$ and $\delta$ do not overlap, then the result of applying $\gamma$ followed
by $\delta$ is the same as the result of applying $\delta$ followed by $\gamma$, and
the corresponding diagram of 2-cells commutes by naturality. If $\gamma$ and $\delta$
do overlap, then we essentially have the following situation:
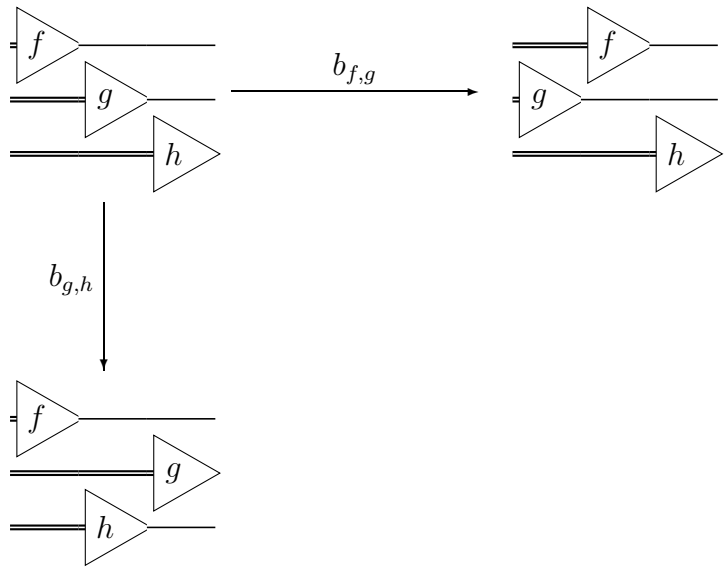
\[
	\braidstyle{xscale=1.2,yscale=0.6}\tikzstyle{braidtriangle}+=[inner sep=0.5pt,xshift=-2pt]
	\begin{diagram}[h=6em,w=8em]
	\begin{braid}{=,=,=}
		\t[f] \c    \c \\
		\c    \t[g] \c \\
		\c    \c    \t[h]
	\end{braid}
	& \rTo^{b_{f,g}} &
	\begin{braid}{=,=,=}
		\c    \t[g] \c \\
		\t[f] \c    \c \\
		\c    \c    \t[h]
	\end{braid}
	\\ \dTo<{b_{g,h}} \\
	\begin{braid}{=,=,=}
		\t[f] \c    \c \\
		\c    \c    \t[h] \\
		\c    \t[g] \c
	\end{braid}
	\end{diagram}
\]
(where we have omitted from the diagram wires which are not connected to one of the three multiarrows that participate in these interchanges.)
This may be completed as shown in the following diagram:
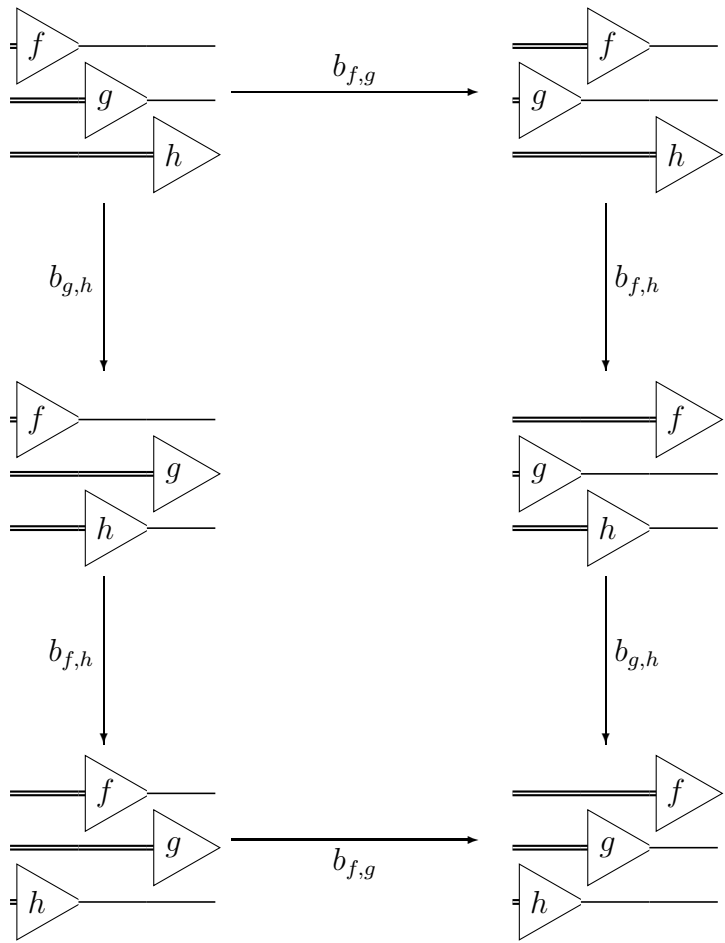
\[
	\braidstyle{xscale=1.2,yscale=0.6}\tikzstyle{braidtriangle}+=[inner sep=0.5pt,xshift=-2pt]
	\begin{diagram}[h=6em,w=8em]
	\begin{braid}{=,=,=}
		\t[f] \c    \c \\
		\c    \t[g] \c \\
		\c    \c    \t[h]
	\end{braid}
	& \rTo^{b_{f,g}} &
	\begin{braid}{=,=,=}
		\c    \t[g] \c \\
		\t[f] \c    \c \\
		\c    \c    \t[h]
	\end{braid}
	\\ \dTo<{b_{g,h}} && \dTo>{b_{f,h}} \\
	\begin{braid}{=,=,=}
		\t[f] \c    \c \\
		\c    \c    \t[h] \\
		\c    \t[g] \c
	\end{braid}
	&&
	\begin{braid}{=,=,=}
		\c    \t[g] \c \\
		\c    \c    \t[h] \\
		\t[f] \c    \c
	\end{braid}
	\\ \dTo<{b_{f,h}} && \dTo>{b_{g,h}} \\
	\begin{braid}{=,=,=}
		\c    \c    \t[h] \\
		\t[f] \c    \c \\
		\c    \t[g] \c
	\end{braid}
	& \rTo_{b_{f,g}} &
	\begin{braid}{=,=,=}
		\c    \c    \t[h]\\
		\c    \t[g] \c \\
		\t[f] \c    \c
	\end{braid}
	\end{diagram}
\]
To see why the corresponding diagram of 2-cells must commute, fill in diagonals as follows:
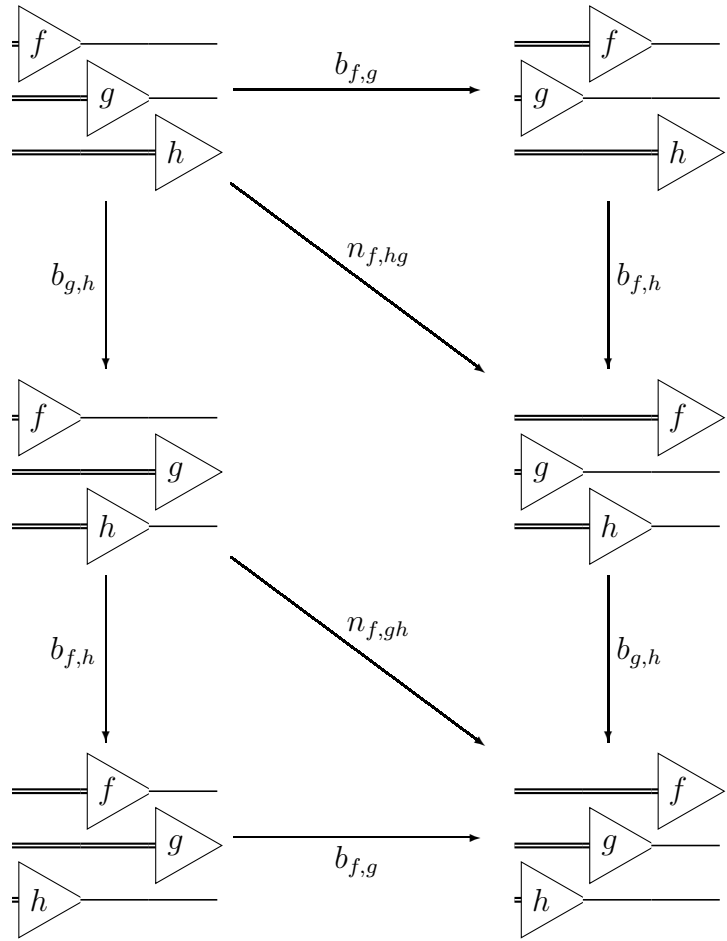
\[
	\braidstyle{xscale=1.2,yscale=0.6}\tikzstyle{braidtriangle}+=[inner sep=0.5pt,xshift=-2pt]
	\begin{diagram}[h=6em,w=8em]
	\begin{braid}{=,=,=}
		\t[f] \c    \c \\
		\c    \t[g] \c \\
		\c    \c    \t[h]
	\end{braid}
	& \rTo^{b_{f,g}} &
	\begin{braid}{=,=,=}
		\c    \t[g] \c \\
		\t[f] \c    \c \\
		\c    \c    \t[h]
	\end{braid}
	\\ \dTo<{b_{g,h}} & \rdTo^{n_{f,hg}} & \dTo>{b_{f,h}} \\
	\begin{braid}{=,=,=}
		\t[f] \c    \c \\
		\c    \c    \t[h] \\
		\c    \t[g] \c
	\end{braid}
	&&
	\begin{braid}{=,=,=}
		\c    \t[g] \c \\
		\c    \c    \t[h] \\
		\t[f] \c    \c
	\end{braid}
	\\ \dTo<{b_{f,h}} & \rdTo^{n_{f,gh}} & \dTo>{b_{g,h}} \\
	\begin{braid}{=,=,=}
		\c    \c    \t[h] \\
		\t[f] \c    \c \\
		\c    \t[g] \c
	\end{braid}
	& \rTo_{b_{f,g}} &
	\begin{braid}{=,=,=}
		\c    \c    \t[h]\\
		\c    \t[g] \c \\
		\t[f] \c    \c
	\end{braid}
	\end{diagram}
\]
where $n_{f,hg}$ denotes the (non-basic) interchange of $f$ with the composite of
$g$ and $h$, and similarly $n_{f,gh}$. The triangles commute by definition, and the
central quadrilateral is another instance of naturality of interchange.

Therefore for every 1-cell $f$, there is a (necessarily unique) corresponding 1-cell
$f'$ in normal form, together with an invertible 2-cell $\gamma_{f}: f\To f'$ built from interchange cells.
Since the system is coherently confluent, and every 2-cell can be represented as a zig-zag of
rewrites, $\gamma_{f}$ is also unique. Hence there is a 2-cell connecting 1-cells $f$ and
$g$ just when $f$ and $g$ have the same normal form, and this 2-cell is unique.

\section{The braided case}
In the braided case the form of the argument is the same, though the details are a little
more involved. The free braided Gray monoid $F_{\beta}(G)$ on a multigraph $G$ can be described in a similar
way to the free ordinary Gray monoid, with the addition of braiding data. Precisely:
\begin{itemize}
    \item A 1-cell of $F_{\beta}(G)$ is the composite of a finite sequence of multiarrow cells
	and crossing cells, which we collectively term \emph{basic 1-cells}. The multiarrow cells
	are as above and:
    \item A braid cell is either a positive crossing $\beta_{\vec X(\vec A,\vec B)\vec Y}$:
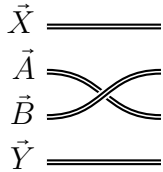
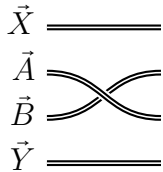
    \[\begin{braid}{=\vec X,=\vec A,=\vec B,=\vec Y}
        \c \s(1,1) \c
    \end{braid}\]
    or a negative crossing $\beta'_{\vec X(\vec A,\vec B)\vec Y}$:
    \[\begin{braid}{=\vec X,=\vec A,=\vec B,=\vec Y}
        \c \s'(1,1) \c
    \end{braid}\]
	As with the interchange cells, we shall typically omit the subscripts corresponding
	to wires that do not participate in the crossing, writing the above crossings as
	$\beta_{\vec A,\vec B}$ and $\beta'_{\vec A,\vec B}$.
    \item A 2-cell is a finite vertical composite of basic 2-cells, where a basic
    2-cell is either a whiskered interchange cell as above, or else a whiskered unit cell
	or braiding cell: the latter two are described below.
    Note that crossings -- as well as multiarrows -- may participate in interchange;
    for example, there is an interchange cell $b_{f,\beta_{C,D}}$ as follows:
    
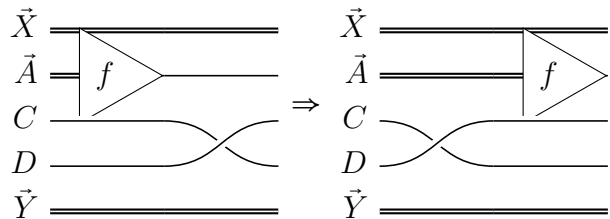
\[
        \begin{braid}{=\vec X,=\vec A,C,D,=\vec Y}
            \c\t[f]\c\c\c \\
            \c\c\s(1,1)\c
        \end{braid}
        \To
        \begin{braid}{=\vec X,=\vec A,C,D,=\vec Y}
            \c\c\s(1,1)\c \\
            \c\t[f]\c\c\c
        \end{braid}
    \]
	\item A unit cell is either $U_{\vec X(\I|\vec A)\vec Y}: \beta_{\vec X(\langle\rangle,\vec A)\vec Y} \To 1$
		or $U_{\vec X(\vec A|\I)\vec Y}: \beta_{\vec X(\vec A,\langle\rangle)\vec Y} \To 1$,
		or the inverse of one of these. In terms of our representation of a 1-cell as a finite sequence
		of basic 1-cells, the target of a unit cell is the empty sequence. So the sequence representing the
		target (of a whiskered unit cell) has one element fewer than the sequence representing the source.
	\item There are two types of braiding cell, overbraiding and underbraiding cells.
	An overbraiding cell
	has the form $B_{\vec X(\vec A,\vec B|\vec C)\vec Y}$:
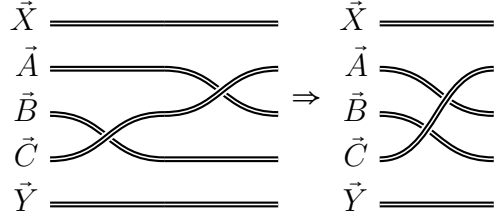
	\[
		\begin{braid}{=\vec X,=\vec A,=\vec B,=\vec C,=\vec Y} \c\c\s(1,1)\c \\ \c\s(1,1)\c\c \end{braid}
		\To
		\begin{braid}{=\vec X,=\vec A,=\vec B,=\vec C,=\vec Y} \c\s(2,1)\c \end{braid}
	\]
	or the inverse $B_{\vec X(\vec A,\vec B|\vec C)\vec Y}^{-1}$. An underbraiding cell
	has the form $B_{\vec X(\vec A|\vec B,\vec C)\vec Y}$:
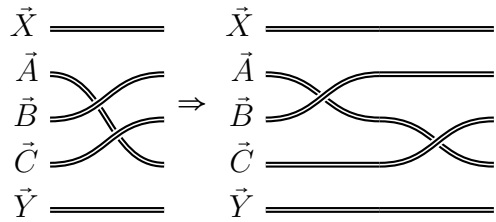
	\[
		\begin{braid}{=\vec X,=\vec A,=\vec B,=\vec C,=\vec Y} \c\s(1,2)\c \end{braid}
		\To
		\begin{braid}{=\vec X,=\vec A,=\vec B,=\vec C,=\vec Y} \c\s(1,1)\c\c \\ \c\c\s(1,1)\c \end{braid}
	\]
	or the inverse. As with the other types of cell, in context we shall usually write
	these simply as $B_{\vec A,\vec B|\vec C}$ and $B_{\vec A|\vec B,\vec C}$.
	\item We also need unit cells $U_{\vec X(\I|\vec A)\vec Y}: \beta_{\vec X(\langle\rangle,\vec A)\vec Y} \To 1_{\vec X\tn\vec A\tn\vec Y}$ and $U_{\vec X(\vec A|\I)\vec Y}: \beta_{\vec X(\vec A,\langle\rangle)\vec Y} \To 1_{\vec X\tn\vec A\tn\vec Y}$.
	\item Finally, there are the 2-cells that give the pseudonaturality of the crossings. We shall take the following as basic:
	\[
	N_{\vec X(f,\vec B)\vec Y}:
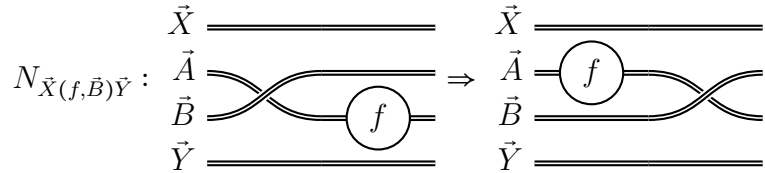
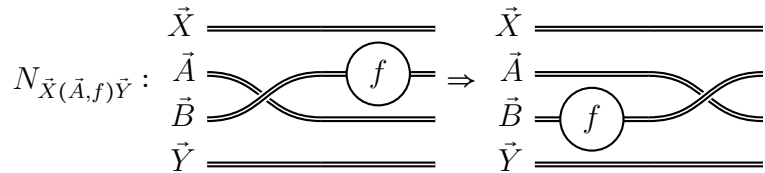
	\begin{braid}{=\vec X,=\vec A,=\vec B,=\vec Y}
		\c\s(1,1)\c \\
		\c\c\f[f]\c
	\end{braid}
	\To
	\begin{braid}{=\vec X,=\vec A,=\vec B,=\vec Y}
		\c\f[f]\c\c \\
		\c\s(1,1)\c
	\end{braid}
	\]
	and
	\[
	N_{\vec X(\vec A,f)\vec Y}:
	\begin{braid}{=\vec X,=\vec A,=\vec B,=\vec Y}
		\c\s(1,1)\c \\
		\c\f[f]\c\c
	\end{braid}
	\To
	\begin{braid}{=\vec X,=\vec A,=\vec B,=\vec Y}
		\c\c\f[f]\c \\
		\c\s(1,1)\c
	\end{braid}
	\]
	where $f$ is a basic 1-cell, i.e.\ a multiarrow or a crossing. These also have inverses.
\end{itemize}
Recall that here we are considering only positive 1-cells, where all the crossings are positive.
(By duality we could equally well take the case where all crossings are negative; it is mixing
the two that introduces additional complexity.)

For rewriting rules, we shall essentially take the interchange, over- and underbraiding, unit, and pseudonaturality cells in their natural (positive) directions. There is a collection of restrictions related to trivial crossings, where one of the `branches' of the crossing is empty. In detail: for the braiding cells $B_{\vec X(\vec A,\vec B|\vec C)\vec Y}$ and $B_{\vec X(\vec A|\vec B,\vec C)\vec Y}$, we require all three sequences $\vec A$, $\vec B$ and $\vec C$ to be non-empty. Similarly, for the pseudonaturality cells $N_{\vec X(f,\vec B)\vec Y}$ we require $\vec B$ to be non-empty, and for $N_{\vec X(\vec A,f)\vec Y}$ we require $\vec A$ to be non-empty. In both types of pseudonaturality cell, if `$f$' denotes a crossing $\beta_{\vec A,\vec B}$, we require $\vec A$ and $\vec B$ to be non-empty. These restrictions should give the reader pause, since we must ensure that every structural 2-cell corresponds to some zig-zag of rewrites. The reason they are admissible is that the excluded braiding and pseudonaturality cells (where some sequence is empty) are all equal to some unit cell or composite of unit cells, by the unit axioms and Propositions~\chref{MonBicats}{prop-braiding-unit-1} and~\chref{MonBicats}{prop-braiding-unit-2}.
\begin{lemma}
	This rewriting system is strongly normalising.
\end{lemma}
\begin{proof}
	As in the non-braided case, we shall define a well-founded partially ordered set of `weightings', and associate a weighting with each (positive) 1-cell; then we shall show that for each rewrite rule the target has a strictly smaller weight than the source.
	
	In the non-braided case, the weightings were simply natural numbers. Here, the set of weightings is taken to be $\mathbb{N}^{4}$ with its lexicographic ordering. Thus to each positive 1-cell we associate four numbers, the \textbf{total overbraid width}, the \textbf{total crossing weight}, the \textbf{total prefix weight} and the \textbf{number of trivial crossings}.
	To define these, we need two auxiliary notions, the \emph{width} and the \emph{weight} of a wire or sequence of wires in a 1-cell. Both these are defined in an iterative way, proceeding (in terms of the diagrams) from left to right; the width (weight) of a sequence of wires is simply the sum of the individual widths (weights).

The width of each wire is initially $1$, i.e. in the identity 1-cell (represented by an empty sequence of basic 1-cells) each wire has a width of $1$. For each multiarrow cell, the width of the output wire is defined to be $1+$ the width of the sequence of input wires. For each crossing cell, the width of each output wire is simply equal to the width of the corresponding input wire. For example, the following diagram shows a 1-cell with every wire annotated with its width.
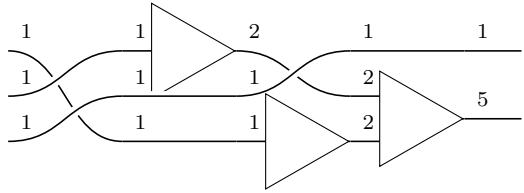
\[\begin{braid}{,,}
	\s<1,1/1> \\
	\t^{1} \c[2]^{1} \\
	\s<2,1> \t^{1} \\
	\c^{1} \T^2/2 \\
	\narrow \c^{1} \C^{5}
\end{braid}\]

The weight of a wire is defined in a related way. Weights are initially zero; for each multiarrow cell the weight of the output wire is $1+$ the weight of the sequence of input wires; for each crossing cell, the weight of each output wire is the sum of the weight and the width of the corresponding input wire. By way of illustration, here is the same 1-cell as above, now annotated with weights:
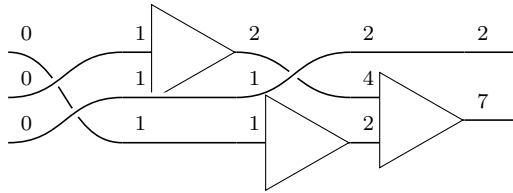
\[\begin{braid}{,,}
	\s<0,0/0> \\
	\t^{1} \c[2]^{1} \\
	\s<2,1> \t^{1} \\
	\c^{2} \T^4/2 \\
	\narrow \c^{2} \C^{7}
\end{braid}\]
\newcommand\Wd{\mathrm{wd}}\newcommand\Wt{\mathrm{wt}}
We denote the width function $\Wd()$, and the weight function $\Wt()$.

Our weighting functions are defined as follows:
\begin{itemize}
	\item The \emph{overbraid width} of a crossing $\beta_{\vec A,\vec B}$ is $2\Wd(\vec X)-1$, and the total overbraid width of a 1-cell is the sum of the overbraid widths of its crossings.
	\item The \emph{crossing weight} of a crossing $\beta_{\vec A,\vec B}$ is $\Wd(\vec A,\vec B)\times\Wt(\vec A,\vec B)$. The total crossing weight of a 1-cell is the sum of the crossing weights of its crossings
	\item The \emph{prefix weight} of a multiarrow or crossing cell is the weight of its prefix. The total prefix weight of a 1-cell is the sum of the prefix weights of all the basic 1-cells.
	\item A crossing $\beta_{\vec A,\vec B}$ is \emph{trivial} if either $\vec A$ or $\vec B$ is empty. The \emph{number of trivial crossings} is the number of trivial crossings.
\end{itemize}

We must verify that every rewrite rule reduces the aggregate weighting.
\begin{itemize}
	\item The overbraiding rule $B_{\vec A,\vec B|\vec C}$ changes the overbraid width of the cells it acts on from $4\Wd(\vec C)-2$ to $2\Wd(\vec C)-1$. The overbraid widths of other crossings are unchanged. Since we are taking $\vec C$ to be non-empty, we know that $2\Wd(\vec C)-1 > 0$, hence the total overbraid width is reduced.
	\item The underbraiding rule $B_{\vec A|\vec B,\vec C}$ changes the overbraid width of the cells it acts on from $2\Wd(\vec B,\vec C)-1$ to $2\Wd(\vec B) + 2\Wd(\vec C)-2 = 2\Wd(\vec B,\vec C)-2$, reducing it by $1$. The overbraid width of other crossings is unaffected.
	\item The pseudonaturality cells do not increase the overbraid width of any crossing. (If a multiarrow cell is being moved across the overbraid part of a crossing, the overbraid width of that crossing is reduced by $1$. Otherwise overbraid widths are unaffected.) In the case of a pseudonaturality cell $N_{f,\vec B}$ or $N_{\vec A,f}$ where `$f$' represents a multiarrow cell, the crossing weight of the affected crossing is clearly reduced (and other crossings are unaffected). The more subtle case is that where `$f$' is another crossing cell. For example, the general case of $N_{\vec V(\beta_{\vec W(\vec A,\vec B)\vec X},\vec Y)\vec Z}$ looks like this:
	
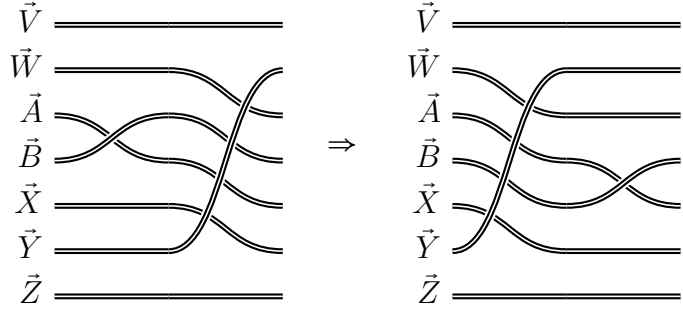
\[
		\begin{braid}{=\vec V,=\vec W,=\vec A,=\vec B,=\vec X,=\vec Y,=\vec Z}
			\c \c \s(1,1) \c \c \c \\
			\c \s(4,1) \c
		\end{braid}
		\quad\To\quad
		\begin{braid}{=\vec V,=\vec W,=\vec A,=\vec B,=\vec X,=\vec Y,=\vec Z}
			\c \s(4,1) \c \\
			\c \c \c \s(1,1) \c \c
		\end{braid}
	\]
	On the left-hand side, the total crossing weight is
	\[
		\Wd(\vec A, \vec B)\times\Wt(\vec A,\vec B) + \Wd(\vec W,\vec A,\vec B,\vec X, \vec Y)\times\bigl(\Wt(\vec W,\vec A,\vec B,\vec X, \vec Y) + \Wd(\vec A,\vec B)\bigr),
	\]
	and on the right-hand side the total crossing weight is
	\[
		\Wd(\vec W,\vec A,\vec B,\vec X, \vec Y)\times\Wt(\vec W,\vec A,\vec B,\vec X, \vec Y)
		+
		\Wd(\vec A,\vec B)\times\bigl(\Wt(\vec A,\vec B) + \Wd(\vec A,\vec B)\bigr).
	\]
	So the difference (left$-$right) is
	\[
		\Wd(\vec W,\vec A,\vec B,\vec X, \vec Y)\times\Wd(\vec A,\vec B) - \Wd(\vec A,\vec B)^{2},
	\]
	which is equal to $\Wd(\vec W,\vec X, \vec Y)\times\Wd(\vec A,\vec B)$. Since we are taking $\vec Y$, $\vec A$ and $\vec B$ to be non-empty, this difference is $>0$. The other type of pseudonaturality cell is handled in a similar way.
	\item The interchange cells do not change the overbraid width nor the crossing weight of any crossing, and (as in the non-braided case) they reduce the total prefix weight.
	\item The unit cells do not affect any of the other measures, but reduce the number of trivial crossings.
\end{itemize}
\end{proof}
\begin{propn}
	The system is coherently confluent.
\end{propn}
\begin{proof}
	We consider the possible ways that two rules could be applied to the same 1-cell.
	In the case where two rules apply to non-overlapping portions of a 1-cell,
	the rules can be applied in either order, and the resulting 2-cells are equal.
	Thus we need to consider conflicts, i.e.\ situations where two rules can apply
	to overlapping portions.
	
	First, consider the interchange rule. There is a class of conflicts where
	some basic 1-cell is moved past a point where some other 2-cell could apply.
	For example, there is a conflict between interchange and underbraiding:
	
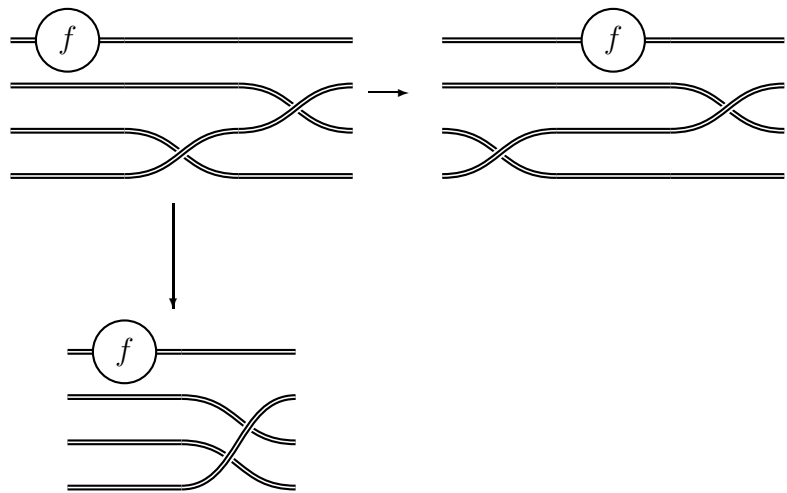
\begin{diagram}[s=5em]
		\begin{braid}{=,=,=,=}
			\f[f] \c[3] \\
			\c \c \s(1,1) \\
			\c \s(1,1) \c
		\end{braid}
 		& \rTo &
		\begin{braid}{=,=,=,=}
			\c \c \s(1,1) \\
			\f[f] \c[3] \\
			\c \s(1,1) \c
		\end{braid}
		\\
		\dTo \\
		\begin{braid}{=,=,=,=}
			\f[f] \c[3] \\
			\c \s(2,1)
		\end{braid}
	\end{diagram}
	where $f$ is either a multiarrow or a crossing.
	This diagram illustrates the style in which we shall show the conflicts. We omit
	extraneous prefix and suffix wires, and leave the arrows unlabelled (since it is
	always clear which rule applies by looking at the source and target).
	This case can of course be completed as
	
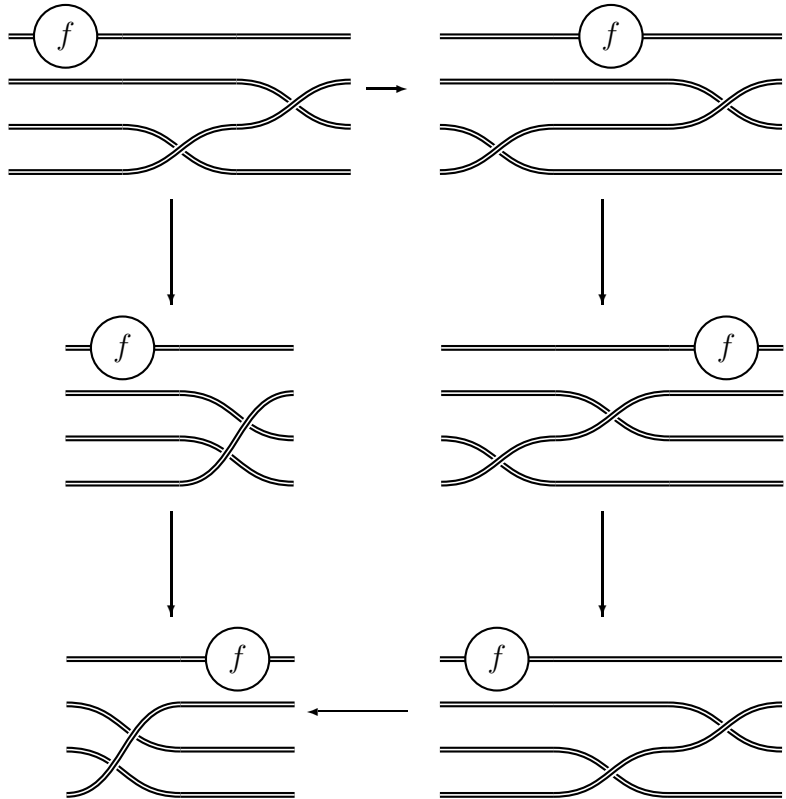
\begin{diagram}[s=5em]
		\begin{braid}{=,=,=,=}
			\f[f] \c[3] \\
			\c \c \s(1,1) \\
			\c \s(1,1) \c
		\end{braid}
 		& \rTo &
		\begin{braid}{=,=,=,=}
			\c \c \s(1,1) \\
			\f[f] \c[3] \\
			\c \s(1,1) \c
		\end{braid}
		\\ \dTo && \dTo \\ 
		\begin{braid}{=,=,=,=}
			\f[f] \c[3] \\
			\c \s(2,1)
		\end{braid}
		&&
		\begin{braid}{=,=,=,=}
			\c \c \s(1,1) \\
			\c \s(1,1) \c \\
			\f[f] \c[3]
		\end{braid}
		\\ \dTo && \dTo \\ 
		\begin{braid}{=,=,=,=}
			\c \s(2,1) \\
			\f[f] \c[3]
		\end{braid}
		& \lTo &
		\begin{braid}{=,=,=,=}
			\f[f] \c[3] \\
			\c \c \s(1,1) \\
			\c \s(1,1) \c
		\end{braid}
	\end{diagram}
	and it is clear that a similar completion is possible for all conflicts of this sort.
	Where, as in this case, there is an obvious reason below for some diagram of 2-cells
	to commute, no comment will be made below. Where the commutativity follows from the
	axioms in the definition of braiding for a monoidal bicategory, we indicate which axiom(s)
	are used.
	Note that interchange can conflict with itself, as in
	
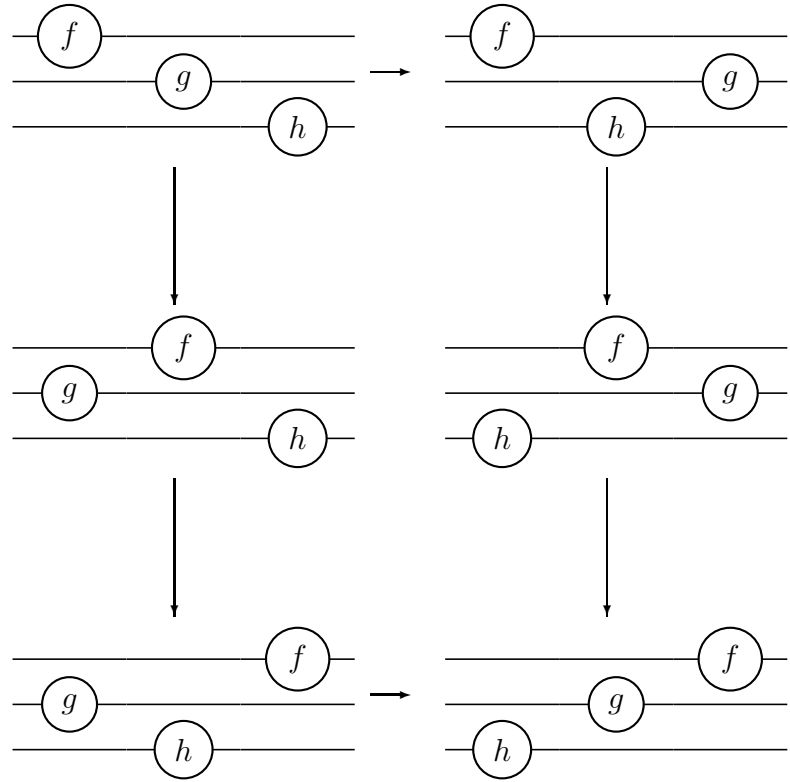
\begin{diagram}[s=5em]
		\begin{braid}{,,}\f[f] \c \c \\ \c \f[g] \c \\ \c \c \f[h]\end{braid}
		& \rTo &
		\begin{braid}{,,}\f[f] \c \c \\ \c \c \f[h] \\ \c \f[g] \c\end{braid}
		\\ \dTo && \dTo \\
		\begin{braid}{,,}\c \f[g] \c \\ \f[f] \c \c \\ \c \c \f[h]\end{braid}
		& & \begin{braid}{,,}\c \c \f[h] \\ \f[f] \c \c \\ \c \f[g] \c\end{braid}
		\\ \dTo && \dTo \\
		\begin{braid}{,,}\c \f[g] \c \\ \c \c \f[h] \\ \f[f] \c \c\end{braid}
		& \rTo & \begin{braid}{,,}\c \c \f[h] \\ \c \f[g] \c \\ \f[f] \c \c\end{braid}
	\end{diagram}
	This is just another instance of the general situation described above, and does
	not require a separate treatment.
	This does not quite exhaust the possible ways that interchange can participate
	in a conflict. One remaining possibility occurs with the underbraiding rule,
	as follows.
	
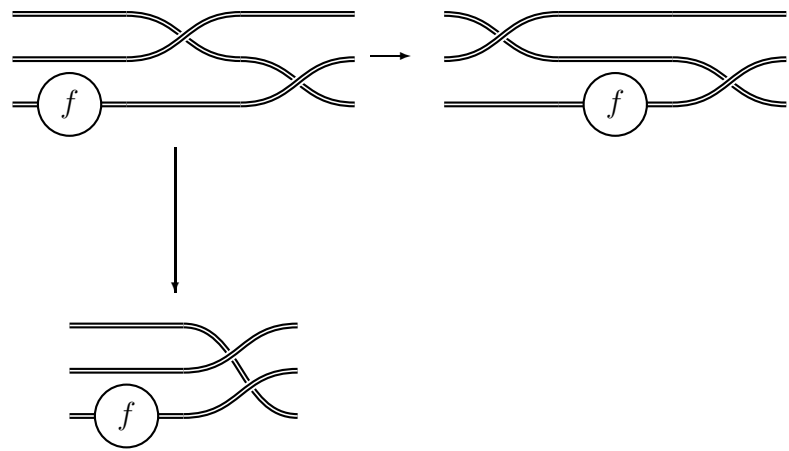
\begin{diagram}[s=5em]
		\begin{braid}{=,=,=}
			\c\c\f[f] \\ \s(1,1) \c \\ \c \s(1,1)
		\end{braid}
		& \rTo &
		\begin{braid}{=,=,=}
			\s(1,1) \c \\ \c\c\f[f] \\ \c \s(1,1)
		\end{braid}
		\\ \dTo \\
		\begin{braid}{=,=,=}
			\c\c\f[f] \\ \s(1,2)
		\end{braid}
	\end{diagram}
	which can be completed as
	
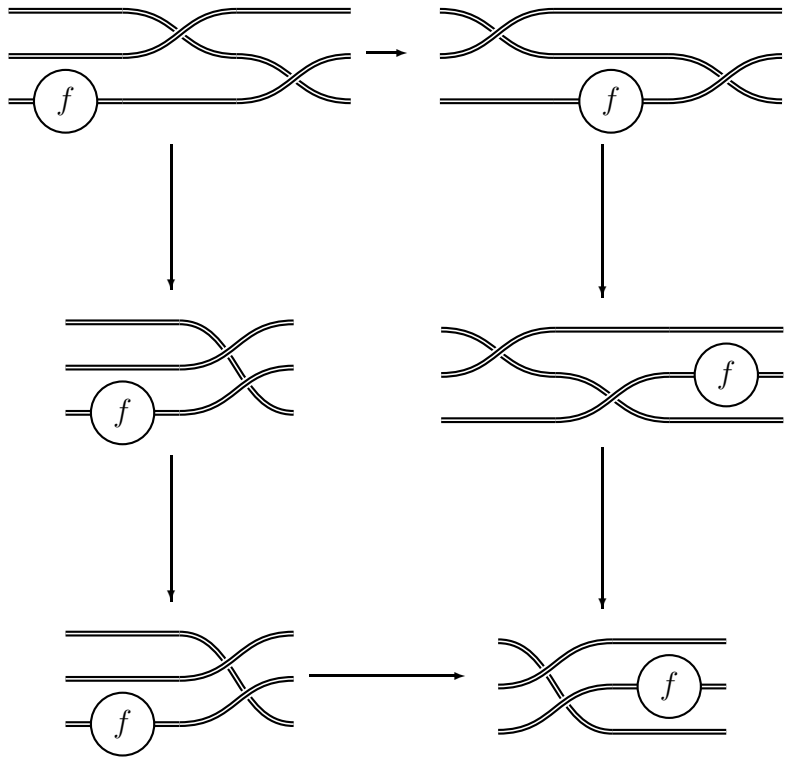
\begin{diagram}[s=5em]
		\begin{braid}{=,=,=}
			\c\c\f[f] \\ \s(1,1) \c \\ \c \s(1,1)
		\end{braid}
		& \rTo &
		\begin{braid}{=,=,=}
			\s(1,1) \c \\ \c\c\f[f] \\ \c \s(1,1)
		\end{braid}
		\\ \dTo && \dTo \\
		\begin{braid}{=,=,=}
			\c\c\f[f] \\ \s(1,2)
		\end{braid}
		&&
		\begin{braid}{=,=,=}
			\s(1,1) \c \\ \c \s(1,1) \\ \c\f[f]\c
		\end{braid}
		\\ \dTo && \dTo \\
		\begin{braid}{=,=,=}
			\c\c\f[f] \\ \s(1,2)
		\end{braid}
		& \rTo &
		\begin{braid}{=,=,=}
			\s(1,2) \\ \c\f[f]\c
		\end{braid}
	\end{diagram}
	The other remaining case involving the interchange rule is where we have
	
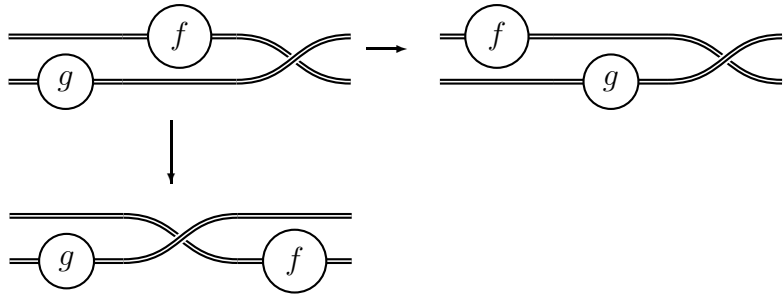
\begin{diagram}[w=5em,h=3em]
		\begin{braid}{=,=} \c \f[g] \\ \f[f] \c \\ \s(1,1) \end{braid}
		& \rTo &
		\begin{braid}{=,=} \f[f] \c \\ \c \f[g] \\ \s(1,1) \end{braid}
		\\ \dTo \\
		\begin{braid}{=,=} \c \f[g] \\ \s(1,1) \\ \c \f[f] \end{braid}
	\end{diagram}
	which can be completed as
	
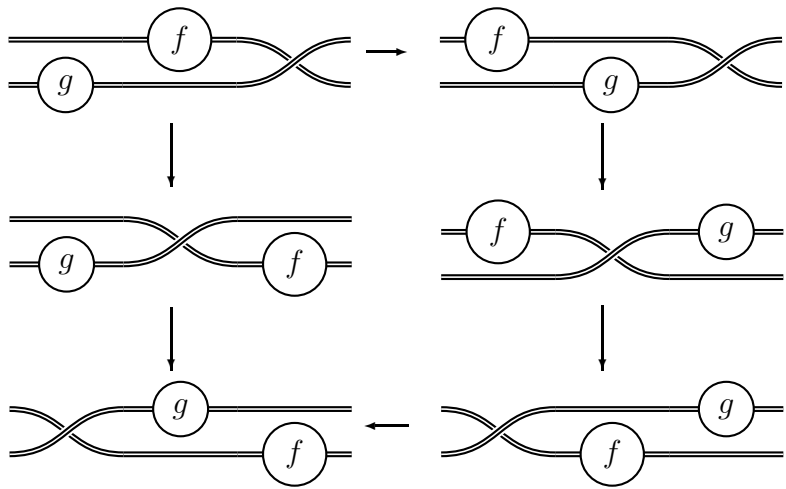
\begin{diagram}[w=5em,h=3em]
		\begin{braid}{=,=} \c \f[g] \\ \f[f] \c \\ \s(1,1) \end{braid}
		& \rTo &
		\begin{braid}{=,=} \f[f] \c \\ \c \f[g] \\ \s(1,1) \end{braid}
		\\ \dTo && \dTo \\
		\begin{braid}{=,=} \c \f[g] \\ \s(1,1) \\ \c \f[f] \end{braid}
		&& \begin{braid}{=,=} \f[f] \c \\ \s(1,1) \\ \f[g] \c \end{braid}
		\\ \dTo && \dTo \\
		\begin{braid}{=,=} \s(1,1) \\ \f[g] \c \\ \c \f[f] \end{braid}
		& \lTo &
		\begin{braid}{=,=} \s(1,1) \\ \c \f[f] \\ \f[g] \c \end{braid}
	\end{diagram}
	The pseudonaturality rules permit another general class of conflicts,
	where a sequence of basic 1-cells that could have some rule applied
	to it could alternatively be moved over or under another bunch of
	strands using the pseudonaturality rule. These conflicts can clearly
	be coherently completed in a natural way, as seen in the following
	illustrative example, which uses the underbraiding rule:
	
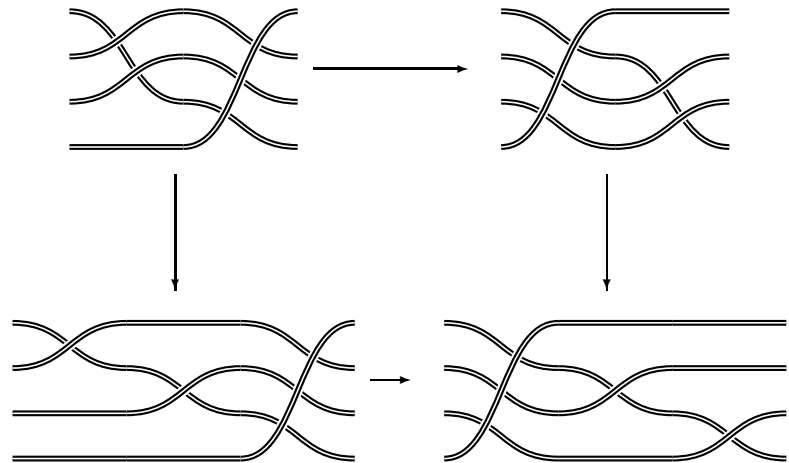
\begin{diagram}[s=5em]
		\begin{braid}{=,=,=,=} \s(1,2) \c \\ \s(3,1) \end{braid}
		& \rTo &
		\begin{braid}{=,=,=,=} \s(3,1) \\ \c \s(1,2) \end{braid}
		\\ \dTo && \dTo \\
		\begin{braid}{=,=,=,=} \s(1,1) \c \c \\ \c\s\c \\ \s(3,1) \end{braid}
		& \rTo &
		\begin{braid}{=,=,=,=} \s(3,1) \\ \c\s\c \\ \c\c\s \end{braid}
	\end{diagram}
	The pseudonaturality rules can also conflict with the over- and underbraiding
	rules in an only slightly more interesting way. With the overbraiding rule, there
	are two such basic possibilities, shown with their completions:
	
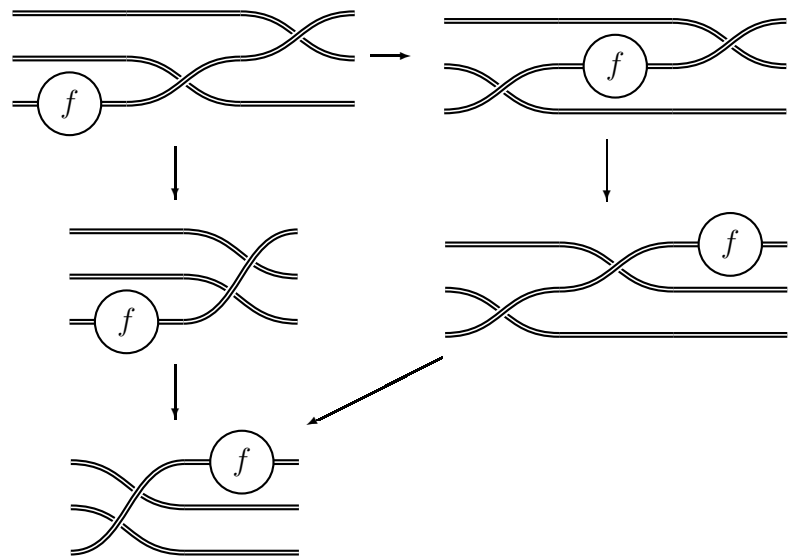
\begin{diagram}[w=5em,h=3.5em]
		\begin{braid}{=,=,=} \c\c\f[f] \\ \c\s \\ \s\c \end{braid}
		& \rTo &
		\begin{braid}{=,=,=} \c\s \\ \c\f[f]\c \\ \s\c \end{braid}
		\\ \dTo && \dTo \\
		\begin{braid}{=,=,=} \c\c\f[f] \\ \s(2,1) \end{braid}
		&& \begin{braid}{=,=,=} \c\s \\ \s\c \\ \f[f]\c\c \end{braid}
		\\ \dTo & \ldTo \\
		\begin{braid}{=,=,=} \s(2,1) \\ \f[f]\c\c \end{braid}
	\end{diagram}
	and
	
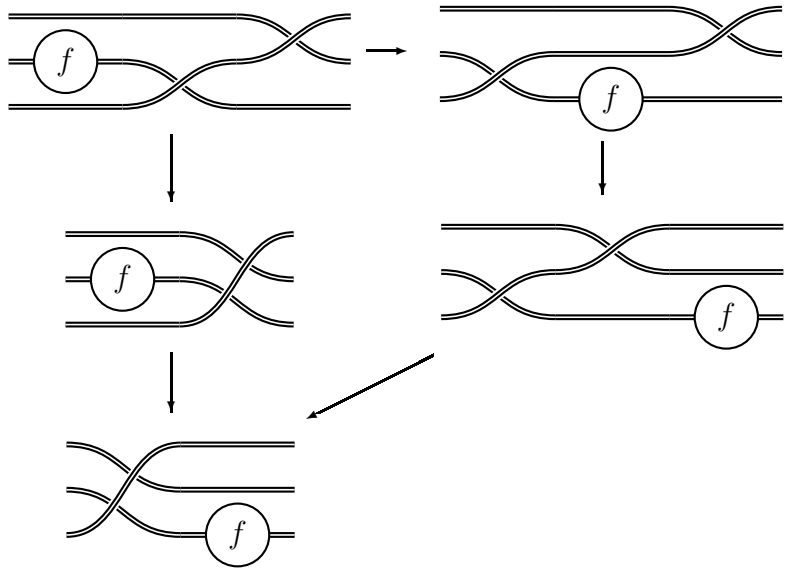
\begin{diagram}[w=5em,h=3.5em]
		\begin{braid}{=,=,=} \c\f[f]\c \\ \c\s \\ \s\c \end{braid}
		& \rTo &
		\begin{braid}{=,=,=} \c\s \\ \c\c\f[f] \\ \s\c \end{braid}
		\\ \dTo && \dTo \\
		\begin{braid}{=,=,=} \c\f[f]\c \\ \s(2,1) \end{braid}
		&& \begin{braid}{=,=,=} \c\s \\ \s\c \\ \c\c\f[f] \end{braid}
		\\ \dTo & \ldTo \\
		\begin{braid}{=,=,=} \s(2,1) \\ \c\c\f[f] \end{braid}
	\end{diagram}
	With the underbraiding rule, there are three, which all follow
	fundamentally the same pattern:
	
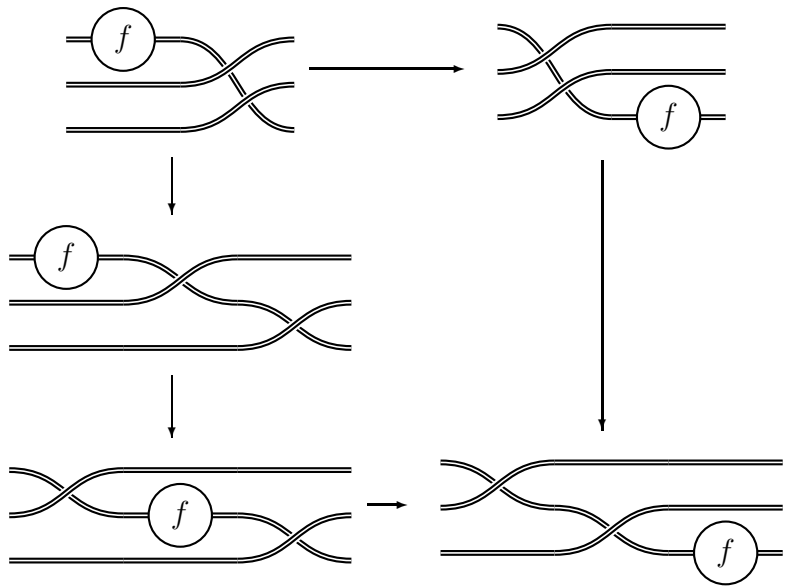
\begin{diagram}[w=5em,h=3.5em]
		\begin{braid}{=,=,=} \f[f]\c\c \\ \s(1,2) \end{braid}
		& \rTo &
		\begin{braid}{=,=,=} \s(1,2) \\ \c\c\f[f] \end{braid}
		\\ \dTo \\
		\begin{braid}{=,=,=} \f[f]\c\c \\ \s\c \\ \c\s \end{braid}
		&& \dTo \\ \dTo \\
		\begin{braid}{=,=,=} \s\c \\ \c\f[f]\c \\ \c\s \end{braid}
		& \rTo &
		\begin{braid}{=,=,=} \s\c \\ \c\s \\ \c\c\f[f] \end{braid}
	\end{diagram}
	and the second:
	
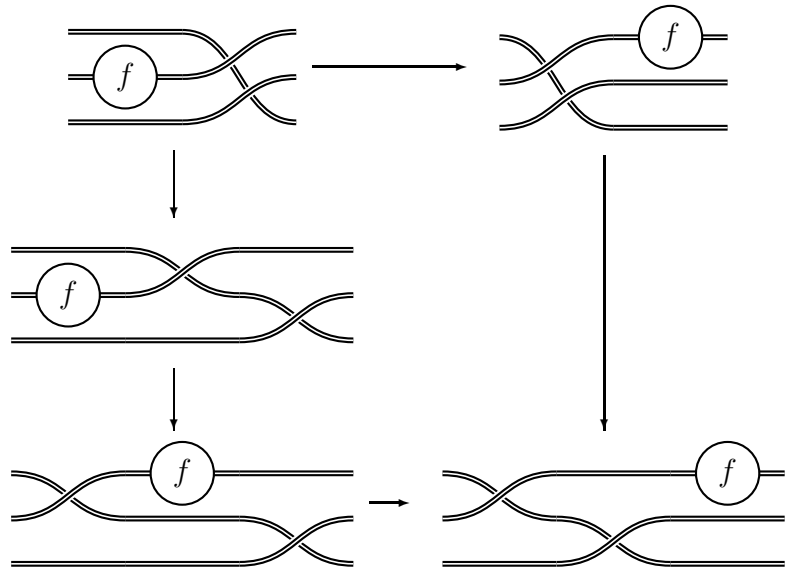
\begin{diagram}[w=5em,h=3.5em]
		\begin{braid}{=,=,=} \c\f[f]\c \\ \s(1,2) \end{braid}
		& \rTo &
		\begin{braid}{=,=,=} \s(1,2) \\ \f[f]\c\c \end{braid}
		\\ \dTo \\
		\begin{braid}{=,=,=} \c\f[f]\c \\ \s\c \\ \c\s \end{braid}
		&& \dTo \\ \dTo \\
		\begin{braid}{=,=,=} \s\c \\ \f[f]\c\c \\ \c\s \end{braid}
		& \rTo &
		\begin{braid}{=,=,=} \s\c \\ \c\s \\ \f[f]\c\c \end{braid}
	\end{diagram}
	and the third:
	
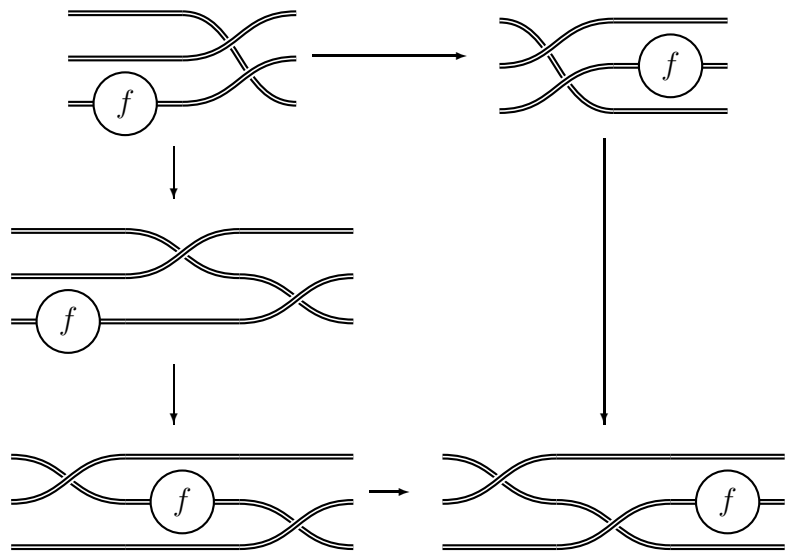
\begin{diagram}[w=5em,h=3.5em]
		\begin{braid}{=,=,=} \c\c\f[f] \\ \s(1,2) \end{braid}
		& \rTo &
		\begin{braid}{=,=,=} \s(1,2) \\ \c\f[f]\c \end{braid}
		\\ \dTo \\
		\begin{braid}{=,=,=} \c\c\f[f] \\ \s\c \\ \c\s \end{braid}
		&& \dTo \\ \dTo \\
		\begin{braid}{=,=,=} \s\c \\ \c\f[f]\c \\ \c\s \end{braid}
		& \rTo &
		\begin{braid}{=,=,=} \s\c \\ \c\s \\ \c\f[f]\c \end{braid}
	\end{diagram}
	There is also a more interesting kind of conflict between the underbraiding
	and pseudonaturality rules, as follows:
	
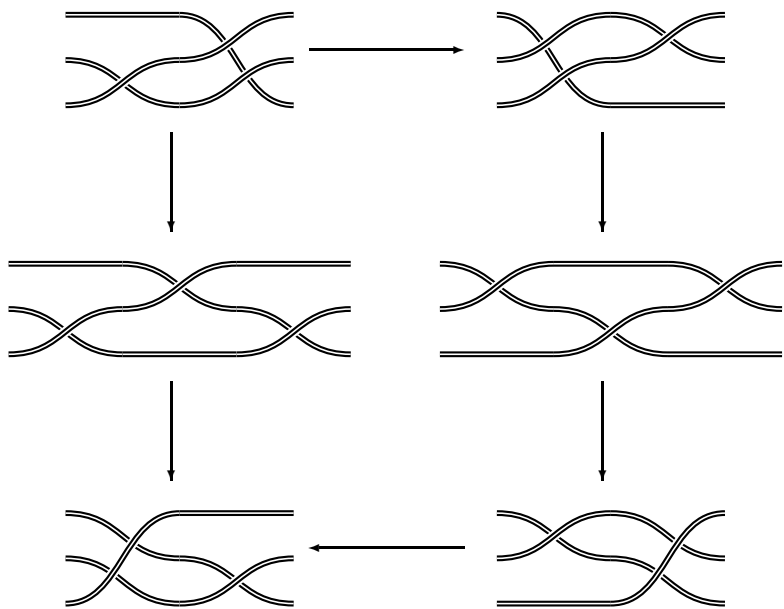
\begin{diagram}[w=5em,h=4em]
		\begin{braid}{=,=,=} \c\s \\ \s(1,2) \end{braid}
		& \rTo &
		\begin{braid}{=,=,=} \s(1,2) \\ \s\c \end{braid}
		\\ \dTo && \dTo \\
		\begin{braid}{=,=,=} \c\s \\ \s\c \\ \c\s \end{braid}
		&& \begin{braid}{=,=,=} \s\c \\ \c\s \\ \s\c \end{braid}
		\\ \dTo && \dTo \\
		\begin{braid}{=,=,=} \s(2,1) \\ \c\s \end{braid}
		& \lTo &
		\begin{braid}{=,=,=} \s\c \\ \s(2,1) \end{braid}
	\end{diagram}
	This diagram of 2-cells commutes by the Yang-Baxter axiom, i.e. the
	fourth axiom in our definition of braiding. In fact the conflict shown
	above is not the most general case of this situation; the general
	case looks like this:
	
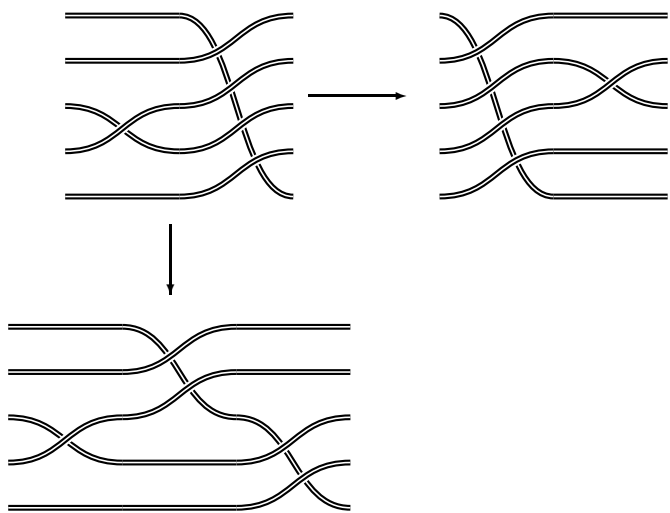
\begin{diagram}[w=5em,h=5em]
		\begin{braid}{=,=,=,=,=} \c\c\s\c \\ \s(1,4) \end{braid}
		& \rTo &
		\begin{braid}{=,=,=,=,=} \s(1,4) \\ \c\s\c\c \end{braid}
		\\ \dTo \\
		\begin{braid}{=,=,=,=,=} \c\c\s\c \\ \s(1,2)\c\c \\ \c\c\s(1,2) \end{braid}
	\end{diagram}
	and can be completed in a similar way, once the underbraiding rule has
	been applied twice to both the resulting 1-cells. It is an easy exercise
	to show that the resulting diagram of 2-cells commutes.

	There is also a non-trivial way in which the overbraiding and underbraiding rules can
	conflict with each other:
	
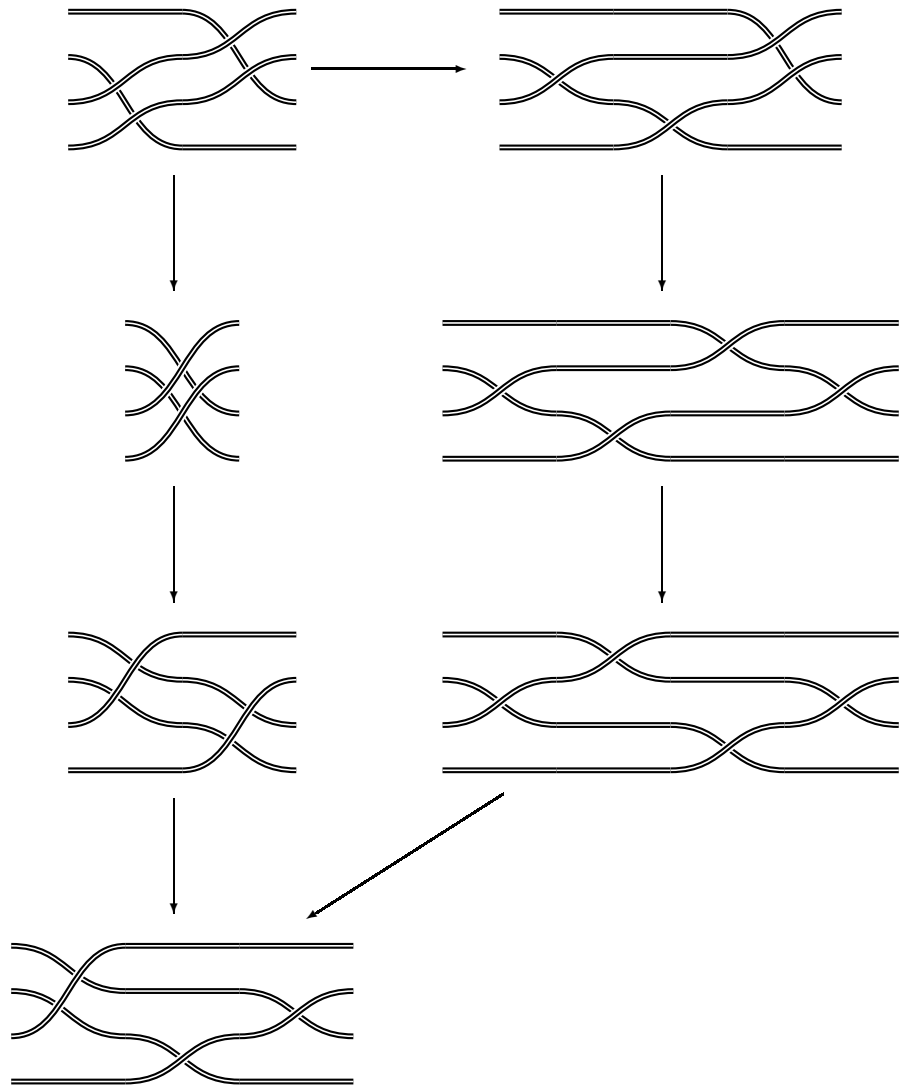
\begin{diagram}[s=5em]
		\begin{braid}{=,=,=,=} \c\s(1,2) \\ \s(1,2)\c \end{braid}
		& \rTo &
		\begin{braid}{=,=,=,=} \c\s\c \\ \c\c\s \\ \s(1,2) \c \end{braid}
		\\ \dTo && \dTo \\
		\begin{braid}{=,=,=,=} \s(2,2) \end{braid}
		&& \begin{braid}{=,=,=,=} \c\s\c \\ \c\c\s \\ \s\c\c \\ \c\s\c \end{braid}
		\\ \dTo && \dTo \\
		\begin{braid}{=,=,=,=} \s(2,1)\c \\ \c\s(2,1) \end{braid}
		&& \begin{braid}{=,=,=,=} \c\s\c \\ \s\c\c \\ \c\c\s \\ \c\s\c \end{braid}
		\\ \dTo & \ldTo \\
		\begin{braid}{=,=,=,=} \s(2,1)\c \\ \c\c\s \\ \c\s\c \end{braid}
	\end{diagram}
	This corresponds to the mysterious third axiom in the definition of braiding.
	Of course the first two steps of the right-hand path could have been
	applied in the other order.

	The final case to consider is conflict between the overbraiding rule and itself,
	or similarly between the underbraiding rule and itself. In the case of the
	overbraiding rule, the situation is this:
	
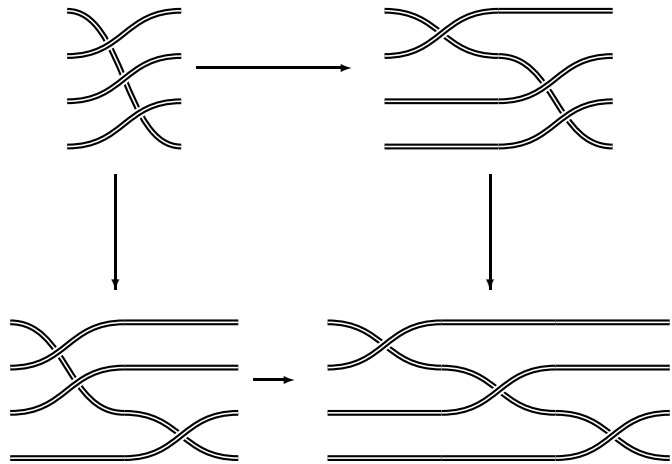
\begin{diagram}[s=5em]
		\begin{braid}{=,=,=,=,=} \s(1,3) \end{braid}
		& \rTo & \begin{braid}{=,=,=,=} \s\c\c \\ \c\s(1,2) \end{braid}
		\\ \dTo && \dTo \\
		\begin{braid}{=,=,=,=} \s(1,2) \c \\ \c\c\s \end{braid}
		& \rTo & \begin{braid}{=,=,=,=} \s\c\c \\ \c\s\c \\ \c\c\s \end{braid}
	\end{diagram}
	which corresponds to the second axiom. The obvious analogue for underbraiding
	corresponds to the first axiom.
	
	Notice that we have set things up in such a way that there are no non-trivial
	conflicts with the unit rules. (The unit axioms were used to justify the decision
	to exclude empty sequences from the domain of application of the other rules.)
	Systematic consideration of the rules shows that we have exhausted the possible
	conflicts, so the proof is complete.
\end{proof}
\end{thesischapter}

\documentclass{robinthesis}
\tikzstyle{every picture}+=[remember picture, >=latex]

\begin{thesischapter}{Language}{A Calculus of Components}
%
\def\rnode#1#2{%
	\tikz[baseline=(#1.base),inner sep=0pt,outer sep=3pt]
		\node(#1){\mathsurround=0pt$\displaystyle#2$};
	}
\section{The problem}
Even with the help of the coherence theorem, it can be extremely
tedious to prove even simple facts about pseudomonoids. The problem
is essentially notational rather than mathematical: in an ordinary
category, we can use commutative diagrams to establish equations
between arrows. It is almost always easier to understand a diagram
than an explicit sequence of equalities between expressions, because
\begin{itemize}
\item The types of the arrows are clear from the diagram.
\item The diagram abstracts away from certain details of the proof:
	if two equations could be applied in either order, the diagram
	does not have to choose an order arbitrarily.
\item The global structure of the whole argument can be seen at a glance.
\end{itemize}
In a bicategory, to prove an equation between 2-cells by similar
means, we should need three-dimensional diagrams. This poses
practical problems -- paper is two-dimensional -- and also requires
the reader to visualise a three-dimensional structure, something
that most humans do not find easy. This technique has sometimes
been attempted nonetheless: by \citet[][Section~3.4]{LackThesis},
for example.
The more usual alternative is to use a sequence of equations
between string or pasting diagrams. As well as using a lot
of paper, such proofs are often difficult to follow.

Now, there is certainly one monoidal bicategory in which this problem
does not arise. Working in the monoidal bicategory $\Cat$, one
typically reasons about natural transformations via their components.
For example, instead of the pasting equation
\[\begin{diagram}[w=4em]
	\\
	\rnode{C}{\C} & \rTo[snake=-1.5em]^{G} & \rnode{D}{\D} \\
	\begin{tikzpicture}[overlay]
		\path[->] (C) edge [out=45, in=135, out looseness=1]
			node [above, pos=0.5] {$F$} node [below, pos=0.5] {$\Downarrow s$} (D);
		\path[->] (C) edge [out=-45, in=-135, out looseness=1]
			node [below, pos=0.5] {$H$} node[above, pos=0.5] {$\Downarrow t$} (D);
	\end{tikzpicture}
\end{diagram}
\quad=\quad
\begin{diagram}[w=4em]
	\\
	\rnode{C}{\C} & \Downarrow u & \rnode{D}{\D} \\
	\begin{tikzpicture}[overlay]
		\path[->] (C) edge [out=30, in=150, out looseness=0.7]
			node [above, pos=0.5] {$F$} (D);
		\path[->] (C) edge [out=-30, in=-150, out looseness=0.7]
			node [below, pos=0.5] {$H$} (D);
	\end{tikzpicture}
\end{diagram}\]
one would typically draw the commutative triangle
\begin{diagram}[vtrianglewidth=1em]
	F(C) && \rTo^{u_{C}} && H(C) \\
	& \rdTo_{s_{C}} && \ruTo_{t_{C}} \\
	&& G(C)
\end{diagram}
for an arbitrary object $C\in\C$, considering the components of
the natural transformations rather than the natural transformations
as a whole.
It turns out that this `calculus of components' is in fact
a valid mode of reasoning in \emph{any} monoidal
bicategory. We prove this by formalising the
component-based reasoning, and showing how the resulting
formal language may be interpreted in a monoidal bicategory.
As a corollary, we show that certain types of theorem
hold in any monoidal bicategory just when they hold in
$\Cat$. For example, the famous coherence theorem for monoidal
categories implies a coherence theorem for general pseudomonoids.

The calculus of components may be used in situations of the
following form: suppose we have a collection of 1-cells of
the form
\[
	A_{1}\tn\cdots\tn A_{n} \to B.
\]
(In a general monoidal bicategory, some particular bracketing
needs to be chosen for the tensor.)
These can be combined by tensor and composition to create composite
1-cells of the same shape (i.e. where the target is a single object
rather than a tensor of objects), as in a multicategory. For example,
if we have 1-cells $F: A\tn B \to C$ and $G: X\tn Y \to B$, there is
a natural composite $A\tn X\tn Y\to C$. In $\Cat$, we could say that
this composite takes the value $F(a,G(x,y))$ on objects $a\in A$,
$x \in X$ and $y\in Y$. Such an expression can be used to describe
the composite in any monoidal bicategory, where we regard the
elements $a$, $x$ and $y$ in a purely formal way.
Now suppose that we have another 1-cell $H: X\tn Y\to B$, and a
2-cell $t: G \To H$. There is a composite with components $F(a,H(x,y))$
(with $a$, $x$ and $y$ again being formal elements of $A$, $X$ and $Y$),
and there is clearly a 2-cell from `$F(a,G(x,y))$' to `$F(a,H(x,y))$',
built from $t$ and $F$. If we were in $\Cat$, this 2-cell would have
components $t_{a,x,y}: F(a,G(x,y)) \to F(a,H(x,y))$; and again, this
description is adequate to describe the equivalent 2-cell in any monoidal
bicategory, provided that we treat the `elements' and `components' in
a purely formal fashion. Furthermore, as we shall see, one can perform
equational reasoning (or equivalently and more importantly diagrammatic
reasoning) with these components precisely as if they \emph{were} components
of ordinary natural transformations.

So the purpose of this chapter is to show how the notational
difficulties associated with monoidal bicategories may, in some cases, be overcome by using a formal
language to specify 1-cells, 2-cells, and equations between 2-cells
in a monoidal bicategory. The syntax of the language is designed in such
a way that a proof using the language closely resembles -- typically,
is formally identical to -- a proof
using categories, functors and natural transformations in the usual
way. In other words, provided that one uses only admissible techniques,
a proof in the 2-monoidal 2-category $\Cat$ may be reinterpreted
as a general proof that applies to any monoidal bicategory. In particular,
we shall be able to prove various general facts about pseudomonoids
simply by reusing the usual proof of the corresponding fact for
monoidal categories. The language could be regarded as a higher-dimensional analogue
of the formal language developed \citet{JayLanguages} for monoidal
categories.

This formal language is not completely general. The fundamental
restriction (mentioned above) is that it may only be used to talk about theories
whose 1-cells are of the form
\[
	A_{1}\tn\cdots\tn A_{n} \to B.
\]
In particular, the braiding on a braided Gray monoid is not of this
form, and we will need an extension of the language to prove facts about
structures that interact with the braiding, such as the braiding of
a braided pseudomonoid. This extended language is developed in
the next chapter.

Most of the formal development is fairly routine. The interpretation of
a theory is defined first for Gray monoids, and then extended to arbitrary
monoidal bicategories using coherence. The only potentially difficult part
is to account for the non-strictness of the interchange law in a Gray monoid.
In an earlier version of this chapter, some rather notationally-intimidating
proofs were needed to account for that; however the coherence results of Chapter~\refchapter{Coh} now save us the trouble.\footnote{In particular, they save \emph{you, the reader} the trouble, the author being unable to spare himself retroactively.}

\section{The language}
The language is used to prove equations between 2-cells that hold in
every model of a given theory.
We require the theory to be presented as a collection
of objects, (multi-)1-cells, 2-cells and equations between 2-cells.

A \emph{theory} $\T$ consists of:
\begin{itemize}
	\item A set $\T_{0}$ of \emph{objects},
	\item For every non-empty sequence $\A_{1}, \ldots, \A_{n}, \CB$ of
		elements of $T_{0}$,
		a set $\T(\A_{1},\ldots,\A_{n};\CB)$
		of \emph{1-cells} $(\A_{1}, \dots, \A_{n})\to \CB$.
		We suppose that the different sets of 1-cells are pairwise disjoint.
	\item Sets of 2-cells and equations, as described below.
\end{itemize}
Before we attempt a formal description of the 2-cells and equations,
we shall introduce the part of the language that describes 1-cells.

\subsection{1-cells}
A 1-cell will be described by a \emph{1-cell sequent}: suppose
we are considering a theory with objects $\A, \CB, \C$ and
1-cells $f: (\A,\CB)\to\C$ and $g:\C\to\C$. Then the arrow
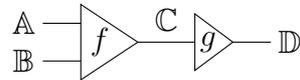
\[\begin{tikzpicture}[scale=0.25]
	\draw (0,-2) -- (0,2) -- (3,0) -- cycle ;
		\node at (1,0) {$f$} ;
	\node at (-3,1)  {$\A$}  ; \draw (-2,1) -- (0,1) ;
	\node at (-3,-1) {$\CB$} ; \draw (-2,-1) -- (0,-1) ;
	\draw (3,0) -- (6,0)
		node[pos=0.5,auto] {$\C$} ;
	\draw (6,-1.5) -- (6,1.5) -- (8,0) -- cycle ;
		\node at (6.7,0) {$g$} ;
	\draw (8,0) -- (10,0) ;
		\node at (11,0) {$\D$} ;
\end{tikzpicture}\]
might be represented by the sequent
\[
	A\in\A, B\in\CB \proves g(f(A,B)) \in \D.
\]
The names in the context -- to the left of the turnstile -- should
be regarded as bound, so for example the sequent
\[
	X\in\A, Y\in\CB \proves g(f(X,Y)) \in \D
\]
is equivalent to the one above, modulo the renaming of bound
variables, and represents the same 1-cell. (We assume
throughout that we have some infinite set of names
on which to draw. Names here will be represented
by upper-case italic letters.)
Formally, the derivation rules for 1-cell sequents are
as follows:
\begin{itemize}
\item The axiom rule: for every $\C\in\T_{0}$,
\[
	\begin{prooftree}
		\justifies A\in\C \proves A\in\C \using (\C)
	\end{prooftree}
\]
\item The application rule: for every $f\in\T(\A_{1},\dots,\A_{n}; \CB)$,
\[
	\begin{prooftree}
		\Gamma_{1}\proves \alpha_{1}\in \A_{1}
		\quad
		\cdots
		\quad
		\Gamma_{n}\proves \alpha_{n}\in \A_{n}
		\justifies
		\Gamma_{1},\dots,\Gamma_{n} \proves f(\alpha_{1},\dots,\alpha_{n})\in\CB
		\using f(\bullet)
	\end{prooftree}
\]
where the sets of names in the $\Gamma_{i}$ are assumed pairwise disjoint.
\end{itemize}
(Whenever several contexts are mentioned together, we shall \emph{always} assume
that their sets of names are pairwise disjoint, usually without explicitly mentioning
this assumption.)
So a 1-cell sequent is simply a formal composite of formal 1-cells.
It is clear (and easily proved by induction) that every derivable 1-cell
sequent has a unique derivation. Also:
\begin{definition}
	Let $A_{1}\in\A_{1}, \cdots, A_{n}\in\A_{n}\proves \beta\in\CB$
	be a derivable 1-cell sequent; and for every $1\leq i\leq n$, let
	$\Gamma_{i} \proves \alpha_{i}\in\A_{i}$ be a derivable 1-cell sequent.
	Then we write $[A_{i} := \alpha_{i}]_{i=1}^{n}$ to denote the multiple
	substitution
	\[
		[A_{1} := \alpha_{1}, \cdots, A_{n}:= \alpha_{n}].
	\]
	On occasion this is abbreviated to $[A_{i} := \alpha_{i}]_{i}$,
	where the value of $n$ is apparent from the context.
	
	Later, we sometimes need to use nested sequences: if
	\[
		(A^{1}_{1}, \cdots, A^{1}_{n_{i}}),
		\dots,
		(A^{n}_{1}, \cdots, A^{n}_{n_{n}})
	\]
	is a nested sequence of objects, with $\alpha^{i}_{j}$ being
	a similarly-numbered nested sequence of 1-cell expressions, then
	\[
		[[A^{i}_{j} := \alpha^{i}_{j}]_{j=1}^{n_{i}}]_{i=1}^{n}
	\]
	denotes the substitution
	\[
		[A^{1}_{1} := \alpha^{1}_{1}, \cdots, A^{1}_{n_{1}}:= \alpha^{1}_{n_{1}},
		\enskip\cdots,\enskip
		A^{n}_{1} := \alpha^{n}_{1}, \cdots, A^{n}_{n_{n}}:= \alpha^{n}_{n_{n}}
		].
	\]
\end{definition}
\begin{lemma}[1-cell substitution]
	If $A_{1}\in\A_{1}, \cdots, A_{n}\in\A_{n}\proves \beta\in\CB$
	is a derivable 1-cell sequent, and for every $1\leq i\leq n$,
	$\Gamma_{i} \proves \alpha_{i}\in\A_{i}$ is a derivable 1-cell sequent,
	with the names in the $\Gamma_{i}$ pairwise disjoint, then
	\[
		\Gamma_{i},\cdots,\Gamma_{n}\proves \beta[A_{i} := \alpha_{i}]_{i=1}^{n}
	\]
	is a derivable 1-cell sequent,
\end{lemma}
\begin{proof}
	Consider a derivation of $\beta$, and make the substitution
	throughout the derivation, as follows. For each sequent
	$\Gamma\proves \gamma\in\C$ in the derivation, replace $\gamma$
	by $\gamma[A_{i} := \alpha_{i}]_{i}$, and replace each name
	$A_{i}\in\A_{i}$ of $\Gamma$ by the contents of $\Gamma_{i}$.
	Each application rule is thus transformed
	into another instance of that application rule, and each axiom
	$A_{i}\in\A_{i} \proves A_{i}\in\A_{i}$
	is transformed into $\Gamma_{i}\proves \alpha_{i}\in\A_{i}$, which is
	derivable by assumption.
\end{proof}

\subsection{2-cells}
Now we can define the 2-cells of $\T$. In addition to the
objects and 1-cells, we have:
for every pair $\Gamma\proves\alpha\in\CB$, $\Gamma\proves\beta\in\CB$
of derivable 1-cell sequents, a set $\T^{\Gamma}_{\CB}[\alpha,\beta]$
of 2-cells. Of course, we take these to be invariant under renaming,
so if $\sigma$ is a permutation of the set of names then we require
\[
	\T^{\Gamma^{\sigma}}_{\CB}[\alpha^{\sigma},\beta^{\sigma}] = \T^{\Gamma}_{\CB}[\alpha,\beta].
\]
Any two \emph{different} sets of 2-cells (i.e.\ sets that are not presumed equal by the above)
are taken to be disjoint.
A 2-cell sequent is of the form $\Gamma\proves \phi: \alpha\to\beta \in\CB$.
The derivation rules for 2-cell sequents are:
\begin{itemize}
\item The identity rule:
	\[\begin{prooftree}
		\Gamma\proves \alpha\in\A
		\justifies
		\Gamma\proves 1_{\alpha}: \alpha\to\alpha\in\A
		\using 1
	\end{prooftree}\]
\item The axiom rule: for every $t\in \T^{\Gamma}_{\CB}[\alpha,\beta]$,
	with $\Gamma = (A_{1}\in\A_{1}, \dots, A_{n}\in\A_{n})$,
	\[\begin{prooftree}
		\Gamma_{1}\proves\gamma_{1}\in\A_{1}
		\quad\cdots\quad
		\Gamma_{n}\proves\gamma_{n}\in\A_{n}
		\justifies
		\Gamma_{1}\cdots\Gamma_{n} \proves t_{\gamma_{1},\dots, \gamma_{n}}
			: \alpha[A_{i} := \gamma_{i}]_{i}
			\to\beta[A_{i} := \gamma_{i}]_{i}
			\in \CB
		\using t_{\bullet} 
	\end{prooftree}\]
	where the sets of names in the $\Gamma_{i}$ are pairwise disjoint.
\item The composition rule:
\[\begin{prooftree}
	\Gamma\proves \phi: \beta\to\gamma \in\A
	\quad
	\Gamma\proves \psi: \alpha\to\beta \in\A
	\justifies
	\Gamma\proves \phi\cdot\psi: \alpha\to\gamma \in\A
	\using \mathrm{comp}
\end{prooftree}\]
\item The 1-cell application rule: for every $f\in\T(\A_{1},\dots,\A_{n};\CB)$,
	\[\begin{prooftree}
		\Gamma_{1}\proves \phi_{1} : \alpha_{1}\to\beta_{1} \in \A_{1}
		\quad \cdots \quad
		\Gamma_{n}\proves \phi_{n} : \alpha_{n}\to\beta_{n} \in \A_{n}
		\justifies
		\Gamma_{1},\cdots,\Gamma_{n}\proves f(\phi_{1},\dots,\phi_{n})
			: f(\alpha_{1},\dots,\alpha_{n}) \to f(\beta_{1},\dots,\beta_{n})
			\in\CB
		\using f(\to)
	\end{prooftree}\]
	The names in the $\Gamma_{i}$ are again required to be pairwise disjoint.
\end{itemize}
Since the conclusion of each rule has a distinct syntactic form, each derivable
2-cell sequent has a unique derivation.
Notice that, if $\Gamma\proves \phi:\alpha\to\beta\in\CB$ is a derivable
2-cell sequent, then $\Gamma\proves\alpha\in\CB$ and $\Gamma\proves\beta\in\CB$
are derivable 1-cell sequents.
We also observe that substituting a 1-cell expression in a 2-cell expression,
or vice versa, results in a derivable 2-cell sequent:
\begin{lemma}[1-in-2 substitution]\label{lemma-1in2}
	Let \( \Delta \proves \phi:\alpha\to\beta \in\C \)
	be a derivable 2-cell sequent, with
	\[
		\Delta = B_{1}\in\CB_{1}, \cdots, B_{n}\in\CB_{n}.
	\]
	For each $1\leq i\leq n$, let
	\[
		\Gamma_{i} \proves \gamma_{i}\in\CB_{i}
	\]
	be a derivable 1-cell sequent. Then the 2-cell sequent
	\[ \Gamma_{1}, \cdots, \Gamma_{n}
		\proves \phi[B_{i}:=\gamma_{i}]_{i}
		: \alpha[B_{i}:=\gamma_{i}]_{i}\to\beta[B_{i}:=\gamma_{i}]_{i}
		\in \C, \]
	is also derivable,
	assuming that the sets of names in the $\Gamma_{i}$ and $\Delta$
	are pairwise disjoint.
\end{lemma}
\begin{proof}
	Consider a derivation of \( \Delta \proves \phi:\alpha\to\beta \in\C \),
	and make the substitution throughout the derivation. In detail: for
	every sequent in the derivation, in the context replace each $B_{i}\in\CB_{i}$
	by $\Gamma_{i}$ and in the expression replace each $B_{i}$ by $\gamma_{i}$.
	For each derivation rule, this
	substitution results in another instance of the same rule, and the
	1-cell substitution lemma ensures that every 1-cell sequent in the
	derivation remains derivable. The conclusion of this new derivation
	is
	\[ \Gamma_{1}, \cdots, \Gamma_{n}
		\proves \phi[B_{i}:=\gamma_{i}]_{i}
		: \alpha[B_{i}:=\gamma_{i}]_{i}\to\beta[B_{i}:=\gamma_{i}]_{i}
		\in \C, \]
	which is therefore derivable.
\end{proof}
\begin{lemma}[2-in-1 substitution]\label{lemma-2in1}
	Let $\Gamma\proves\gamma\in\C$ be a derivable 1-cell sequent, for
	some \[ \Gamma = (A_{1}\in\A_{1},\dots, A_{n}\in\A_{n}). \]
	If the 2-cell sequents
	\[
	\Delta_{1}\proves \phi_{1} : \alpha_{1}\to\beta_{1} \in \A_{1}
	\quad \cdots \quad
	\Delta_{n}\proves \phi_{n} : \alpha_{n}\to\beta_{n} \in \A_{n}
	\]
	are all derivable, then so is
	\[
	\Delta_{1},\cdots,\Delta_{n}\proves \gamma[A_{i} := \phi_{i}]_{i}
		: \gamma[A_{i} := \alpha_{i}]_{i} \to \gamma[A_{i} := \beta_{i}]_{i}
		\in\CB.
	\]
	The names in the $\Delta_{i}$ must be pairwise disjoint, of course.
\end{lemma}
\begin{proof}
	\def\<#1>{\langle#1\rangle}
	This is simply a matter of repeatedly applying the 1-cell application
	rule. Formally,
	the proof is by induction on $\gamma$.
	If $\gamma$ is a constant, i.e.\ $\Gamma$ is the singleton $C\in\C$ and $\gamma$
	is just $C$, then the conclusion is equal to the hypothesis and the claim
	is trivial. Suppose, then, that $\gamma = f(\gamma_{1}, \dots, \gamma_{m})$ for some
	$m\in\N$, 1-cell symbol $f\in\T(\CB_{1}, \dots, \CB_{m}; \C)$ and
	1-cell expressions $\gamma_{i}$.
	
	Clearly the context $\Gamma$ must be equal to
	$(\Gamma_{1}, \cdots, \Gamma_{m})$, where each $\Gamma_{i}$ contains
	the names used in $\gamma_{i}$.
	We want to reindex the names so that it is clear to which $\gamma_{i}$
	each belongs, for which we introduce a simple technical device that is
	also used in several other proofs below.
	For each $1\leq i\leq m$, let $m_{i}$ be the number of names in
	$\Gamma_{i}$. Then, given $1\leq i \leq m$ and $1\leq j\leq m_{i}$,
	define $\<i,j>$ to be $j+\sum_{r<i}m_{r}$. The effect of this definition
	is that
	\[
		\Gamma_{i} = (A_{\<i,1>} \in \A_{\<i,1>}, \cdots, A_{\<i,m_{i}>} \in \A_{\<i,m_{i}>})
	\]
	or, in abbreviated notation, that
	\[
		\Gamma_{i} = (A_{\<i,j>} \in \A_{\<i,j>})_{j=1}^{m_{i}}
	\]
	The inductive hypothesis implies that, for each $1\leq i\leq m$, the
	2-cell sequent
	\[
	\Gamma_{i}\proves \gamma_{i}[A_{\<i,j>} := \phi_{\<i,j>}]_{j=1}^{m_{i}}
		: \gamma_{i}[A_{\<i,j>} := \alpha_{\<i,j>}]_{j=1}^{m_{i}}
		\to \gamma_{j}[A_{\<i,j>} := \beta_{\<i,j>}]_{j=1}^{m_{i}}
		\in\CB_{i}.
	\]
	is derivable. Now apply the 1-cell application rule to these sequents and
	the 1-cell symbol $f$, and the conclusion follows.
\end{proof}

\subsection{Equations}
The final ingredient that we need to complete our description
of $\T$ is a collection of equations between 2-cells.
Formally, we declare that for each pair
$\Gamma\proves\alpha\in\A$ and $\Gamma\proves\beta\in\A$ of
1-cell sequents, there is a set that we denote
$\T_{=}(\Gamma\proves \alpha\to\beta\in\A)$. This set consists
of pairs $(\phi,\psi)$,
where $\Gamma\proves\phi:\alpha\to\beta\in\A$ and
$\Gamma\proves\psi:\alpha\to\beta\in\A$ are derivable
2-cell sequents.
We also require the collection of equations to be closed under renaming,
so if $\sigma$ is a permutation of the set of names then
\[
	(\phi,\psi)\in\T_{=}(\Gamma\proves \alpha\to\beta\in\A)
	\iff 
	(\phi^{\sigma},\psi^{\sigma})\in\T_{=}(\Gamma^{\sigma}\proves \alpha^{\sigma}\to\beta^{\sigma}\in\A).
\]
In fact this restriction is not strictly necessary, since the
axiom rule for equations (below) permits any 1-in-2 substitution instance
of an equation to be used, but we retain it for the sake of
consistency.

The derivation rules for equations are as follows.
There are three rules expressing
the fact that any equality worth the name should be
reflexive, symmetric, and transitive:
\begin{itemize}
	\item For every derivable 2-cell sequent
	$\Gamma\proves\phi:\alpha\to\beta\in\A$, we have
	reflexivity:
	\[\begin{prooftree}
		\justifies
		\Gamma\proves\phi=\phi:\alpha\to\beta\in\A
		\using=_{r}
	\end{prooftree}\]
	\item symmetry:
	\[\begin{prooftree}
		\Gamma\proves\phi=\psi:\alpha\to\beta\in\A
		\justifies
		\Gamma\proves\psi=\phi:\alpha\to\beta\in\A
		\using=_{s}
	\end{prooftree}\]
	\item and transitivity:
	\[\begin{prooftree}
		\Gamma\proves\psi=\chi:\alpha\to\beta\in\A
		\quad
		\Gamma\proves\phi=\psi:\alpha\to\beta\in\A
		\justifies
		\Gamma\proves\phi=\chi:\alpha\to\beta\in\A
		\using=_{t}
	\end{prooftree}\]
\end{itemize}
There is an axiom rule for equations, which allows us to use any 1-in-2 substitution
instance of an equation axiom:
\begin{itemize}
	\item For every $(\Delta,\A,\alpha,\beta,\phi,\psi)\in\T_{=}$,
	with $\Delta = (B_{1}\in\CB_{1}, \cdots, B_{n}\in\CB_{n})$,
	\[\begin{prooftree}
		\Gamma_{1} \proves \gamma_{1}\in\CB_{1}
		\quad\cdots\quad
		\Gamma_{n} \proves \gamma_{n}\in\CB_{n}
		\justifies
		\Gamma_{1}\cdots\Gamma_{n}
			\proves \phi[B_{i}:=\gamma_{i}]_{i} = \psi[B_{i}:=\gamma_{i}]_{i}
			: \alpha[B_{i}:=\gamma_{i}]_{i} \to \beta[B_{i}:=\gamma_{i}]_{i}
			\in\A
		\using (\phi=\psi)
	\end{prooftree}\]
\end{itemize}
The next two rules express the idea that the 2-cell construction rules
should preserve equality:
\begin{itemize}
\item The composition rule preserves equality:
\[\begin{prooftree}
	\Gamma\proves \phi=\phi': \beta\to\gamma \in\A
	\quad
	\Gamma\proves \psi=\psi': \alpha\to\beta \in\A
	\justifies
	\Gamma\proves \phi\cdot\psi = \phi'\cdot\psi': \alpha\to\gamma \in\A
	\using\mathrm{comp}
\end{prooftree}\]
\item The 1-cell application rule preserves equality: for every
	$f\in\T(\A_{1},\dots,\A_{n};\CB)$,
	\[\begin{prooftree}
		\Gamma_{1}\proves \phi_{1}=\phi'_{1} : \alpha_{1}\to\beta_{1} \in \A_{1}
		\quad \cdots \quad
		\Gamma_{n}\proves \phi_{n}=\phi'_{n} : \alpha_{n}\to\beta_{n} \in \A_{n}
		\justifies
		\Gamma_{1},\cdots,\Gamma_{n}
			\proves f(\phi_{1},\dots,\phi_{n}) = f(\phi'_{1},\dots,\phi'_{n})
			: f(\alpha_{1},\dots,\alpha_{n}) \to f(\beta_{1},\dots,\beta_{n})
			\in\CB
		\using f(\to)
	\end{prooftree}\]
\end{itemize}
Next, there are equations expressing the fact that 1-cell application
preserves identities and composition:
\begin{itemize}
\item 1-cell application preserves identities: for every
	$f\in\T(\A_{1},\dots,\A_{n};\CB)$,
	\[\begin{prooftree}
		\Gamma_{1}\proves \alpha_{1}\in \A_{1}
		\quad\cdots\quad
		\Gamma_{n}\proves \alpha_{n}\in \A_{n}
		\justifies
		\Gamma_{1},\cdots,\Gamma_{n} \proves
			f(1_{\alpha_{1}}, \dots, 1_{\alpha_{n}}) = 1_{f(\alpha_{1}, \dots, \alpha_{n})}
			\in\CB
		\using f(1)
	\end{prooftree}\]
\item 1-cell application preserves composition: for every
	$f\in\T(\A_{1},\dots,\A_{n};\CB)$,
	\[\scalebox{0.75}{\begin{prooftree}
		\Gamma_{1}\proves\phi_{1}: \beta_{1}\to\gamma_{1}\in\A_{1}
			\enskip \Gamma_{1}\proves\psi_{1}: \alpha_{1}\to\beta_{1}\in\A_{n}
		\quad\dots\quad
		\Gamma_{n}\proves\phi_{n}: \beta_{n}\to\gamma_{n}\in\A_{n}
			\enskip \Gamma_{n}\proves\psi_{n}: \alpha_{n}\to\beta_{n}\in\A_{n}
		\justifies
		\Gamma_{1},\cdots,\Gamma_{n}
		\proves f(\phi_{1},\dots,\phi_{n})\cdot f(\psi_{1},\dots,\psi_{n})
			= f(\phi_{1}\cdot\psi_{1},\dots,\phi_{n}\cdot\psi_{n})
			: f(\alpha_{1}, \dots, \alpha_{n})
			\to f(\gamma_{1},\dots, \gamma_{n})
			\in\CB
		\using f(\cdot)
	\end{prooftree}}\]
\end{itemize}
The identity 2-cells should be units for composition:
\begin{itemize}
\item Left identity:
	\[\begin{prooftree}
		\Gamma\proves \phi:\alpha\to\beta\in\CB
		\justifies
		\Gamma\proves 1_{\beta}\cdot\phi = \phi: \alpha\to\beta\in\CB
		\using l_{\beta}
	\end{prooftree}\]
\item Right identity:
	\[\begin{prooftree}
		\Gamma\proves \phi:\alpha\to\beta\in\CB
		\justifies
		\Gamma\proves \phi\cdot1_{\alpha} = \phi: \alpha\to\beta\in\CB
		\using r_{\beta}
	\end{prooftree}\]
\end{itemize}
Finally, we have:
\begin{itemize}
\item The naturality rule:
	for every $t\in \T^{\Gamma}_{\CB}[\alpha,\beta]$,
	with $\Gamma = (A_{1}\in\A_{1}, \dots, A_{n}\in\A_{n})$,
	\begin{mspill}\begin{prooftree}
		\Gamma_{1}\proves\phi_{1}:\gamma_{1}\to\delta_{1}\in\A_{1}
		\quad\cdots\quad
		\Gamma_{n}\proves\phi_{n}:\gamma_{n}\to\delta_{n}\in\A_{n}
		\justifies
		\Gamma_{1}\cdots\Gamma_{n}
			\proves \beta[A_{i} := \phi_{i}]_{i} \cdot t_{\gamma_{1},\dots, \gamma_{n}}
				= t_{\delta_{1},\dots, \delta_{n}} \cdot \alpha[A_{i} := \phi_{i}]_{i}
				: \alpha[A_{i} := \gamma_{i}]_{i} \to\beta[A_{i}:=\delta_{i}]_{i} \in \CB
		\using t_{\natural} 
	\end{prooftree}\end{mspill}
	where the sets of names in the $\Gamma_{i}$ are pairwise disjoint.
\end{itemize}
The following proposition is just a sanity check of the equation rules.
\begin{propn}
	If \[\Gamma\proves\phi=\psi: \alpha\to\beta\in\CB\] is a derivable equation
	sequent, then $\Gamma\proves\phi: \alpha\to\beta\in\CB$
	and $\Gamma\proves\psi: \alpha\to\beta\in\CB$
	are derivable 2-cell sequents.
\end{propn}
\begin{proof}
	An easy induction over the derivation, using
	Lemma~\ref{lemma-1in2} for the axiom rule and
	Lemmas~\ref{lemma-1in2} and~\ref{lemma-2in1} for the naturality rule.
\end{proof}

\section{Interpretation in a Gray monoid}
To interpret these sequents in a target monoidal bicategory $\B$, we need
an interpretation, or model, of $\T$. This section defines the interpretation
in a Gray monoid, so suppose for now that $\B$ is a Gray monoid.
A model $v:\T\to\B$ consists of:
\begin{itemize}
	\item for every $\A\in\T_{0}$, an object $v(\A)$ in $\B$;
	\item for every $f\in\T(\A_{1},\dots,\A_{n};\CB)$, a 1-cell
	\[
		v(f): v(\A_{1})\tn\cdots\tn v(\A_{n})\to v(\CB);
	\]
	in $\B$.
	\item for every $t\in\T^{\Gamma}_{\CB}[\alpha, \beta]$, a 2-cell
	\[
		v(t): \llbracket \Gamma\proves\alpha\in\CB \rrbracket_{v} \To
			\llbracket \Gamma\proves\beta\in\CB \rrbracket_{v}
	\]
	where the function $\llbracket-\rrbracket_{v}$ is as defined for 1-cell
	sequents below,
	\item such that for every $(\Gamma,\alpha,\beta,\phi,\psi)\in\T_{=}$
	we have
	\[
		\llbracket \Gamma\proves\phi:\alpha\to\beta \rrbracket_{v}
		=
		\llbracket \Gamma\proves\psi:\alpha\to\beta \rrbracket_{v}
	\]
	where the function $\llbracket-\rrbracket_{v}$ is as defined for 2-cell
	sequents below.
\end{itemize}
If $\Gamma = (A_{1}\in\A_{1}, \cdots, A_{n}\in\A_{n})$ is a context, we write
$v(\Gamma)$ as an abbreviation for $v(\A_{1})\tn\cdots\tn v(\A_{n})$.
\subsection{Interpretation of 1-cells}
The semantic interpretation $\semint-$ of a 1-cell
derivation is defined by induction:
\begin{itemize}
\item $\Bigl\llbracket	\begin{prooftree}
	\justifies A\in\C \proves A\in\C \using (\C)
\end{prooftree} \Bigr\rrbracket_{v} = 1_{v(\C)}$,
\item $\left\llbracket\begin{prooftree}
		\[[\pi_{1}]\justifies\Gamma_{1}\proves \alpha_{1}\in \A_{1}\]
		\quad\cdots\quad
		\[[\pi_{n}]\justifies\Gamma_{n}\proves \alpha_{n}\in \A_{n}\]
		\justifies
		\Gamma_{1},\dots,\Gamma_{n} \proves f(\alpha_{1},\dots,\alpha_{n})\in\CB
		\using f(\bullet)
	\end{prooftree}\right\rrbracket_{v}$
	\newline\vskip1ex\strut\hfil$= v(f)\o\Bigl(
	\biggl\llbracket\begin{prooftree}
		[\pi_{1}]\justifies\Gamma_{1}\proves \alpha_{1}\in \A_{1}
	\end{prooftree}\biggr\rrbracket_{v}
	\tn\cdots\tn
	\biggl\llbracket\begin{prooftree}
		[\pi_{n}]\justifies\Gamma_{n}\proves \alpha_{n}\in \A_{n}
	\end{prooftree}\biggr\rrbracket_{v}
	\Bigr)$.
\end{itemize}
Since a derivable 1-cell sequent has a unique derivation, we may
regard the input to the interpretation function as a sequent rather
than a derivation. Clearly $\semint{\Gamma\proves \alpha\in\CB}$
is always a 1-cell $v(\Gamma)\to v(\CB)$.
Where $\Gamma$ and $\CB$ are evident from the context, we shall
abbreviate $\semint{\Gamma\proves \alpha\in\CB}$ to $\semint{\gamma}$.

\subsection{Semantics of substitution}\label{s-semantics-of-substitution}
In order to define the interpretation of 2-cells (and, later, to
show soundness) we shall need a careful analysis of the semantics
of 1-cell substitution. The reason this is non-trivial is that the
interchange law of a Gray monoid does not hold on the nose: there
is some work to be done to account for this.

\begin{definition}
	Given arrows $f_{1}, \dots, f_{n}$ of a Gray monoid, we write
	$\Tn_{i=1}^{n}f_{i}$ to mean \[f_{1}\tn\cdots\tn f_{n}.\]
\end{definition}
\begin{definition}
	Given arrows $f_{1}, \dots, f_{n}$ and $g_{1}, \dots, g_{n}$ of
	a Gray monoid, let $\Ic_{i=1}^{n}(f_{i},g_{i})$ denote the
	interchange isomorphism
	\[
		\Tn_{i=1}^{n}f_{i}\o\Tn_{i=1}^{n}g_{i}\to\Tn_{i=1}^{n}(f_{i}\o g_{i}).
	\]
	If $((f^i_j)_{j=1}^{n_{i}})_{i=1}^{n}$
	and $((g^i_j)_{j=1}^{n_{i}})_{i=1}^{n}$
	are nested sequences of 1-cells, then the interchange isomorphism
	\[
		\Tn_{i=1}^{n}\Tn_{j=1}^{n^{i}_{j}}f^{i}_{j}\o\Tn_{i=1}^{n}\Tn_{j=1}^{n^{i}_{j}}g^{i}_{j}
		\to \Tn_{i=1}^{n}\Tn_{j=1}^{n^{i}_{j}}(f^{i}_{j}\o g^{i}_{j})
	\]
	is denoted $(\Ic_{j=1}^{n_{i}})_{i=1}^{n}(f^i_j,g^i_j)$.
\end{definition}
\begin{definition}
	Now let $\Gamma\proves \beta\in\C$ be a derivable 1-cell sequent, where
	\[\Gamma = (B_{i}\in\CB_{i})_{i=1}^{n},\] and for each $1\leq i\leq n$
	let $\Delta_{i}\proves \alpha_{i}\in\CB_{i}$ be a derivable 1-cell sequent.
	We shall explicitly define a `normalisation' isomorphism
	\[
		\norm_{\C}^{\Gamma}((\alpha_{i})_{i=1}^{n},\beta)
			: \semint{\beta}\o\Tn_{i=i}^{n}\semint{\alpha_{i}}
			\rTo^{\cong}
			\semint{\beta[B_{i}:=\alpha_{i}]_{i=1}^{n}}
	\]
	The definition is by recursion over $\beta$. If $\beta = B_{1}$, then
	we use the identity. So suppose that $\beta = f(\beta_{1}, \dots, \beta_{m})$
	for some natural number $m$ and 1-cell symbol $f\in\T(\C_{1},\cdots,\C_{m};\D)$.
	For each $1\leq i\leq m$, let $m_{i}$ denote the
	number of names that occur in $\beta_{i}$. Let $B^{i}_{j}$
	denote $B_{j+\sum_{r<i}m_{r}}$ and $\alpha^{i}_{j}$ denote
	$\alpha_{j+\sum_{r<i}m_{r}}$. Thus $(B_{i})_{i=1}^{n}$ is divided
	into the nested sequence $((B^{i}_{j})_{j=1}^{m_{i}})_{i=1}^{m}$,
	where the inner sequence $(B^{i}_{j})_{j=1}^{m_{i}}$ contains just
	the names that occur in $\beta_{i}$. In a similar way, divide the
	context $\Gamma$ into $(\Gamma_{1}, \cdots, \Gamma_{m})$ so that
	each $\Gamma_{i}$ contains the names that occur in $\beta_{i}$,
	i.e.\[\Gamma_{i} = (B^{i}_{j}\in\B^{i_{j}})_{j=1}^{m_{i}},\] where
	the nested sequence $(\B^{i}_{j})$ is defined in the obvious way.

	Now, define $\norm^{\Gamma}_{\D}((\alpha_{i})_{i=1}^{n},\beta)$
	to be the composite
	\begin{diagram}
		v(f) \o \Tn_{i=1}^{m}\semint{\beta_{i}} \o \Tn_{i=1}^{m}\Tn_{j=1}^{m_{i}}\semint{\alpha^{i}_{j}}
		\\
		\dTo>{v(f) \o \Ic_{i=1}^{m}( \semint{\beta_{i}}, \Tn_{j=1}^{m_{j}}\semint{\alpha^{i}_{j}} )}
		\\
		v(f) \o \Tn_{i=1}^{m}( \semint{\beta_{i}} \o \Tn_{j=1}^{m_{i}}\semint{\alpha^{i}_{j}} )
		\\
		\dTo>{v(f) \o \Tn_{i=i}^{m} \norm^{\Gamma_{i}}_{\C_{i}}((\alpha^{i}_{j})_{j=1}^{m_{i}}, \beta_{i})}
		\\
		v(f) \o \Tn_{i=1}^{m}\semint{\beta_{i}[B^{i}_{j} := \alpha^{i}_{j}]_{j=1}^{m_{i}}}
	\end{diagram}
	This completes the recursive definition.
\end{definition}
We shall need the fact that the $\norm$ operation has a kind of
associativity property, described in the following Proposition.
This is absolutely fundamental, since it is used to show that Gray monoids
have enough coherence for our language to be soundly interpretable.
Fortunately it follows immediately from the coherence theorem of
Chapter~\refchapter{Coh}.
\begin{propn}[Double norm]\label{prop-double-norm}
	Let $\Gamma\proves\gamma\in\D$ be a derivable 1-cell sequent,
	where $\Gamma = (C_{i}\in\C_{i})_{i=1}^{n}$. For each $1\leq i\leq n$,
	let $\Delta_{i}\proves\beta_{i} \in\C_{i}$ be a derivable 1-cell sequent,
	where $\Delta_{i} = (B^{i}_{j}\in\CB^{i}_{j})_{j=1}^{n_{i}}$. For each $1\leq i\leq n$
	and $1\leq j\leq n_{i}$, let $\Xi^{i}_{j}\proves\alpha^{i}_{j}\in\CB^{i}_{j}$
	be a derivable 1-cell sequent. Then the diagram
	\[\scalebox{0.75}{\begin{diagram}
		\semint{\gamma}
			\o \Tn_{i=1}^{n}\semint{\beta_{i}}\o\Tn_{i=1}^{n}\Tn_{j=1}^{n_{i}}\semint{\alpha^{i}_{j}}
			& \rTo^{\semint{\gamma}
				\o \Ic_{i=1}^{n}(\semint{\beta_{i}}, \Tn_{j=1}^{n_{i}}\semint{\alpha^{i}_{j}})}
			& \semint{\gamma} \o \Tn_{i=1}^{n}( \semint{\beta_{i}}
				\o \Tn_{j=1}^{n_{i}}\semint{\alpha^{i}_{j}} ) \\
		\dTo[uppershortfall=-3pt,lowershortfall=0pt]>{\norm^{\Gamma}_{\D}\bigl((\beta_{i})_{i=1}^{n}, \gamma\bigr)
		 	\o \Tn_{i=1}^{n}\Tn_{j=1}^{n_{i}}\semint{\alpha^{i}_{j}}}
			& \ruTo[crab=-2em,shortfall=0pt](0,2)<{\semint{\gamma}
				\o \Tn_{i=1}^{n}\norm^{\Delta_{i}}_{\C_{i}}\bigl((\alpha^{i}_{j})_{j=1}^{n_{i}}, \beta_{i}\bigr)} \\
		\semint{\gamma[C_{i} := \beta_{i}]_{i=1}^{n}}
			\o \Tn_{i=1}^{n}\Tn_{j=1}^{n_{i}}\semint{\alpha^{i}_{j}}
			&& \semint{\gamma} \o \Tn_{i=1}^{n}\semint{\beta_{i}[B^{i}_{j} := \alpha^{i}_{j}]_{j=1}^{n_{i}}}
			\\
		\dTo[uppershortfall=-3pt]>{
			\norm^{(\Delta_{i})_{i=1}^{n}}_{\D}\bigl(((\alpha^{i}_{j})_{j=1}^{n_{i}})_{i=1}^{n},
				\gamma[C_{i} := B_{i}]_{i=1}^{n}\bigr)}
			& \ruTo[crab=-2em,shortfall=0pt](0,2)<{\norm^{\Gamma}_{\D}\bigl((\beta_{i}[B^{i}_{j} := \alpha^{i}_{j}]_{j=1}^{n_{i}})_{i=1}^{n}, \gamma\bigr)} \\
		\semint{\gamma[C_{i} := \beta_{i}]_{i=1}^{n}[[B^{i}_{j} := \alpha^{i}_{j}]_{j=1}^{n_{i}}]_{i=1}^{n}}
			& \rTo_{=}
			& \semint{\gamma[C_{i} := \beta_{i}[B^{i}_{j} := \alpha^{i}_{j}]_{j=1}^{n_{i}}]_{i=1}^{n}}
	\end{diagram}}\]
	commutes.
\end{propn}
\begin{proof}
	Immediate from \chref{Coh}{thm-coh}.
\end{proof}

\subsection{Interpretation of 2-cells}\label{s-interpretation-of-2-cells}
The semantic interpretation of a 2-cell derivation is also defined by induction.
As for 1-cells, we define the interpretation on derivations, but since each
derivable 2-cell sequent has a unique derivation the interpretation may be
applied directly to derivable sequents. Also, we sometimes omit the type
information when it is obvious from the context.
\begin{itemize}
\item
	$\ssemint[bigg]{v}{\begin{prooftree}
		\Gamma\proves \alpha\in\A
		\justifies
		\Gamma\proves 1_{\alpha}: \alpha\to\alpha\in\A
		\using 1
	\end{prooftree}}
	= 1_{\semint{\Gamma\proves \alpha\in\A}}$,
\item
	$\ssemint[bigg]{v}{\begin{prooftree}
		\[[\pi_{1}]\justifies\Gamma\proves \phi: \beta\to\gamma \in\A\]
		\quad
		\[[\pi_{2}]\justifies\Gamma\proves \psi: \alpha\to\beta \in\A\]
		\justifies
		\Gamma\proves \phi\cdot\psi: \alpha\to\gamma \in\A
		\using \mathrm{comp}
	\end{prooftree}} = \semint{\phi}\cdot\semint{\psi}$.
\item $\ssemint[Bigg]{v}{\begin{prooftree}
		\Gamma_{1}\proves\gamma_{1}\in\A_{1}
		\quad\cdots\quad
		\Gamma_{n}\proves\gamma_{n}\in\A_{n}
		\justifies
		\Gamma_{1}\cdots\Gamma_{n} \proves t_{\gamma_{1},\dots, \gamma_{n}}
			: \alpha[A_{i} := \gamma_{i}]_{i} \to\beta[A_{i} := \gamma_{i}]_{i}
			\in \CB
		\using t_{\bullet} 
	\end{prooftree}}$
	\newline\vskip0ex
	is defined to be the composite
	\begin{diagram}
		\semint{\alpha[A_{i} := \gamma_{i}]_{i}}
			&& \semint{\beta[A_{i} := \gamma_{i}]_{i}}. \\
		\dTo<{\norm^{-1}} && \uTo>{\norm} \\
		\semint{\alpha}\o\Tn_{i=1}^{n}\semint{\gamma_{i}}
			&\rTo^{v(t)\o\Tn_{i=1}^{n}\semint{\gamma_{i}}}
				&\semint{\beta}\o\Tn_{i=1}^{n}\semint{\gamma_{i}}		
	\end{diagram}
\item $\ssemint[stretch]{v}{\begin{prooftree}
		\[[\pi_{1}]\justifies\Gamma_{1}\proves \phi_{1} : \alpha_{1}\to\beta_{1} \in \A_{1}\]
		\quad \cdots \quad
		\[[\pi_{n}]\justifies\Gamma_{n}\proves \phi_{n} : \alpha_{n}\to\beta_{n} \in \A_{n}\]
		\justifies
		\Gamma_{1},\cdots,\Gamma_{n}\proves f(\phi_{1},\dots,\phi_{n})
			: f(\alpha_{1},\dots,\alpha_{n}) \to f(\beta_{1},\dots,\beta_{n})
			\in\CB
		\using f(\to)
	\end{prooftree}}$
	\newline\vskip1ex\strut\hfil$=
	v(f) \o \Tn_{i=1}^{n}\semint{\phi_{i}}$.
\end{itemize}

\begin{propn}[Semantics of 1-in-2 substitution]\label{prop-sem-1in2}
	Let
	\[
		\Gamma_{1} \proves \gamma_{1}\in\CB_{1}
		\quad\cdots\quad
		\Gamma_{n} \proves \gamma_{n}\in\CB_{n}
	\]
	be derivable 1-cell sequents, and let
	\[
		\Delta\proves \phi: \alpha\to\beta\in\C
	\]
	be a derivable 2-cell sequent with
	$\Delta = (B_{1}\in\CB_{1}, \cdots, B_{n}\in\CB_{n})$.
	Then
	\[
		\semint{
			\Gamma_{1}\cdots\Gamma_{n}
				\proves \phi[B_{i}:=\gamma_{i}]_{i}
				: \alpha[B_{i}:=\gamma_{i}]_{i} \to \beta[B_{i}:=\gamma_{i}]_{i}
				\in\C
		}
	\]
	is equal to the composite
	\begin{diagram}
		\semint{\alpha[B_{i} := \gamma_{i}]_{i}}
			&& \semint{\beta[B_{i} := \gamma_{i}]_{i}}. \\
		\dTo<{\norm^{-1}} && \uTo>{\norm} \\
		\semint{\alpha}\o\Tn_{i=1}^{n}\semint{\gamma_{i}}
			&\rTo^{\semint{\phi}\o\Tn_{i=1}^{n}\semint{\gamma_{i}}}
				&\semint{\beta}\o\Tn_{i=1}^{n}\semint{\gamma_{i}}		
	\end{diagram}
\end{propn}
\begin{proof}
	\def\<#1>{\langle#1\rangle}
	The proof is by induction over the derivation of $\phi$:
	the non-trivial cases are 1-cell application and the axiom rule.
	For 1-cell application, let $\phi = f(\phi_{1}, \dots, \phi_{m})$.
	As before, let $m_{i}$ be the number of names occurring in $\phi_{i}$,
	and let $\<i,j> = j+\sum_{r<i}m_{i}$.
	Now consider the diagram
	\begin{diagram}[labelstyle=\scriptstyle]
		v(f)\o\Tn_{i=1}^{m}\semint{\alpha_{i}[B_{\<i,j>} := \gamma_{\<i,j>}]_{j=1}^{m_{i}}}
		& \rTo^{ v(f)\o\Tn_{i=1}^{m}\semint{\phi_{i}[B_{\<i,j>} := \gamma_{\<i,j>}]_{j=1}^{m_{i}}} }
		& v(f)\o\Tn_{i=1}^{m}\semint{\beta_{i}[B_{\<i,j>} := \gamma_{\<i,j>}]_{j=1}^{m_{i}}} \\
		\dTo<{v(f) \o \Tn_{i=1}^{m}\norm^{-1}((\gamma_{\<i,j>})_{j}, \alpha_{i})}
		&& \uTo<{v(f) \o \Tn_{i=1}^{m}\norm((\gamma_{\<i,j>})_{j}, \beta_{i})} \\
		v(f)\o\Tn_{i=1}^{m}\bigl(
				\semint{\alpha_{i}} \o \Tn_{j=1}^{m_{i}}\semint{\gamma_{\<i,j>}}
			\bigr)
			& \rTo^{v(f)\o\Tn_{i=1}^{m}\bigl(
				\semint{\phi_{i}} \o \Tn_{j=1}^{m_{i}}\semint{\gamma_{\<i,j>}}
			\bigr)}
			& v(f)\o\Tn_{i=1}^{m}\bigl(
				\semint{\beta_{i}} \o \Tn_{j=1}^{m_{i}}\semint{\gamma_{\<i,j>}}
			\bigr) \\
			\dTo<{v(f)\o \Ic_{i=1}^{m}(
					\semint{\alpha_{i}}, \Tn_{j=1}^{m_{i}}\semint{\gamma_{\<i,j>}}
				)^{-1}}
				&& \uTo<{v(f)\o \Ic_{i=1}^{m}(
					\semint{\beta_{i}}, \Tn_{j=1}^{m_{i}}\semint{\gamma_{\<i,j>}}
				)} \\
			v(f) \o \Tn_{i=1}^{m}\semint{\alpha_{i}}
					\o \Tn_{i=1}^{m}\Tn_{j=1}^{m_{i}}\semint{\gamma_{\<i,j>}}
				& \rTo_{v(f) \o \Tn_{i=1}^{m}\semint{\phi_{i}}
					\o \Tn_{i=1}^{m}\Tn_{j=1}^{m_{i}}\semint{\gamma_{\<i,j>}}}
				& v(f) \o \Tn_{i=1}^{m}\semint{\beta_{i}}
					\o \Tn_{i=1}^{m}\Tn_{j=1}^{m_{i}}\semint{\gamma_{\<i,j>}}
	\end{diagram}
	where the upper square commutes by definition, and the lower square
	commutes by naturality of $\Ic$. Thus the outside commutes: the left-hand
	vertical edge is equal, by definition, to $\norm((\gamma_{i})_{i=1}^{n}, \alpha)^{-1}$,
	and similarly the right-hand edge is equal to $\norm((\gamma_{i})_{i=1}^{n}, \beta)$.
	The top edge is equal, again by definition, to $\semint{\phi[B_{i} := \gamma_{i}]_{i=1}^{n}}$,
	and the lower edge to $\semint{\phi}\o\Tn_{i=1}^{n}\semint{\gamma_{\<i,j>}}$.
	
	For the axiom rule, let $\phi = t_{\delta_{1}, \dots, \delta_{m}}$,
	where $t\in\T^{(A_{i}\in\A_{i})_{i=1}^{m}}_{\C}[\lambda, \mu]$.
	Note that \[
		\alpha = \lambda[A_{i} := \delta_{i}]_{i=1}^{m}
		\mbox{\quad and\quad}
		\beta = \lambda[A_{i} := \delta_{i}]_{i=1}^{l}.
	\]
	Let $m_{i}$ be the number of names occurring in $\delta_{i}$,
	and let $\<i,j> = j+\sum_{r<i}m_{i}$. Note in particular that
	\[
		\Tn_{i=1}^{n}\semint{\gamma_{i}} = \Tn_{i=1}^{m}\Tn_{j=1}^{m_{i}}\semint{\gamma_{\<i,j>}}.
	\]	
	Now consider the diagram
	\begin{diagram}
		\semint{
			\lambda[A_{i} := \delta_{i}[B_{\<i,j>} := \gamma_{\<i,j>}]_{j=1}^{m_{i}}]_{i=1}^{m}
		}
			& \rTo^{\semint{\phi[B_{i} := \gamma_{i}]_{i=1}^{n}}}
			& \semint{
				\mu[A_{i} := \delta_{i}[B_{\<i,j>} := \gamma_{\<i,j>}]_{j=1}^{m_{i}}]_{i=1}^{m}
			} \\
		\dTo<{\norm^{-1}} && \uTo>{\norm} \\
		\semint{\lambda} \o \Tn_{i=1}^{m}\semint{\delta_{i}[B_{\<i,j>} := \gamma_{\<i,j>}]_{j=1}^{m_{i}}}
			& \rTo^{v(t) \o \Tn_{i=1}^{m}(\cdots)}
			& \semint{\mu}
				\o \Tn_{i=1}^{m}\semint{\delta_{i}[B_{\<i,j>} := \gamma_{\<i,j>}]_{j=1}^{m_{i}}} \\
		\dTo<{\semint{\lambda}\o\Tn_{i}\norm^{-1}} && \uTo>{\semint{\mu}\o\Tn_{i=1}^{m}\norm} \\
		\semint{\lambda} \o \Tn_{i=1}^{m}\bigl(
				\semint{\delta_{i}} \o \Tn_{j=1}^{m_{i}}\semint{\gamma_{\<i,j>}}
			\bigr)
			& \rTo^{v(t) \o \Tn_{i=1}^{m}(\cdots)}
			& \semint{\mu} \o \Tn_{i=1}^{m}\bigl(
				\semint{\delta_{i}} \o \Tn_{j=1}^{m_{i}}\semint{\gamma_{\<i,j>}}
			\bigr) \\
			\dTo<{\semint{\lambda}\o\Ic^{-1}} && \uTo>{\semint{\mu}\o\Ic} \\
			\semint{\lambda} \o \Tn_{i=1}^{m}
				\semint{\delta_{i}} \o \Tn_{i=1}^{m}\Tn_{j=1}^{m_{i}}\semint{\gamma_{\<i,j>}}
			& \rTo_{v(t)\o (\cdots)}
			& \semint{\mu} \o \Tn_{i=1}^{m}
				\semint{\delta_{i}} \o \Tn_{i=1}^{m}\Tn_{j=1}^{m_{i}}\semint{\gamma_{\<i,j>}}
	\end{diagram}
	The upper square commutes by definition of $\semint{\phi[B_{i} := \gamma_{i}]_{i=1}^{n}}$,
	and the other two squares commute by the 2-categorical interchange law, hence
	the outside commutes.
	Now, applying the Double norm result (Proposition~\ref{prop-double-norm})
	to the vertical sides gives that the outside of the following diagram commutes:
	\begin{diagram}
		\semint{\alpha[B_{i} := \gamma_{i}]_{i=1}^{n}}
			& \rTo^{\semint{\phi[B_{i} := \gamma_{i}]_{i=1}^{n}}}
			& \semint{\beta[B_{i} := \gamma_{i}]_{i=1}^{n}} \\
		\dTo<{\norm^{-1}} && \uTo>{\norm} \\
		\semint{\alpha}\o\Tn_{i=1}^{n}\semint{\gamma_{i}}
			& \rTo^{\semint{\phi} \o \Tn_{i=1}^{n}\semint{\gamma_{i}}}
			& \semint{\beta}\o\Tn_{i=1}^{n}\semint{\gamma_{i}} \\
		\dTo<{\norm^{-1}\o\Tn_{i=1}^{n}\semint{\gamma_{i}}}
			&& \uTo>{\norm\o\Tn_{i=1}^{n}\semint{\gamma_{i}}} \\
		\semint{\lambda} \o \Tn_{i=1}^{m}\semint{\delta_{i}} \o \Tn_{i=1}^{n}\semint{\gamma_{i}}
			& \rTo_{v(t) \o \cdots}
			& \semint{\mu} \o \Tn_{i=1}^{m}\semint{\delta_{i}} \o \Tn_{i=1}^{n}\semint{\gamma_{i}}
	\end{diagram}
	The lower square also commutes, by definition of $\semint{\phi}$, hence the upper
	square commutes as required.
\end{proof}

\begin{propn}[Semantics of 2-in-1 substitution]\label{prop-sem-2in1}
	Let $\Gamma\proves\gamma\in\C$ be a derivable 1-cell sequent, for
	some \[ \Gamma = (A_{1}\in\A_{1},\dots, A_{n}\in\A_{n}). \]
	Let
	\[
	\Delta_{1}\proves \phi_{1} : \alpha_{1}\to\beta_{1} \in \A_{1}
	\quad \cdots \quad
	\Delta_{n}\proves \phi_{n} : \alpha_{n}\to\beta_{n} \in \A_{n}
	\]
	be derivable 2-cell sequents. Then
	\[
		\semint{
		\Delta_{1},\cdots,\Delta_{n}\proves \gamma[A_{i} := \phi_{i}]_{i}
			: \gamma[A_{i} := \alpha_{i}]_{i} \to \gamma[A_{i} := \beta_{i}]_{i}
			\in\CB.
		}
	\]
	is equal to the composite
	\begin{diagram}
		\semint{\gamma[A_{i} := \alpha_{i}]_{i}}
			&& \semint{\gamma[A_{i} := \beta_{i}]_{i}}. \\
		\dTo<{\norm^{-1}} && \uTo>{\norm} \\
		\semint{\gamma}\o\Tn_{i=1}^{n}\semint{\alpha_{i}}
			&\rTo^{\semint{\gamma}\o\Tn_{i=1}^{n}\semint{\phi_{i}}}
				&\semint{\gamma}\o\Tn_{i=1}^{n}\semint{\beta_{i}}		
	\end{diagram}
\end{propn}
\begin{proof}
	\def\<#1>{\langle#1\rangle}
	Compared with the previous Proposition, this one is easy to prove!
	The proof is by induction over $\gamma$. If $\gamma$ is a constant
	then it is trivial, so let $\gamma = f(\gamma_{1}, \dots, \gamma_{m})$
	for some $f\in\T(\CB_{1}, \cdots, \CB_{m}; \C)$. For each $1\leq i\leq m$,
	let $m_{i}$ be the number of names in $\gamma_{i}$, and let $\<i,j>=
	j+\sum_{r<i}m_{r}$. We have to show that the outside of the diagram
	\begin{diagram}
		v(f) \o \Tn_{i=1}^{m}\semint{\gamma_{i}[A_{\<i,j>} := \alpha_{\<i,j>}]_{j=1}^{m_{i}}}
			& \rTo^{v(f) \o \Tn_{i=1}^{m}\semint{\gamma_{i}[A_{\<i,j>} := \phi_{\<i,j>}]_{j=1}^{m_{i}}}}
			& v(f) \o \Tn_{i=1}^{m}\semint{\gamma_{i}[A_{\<i,j>} := \beta_{\<i,j>}]_{j=1}^{m_{i}}} \\
		\dTo<{v(f) \o \Tn_{i=1}^{m} \norm^{-1}} && \uTo>{v(f) \o \Tn_{i=1}^{m} \norm} \\
		v(f) \o \Tn_{i=1}^{m}\bigl(
				\semint{\gamma_{i}}
				\o \Tn_{j=1}^{m_{i}}\semint{\alpha_{\<i,j>}}
			\bigr)
			& \rTo^{
				v(f) \o \Tn_{i=1}^{m}\bigl(
					\semint{\gamma_{i}}
					\o \Tn_{j=1}^{m_{i}}\semint{\phi_{\<i,j>}}
				\bigr)
			}
			& v(f) \o \Tn_{i=1}^{m}\bigl(
					\semint{\gamma_{i}}
					\o \Tn_{j=1}^{m_{i}}\semint{\beta_{\<i,j>}}
				\bigr) \\
		\dTo<{v(f) \o \Ic^{-1}} && \uTo>{v(f) \o \Ic} \\
		v(f) \o \Tn_{i=1}^{m}\semint{\gamma_{i}}
				\o \Tn_{i=1}^{m}\Tn_{j=1}^{m_{i}}\semint{\alpha_{\<i,j>}}
				& \rTo_{
				v(f) \o \Tn_{i=1}^{m}\semint{\gamma_{i}}
					\o \Tn_{i=1}^{m}\Tn_{j=1}^{m_{i}}\semint{\phi_{\<i,j>}}
				}
			& v(f) \o \Tn_{i=1}^{m}\semint{\gamma_{i}}
				\o \Tn_{i=1}^{m}\Tn_{j=1}^{m_{i}}\semint{\beta_{\<i,j>}}
	\end{diagram}
	commutes. The upper square commutes by the inductive hypothesis,
	and the lower square by naturality of $\Ic$, hence the outside
	commutes as required.
\end{proof}

\section{Soundness}
To make use of the equation rules, we need to show that the derivable equations
hold in every model.
\begin{propn}[Soundness]
	If
	\[
		\Gamma\proves\phi=\psi: \alpha\to\beta \in\CB
	\]
	is a derivable equation sequent and $v$ is an interpretation
	then
	\[
		\llbracket \Gamma\proves\phi: \alpha\to\beta \in\CB \rrbracket_{v}
		=
		\llbracket \Gamma\proves\psi: \alpha\to\beta \in\CB \rrbracket_{v}.
	\]
\end{propn}
\begin{proof}
	We consider in turn each of the derivation rules for equations.
	\begin{itemize}
	\item The reflexivity, symmetry and transitivity rules need no comment.
	\item For the axiom rule, let $(\phi,\psi)\in\T_{=}(\Delta\proves\alpha\to\beta\in\A)$ and let
		\[
			\Gamma_{1} \proves \gamma_{1}\in\CB_{1}
			\quad\cdots\quad
			\Gamma_{n} \proves \gamma_{n}\in\CB_{n}
		\]
		be derivable 1-cell sequents. We must show that
		\begin{mmulti}
			\semint{
				\Gamma_{1}\cdots\Gamma_{n}
					\proves \phi[B_{i}:=\gamma_{i}]_{i}
					: \alpha[B_{i}:=\gamma_{i}]_{i} \to \beta[B_{i}:=\gamma_{i}]_{i}
			}
			\\ =
			\semint{
				\Gamma_{1}\cdots\Gamma_{n}
					\proves \psi[B_{i}:=\gamma_{i}]_{i}
					: \alpha[B_{i}:=\gamma_{i}]_{i} \to \beta[B_{i}:=\gamma_{i}]_{i}
			}.
		\end{mmulti}
		By definition of model, we already know that
		\(
			\semint{\phi}
			=
			\semint{\psi},
		\)
		hence
		\[
			\semint{\phi}
				\o
				\Tn_{i=1}^{n}
					\semint{\gamma_{i}}
			=
			\semint{\psi}
				\o
				\Tn_{i=1}^{n}
					\semint{\gamma_{i}}
		\]
		which, by Proposition~\ref{prop-sem-1in2}, implies that
		\[
			\semint{
				\phi[
					B_{i} := \gamma_{i}
				]_{i=1}^{n}
			}
			=
			\semint{
				\psi[
					B_{i} := \gamma_{i}
				]_{i=1}^{n}
			}
		\]
		as required.
	\item Since both horizontal and vertical composition preserve equality
		in a 2-category (or indeed in a bicategory), the rules expressing
		that composition and 1-cell application preserve equality are sound.
	\item Let $f\in\T(\A_{1},\dots,\A_{n};\CB)$ and let
		\[
			\Gamma_{1}\proves \alpha_{1}\in \A_{1}
			\quad\cdots\quad
			\Gamma_{n}\proves \alpha_{n}\in \A_{n}
		\]
		be derivable 1-cell sequents. Then
		\begin{mmulti}[.5em]
			\semint{f(1_{\alpha_{1}},\dots,1_{\alpha_{n}})}
			\\= v(f)\o\Bigl(
				\semint{1_{\alpha_{1}}}
				\tn\cdots\tn
				\semint{1_{\alpha_{n}}}
			\Bigr)
			\\= v(f)\o\Bigl(
				1_{\semint{\alpha_{1}}}
				\tn\cdots\tn
				1_{\semint{\alpha_{n}}}
			\Bigr)
			\\= v(f)\o1_{
				\semint{\alpha_{1}}
				\tn\cdots\tn
				\semint{\alpha_{n}}
			}
			\\= 1_{
				v(f)\o(
					\semint{\alpha_{1}}
					\tn\cdots\tn
					\semint{\alpha_{n}}
				)
			}
			\\= \semint{
					1_{f(\alpha_{1}, \dots, \alpha_{n})}
			}
		\end{mmulti}
		showing that 1-cell application preserves identities.
	\item Let $f\in\T(\A_{1},\dots,\A_{n};\CB)$ and let
		\[
			\Gamma_{1}\proves\phi_{1}: \beta_{1}\to\gamma_{1}\in\A_{1}
			\quad\cdots\quad
			\Gamma_{n}\proves\phi_{n}: \beta_{n}\to\gamma_{n}\in\A_{n}
		\]
		and
		\[
			\Gamma_{1}\proves\psi_{1}: \alpha_{1}\to\beta_{1}\in\A_{1}
			\quad\cdots\quad
			\Gamma_{n}\proves\psi_{n}: \alpha_{n}\to\beta_{n}\in\A_{n}
		\]
		be derivable 2-cell sequents. Now, we have
		\begin{mmulti}[1em]
			\semint{
			\Gamma_{1},\cdots,\Gamma_{n}
				\proves f(\phi_{1},\dots,\phi_{n})\cdot f(\psi_{1},\dots,\psi_{n})
				\in\CB
			}
			\\=
			\semint{\Gamma_{1},\cdots,\Gamma_{n}
				\proves f(\phi_{1},\dots,\phi_{n})\in\CB}
			\cdot
			\semint{\Gamma_{1},\cdots,\Gamma_{n}
				\proves f(\psi_{1},\dots,\psi_{n})\in\CB}
			\\=
			\bigl( v(f)\o (
				\Tn_{i=1}^{n}\semint{\phi_{i}}
			) \bigr)
			\cdot
			\bigl( v(f)\o (
				\Tn_{i=1}^{n}\semint{\psi_{i}}
			) \bigr)
			\\=
			v(f)\o\bigl(
				\Tn_{i=1}^{n}\semint{\phi_{i}}
				\cdot
				\Tn_{i=1}^{n}\semint{\psi_{i}}
			\bigr)
			\\=
			v(f)\o\Tn_{i=1}^{n}\bigl(
				\semint{\phi_{i}}\cdot\semint{\psi_{i}}
			\bigr)
			\\=
			v(f)\o\Tn_{i=1}^{n}\bigl(
				\semint{\phi_{i}\cdot\psi_{i}}
			\bigr)
			\\=
			\semint{f(\phi_{1}\cdot\psi_{1},\dots,\phi_{n}\cdot\psi_{n})}
		\end{mmulti}
		showing that 1-cell application does indeed preserve composition.
	\item The left and right identity rules are obviously sound,
		since $\semint{1_{\alpha}} = 1_{\semint{\alpha}}$ is a strict unit for
		vertical composition in a Gray monoid.
	\item For the naturality rule, let $t\in \T^{\Gamma}_{\CB}[\alpha,\beta]$
	where
	\[
		\Gamma = (A_{1}\in\A_{1}, \cdots, A_{n}\in\A_{n}),
	\]
	and let
	\[
		\Gamma_{1}\proves\phi_{1}:\gamma_{1}\to\delta_{1}\in\A_{1}
		\quad\cdots\quad
		\Gamma_{n}\proves\phi_{n}:\gamma_{n}\to\delta_{n}\in\A_{n}
	\]
	be derivable 2-cell sequents. We wish to show that
	\[
		\semint{\beta[A_{i} := \phi_{i}]_{i} \cdot t_{\gamma_{1},\dots, \gamma_{n}}}
		=
		\semint{t_{\delta_{1},\dots, \delta_{n}} \cdot \alpha[A_{i}:=\phi_{i}]_{i}},
	\]
	i.e. that the diagram
	\begin{diagram}
		\semint{
			\alpha[
				A_{i} := \gamma_{i}
			]_{i}
		}
		& \rTo^{
			\semint{
				t_{\gamma_{1}, \dots, \gamma_{n}}
			}
		}
		& \semint{
			\beta[
				A_{i} := \gamma_{i}
			]_{i}
		}
		\\
		\dTo<{
			\semint{
				\alpha[
					A_{i} := \phi_{i}
				]_{i}
			}
		}
		&& \dTo>{
			\semint{
				\beta[
					A_{i} := \phi_{i}
				]_{i}
			}
		}
		\\
		\semint{
			\alpha[
				A_{i} := \delta_{i}
			]_{i}
		}
		& \rTo_{
			\semint{
				t_{\delta_{1}, \dots, \delta_{n}}
			}
		}
		& 
		\semint{
			\beta[
				A_{i} := \delta_{i}
			]_{i}
		}
	\end{diagram}
	commutes. Expanding the definitions of these four arrows gives the diagram
	\begin{diagram}
		\semint{\alpha} \o \Tn_{i=1}^{n}\semint{\gamma_{i}}
		& \rTo^{
			v(t) \o \Tn_{i=1}^{n}\semint{\gamma_{i}}
		}
		& \semint{\beta} \o \Tn_{i=1}^{n}\semint{\gamma_{i}}
		\\ \uTo<{\norm^{-1}} && \dTo>{\norm} \\
		\semint{
			\alpha[
				A_{i} := \gamma_{i}
			]_{i}
		}
		&& \semint{
			\beta[
				A_{i} := \gamma_{i}
			]_{i}
		}
		\\ \dTo<{\norm^{-1}} && \dTo>{\norm^{-1}} \\
		\semint{\alpha} \o \Tn_{i=1}^{n}\semint{\gamma_{i}}
			&& \semint{\beta} \o \Tn_{i=1}^{n}\semint{\gamma_{i}}
		\\
		\dTo<{
			\semint{\alpha} \o \Tn_{i=1}^{n}\semint{\phi_{i}}
		}
		&& \dTo>{
			\semint{\beta} \o \Tn_{i=1}^{n}\semint{\phi_{i}}
		}
		\\
		\semint{\alpha} \o \Tn_{i=1}^{n}\semint{\delta_{i}}
		&& \semint{\beta} \o \Tn_{i=1}^{n}\semint{\delta_{i}}
		\\ \dTo<{\norm} && \dTo>{\norm} \\
		\semint{
			\alpha[
				A_{i} := \delta_{i}
			]_{i}
		}
		&& \semint{
			\beta[
				A_{i} := \delta_{i}
			]_{i}
		}
		\\ \dTo<{\norm^{-1}} && \uTo>{\norm} \\
		\semint{\alpha} \o \Tn_{i=1}^{n}\semint{\gamma_{i}}
		& \rTo_{
			v(t) \o \Tn_{i=1}^{n}\semint{\delta_{i}}
		}
		& \semint{\beta} \o \Tn_{i=1}^{n}\semint{\delta_{i}}
	\end{diagram}
	Cancelling the $\norm$ arrows gives
	\begin{diagram}
		\semint{\alpha} \o \Tn_{i=1}^{n}\semint{\gamma_{i}}
		& \rTo^{
			v(t) \o \Tn_{i=1}^{n}\semint{\gamma_{i}}
		}
		& \semint{\beta} \o \Tn_{i=1}^{n}\semint{\gamma_{i}}
		\\
		\dTo<{
			\semint{\alpha} \o \Tn_{i=1}^{n}\semint{\phi_{i}}
		}
		&& \dTo>{
			\semint{\beta} \o \Tn_{i=1}^{n}\semint{\phi_{i}}
		}
		\\
		\semint{\alpha} \o \Tn_{i=1}^{n}\semint{\gamma_{i}}
		& \rTo_{
			v(t) \o \Tn_{i=1}^{n}\semint{\delta_{i}}
		}
		& \semint{\beta} \o \Tn_{i=1}^{n}\semint{\delta_{i}}
	\end{diagram}
	which clearly commutes, by the interchange law for composition in a 2-category.
	Hence the naturality rule is sound.
	\end{itemize}
\end{proof}

\section{Interpretation in a general monoidal bicategory}
We have shown how the language may be soundly interpreted in a Gray monoid.
In this section, we use coherence to interpret the language soundly
in any monoidal bicategory.

\newcommand\ee{\mathbf{e}}
\newcommand\ff{\mathbf{f}}
Let $\B$ be a monoidal bicategory, and let $\Gr(\B)$ be the
corresponding Gray monoid, as defined by \citet[][Chapter~10]{GurskiThesis},
with monoidal biequivalences
\[
	\B \pile{\rTo^{\ff}\\\lTo_{\ee}} \Gr(\B).
\]
We briefly recall the definitions of $\Gr(\B)$ and $\ee$:
\begin{itemize}
	\item an object of $\Gr(\B)$ is a finite sequence of objects of $\B$,
		which we regard as a formal tensor of those objects. So the tensor
		product of two objects is just their concatenation as lists.
	\item a `basic 1-cell' of $\Gr(\B)$
	\[
		\langle f,x,y,\alpha,\beta \rangle:
		(X_{i})_{i=1}^{x} (A_{i})_{i=1}^{m} (Y_{i})_{i=1}^{y}
		\to
		(X_{i})_{i=1}^{x} (B_{i})_{i=1}^{n} (Y_{i})_{i=1}^{y}
	\]
	consists of:
	\begin{itemize}
		\item a bracketing\footnote{
			By \emph{bracketing} we mean what Gurski calls a `choice of association'
			(his Definition~10.3.1). This can include copies of the unit object:
			for example, $((\I\tn A)\tn B)\tn \I$ is a bracketing of $(A,B)$.
		} $\alpha$ of the subsequence $(A_{i})_{i=1}^{m}$,
		\item a bracketing $\beta$ of the subsequence $(B_{i})_{i=1}^{n}$,
		\item a 1-cell $f: \Tn_{\alpha}A_{i} \to \Tn_{\beta}B_{i}$ in $\B$,
			where $\Tn_{\alpha}$ denotes the iterated tensor according to the
			bracketing $\alpha$.
	\end{itemize}
	\item Let us define the \emph{canonical tensor} of a sequence of objects or 1-cells to
	associate to the right, for example $\Tn(A,B,C,D) = A\tn(B\tn(C\tn D))$.
	On basic 1-cells, $\ee(\langle h,x,y,\alpha,\beta \rangle)$ is defined to
		be a composite of three 1-cells:
		\begin{diagram}
			\Tn((X_{i})_{i=1}^{x} (A_{i})_{i=1}^{m} (Y_{i})_{i=1}^{y}) \\
			\dTo>{\cong} \\
			(\Tn_{i=1}^{x}X_{i})\tn\bigl((\Tn_{\alpha}A_{i})\tn(\Tn_{i=1}^{y}Y_{i})\bigr) \\
			\dTo>{(\Tn_{i=1}^{x}X_{i})\tn h \tn(\Tn_{i=1}^{y}Y_{i})} \\
			(\Tn_{i=1}^{x}X_{i})\tn\bigl((\Tn_{\beta}B_{i})\tn(\Tn_{i=1}^{y}Y_{i})\bigr) \\
			\dTo>{\cong} \\
			\Tn((X_{i})_{i=1}^{x} (B_{i})_{i=1}^{n} (Y_{i})_{i=1}^{y})
		\end{diagram}
		where the first and third arrows are associator 1-cells.%
			(Gurski assumes that some associator
			has been chosen for each pair of bracketings. We shall further assume
			that each chosen associator has minimal complexity, where the complexity of
			an associator is the number of basic structural 1-cells it is built
			from. In addition we assume that, if $\alpha$ and $\beta$ are bracketings,
			the chosen associator $\beta\to\alpha$ is the inverse of the
			chosen associator $\alpha\to\beta$. This is clearly consistent with the
			previous requirement.)
	\item a 1-cell of $\Gr(\B)$ is a composable finite sequence of basic 1-cells.
	\item On 1-cells, $\ee(f_{1},  \dots, f_{n})$ is defined to be
	\[
		e(f_{n}) \o \cdots \o e(f_{1}).
	\]
	\item a 2-cell $f \To g$ is a 2-cell $\ee(f) \To \ee(g)$ in $\B$.
	\item $\ee$ is the identity on 2-cells.
	\item Horizontal composition of 2-cells in $\Gr(\B)$ is non-trivial, and
		inserts the structural isomorphisms necessary to make the types match.
\end{itemize}
Our primary interest in $\ee$ is that it serves as a machine for inserting
structural 1-cells and 2-cells where appropriate. Unfortunately for us,
the very simplicity of its definition means that it will often insert
structural 1-cells unnecessarily. This is no bar to defining an interpretation
in a general monoidal bicategory, but it would make the resulting interpretation
more complicated than necessary. Therefore it is convenient to introduce
a new operation $\ee'$ on the 1-cells of $\Gr(\B)$, which produces a
1-cell in $\B$ that is isomorphic to, but generally simpler than, the
one produced by $\ee$. We will also explicitly define an invertible
2-cell $\epsilon(h): \ee(h) \To \ee'(h)$, for every 1-cell $h$ of $\Gr(\B)$.

The first difference between $\ee'$ and $\ee$ is that we redefine the value of
$\ee'$ on a basic 1-cell $\langle h,x,y,\alpha,\beta \rangle$. We suppress the
first and/or third (associator) arrow when its source and target are equal.
In this case our minimal-complexity criterion ensures that these associators
are identities, hence there is an obvious invertible 2-cell
\[
	\epsilon(\langle h,x,y,\alpha,\beta \rangle):
		\ee(\langle h,x,y,\alpha,\beta \rangle) \To \ee'(\langle h,x,y,\alpha,\beta \rangle)
\]
built from the unit 2-cells of the bicategory $\B$.
For example, let $A$ and $B$ be objects
of $\B$, and consider the one-element sequences $(A)$ and $(B)$, which are
objects of $\Gr(\B)$. Given $h: A\to B$ in $\B$, there is a basic 1-cell
$\langle h,0,0,(-),(-) \rangle: (A)\to(B)$ in $\Gr(B)$; in fact this is
precisely $\ff(h)$, according to Gurski's triequivalence $\ff$. Now, applying
$\ee$ to this basic 1-cell gives the composite
\[
	A \rTo^{1} A \rTo^{h} B \rTo^{1} B,
\]
whereas applying $\ee'$ just gives $h: A\to B$.

The second case in which $\ee$ may introduce unnecessary 1-cells comes
when two basic 1-cells are juxtaposed within a 1-cell of $\Gr(\B)$.
Suppose, for example, that we have a 1-cell containing
$\langle h,x,y,\alpha,\beta \rangle$ followed by
$\langle h',x',y',\alpha',\beta' \rangle$. Applying $\ee$ to this pair
gives a composite of six 1-cells, the third and fourth of which are:
\begin{diagram}
	(\Tn_{i=1}^{x}X_{i})\tn\bigl((\Tn_{\beta}B_{i})\tn(\Tn_{i=1}^{n}B_{i})\bigr) \\
	\dTo>{\cong} \\
	\Tn((X_{i})_{i=1}^{x} (B_{i})_{i=1}^{n} (Y_{i})_{i=1}^{y}) \\
	= \Tn((X'_{i})_{i=1}^{x'} (B'_{i})_{i=1}^{n'} (Y'_{i})_{i=1}^{y'}) \\
	\dTo>{\cong} \\
	(\Tn_{i=1}^{x'}X'_{i})\tn\bigl((\Tn_{\alpha'}B'_{i})\tn(\Tn_{i=1}^{y'}Y'_{i})\bigr)
\end{diagram}
We define $\ee'$ so that:
\begin{itemize}
	\item If these associators are mutually inverse, both are suppressed.
	\item Otherwise, this composite is replaced by the relevant chosen
		associator of minimal complexity.
\end{itemize}
And we define $\epsilon$ in the obvious way.

Finally, we define $\ee'$ on the 2-cells of $\Gr(\B)$. If $\xi: (h_{i}) \to (k_{i})$
is a 2-cell in $\Gr(\B)$, let $\ee'(\xi)$ be the composite
\[
	\ee'((h_{i})) \rTo^{\epsilon^{-1}((h_{i}))} \ee((h_{i}))
		\rTo^{\ee(\xi)} \ee((k_{i}))
		\rTo^{\epsilon((k_{i}))} \ee'((k_{i})).
\]

We shall say in stages what it means to give an interpretation $v: \T\to\B$,
and use $v$ to simultaneously define:
\begin{itemize}
	\item an interpretation $v_{\Gr}: \T \to \Gr(\B)$, and
	\item the semantic function $\semint{-}$.
\end{itemize}
The interpretation $v$ is given as follows:
\begin{itemize}
	\item To give the object part of $v$, one gives, for every
	object $\A$ of $\T$, an object $v(\A)$ of $\B$. Using
	this, we define the object part of $v_{\Gr}$ by letting
	$v_{\Gr}(\A) = \ff(v(\A))$.
	\item To give the 1-cell part of $v$, one gives,
	for every $h\in\T(\A_{1}, \dots, \A_{n}; \CB)$,
	a 1-cell $v(h): \ee(\Tn_{i=1}^{n}v_{\Gr}(\A_{i})) \to v(\CB)$.
	Using this, we define $v_{\Gr}(h)$ to be the basic 1-cell
	\[
		(v_{\Gr}(\A_{i}))_{i=1}^{n} \to v_{\Gr}(\CB)
	\]
	for which $\ee'(v_{\Gr}(h)) = v(h)$.
	\item The interpretation $\semint{\gamma}$ of a 1-cell is
	defined to be $\ee'(\semint[v_{\Gr}]{\gamma})$.
	\item To give the 2-cell part of $v$, one gives,
	for every $t \in \T^{\Gamma}_{\CB}(\alpha, \beta)$,
	a 2-cell
	\[
		v(t): \semint{\alpha} \to \semint{\beta}
	\] in $\B$.
	Since $\ee$ is a bijection on 2-cells, so is $\ee'$, hence there is a unique 2-cell
	\[
		v_{\Gr}(t): \semint[v_{\Gr}]{\alpha} \to \semint[v_{\Gr}]{\beta}
	\]
	for which $\ee'(v_{\Gr}(t)) = v(t)$.
	\item The interpretation $\semint{\phi}$ of a 2-cell is
	defined to be $\ee(\semint[v_{\Gr}]{\phi})$.
	\item Additionally we demand that, for every
	equation $(\phi, \psi) \in \T_{=}(\Gamma\proves\alpha\to\beta\in\CB)$,
	$\semint{\phi} = \semint{\psi}$. Note that this is so if and only if
	$\semint[v_{\Gr}]{\phi} = \semint[v_{\Gr}]{\psi}$.
	\item Since the interpretation $\semint[v_{\Gr}]{-}$ is known to be sound,
	the interpretation $\semint{-}$ is also sound.
\end{itemize}

\section{Towards a braided extension}\label{s-braided}
Thanks to Theorem~\chref{Coh}{thm-coh-braiding}, there is no serious
obstacle to defining a version of the language applicable to braided
monoidal bicategories. With each context one could associate a
positive braid on that context, and arrange matters so that the
associated expression (on the right of the turnstile) uses the variables
in the order that they `emerge' from the braid. This could then be used
to treat braided structures, such as braided pseudomonoids, via components.

Only lack of time has prevented us from developing the braided extension
in full detail here.

\end{thesischapter}

\documentclass{robinthesis}

\begin{thesischapter}{Psmon}{Pseudomonoids}
\renewcommand\ss{\mathfrak{s}}
A monoid, in a monoidal category, consists of an object $A$ equipped
with a unit $u: I\to A$ and a multiplication $m: A\tn A\to A$, satisfying
the obvious unit and associativity axioms. In a monoidal \emph{bicategory},
the corresponding notion is that of a pseudomonoid, where the unit and
associativity laws hold, not on the nose, but up to coherent isomorphism.
The primeval example is of course that of monoidal categories, which are
pseudomonoids in the monoidal bicategory $\Cat$; though for present
purposes we are particularly interested in promonoidal categories,
which are pseudomonoids in $\Prof$.

Little has been published about pseudomonoids per se, though the
definition is given by \citet{MonBicat}. However, the
equivalent notion of \emph{pseudomonad} (in a tricategory or Gray-category)
has received more attention:
from \citet{MarmolejoPseudomonads, MarmolejoDistributive, MarmolejoDistributiveII, LackPseudomonads, TanakaThesis},
among others.

\begin{definition} 
	A pseudomonoid $\C$ in a monoidal bicategory $\B$ is a normal
	pseudofunctor $1\to\B$. More concretely, it consists of an object $\C$,
	1-cells
	\[\begin{array}l
		J: \I\to\C,\\
		P:\C\tn\C\to\C,
	\end{array}\]
	and invertible 2-cells
	\[\begin{array}{c@{\qquad}c}
		\multicolumn2c{\begin{diagram}
			\C\tn(\C\tn\C) && \rTo^{a_{\C,\C,\C}} && (\C\tn\C)\tn\C
			\\
			\dTo<{\C\tn P}
			&& \begin{array}c\To\\[-4pt]\aa\end{array}
			&& \dTo>{P\tn\C}
			\\
			\C\tn\C & \rTo_{P} & \C & \lTo_{P} & \C\tn\C
		\end{diagram}}
		\\[6em]
		\begin{diagram}[vtrianglewidth=1.5em,tight]
			\I\tn\C &&\rTo^{J\tn\C}&&\C\tn\C\\
			&\rdTo[snake=-1ex]<{l_\C}
				&\raise1ex\hbox{$\begin{array}c\Rightarrow\\[-5pt]\ll\end{array}$}%
				&\ldTo[snake=1ex]>{P}\\
			&&\C
		\end{diagram}
		&
		\begin{diagram}[vtrianglewidth=1.5em,tight]
			\C\tn \I &&\rTo^{\C\tn J} && \C\tn\C\\
			&\rdTo[snake=-1ex]<{r_\C}
				&\raise1ex\hbox{$\begin{array}c\Rightarrow\\[-5pt]\rr\end{array}$}%
				&\ldTo[snake=1ex]>{P}\\
			&&\C
		\end{diagram}
	\end{array}\]
	subject to the two equations below (stated in the Gray monoid setting).
	Since the 2-cells are assumed to be invertible, we shall permit ourselves
	to omit the arrow. Also, here and elsewhere, we write $\C^{2}$ for $\C\tn\C$.
	\begin{equation}\label{eq-lra}
	\begin{diagram}
		\C\tn\I\tn\C\\
		\dTo<1 &\hbox to0pt{\hss$\C\tn\ll$}\rdTo(2,1)^{\C\tn J\tn\C} & \C\tn\C\tn\C\\
		\C\tn\C & \ldTo(2,1)_{\C\tn P} &\dTo>{P\tn\C}\\
		\dTo<P & \raise1.5em\hbox{$\aa$}& \C\tn\C\\
		\C&\ldTo(2,1)_P
	\end{diagram}
	\qquad=\qquad
	\begin{diagram}
		\C\tn\I\tn\C\\
		\dTo<1 &\hbox to0pt{\hss$\rr\tn\C$}\rdTo(2,1)^{\C\tn J\tn\C} & \C\tn\C\tn\C\\
		\C\tn\C & \ldTo(2,1)_{P\tn\C}\\
		\dTo<P\\
		\C
	\end{diagram}
	\end{equation}
	and
	\begin{equation}\label{eq-aa}
		\begin{diagram}[s=2.2em,labelstyle=\scriptstyle,tight]
			&&\C^3\\
			&\ruTo^{\C^2\tn P}&\dTo[snake=-5pt]<{P\tn\C}&\rdTo^{\C\tn P}\\
			\C^4 &\sim& \C^2 &\mathop{\Leftarrow}\limits_{\;\;\;\aa}& \C^2\\
			\dTo<{P\tn\C^2}&\ruTo_{\C\tn P} && \rdTo_P & \dTo>{P}\\
			\C^3 && \Arr\Downarrow\aa && \C\\
			&\rdTo_{P\tn\C}&&\ruTo>{P}\\
			&&\C^2
		\end{diagram}
		\qquad=\qquad
		\begin{diagram}[s=2.2em,labelstyle=\scriptstyle,tight]
			&&\C^3\\
			&\ruTo^{\C^2\tn P}&&\rdTo^{\C\tn P}\\
			\C^4 && \Arr\Downarrow{\C\tn \aa} && \C^2\\
			\dTo<{P\tn\C^2}&\rdTo^{\C\tn P\tn\C} && \ruTo^{\C\tn P} & \dTo>{P}\\
			\C^3 &\mathop{\Leftarrow}\limits_{\;\;\;\aa\tn\C}& \C^3 &\Arr\Swarrow{\scriptstyle\!\!\!\aa}& \C\\
			&\rdTo_{P\tn\C}&\dTo[snake=5pt]>{\!\!P\tn\C}&\ruTo>{P}\\
			&&\C^2
		\end{diagram}
	\end{equation}
\end{definition}
The first of these equations corresponds to the triangle axiom relating
$\alpha$, $\lambda$ and $\rho$, and the second to Mac Lane's pentagon
axiom.

\section{Some facts about pseudomonoids}
If we express the results of \cite{KellyML} in the language
of general pseudomonoids, we obtain the three equations below.
In the following sections, we shall prove that they hold in
general, by showing how Kelly's argument can be applied,
via the calculus of components, to any pseudomonoid $\C$.
\begin{equation}\label{eq-lla}
\begin{diagram}
	\I\tn\C^2 & \rTo^{J\tn\C^2} &\C^3& \rTo^{\C\tn P} & \C^2\\
	&\rdTo(2,2)<1\raise1em\hbox to0pt{$\ll\tn\C$\hss} & \dTo[snake=.5em]>{P\tn\C} &\aa&\dTo>P\\
	&&\C^2 &\rTo_P &\C
\end{diagram}
\quad=\quad
\begin{diagram}
	\I\tn\C^2 &\rTo^{J\tn\C^2} & \C^3\\
	\dTo<{\I\tn P} &\sim & \dTo>{\C\tn P}\\
	\I\tn\C &\rTo^{J\tn\C} & \C^2\\
	&\rdTo(2,2)_1\raise0.5em\hbox to0pt{\hskip0.5em$\ll$\hss} &\dTo>P\\
	&&\C
\end{diagram}
\end{equation}
\begin{equation}\label{eq-rra}
\begin{diagram}
	\C^2\tn\I & \rTo^{\C^2\tn J} &\C^3& \rTo^{P\tn\C} & \C^2\\
	&\rdTo(2,2)<1\raise1em\hbox to0pt{$\C\tn\rr$\hss} & \dTo[snake=.5em]>{\C\tn P} &\aa&\dTo>P\\
	&&\C^2 &\rTo_P &\C
\end{diagram}
\quad=\quad
\begin{diagram}
	\C^2\tn\I &\rTo^{\C^2\tn J} & \C^3\\
	\dTo<{P\tn\I} &\sim & \dTo>{P\tn\C}\\
	\C\tn\I &\rTo^{\C\tn J} & \C^2\\
	&\rdTo(2,2)_1\raise0.5em\hbox to0pt{\hskip0.5em$\rr$\hss} &\dTo>P\\
	&&\C
\end{diagram}
\end{equation}
\begin{equation}\label{eq-lr}
\begin{diagram}
	\I\tn\I\\
	\dTo<{J\tn\I} & \rdTo(2,1)^{\I\tn J} & \I\tn\C\\
	\C\tn\I &\raise1em\hbox{$\sim$}&\dTo>{J\tn\C}\\
	\dTo<1 & \hbox to0pt{\hss$\rr$\hskip0.5em}\rdTo(2,1)^{\C\tn J}& \C\tn\C\\
	\C&\ldTo(2,1)_P
\end{diagram}
\quad=\quad
\begin{diagram}
	\I\tn\I\\
	\dTo<{J\tn\I} & \rdTo(2,1)^{\I\tn J} & \I\tn\C\\
	\C\tn\I &\ldTo(2,3)^1&\dTo>{J\tn\C}\\
	\dTo<1 & \raise1em\hbox to0pt{\hskip1em$\ll$\hss}& \C\tn\C\\
	\C&\ldTo(2,1)_P
\end{diagram}
\end{equation}
%
%
\citet[section~3.4]{LackThesis} describes an interesting geometrical way to
prove these equations, using certain four-dimensional diagrams.\footnote{
	Lack is working in the slightly more general
	context of enriched bicategories: a pseudomonoid is an enriched
	bicategory with one object.}
\citet[][Proposition~8.1]{MarmolejoPseudomonads} gives a more down-to-earth version of the argument, using pasting diagrams. Instead we will show how they follow from Kelly's proof for ordinary monoidal categories, by the calculus of components.

\section{Pseudomonoids via the calculus of components}
Next we see how the language of components may be used to reason about pseudomonoids.
We will define the theory $\M$ of pseudomonoids, and show that reasoning
in the formal language corresponds precisely to the usual modes of reasoning
about monoidal categories, and that a model of $\M$ is precisely a pseudomonoid.
The theory concerns
a single object $\C$, so $\M_{0} = \{\C\}$. There are two basic 1-cells,
$J:\I\to\C$ and $P:\C\tn\C\to\C$, so formally we have $\M(;\C) = \{J\}$
and $\M(\C,\C;\C):=\{P\}$. There are six basic 2-cells, corresponding to
$\aa$, $\ll$, $\rr$ and their inverses. The 2-cell $\aa$ goes from
\[
	A\in\C, B\in\C, C\in\C \proves P(A,P(B,C))\in\C
\]
to
\[
	A\in\C, B\in\C, C\in\C \proves P(P(A,B),C)\in\C;
\]
formally, for every three distinct names $A$, $B$ and $C$ we have
\[\M^{(A\in\C,B\in\C,C\in\C)}_{\C}[P(A,P(B,C)),P(P(A,B),C)] = \{\aa\}.\]
To make the notation appear more familiar, we shall write $A\ast B$ to
mean $P(A,B)$, and $I$ to mean $J()$. Thus $\aa$ is a 2-cell with components
\[
	\aa_{A,B,C}: A\ast (B\ast C) \to (A\ast B)\ast C
\]
for $A$, $B$, $C\in\C$.
In fact we want $\aa$ to be an invertible 2-cell, which formally
means that there is another 2-cell $\aa^{-1}$ with components
\[
	\aa^{-1}_{A,B,C}: (A\ast B)\ast C \to A\ast (B\ast C)
\]
such that $\aa_{A,B,C}\cdot\aa^{-1}_{A,B,C}$ and
$\aa^{-1}_{A,B,C}\cdot\aa_{A,B,C}$ are identities.
In terms of the formal definition of the theory $\M$, this means
\[\M^\Gamma_{\C}[P(P(A,B),C),P(A,P(B,C))] = \{\aa^{-1}\},\]
\[(\Gamma,\C,P(P(A,B),C),P(P(A,B),C),\aa^{-1}_{A,B,C}\cdot\aa_{A,B,C},
1_{P(P(A,B),C)})\in\M_{=},\]
\[(\Gamma,\C,P(A,P(B,C)),P(A,P(B,C)),\aa_{A,B,C}\cdot\aa^{-1}_{A,B,C},
1_{P(A,P(B,C))})\in\M_{=},\]
where $\Gamma=(A\in\C,B\in\C,C\in\C)$.

In the same way, we want $\ll$ to be an invertible 2-cell with
components
\[
	\ll_{A}: I\ast A \to A
\]
for $A\in\C$, and $\rr$ to be an invertible 2-cell with
components
\[
	\rr_{A}: A\ast I \to A
\]
for $A\in\C$. (Formally, the theory $\M$ contains 2-cells $\ll$,
$\ll^{-1}$, $\rr$ and $\rr^{-1}$, and four equations expressing
that $\ll^{-1}$ is inverse to $\ll$ and $\rr^{-1}$ is inverse to $\rr$.)

Finally, we have the pentagon and triangle axioms. Consider the
pentagon:
\begin{mspill}\begin{diagram}
  A\ast (B\ast (C\ast D))
	  &\rTo^{\aa_{A,B,C\ast D}}&(A\ast B)\ast (C\ast D)
	  &\rTo^{\aa_{A\ast B,C,D}} & \bigl(((A\ast B)\ast C)\ast D\bigr)
  \\
  &\rdTo[snake=-1em](1,2)<{A\ast \aa_{B,C,D}} &
	  & \ruTo[snake=1em](1,2)>{\aa_{A,B,C}\ast D}
  \\
  & \spleft{A\ast ((B\ast C)\ast D)}
	  & \rTo_{\aa_{A,B\ast C,D}}
	  & \spright{(A\ast (B\ast C))\ast D}
\end{diagram}\end{mspill}
Of course, this diagram is just a convenient way of writing the equation
\[
	\aa_{A\ast B,C,D} \cdot \aa_{A,B\ast C,D}
	=
	P(\aa_{A,B,C},D) \cdot \aa_{A,B\ast C,D} \cdot (A\ast\aa_{B,C,D})
\]
which we take as a formal equation of $\M$. Thus any model of $\M$
must satisfy
\[
	\semint{\aa_{A\ast B,C,D} \cdot \aa_{A,B\ast C,D}}
	=
	\semint{P(\aa_{A,B,C},D) \cdot \aa_{A,B\ast C,D} \cdot (A\ast\aa_{B,C,D})}.
\]
Similarly the triangle:
\begin{diagram}
        A\ast (I\ast C) &\rTo^{\aa_{A,I,C}}& (A\ast I)\ast C\\
        &\rdTo[snake=-1ex](1,2)<{A\ast\ll_C}\ldTo[snake=1ex](1,2)>{\rr_A\ast C}\\
        &A\ast C
\end{diagram}
represents the equation
\[
	(\rr_A\ast C) \cdot \aa_{A,I,C}  =  A\ast\ll_C,
\]
hence a model of $\M$ must also satisfy
\[
	\semint{(\rr_A\ast C) \cdot \aa_{A,I,C}}  =  \semint{A\ast\ll_C}.
\]
Now, let us consider an interpretation of $\M$ in a Gray monoid $\B$. We shall
omit $v()$, writing just $\C$ instead of $v(\C)$, and $P$ instead of $v(P)$, etc.
So an interpretation of $\M$ consists of an object $\C$,
1-cells $J: \I \to \C$ and $P:\C\tn\C\to\C$,
and invertible 2-cells
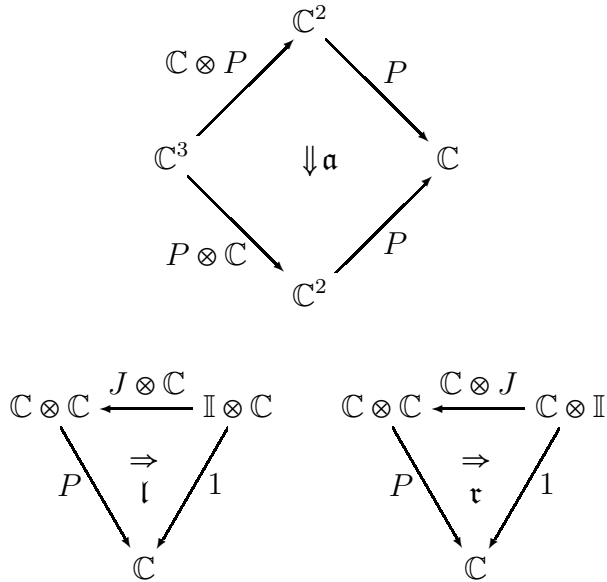
\[\begin{array}{c@{\qquad}c}
	\multicolumn2c{	\begin{diagram}[s=2.2em,tight]
		&& \C^2\\
		&\ruTo^{\C\tn P} && \rdTo^P\\
		\C^3 && \Arr\Downarrow\aa && \C\\
		&\rdTo_{P\tn\C}&&\ruTo>{P}\\
		&&\C^2
	\end{diagram}
}
	\\[6em]
	\begin{diagram}[vtrianglewidth=1.5em,tight]
		\C\tn\C &&\lTo^{J\tn\C}&&\I\tn\C\\
		&\rdTo[snake=-1ex]<{P}
			&\raise1ex\hbox{$\begin{array}c\Rightarrow\\[-5pt]\ll\end{array}$}%
			&\ldTo[snake=1ex]>{1}\\
		&&\C
	\end{diagram}
	&
	\begin{diagram}[vtrianglewidth=1.5em,tight]
		\C\tn \C &&\lTo^{\C\tn J} && \C\tn\I\\
		&\rdTo[snake=-1ex]<{P}
			&\raise1ex\hbox{$\begin{array}c\Rightarrow\\[-5pt]\rr\end{array}$}%
			&\ldTo[snake=1ex]>{1}\\
		&&\C
	\end{diagram}
\end{array}\]
where we have written $\C^{2}$ as an abbreviation for $\C\tn\C$, etc.,
subject to equations corresponding to the pentagon and triangle conditions.
Let us first consider the pentagon equation
\[
	\semint{\aa_{A\ast B,C,D} \cdot \aa_{A,B,C\ast D}}
	=
	\semint{(\aa_{A,B,C}\ast D) \cdot \aa_{A,B\ast C,D} \cdot (A\ast\aa_{B,C,D})}.
\]
equivalently
\[
	\semint{\aa_{A\ast B,C,D}} \cdot \semint{\aa_{A,B,C\ast D}}
	=
	\semint{(\aa_{A,B,C}\ast D)} \cdot \semint{\aa_{A,P(B,C),D}} \cdot \semint{P(A,\aa_{B,C,D})}.
\]
By definition, $\semint{\aa_{A,B, C\ast D}}: \semint{A\ast(B\ast(C\ast D))}
	\To \semint{(A\ast B)\ast(C\ast D)}$ is
\begin{diagram}
	\semint{A\ast(B\ast(C\ast D))} && \semint{(A\ast B)\ast(C\ast D)} \\
	\dTo<{\norm^{-1}} && \uTo>{\norm} \\ 
	\semint{A\ast(B\ast X)} \o (\C\tn\C\tn P)
	& \rTo_{\aa \o (\C\tn\C\tn P)}
	& \semint{(A\ast B)\ast X} \o (\C\tn\C\tn P)
\end{diagram}
On the left we have
\begin{mmulti}
	\semint{
	A\ast(B\ast(C\ast D))
	}
	\\= P \o (\C\tn\semint{
	B\ast(C\ast D)
	})
	\\= P \o (\C\tn(
		P \o (\C\tn\semint{
			C\ast D
		}
	)))
	\\= P \o (\C\tn(
		P \o (\C\tn P)
	))
	\\= P \o (\C\tn P) \o (\C\tn\C\tn P)
\end{mmulti}
and the $\norm^{-1}$ map is just the identity. The reason is
essentially that every occurrence of $P$ has one argument equal
to a constant. Thus the right-hand side is more interesting: we have
\[
	\semint{
	(A\ast B)\ast(C\ast D)
	}
	=
	P \o (P\tn P)
\]
and the $\norm$ map is the isomorphism
\[
	P \o (P\tn\C) \o (\C\tn\C\tn P) \to P \o (P\tn P).
\]
So this 2-cell is
{\def\rnode#1#2
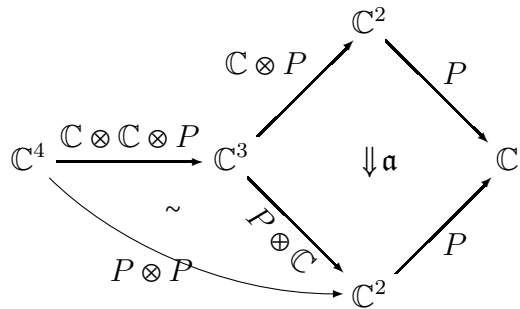
{%
	\tikz[baseline=(#1.base),inner sep=0pt,outer sep=3pt]
		\node(#1){\mathsurround=0pt$\displaystyle#2$};
	}
\begin{diagram}[s=2.2em]
	&& && \C^2\\
	&& &\ruTo^{\C\tn P} && \rdTo^P\\
	\rnode{left}{\C^{4}}&\rTo^{\C\tn \C\tn P}& \C^3 && \Arr\Downarrow\aa && \C\\
	&& &\rdTo[hug]_{P\tn\C}&&\ruTo>{P}\\
	&& &&\rnode{bottom}{\C^2}
	\begin{tikzpicture}[overlay]
		\path[->] (left) edge [out=-45, in=180, out looseness=0.7]
			node [left, pos=0.6] {$P\tn P$} node [above=1em, pos=0.5] {$\sim$} (bottom);
	\end{tikzpicture}
\end{diagram}
In a similar way, one may calculate that
\[
	\semint{\aa_{P(A,B),C,D}}
\]
is equal to
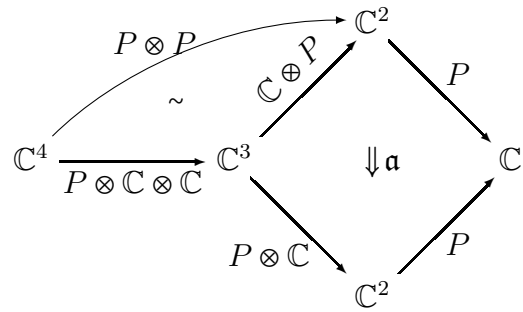
\begin{diagram}[s=2.2em]
	&& && \rnode{top}{\C^2}\\
	&& &\ruTo[hug]^{\C\tn P} && \rdTo^P\\
	\rnode{left}{\C^{4}}&\rTo_{P\tn\C\tn\C}& \C^3 && \Arr\Downarrow\aa && \C\\
	&& &\rdTo_{P\tn\C}&&\ruTo>{P}\\
	&& &&\rnode{bottom}{\C^2}
	\begin{tikzpicture}[overlay]
		\path[->] (left) edge [in = 180, out looseness=0.7]
			node [left, pos=0.6] {$P\tn P$} node [below=1em, pos=0.5] {$\sim$} (top);
	\end{tikzpicture}
\end{diagram}}
Note that, according to the definition of Gray monoid, in fact
\[
	P\tn P = (\C\tn P)\o(P\tn\C\tn\C),
\]
and this interchange cell is the identity.
Thus the composite
\[
	\semint{\aa_{A\ast B,C,D} \cdot \aa_{A,B,C\ast D}}
	= \semint{\aa_{A\ast B,C,D}} \cdot \semint{\aa_{A,B,C\ast D}}
\]
is
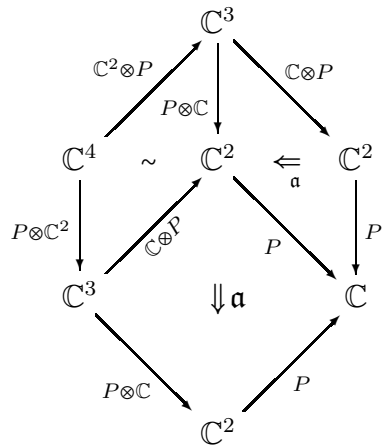
\begin{diagram}[s=2.2em,labelstyle=\scriptstyle,tight]
	&&\C^3\\
	&\ruTo^{\C^2\tn P}&\dTo[snake=-5pt]<{P\tn\C}&\rdTo^{\C\tn P}\\
	\C^4 &\sim& \C^2 &\mathop{\Leftarrow}\limits_{\;\;\;\aa}& \C^2\\
	\dTo<{P\tn\C^2}&\ruTo[hug]_{\C\tn P} && \rdTo_P & \dTo>{P}\\
	\C^3 && \Arr\Downarrow\aa && \C\\
	&\rdTo_{P\tn\C}&&\ruTo>{P}\\
	&&\C^2
\end{diagram}
Calculating
\[
	\semint{(\aa_{A,B,C}\ast D) \cdot \aa_{A,B\ast C,D} \cdot (A\ast\aa_{B,C,D})}.
\]
is less interesting, since all the $\norm$ maps are identities, so
it is just
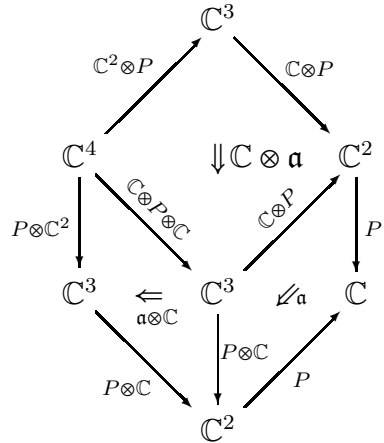
\begin{diagram}[s=2.2em,labelstyle=\scriptstyle,tight]
	&&\C^3\\
	&\ruTo^{\C^2\tn P}&&\rdTo^{\C\tn P}\\
	\C^4 && \Arr\Downarrow{\C\tn \aa} && \C^2\\
	\dTo<{P\tn\C^2}&\rdTo[hug]^{\C\tn P\tn\C} && \ruTo[hug]^{\C\tn P} & \dTo>{P}\\
	\C^3 &\mathop{\Leftarrow}\limits_{\;\;\;\aa\tn\C}& \C^3 &\Arr\Swarrow{\scriptstyle\!\!\!\aa}& \C\\
	&\rdTo_{P\tn\C}&\dTo[snake=5pt]>{\!\!P\tn\C}&\ruTo>{P}\\
	&&\C^2
\end{diagram}
Thus the pentagon axiom, in a Gray monoid, is precisely
equation~\pref{eq-aa}.
In a similar way, the triangle axiom corresponds
to equation~\pref{eq-lra}.
%
The interpretation in a general monoidal bicategory works in a
similar way; the monoidal biequivalence $\mathbf{e}'$ causes structural
1-cells and 2-cells to be inserted where necessary. Let us consider
the 2-cell $\aa$: its interpretation needs to be a 2-cell
\[
	\semint{A\ast (B\ast C)} \to \semint{(A\ast B)\ast C},
\]
and $\semint{A\ast (B\ast C)}$ is simply
\[
	\C\tn(\C\tn\C)
	\rTo^{\C\tn P}
	\C\tn\C
	\rTo^{P}
	\C
\]
whereas $\semint{(A\ast B)\ast C}$ is
\[
	\C\tn(\C\tn\C)
	\rTo^{a_{\C,\C,\C}}
	(\C\tn\C)\tn\C
	\rTo^{P\tn\C}
	\C\tn\C
	\rTo^{P}
	\C,
\]
hence $\aa$ should be a 2-cell
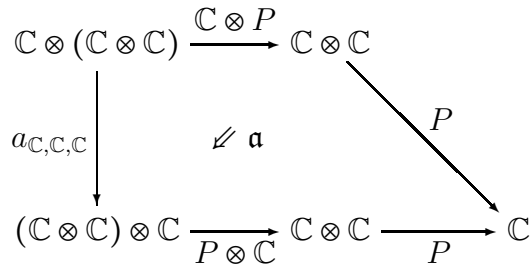
\begin{diagram}
		\C\tn(\C\tn\C)
		& \rTo^{\C\tn P} & \C\tn\C
		\\
		\dTo<{a_{\C,\C,\C}}
		& \Swarrow\aa
		&& \rdTo^{P}
		\\
		(\C\tn\C)\tn\C
		& \rTo_{P\tn\C} & \C\tn\C
		& \rTo_{P} & \C
\end{diagram}
The axioms also sport structural 2-cells. For example, the pentagon equation becomes
the requirement that
{\def\rnode#1#2
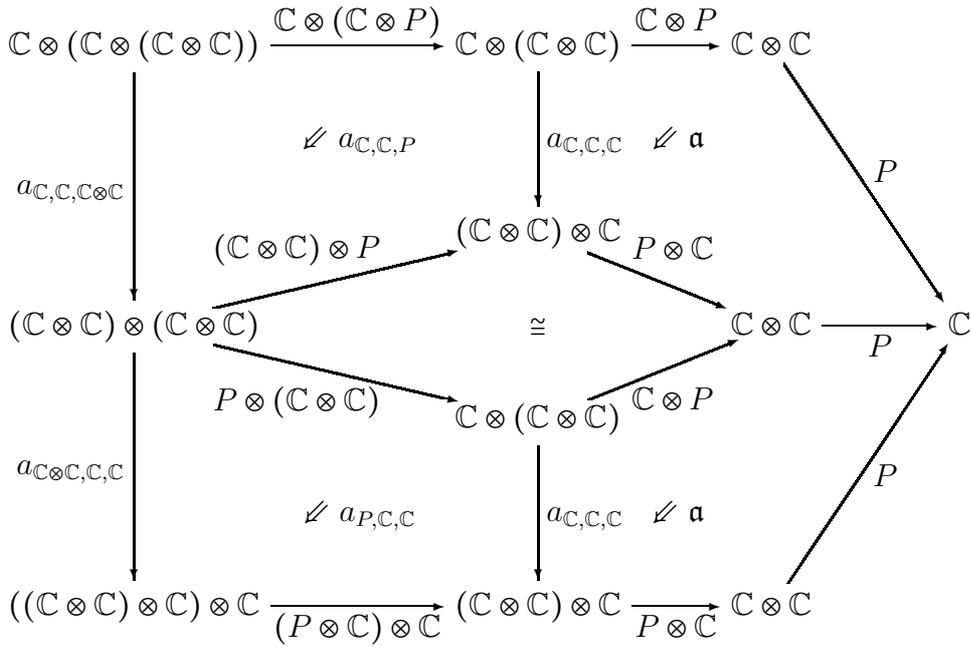
{%
	\tikz[baseline=(#1.base),inner sep=0pt,outer sep=3pt]
		\node(#1){\mathsurround=0pt$\displaystyle#2$};
	}
\begin{mspill}\begin{diagram}
	\C\tn(\C\tn(\C\tn\C))
	& \rTo^{\C\tn(\C\tn P)} & \C\tn(\C\tn\C)
	& \rTo^{\C\tn P} & \C\tn\C
	\\
	\dTo[snake=-1.5em]<{a_{\C,\C,\C\tn\C}}
	& \Swarrow a_{\C,\C,P}
	& \dTo>{a_{\C,\C,\C}}
	& \Swarrow\aa
	&& \rdTo(2,3)^P
	\\
	&& (\C\tn\C)\tn\C
	\\
	(\C\tn\C)\tn(\C\tn\C)
	& \ruTo(2,1)^{(\C\tn\C)\tn P}
	& \cong
	& \rdTo(2,1)^{P\tn\C}
	& \C\tn\C
	& \rTo_P & \C
	\\
	\dTo[snake=-1.5em]<{a_{\C\tn\C,\C,\C}}
	& \rdTo(2,1)_{P\tn(\C\tn\C)}
	&\C\tn(\C\tn\C)
	& \ruTo(2,1)_{\C\tn P}
	&& \ruTo(2,3)_P
	\\
	& \Swarrow a_{P,\C,\C}
	& \dTo>{a_{\C,\C,\C}}
	& \Swarrow\aa
	\\
	((\C\tn\C)\tn\C)\tn\C
	& \rTo_{(P\tn\C)\tn\C} & (\C\tn\C)\tn\C
	& \rTo_{P\tn\C} & \C\tn\C
\end{diagram}\end{mspill}
must be equal to
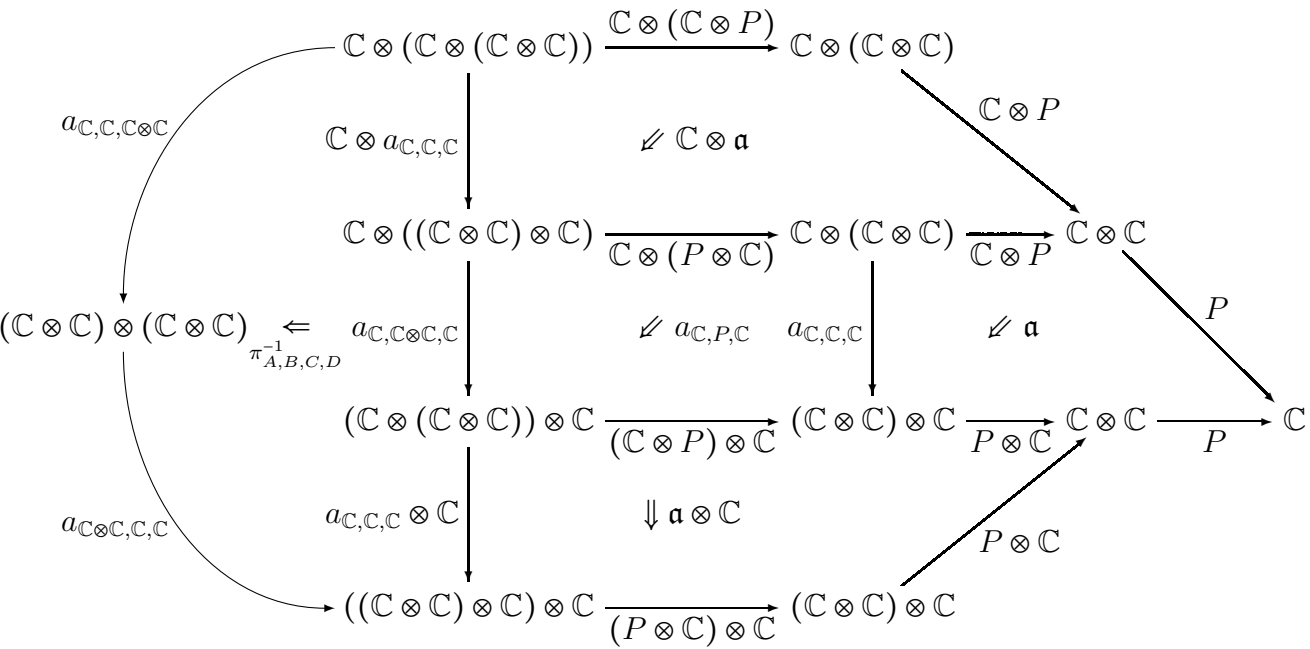
\begin{diagram}
	&& \rnode{t}{\C\tn(\C\tn(\C\tn\C))}
	& \rTo^{\C\tn(\C\tn P)} & \C\tn(\C\tn\C)
	\\
	& 
	& \dTo<{\C\tn a_{\C,\C,\C}}
	& \Swarrow\C\tn \aa
	&& \rdTo^{\C\tn P}
	\\
	&& \C\tn ((\C\tn\C)\tn\C)
	& \rTo_{\C\tn(P\tn\C)} & \C\tn(\C\tn\C)
	& \rTo_{\C\tn P} & \C\tn\C
	\\
	\rnode{l}{(\C\tn\C)\tn(\C\tn\C)} & \Left_{\pi_{A,B,C,D}^{-1}}
	& \dTo<{a_{\C,\C\tn\C,\C}}
	& \Swarrow a_{\C,P,\C}
	& \dTo<{a_{\C,\C,\C}}
	& \Swarrow\aa
	&& \rdTo^{P}
	\\
	& 
	& (\C\tn(\C\tn\C))\tn\C
	& \rTo_{(\C\tn P)\tn\C}
	& (\C\tn\C)\tn\C
	& \rTo_{P\tn\C}
	& \C\tn\C & \rTo_P & \C
	\\
	&& \dTo<{a_{\C,\C,\C}\tn\C}
	& \Downarrow\aa\tn\C
	&& \ruTo_{P\tn\C}
	\\
	&& \rnode{b}{((\C\tn\C)\tn\C)\tn\C} & \rTo_{(P\tn\C)\tn\C} & (\C\tn\C)\tn\C
	\begin{tikzpicture}[overlay]
		\path[->] (t) edge [out=180, in=90]
			node [left] {$a_{\C,\C,\C\tn\C}$} (l);
		\path[->] (l) edge [out=-90, in=180]
			node [left] {$a_{\C\tn\C,\C,\C}$} (b); 
	\end{tikzpicture}
\end{diagram}
and the triangle equation becomes the requirement that
\begin{diagram}
	\C\tn(\I\tn\C) & \rTo^{a_{\C,\I,\C}} & \rnode{t}{(\C\tn\I)\tn\C} \\
	\dTo<{\C\tn(J\tn\C)} & \Arr\Nearrow{a_{\C,J,\C}} & \dTo>{(\C\tn J)\tn\C} \\
	\C\tn(\C\tn\C) & \rTo_{a_{\C,\C,\C}} & (\C\tn\C)\tn\C & \Right_{\rr\tn\C} \\
	\dTo<{\C\tn P} & \Arr\Nearrow{\aa} && \rdTo_{P\tn\C} \\
	\C\tn\C & \rTo_{P} & \C & \lTo_{P} & \rnode{b}{\C\tn\C}
	\begin{tikzpicture}[overlay]
		\path[->] (t) edge [out=0, in=90]
			node [right] {$r_{\C}\tn\C$} (b);
	\end{tikzpicture}
\end{diagram}
be equal to
\begin{diagram}[h=2.2em]
	& & \rnode{CIC}{\C\tn(\I\tn\C)} \\
	& \ldTo^{\C\tn (J\tn\C)} && \rdTo^{a_{\C,\I,\C}} \\
	\C^{3} & \rlap{$\Right_{\rr\tn\C}$} & \dTo~{\C\tn l_{\C}}
		& \llap{$\Right_{\mu_{\C,\C}}$} & (\C\tn\I)\tn\C\\
	& \rdTo_{\C\tn P} && \ldTo_{r_{\C}\tn\C} \\
	& & \rnode{CC}{\C\tn\C} \\
	& & \dTo<{P} \\
	& & \C 
\end{diagram}}
(In fact the `raw' version of the equation, as it emerges from the
interpretation, has $\mu^{-1}$ on the left-hand side, rather than
$\mu$ on the right as we have written.)

\section{Calculating in the theory of pseudomonoids}
Now we may use the language to prove various facts about
pseudomonoids, essentially using the formal interpretation of the
language as a translation tool that allows us to transfer
proofs from the familiar setting of monoidal categories.
As a first example, consider the simple fact that
\[
	\ll_{I\ast A} = I\ast\ll_{A}: I\ast(I\ast A) \to I\ast A.
\]
The usual proof runs as follows: since $\ll$ is a natural transformation,
we have a naturality square
\begin{diagram}
	I\ast (I\ast A) & \rTo^{I\ast\ll_{A}} & I\ast A \\
	\dTo<{\ll_{I\ast A}} &\natural& \dTo>{\ll_{A}} \\
	I\ast A & \rTo_{\ll_{A}} & A
\end{diagram}
which, since $\ll_{A}$ is invertible, implies the claim.
The formal proof in our language is precisely the same:
the naturality square is an instance of the naturality
axiom; we then compose with $\ll^{-1}_{A}$ (using the axiom
that composition preserves equality), then use our
axiom relating $\ll$ and $\ll^{-1}$ to derive
\[
	1_{A} \cdot \ll_{I\ast A} = 1_{A} \cdot I\ast \ll_{A},
\]
and finally use the identity axiom (and symmetry and transitivity)
to conclude that
\[
	\ll_{I\ast A} = I\ast\ll_{A}: I\ast(I\ast A) \to I\ast A
\]
as required. In a Gray monoid, this shows that
{\def\rnode#1#2{%
	\tikz[baseline=(#1.base),inner sep=0pt,outer sep=3pt]
		\node(#1){\mathsurround=0pt$\displaystyle#2$};
	}
\begin{diagram}[h=1.5em]
	\\\ \\
	&&& \sim\\
	\rnode{l}{\C} & \rTo_{J\tn\C} & \C\tn\C & \rTo_{P} & \rnode{l2}{\C} & \rTo_{J\tn\C}
		& \rnode{r}{\C\tn\C} & \rTo_{P} & \rnode{r2}{\C} \\
	&&&&&& \Downarrow\ll \\
	\begin{tikzpicture}[overlay]
		\path[->] (l) edge [out=30, in=150]
			node [above] {$J\tn(P\o(J\tn\C))$} (r);
		\path[->] (l2) edge [out=-60, in=-120]
			node [below] {$1$} (r2);
	\end{tikzpicture}
\end{diagram}
is equal to
\begin{diagram}[h=1.5em]
	\\\ \\
	&&& \sim\\
	\rnode{l}{\C} & \rTo_{J\tn\C} & \C\tn\C & \rTo_{P} & \rnode{r2}{\C} & \rTo_{J\tn\C}
		& \rnode{r}{\C\tn\C} & \rTo_{P} & \C \\
	&& \Downarrow\ll \\
	\begin{tikzpicture}[overlay]
		\path[->] (l) edge [out=30, in=150]
			node [above] {$J\tn(P\o(J\tn\C))$} (r);
		\path[->] (l) edge [out=-60, in=-120]
			node [below] {$1$} (r2);
	\end{tikzpicture}
\end{diagram}
Since we may cancel the common $\norm$ cell, this can equivalently
be stated as:
\begin{diagram}[h=1.5em]
	\rnode{l}{\C} & \rTo_{J\tn\C} & \C\tn\C & \rTo_{P} & \rnode{l2}{\C} & \rTo_{J\tn\C}
		& \rnode{r}{\C\tn\C} & \rTo_{P} & \rnode{r2}{\C} \\
	&&&&&& \Downarrow\ll \\
	\begin{tikzpicture}[overlay]
		\path[->] (l2) edge [out=-60, in=-120]
			node [below] {$1$} (r2);
	\end{tikzpicture}
\end{diagram}
is equal to
\begin{diagram}[h=1.5em]
	\rnode{l}{\C} & \rTo_{J\tn\C} & \C\tn\C & \rTo_{P} & \rnode{r2}{\C} & \rTo_{J\tn\C}
		& \rnode{r}{\C\tn\C} & \rTo_{P} & \C \\
	&& \Downarrow\ll \\
	\begin{tikzpicture}[overlay]
		\path[->] (l) edge [out=-60, in=-120]
			node [below] {$1$} (r2);
	\end{tikzpicture}
\end{diagram}}
For a more substantial application, consider the triangle \pref{eq-lla}:
\begin{diagram}[vtriangleheight=3em]
	I\ast(A\ast B) && \rTo^{\aa_{I,A,B}} && (I\ast A)\ast B \\
	& \rdTo_{\ll_{A\ast B}} && \ldTo_{\ll_{A}\ast B} \\
	&& A\ast B
\end{diagram}
The proof of this, as given by \citet{KellyML} for monoidal categories,
runs as follows. Consider the diagram
\begin{mspill}\begin{diagram}[midvshaft,hug]
	\bigl(A,(I,(C,D))\bigr)
		&\rTo^{\aa_{A,I,(C\ast D)}} & (A\ast I)\ast (C\ast D)
		&\rTo^{\aa_{(A\ast I),C,D}} & ((A\ast I)\ast C)\ast D
	\\
	& \rdTo(1,2)^{(A,\ll_{(C,D)})} \ldTo(1,2)_{\rr_{A}\ast (C,D)}
		& \rlap{\qquad$\natural$} &\ldTo(1,2)^{(\rr_{A}\ast C)\ast D}
	\\
		& A\ast (C\ast D) & \rTo_{\aa_{A,C,D}} & (A\ast C)\ast D
	\\
	\dTo<{A\ast\aa_{I,C,D}}>{\qquad?}
		\ruTo(1,2)_{A\ast(\ll_{C}\ast D)}
		&& \natural &&
		\luTo(1,2)_{(A\ast\ll_{C})\ast D}
		\uTo>{\aa_{A,I,C}\ast D}
	\\
	A\ast((I\ast C)\ast D)
		&& \rTo_{\aa_{A,I\ast C,D}}
		&& (A\ast(I\ast C))\ast D
\end{diagram}\end{mspill}
The outside commutes by the pentagon axiom.
The quadrilaterals commute by naturality, and the unmarked triangles
by the triangle axiom. Since $\aa_{A,C,D}$ is invertible, it follows
that the triangle marked `$?$' commutes.
Now set $A = I$, and use the naturality and invertibility of $\ll$ to
conclude that the required triangle commutes.
It's easy to see that all this reasoning is formalisable in the language,
hence
\[
	\semint{\aa_{I, A, B} \cdot \ll_{A\ast B}}
	=
	\semint{\ll_{A}\ast B}
\]
in any model.
By a dual argument,
\[
	\semint{\rr_{A\ast B} \cdot \aa_{A,B,I}} = \semint{A\ast\rr_{B}}
\]
as well. Also, we can show that $\semint{\ll_{I}} = \semint{\rr_{I}}$,
as follows:
\[
	(\ll_{I}\ast A) \cdot \aa_{I,I,A} = \ll_{I\ast A}
\]
by the triangle we proved above, which is equal to $I\ast \ll_{A}$
by the naturality argument at the start of this section, which in
turn is equal to $(\rr_{I}\ast A) \cdot \aa_{I,I,A}$ by the
triangle axiom. Since $\aa_{I,I,A}$ is invertible, we have
that $\ll_{J()}\ast A = \rr_{J()}\ast A$. Then use the invertibility
and naturality of $\rr$ to conclude $\ll_{I} = \rr_{I}$, as
required.

Again, this reasoning may easily be formalised in the language,
and as promised the language allows us to transfer proofs from
the setting of ordinary monoidal categories to that of arbitrary
pseudomonoids.

\section{Braided pseudomonoids}\label{s-braided}
In the context of a braided monoidal bicategory $\B$, we can define
what it means to have a braiding on a pseudomonoid in $\B$. Observe
that it is not possible to define \emph{symmetric} pseudomonoids
in this setting: there is simply no way to express the desired
equation. To define symmetric pseudomonoids in general, one needs
some additional structure on the braiding of the monoidal bicategory:
this additional structure is called a \emph{syllepsis}, and consists
of an invertible modification between the identity transformation
on the tensor pseudofunctor and the transformation with components
\[
	A\tn B \rTo^{s_{A,B}} B\tn A \rTo^{s_{B,A}} A\tn B,
\]
subject to coherence conditions. However, all the general facts
that we need concerning symmetric promonoidal categories are true
more generally in the braided case, hence we have no need to
consider symmetry explicitly in the abstract setting. 
\begin{definition} 
	Let $\C$ be a pseudomonoid in the braided monoidal bicategory $\B$.
	A \defn{braiding} for $\C$ is a 2-cell $\ss$:
	\begin{diagram}
		\C\tn\C &\rTo^{s_{{\C,\C}}}&\C\tn\C\\
		&\rdTo[snake=-1ex](1,2)<{P}
			\raise1ex\hbox{$\begin{array}c\Rightarrow\\[-5pt]\ss\end{array}$}%
			\ldTo[snake=1ex](1,2)>{P}\\
		&\C
	\end{diagram}
	subject to two equations, which (in a Gray monoid) are as follows:
	\begin{equation}\label{eq-sa-left}
		\begin{array}{l}
		\begin{diagram}
			&&\rnode{tl}{\C^{3}} & \lTo^{s_{\C,\C^{2}}} & \rnode{tr}{\C^{3}} \\
			&&\dTo<{P\tn \C} & s_{\C,P} & \dTo>{\C\tn P} \\
			&&\C^{2} & \lTo^{{s_{\C,\C}}} & \C^{2} \\
			&\aa && \raise 1em\hbox{$\ss$} \rdTo(1,2)_{P} \ldTo(1,2)_{P} && \aa \\
			\rnode{bl}{\C^{2}} && \rTo_{P} & \C & \lTo_{P} && \rnode{br}{\C^{2}}
			\nccurve[angleA=180,angleB=90]{->}{tl}{bl}\Bput{\C\tn P}
			\nccurve[angleA=0,angleB=90]{->}{tr}{br}\Aput{P\tn \C}
		\end{diagram}
		\\
		\multicolumn 1r{\quad=\quad
		\begin{diagram}
			&&\rnode{tl}{\C^{3}} & \lTo^{s_{\C,\C^{2}}} & \rnode{tr}{\C^{3}} \\
			&&& \raise 1em\hbox{$S_{\C|\C,\C}$}
				\luTo(1,2)_{\C\tn s_{\C,\C}}
				\ldTo(1,2)_{s_{\C,\C}\tn\C} \\
			&\C\tn\ss && \C^{3} && \ss\tn\C \\
			&&\ldTo(3,2)_{\C\tn P} & \aa & \rdTo(3,2)_{P\tn \C} \\
			\rnode{bl}{\C^{2}} && \rTo_{P} & \C & \lTo_{P} && \rnode{br}{\C^{2}}
			\nccurve[angleA=180,angleB=90]{->}{tl}{bl}\Bput{\C\tn P}
			\nccurve[angleA=0,angleB=90]{->}{tr}{br}\Aput{P\tn \C}
		\end{diagram}}
		\end{array}
	\end{equation}
	and
	\begin{equation}\label{eq-sa-right}
		\begin{array}{l}
		\begin{diagram}
			&&\rnode{tl}{\C^{3}} & \rTo^{s_{\C^{2},\C}} & \rnode{tr}{\C^{3}} \\
			&&\dTo<{P\tn \C} & s_{P,\C} & \dTo>{\C\tn P} \\
			&&\C^{2} & \rTo^{{s_{\C,\C}}} & \C^{2} \\
			&\aa && \raise 1em\hbox{$\ss$} \rdTo(1,2)_{P} \ldTo(1,2)_{P} && \aa \\
			\rnode{bl}{\C^{2}} && \rTo_{P} & \C & \lTo_{P} && \rnode{br}{\C^{2}}
			\nccurve[angleA=180,angleB=90]{->}{tl}{bl}\Bput{\C\tn P}
			\nccurve[angleA=0,angleB=90]{->}{tr}{br}\Aput{P\tn \C}
		\end{diagram}
		\\
		\multicolumn 1r{\quad=\quad
		\begin{diagram}
			&&\rnode{tl}{\C^{3}} & \rTo^{s_{\C^{2},\C}} & \rnode{tr}{\C^{3}} \\
			&&& \raise 1em\hbox{$S_{\C,\C|\C}$}
				\rdTo(1,2)_{\C\tn s_{\C,\C}}
				\ruTo(1,2)_{s_{\C,\C}\tn\C} \\
			&\C\tn\ss && \C^{3} && \ss\tn\C \\
			&&\ldTo(3,2)_{\C\tn P} & \aa & \rdTo(3,2)_{P\tn \C} \\
			\rnode{bl}{\C^{2}} && \rTo_{P} & \C & \lTo_{P} && \rnode{br}{\C^{2}}
			\nccurve[angleA=180,angleB=90]{->}{tl}{bl}\Bput{\C\tn P}
			\nccurve[angleA=0,angleB=90]{->}{tr}{br}\Aput{P\tn \C}
		\end{diagram}}
		\end{array}
	\end{equation}
\end{definition}
\begin{definition}
	A \emph{braided pseudomonoid} is a pseudomonoid
	equipped with a braiding.
\end{definition}
Observe that, if $\ss$ is a braiding for $\C$ with respect to the
monoidal bicategory braiding $s$, then the inverse of the right mate of $\ss$,
with respect to $s_{\C,\C}$ and $1_{\C}$,
is a braiding with respect to $s^{*}$. We shall denote this dual braiding
as $\ss^{*}$. (Note that, by Lemma~\chref{Bicats}{lemma-adjeq-twisted},
$\ss^{*}$ is also the right mate of the inverse of $\ss$.)

\section{Another approach to braided pseudomonoids}\label{s-braided-facts}
This section concerns the equation
\begin{equation}\label{eq-lrs}
	\begin{diagram}[s=2.5em,tight]
		&&\I\tn \C \\
		&\ldTo^{J\tn \C} && \luTo^{s_{\C,\I}}\\
		\C^{2} & \hbox to0pt{\hskip 4pt$\ll$\hss} & \dTo>1
			& \hskip-4pt U_{\C|\I}
			& \C\tn \I \\
		&\rdTo_{P} && \ldTo_{1} \\
		&&\C
	\end{diagram}
	\qquad=\qquad
	\begin{diagram}[s=2.5em,tight]
		\rnode{CC}{\C^{2}} & \lTo^{J\tn\C}& \I\tn \C &
			\lTo^{s_{\C,\I}}& \rnode{CI}{\C\tn \I} \\
		&\luTo_{s_{\C,\C}} &s_{\C,J}& \ldTo_{\C\tn J} \\
		&&\C^{2} \\
		&\raise 2em\hbox{$\ss$} &\dTo>P& \raise 2em\hbox{$\rr$} \\
		&&\rnode{C}{\C}
		\ncarc[arcangle=-45]{->}{CC}{C}\Bput{P}
		\ncarc[arcangle=45]{->}{CI}{C}\Aput{1}
	\end{diagram}
\end{equation}
We'll first show that this equation holds in every braided pseudomonoid,
and then we'll show that, in the presence of axioms \pref{eq-aa}, \pref{eq-sa-left}
and \pref{eq-sa-right}, the equations \pref{eq-lla} and \pref{eq-lrs} together
imply \pref{eq-lra}. This gives a useful alternative axiomatisation of braided
pseudomonoids.

In the case of ordinary braided monoidal categories, this equation corresponds
to the triangle:
\begin{diagram}[vtrianglewidth=1em]
	I\tn A && \rTo^{\sigma_{I,A}} && A\tn I \\
	&\rdTo_{\lambda_{A}} && \ldTo_{\rho_{A}} \\
	&&A
\end{diagram}
and the alternative axiomatisation consists of the ordinary pentagon
and hexagon equations together with this triangle and the triangle
\begin{diagram}[vtrianglewidth=1em]
	I\tn(A\tn B) && \rTo^{\alpha} && (I\tn A)\tn B \\
	&\rdTo[snake=-1ex]_{\lambda_{A\tn B}}
		&& \ldTo[snake=1ex]_{\lambda_{A}\tn B} \\
	&&A\tn B
\end{diagram}
The advantage of this axiomatisation, in the general case just
as in the case of ordinary monoidal categories, is that $\rho$ appears
just once; hence it may be eliminated from the data and defined
in terms of $\sigma$ and $\lambda$.

In fact this can be proved quite easily using the braided extension
of the calculus of components that was mentioned briefly in Section~\chref{Language}{s-braided}. Sadly time constraints have made
it impossible to incorporate this improvement in adequate detail, so
instead we give a direct proof using pasting diagrams. This may serve
at least to indicate what a dramatic simplification is made possible
by component-based reasoning.

\begin{remark}
	Notice that, if we can prove that this equation holds of every braided
	pseudomonoid, then in particular it holds of the dual braiding $\ss^{*}$,
	so we have
	\[
		\begin{diagram}[s=2.5em,tight]
			&&\I\tn \C \\
			&\ldTo^{J\tn \C} && \luTo^{s^{*}_{\C,\I}}\\
			\C^{2} & \hbox to0pt{\hskip 4pt$\ll$\hss} & \dTo>1
				& \hskip-4pt U^{*}_{\C|\I}
				& \C\tn \I \\
			&\rdTo_{P} && \ldTo_{1} \\
			&&\C
		\end{diagram}
		\qquad=\qquad
		\begin{diagram}[s=2.5em,tight]
			\rnode{CC}{\C^{2}} & \lTo^{J\tn\C}& \I\tn \C &
				\lTo^{s^{*}_{\C,\I}}& \rnode{CI}{\C\tn \I} \\
			&\luTo_{s^{*}_{\C,\C}} & s^{*}_{\C,J} & \ldTo_{\C\tn J} \\
			&&\C^{2} \\
			&\raise 2em\hbox{$\ss^{*}$} &\dTo>P& \raise 2em\hbox{$\rr$} \\
			&&\rnode{C}{\C}
			\ncarc[arcangle=-45]{->}{CC}{C}\Bput{P}
			\ncarc[arcangle=45]{->}{CI}{C}\Aput{1}
		\end{diagram}
	\]
	Taking mates then gives
	\begin{equation}\label{eq-lrs'}
		\begin{diagram}[s=2.5em,tight]
			&&\I\tn \C \\
			&\ldTo^{J\tn \C} && \rdTo^{s_{\I,\C}}\\
			\C^{2} & \hbox to0pt{\hskip 4pt$\ll$\hss} & \dTo>1
				& \hskip-4pt U_{\I|\C}
				& \C\tn \I \\
			&\rdTo_{P} && \ldTo_{1} \\
			&&\C
		\end{diagram}
		\qquad=\qquad
		\begin{diagram}[s=2.5em,tight]
			\rnode{CC}{\C^{2}} & \lTo^{J\tn\C}& \I\tn \C &
				\rTo^{s_{\I,\C}}& \rnode{CI}{\C\tn \I} \\
			&\rdTo_{s_{\C,\C}} &s_{J,\C}& \ldTo_{\C\tn J} \\
			&&\C^{2} \\
			&\raise 2em\hbox{$\ss$} &\dTo>P& \raise 2em\hbox{$\rr$} \\
			&&\rnode{C}{\C}
			\ncarc[arcangle=-45]{->}{CC}{C}\Bput{P}
			\ncarc[arcangle=45]{->}{CI}{C}\Aput{1}
		\end{diagram}
	\end{equation}
	So a proof of \pref{eq-lrs} will also establish \pref{eq-lrs'}. This is
	an example of how the duality principle can be used to establish facts
	about the braiding $\ss$, not only about the dual braiding $\ss^{*}$.
\end{remark}

These equations generalise the one of \citet[Prop.~2.1, part~1]{BTC},
and indeed the essence of the lengthy argument here is contained in the
two-line sketch proof therein. The proof is originally due to \citet{KellyML}
-- that he considered a symmetry, rather than a braiding, does not affect
the proof.

At the end of the proof, we shall need to appeal to the following lemma.
It corresponds to the fact that, in a monoidal category,
the functors $I\tn-$ and $-\tn I$ are faithful.
\begin{lemma}\label{lemma-faithful}
	Let $A$ be some object of $\B$, let $f$, $g: A\to\C$
	and let $\gamma$, $\delta: f\To g$. If
	\[\hbox{\vrule height 3em depth 2em width 0pt}
	\begin{diagram}
		\rnode{CA}{\C\times A} & \Downarrow{\scriptstyle\C\times\gamma} & \rnode{CC}{\C\times\C} &\rTo^P &\C
		\ncarc{->}{CA}{CC}\Aput{\C\times f}
		\ncarc{<-}{CC}{CA}\Aput{\C\times g}
	\end{diagram}
	\quad=\quad
	\begin{diagram}
		\rnode{CA}{\C\times A} & \Downarrow{\scriptstyle\C\times\delta} & \rnode{CC}{\C\times\C} &\rTo^P &\C
		\ncarc{->}{CA}{CC}\Aput{\C\times f}
		\ncarc{<-}{CC}{CA}\Aput{\C\times g}
	\end{diagram}
	\]
	then $\gamma = \delta$.
	Dually, if
	\[\hbox{\vrule height 3em depth 2em width 0pt}
	\begin{diagram}
		\rnode{AC}{A\times\C} & \Downarrow{\scriptstyle\gamma\times\C} & \rnode{CC}{\C\times\C} &\rTo^P &\C
		\ncarc{->}{AC}{CC}\Aput{f\times\C}
		\ncarc{<-}{CC}{AC}\Aput{g\times\C}
	\end{diagram}
	\quad=\quad
	\begin{diagram}
		\rnode{AC}{A\times\C} & \Downarrow{\scriptstyle\delta\times\C} & \rnode{CC}{\C\times\C} &\rTo^P &\C
		\ncarc{->}{AC}{CC}\Aput{f\times\C}
		\ncarc{<-}{CC}{AC}\Aput{g\times\C}
	\end{diagram}
	\]
	then $\gamma=\delta$.
\end{lemma}
\begin{proof}
	Easy, via the calculus of components.
\end{proof}

\begin{propn}\label{prop-lrs}
	Equation~\pref{eq-lrs} holds in any braided pseudomonoid,
	hence (by the discussion above) so does~\pref{eq-lrs'}.
\end{propn}
\begin{proof}
	\diagramstyle[hug]
	Consider the 2-cell
	
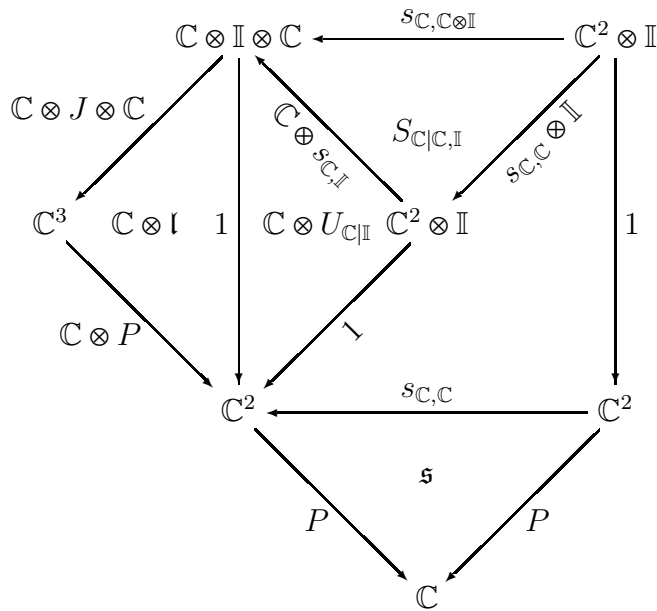
\begin{diagram}[hug,s=3em,tight]
		&&\C\tn \I\tn \C && \lTo^{s_{\C,\C\tn \I}} && \C^{2}\tn \I \\
		&\ldTo[nohug,snake=-1ex]^{\C\tn J\tn \C}
			&& \luTo_{\C\tn s_{\C,\I}}
			& S_{\C|\C,\I}
			& \ldTo_{s_{\C,\C}\tn \I} \\
		\C^{3} & \C\tn\ll & \dTo<1 & \hskip-1em\C\tn U_{\C|\I} & \C^{2}\tn \I
			&& \dTo>1 \\
		&\rdTo[nohug,snake=-1ex]_{\C\tn P} &&\ldTo_{1} \\
		&& \C^{2} && \lTo^{s_{\C,\C}} && \C^{2} \\
		&&& \rdTo[nohug]_{P} & \raise 1em\hbox{$\ss$} & \ldTo[nohug]_{P} \\
		&&&& \C
	\end{diagram}
	By Proposition~\chref{MonBicats}{prop-braiding-unit-1}, this is equal to
	\begin{diagram}[nohug,s=2em,tight]
		&&\C\tn \I\tn \C && \lTo^{s_{\C,\C\tn \I}} && \C^{2}\tn \I \\
		&\ldTo[snake=-1ex]^{\C\tn J\tn \C}
			&&
			&
			& \\
		\C^{3} & \C\tn\ll & \dTo>1 &&
			&& \dTo>1 \\
		&\rdTo[snake=-1ex]_{\C\tn P} \\
		&& \C^{2} && \lTo^{s_{\C,\C}} && \C^{2} \\
		&&& \rdTo[nohug]_{P} & \raise 1em\hbox{$\ss$} & \ldTo[nohug]_{P} \\
		&&&& \C
	\end{diagram}
 	which is equal, by equation \pref{eq-lra}, to
	\begin{diagram}[nohug,s=2em,tight]
		&&\C\tn \I\tn \C && \lTo^{s_{\C,\C\tn \I}} && \C^{2}\tn \I \\
		&\ldTo[snake=-1ex]^{\C\tn J\tn \C}
			&&
			&
			& \\
		\C^{3} & \rr\tn\C & \dTo>1 &&
			&& \dTo>1 \\
		&\rdTo[hug]_{P\tn\C} \\
		\dTo<{\C\tn P} && \C^{2} && \lTo^{s_{\C,\C}} && \C^{2} \\
		&\aa && \rdTo[nohug]_{P} & \raise 1em\hbox{$\ss$} & \ldTo[nohug]_{P} \\
		\C^{2} &&\rTo_{P}&& \C
	\end{diagram}
	Since $s$ is pseudo-natural, this is equal to
	\begin{diagram}
		&&\C\tn \I\tn \C & \lTo^{s_{\C,\C\tn \I}} & \C^{2}\tn \I \\
		&\ldTo^{\C\tn J\tn \C}
			& s_{\C,\C\tn J}
			& \ldTo^{\C^{2}\tn J}
			& \\
		\C^{3} & \lTo^{s_{\C,\C^{2}}} & \C^{3} & \C\tn\rr
			& \dTo>1 \\
		&\rdTo_{P\tn\C} & s_{\C,P} & \rdTo_{\C\tn P} \\
		\dTo<{\C\tn P} && \C^{2} & \lTo^{s_{\C,\C}} & \C^{2} \\
		&\aa && \rdTo[nohug](1,2)_{P} \raise 1em\hbox{$\ss$} \ldTo[nohug](1,2)_{P} \\
		\C^{2} &&\rTo_{P}& \C
	\end{diagram}
	which, by equation \pref{eq-rra}, equals
	\begin{diagram}
		&&&&\C\tn \I\tn \C&\lTo^{s_{\C,\C\tn \I}} & \rnode{CCI}{\C^{2}\tn \I} \\
		&&&\ldTo^{\C\tn J\tn \C} & s_{\C,\C\tn J} & \ldTo_{\C^{2}\tn J} \\
		&&\rnode{tl}{\C^{3}} & \lTo^{s_{\C,\C^{2}}} & \rnode{tr}{\C^{3}}
			&& \hbox to0pt{\quad$\sim$\hss} \\
		&&\dTo<{P\tn \C} & s_{\C,P} & \dTo>{\C\tn P} \\
		&&\C^{2} & \lTo^{{s_{\C,\C}}} & \C^{2} &&&& \rnode{CI}{\C\tn \I} \\
		&\aa && \raise 1em\hbox{$\ss$} \rdTo[nohug](1,2)_{P} \ldTo[nohug](1,2)_{P}
			&& \aa && \ldTo_{\C\tn J}\\
		\rnode{bl}{\C^{2}} && \rTo_{P} & \rnode{C}{\C} & \lTo_{P} && \rnode{br}{\C^{2}} \\
		&&&&&&\rr \\
		\nccurve[angleA=180,angleB=90]{->}{tl}{bl}\Bput{\C\tn P}
		\nccurve[angleA=0,angleB=90]{->}{tr}{br}\Aput{P\tn \C}
		\nccurve[angleA=-20,angleB=90]{->}{CCI}{CI}\Aput{P\tn \I}
		\nccurve[angleA=270,angleB=-45,ncurv=1.5]{->}{CI}{C}\Aput{1}
	\end{diagram}
	By \pref{eq-sa-left}, this is equal to
	\begin{diagram}
		&&&&\C\tn \I\tn \C&\lTo^{s_{\C,\C\tn \I}} & \rnode{CCI}{\C^{2}\tn \I} \\
		&&&\ldTo^{\C\tn J\tn \C} & s_{\C,\C\tn J} & \ldTo_{\C^{2}\tn J} \\
		&&\rnode{tl}{\C^{3}} & \lTo^{s_{\C,\C^{2}}} & \rnode{tr}{\C^{3}}
			&& \hbox to0pt{\quad$\sim$\hss} \\
		&&& \raise 1em\hbox{$S_{\C|\C,\C}$}
			\luTo[nohug](1,2)_{\C\tn s_{\C,\C}}
			\ldTo[nohug](1,2)_{s_{\C,\C}\tn\C} \\
		&\C\tn\ss && \C^{3} && \ss\tn\C &&& \rnode{CI}{\C\tn \I} \\
		&&\ldTo(3,2)_{\C\tn P} & \aa & \rdTo(3,2)_{P\tn \C}
				&&& \ldTo_{\C\tn J}\\
		\rnode{bl}{\C^{2}} && \rTo_{P} & \rnode{C}{\C} & \lTo_{P}
			&& \rnode{br}{\C^{2}} \\
		&&&&&&\rr \\
		\nccurve[angleA=180,angleB=90]{->}{tl}{bl}\Bput{\C\tn P}
		\nccurve[angleA=0,angleB=90]{->}{tr}{br}\Aput{P\tn \C}
		\nccurve[angleA=-20,angleB=90]{->}{CCI}{CI}\Aput{P\tn \I}
		\nccurve[angleA=270,angleB=-45,ncurv=1.5]{->}{CI}{C}\Aput{1}
	\end{diagram}
	which, since $\tn$ is a pseudo-functor, is equal to
	
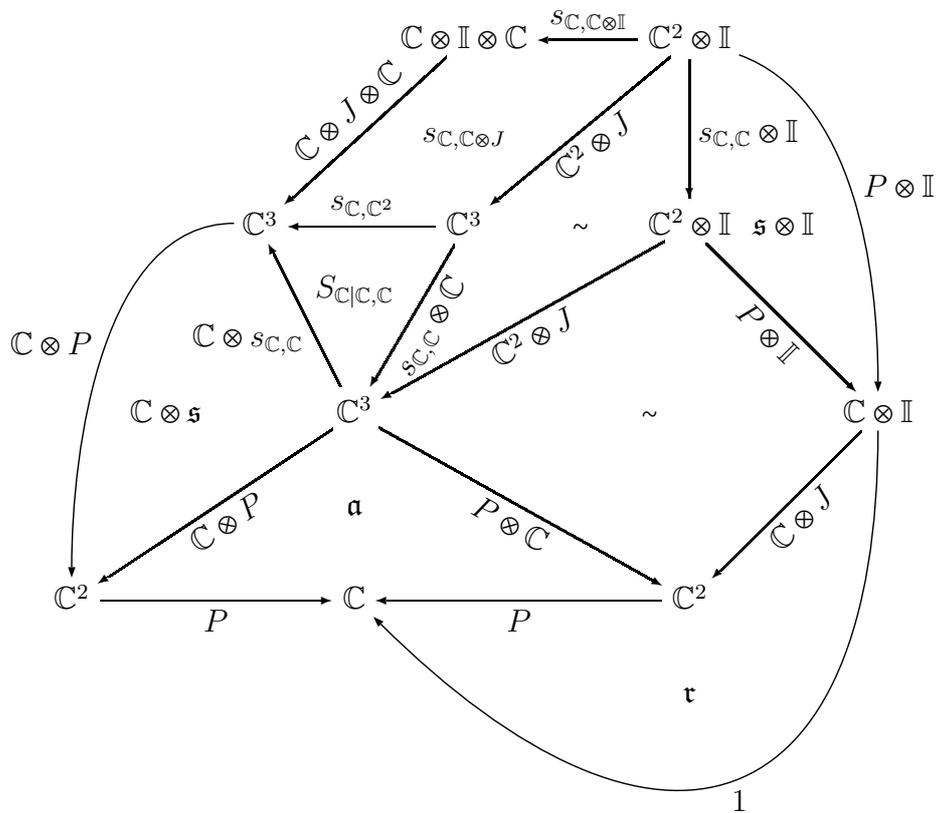
\begin{diagram}
		&&&&\C\tn \I\tn \C&\lTo^{s_{\C,\C\tn \I}} & \rnode{CCI}{\C^{2}\tn \I} \\
		&&&\ldTo^{\C\tn J\tn \C} & s_{\C,\C\tn J} & \ldTo_{\C^{2}\tn J}
		 	& \dTo>{s_{\C,\C}\tn\I} \\
		&&\rnode{tl}{\C^{3}} & \lTo^{s_{\C,\C^{2}}} & \rnode{tr}{\C^{3}}
			& \sim & \C^{2}\tn \I & \ss\tn\I \\
		&&& \raise 1em\hbox{$S_{\C|\C,\C}$}
			\luTo[nohug](1,2)_{\C\tn s_{\C,\C}}
			\ldTo(1,2)_{s_{\C,\C}\tn\C}
			&& \ldTo(3,2)_{\C^{2}\tn J}
			&& \rdTo_{P\tn \I} \\
		&\C\tn\ss && \C^{3} &&& \hbox to 0pt{\hss$\sim$\quad} && \rnode{CI}{\C\tn \I} \\
		&&\ldTo(3,2)_{\C\tn P} & \aa & \rdTo(3,2)_{P\tn \C}
				&&& \ldTo_{\C\tn J}\\
		\rnode{bl}{\C^{2}} && \rTo_{P} & \rnode{C}{\C} & \lTo_{P}
			&& \rnode{br}{\C^{2}} \\
		&&&&&&\rr \\
		\nccurve[angleA=180,angleB=90]{->}{tl}{bl}\Bput{\C\tn P}
		\nccurve[angleA=-20,angleB=90]{->}{CCI}{CI}\Aput{P\tn \I}
		\nccurve[angleA=270,angleB=-45,ncurv=1.5]{->}{CI}{C}\Aput{1}
	\end{diagram}
	Since $S$ is a modification, this in turn is equal to
	
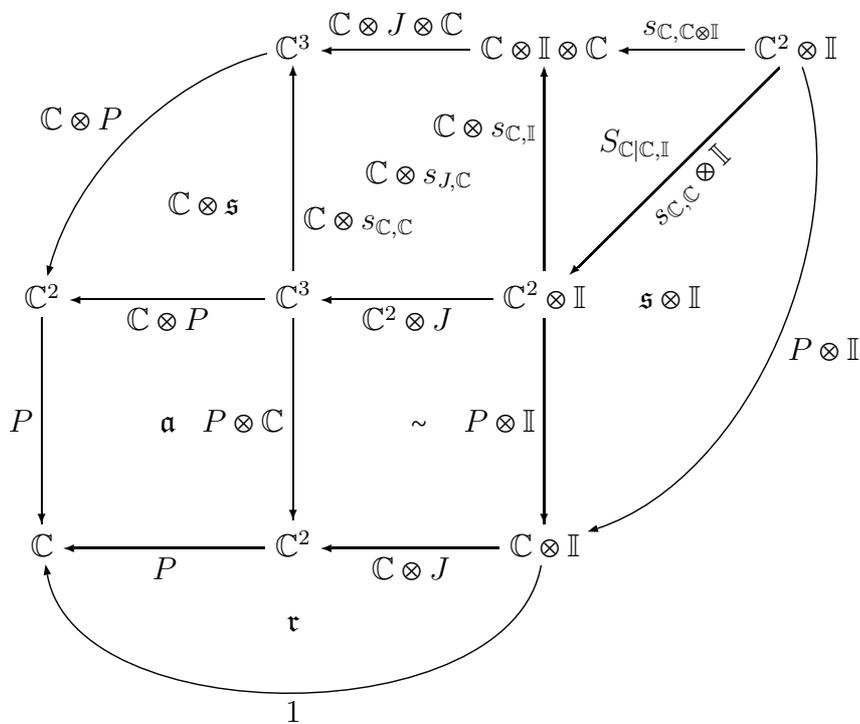
\begin{diagram}[s=4em,tight]
		&& \rnode{CCC}{\C^{3}} & \lTo^{\C\tn J\tn \C} & \C\tn \I\tn \C
			& \lTo^{s_{\C,\C\tn \I}} & \rnode{CCI}{\C^{2}\tn \I} \\
		&\raise -1em\rlap{$\C\tn\ss$} & \uTo[snake=-1.5em]>{\C\tn s_{\C,\C}}
			& \C\tn s_{J,\C}
			& \uTo[snake=1.5em]<{\C\tn s_{\C,\I}}
			& \raise1em\llap{$S_{\C|\C,\I}$}\ldTo_{s_{\C,\C}\tn\I} \\
		\rnode{CC}{\C^{2}} & \lTo_{\C\tn P} & \C^{3} & \lTo_{\C^{2}\tn J}
			& \C^{2}\tn \I & \ss\tn\I \\
		\dTo<{P} & \aa & \dTo<{P\tn \C} & \sim & \dTo<{P\tn \I} \\
		\rnode{C}{\C} & \lTo_{P} & \C^{2} & \lTo_{\C\tn J} & \rnode{CI}{\C\tn \I} \\
		&&\raise1.6em\hbox{$\rr$}
		\ncarc[arcangle=80,ncurv=.7]{->}{CI}{C} \Aput{1}
		\ncarc[arcangle=-30]{->}{CCC}{CC} \Bput{\C\tn P}
		\ncarc[arcangle=45]{->}{CCI}{CI} \Aput{P\tn \I}
	\end{diagram}
	which by \pref{eq-rra} is equal to
	
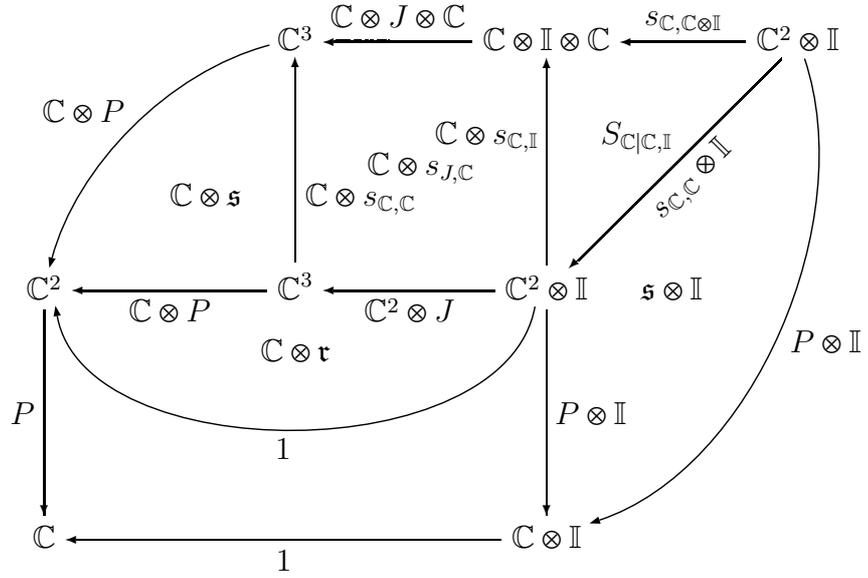
\begin{diagram}[s=4em,tight]
		&& \rnode{CCC}{\C^{3}} & \lTo^{\C\tn J\tn \C} & \C\tn \I\tn \C
			& \lTo^{s_{\C,\C\tn \I}} & \rnode{CCI}{\C^{2}\tn \I} \\
		&\raise -1em\rlap{$\C\tn\ss$} & \uTo[snake=-1em]>{\C\tn s_{\C,\C}} & \C\tn s_{J,\C}
			& \uTo[snake=1em]<{\C\tn s_{\C,\I}}
			& \raise1em\llap{$S_{\C|\C,\I}$}\ldTo_{s_{\C,\C}\tn\I} \\
		\rnode{CC}{\C^{2}} & \lTo_{\C\tn P} & \C^{3} & \lTo_{\C^{2}\tn J}
			& \rnode{mid}{\C^{2}\tn \I} & \ss\tn\I \\
		\dTo<{P} & & \raise2em\hbox{$\C\tn\rr$} & & \dTo>{P\tn \I} \\
		\rnode{C}{\C} & & \lTo_{1} & & \rnode{CI}{\C\tn \I} \\
		\ncarc[arcangle=80,ncurv=.7,offset=-3pt]{->}{mid}{CC} \aput(0.517){1}
		\ncarc[arcangle=-30]{->}{CCC}{CC} \Bput{\C\tn P}
		\ncarc[arcangle=45]{->}{CCI}{CI} \Aput{P\tn \I}
	\end{diagram}
	If we compare this with the diagram we began with, and
	cancel the invertible 2-cells $S_{\C|\C,\I}$ and $\ss=\ss\tn\I$,
	we have that
	
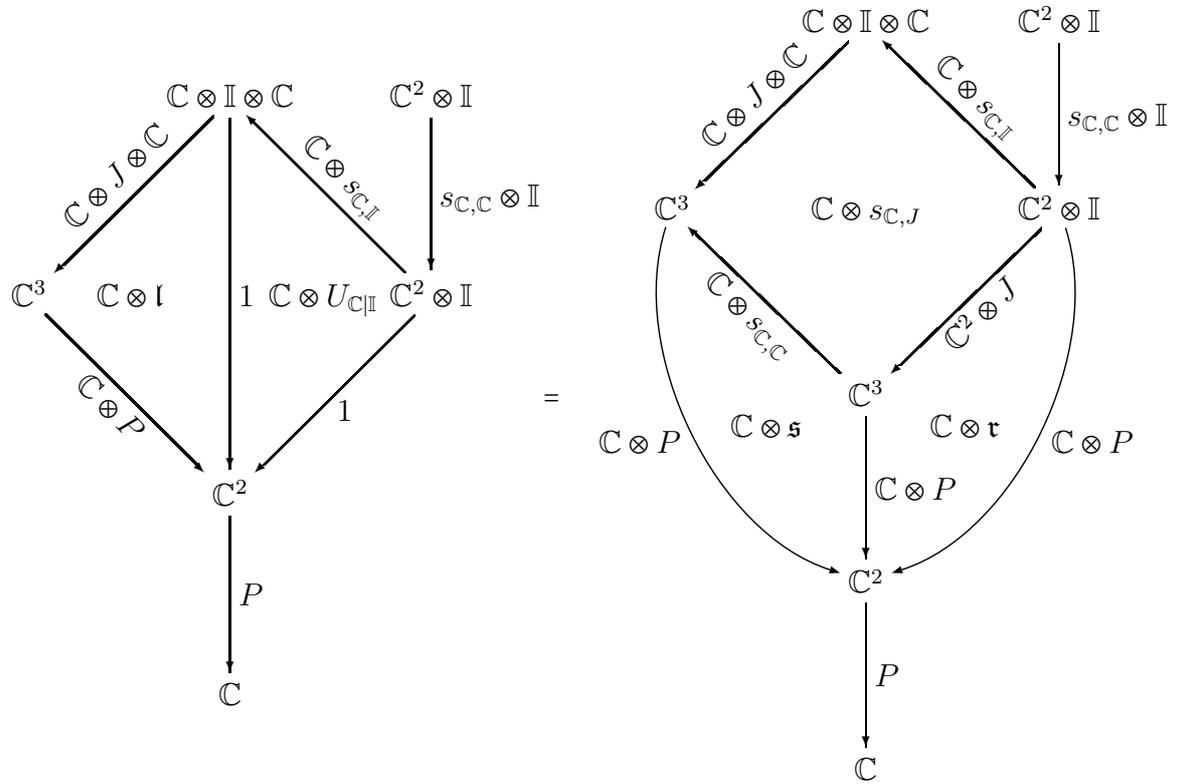
\[
	\begin{diagram}[s=3.2em,tight]
		&&\C\tn \I\tn \C && \C^{2}\tn \I \\
		&\ldTo^{\C\tn J\tn \C}
			&& \luTo^{\C\tn s_{\C,\I}}
			& \dTo_{s_{\C,\C}\tn \I} \\
		\C^{3} & \C\tn\ll & \dTo>1 & \hbox to 3em{\hss$\C\tn U_{\C|\I}$}
			& \C^{2}\tn \I \\
		&\rdTo_{\C\tn P} &&\ldTo[nohug]_{1} \\
		&& \C^{2} \\
		&& \dTo[nohug]_{P} \\
		&& \C
	\end{diagram}
	\qquad=\qquad
	\begin{diagram}
		&&\C\tn\I\tn \C && \C^{2}\tn \I\\
		&\ldTo^{\C\tn J\tn \C} && \luTo^{\C\tn s_{\C,\I}} & \dTo>{s_{\C,\C}\tn\I} \\
		\rnode{CCC}{\C^{3}} && \C\tn s_{\C,J} && \rnode{CCI}{\C^{2}\tn \I} \\
		&\luTo_{\C\tn s_{\C,\C}} && \ldTo_{\C^{2}\tn J} \\
		&&\C^{3} \\
		&\raise 2em\hbox{$\C\tn\ss$} &\dTo>{\C\tn P}& \raise 2em\hbox{$\C\tn\rr$} \\
		&&\rnode{CC}{\C^{2}} \\
		&&\dTo>P \\
		&&\C
		\ncarc[arcangle=-45]{->}{CCC}{CC}\Bput{\C\tn P}
		\ncarc[arcangle=45]{->}{CCI}{CC}\Aput{\C\tn P}
	\end{diagram}
	\]
	and since $s_{\C,\C}$ is an equivalence, it follows that
	\[
	\begin{diagram}[s=3.2em,tight]
		&&\C\tn \I\tn \C \\
		&\ldTo^{\C\tn J\tn \C}
			&& \luTo^{\C\tn s_{\C,\I}} \\
		\C^{3} & \C\tn\ll & \dTo>1 & \hbox to 3em{\hss$\C\tn U_{\C|\I}$}
			& \C^{2}\tn \I \\
		&\rdTo_{\C\tn P} &&\ldTo[nohug]_{1} \\
		&& \C^{2} \\
		&& \dTo[nohug]_{P} \\
		&& \C
	\end{diagram}
	\qquad=\qquad
	\begin{diagram}
		\rnode{CCC}{\C^{3}} &\lTo^{\C\tn J\tn \C}& \C\tn s_{\C,J}
			& \lTo^{\C\tn s_{\C,\I}} & \rnode{CCI}{\C^{2}\tn \I} \\
		&\luTo_{\C\tn s_{\C,\C}} & \raise1ex\hbox{$\C\tn s_{\C,J}$}
			& \ldTo_{\C^{2}\tn J} \\
		&&\C^{3} \\
		&\raise 2em\hbox{$\C\tn\ss$} &\dTo>{\C\tn P}& \raise 2em\hbox{$\C\tn\rr$} \\
		&&\rnode{CC}{\C^{2}} \\
		&&\dTo>P \\
		&&\C
		\ncarc[arcangle=-45]{->}{CCC}{CC}\Bput{\C\tn P}
		\ncarc[arcangle=45]{->}{CCI}{CC}\Aput{\C\tn P}
	\end{diagram}
	\]
	Now Lemma~\ref{lemma-faithful} yields the claim.
\end{proof}

\begin{propn}
	Let the object $\C$, the 1-cells $P$ and $J$,
	and the 2-cells $\aa$, $\ll$, $\rr$ and $\ss$ be given, as in
	the definition of braided pseudomonoid. Suppose that equations \pref{eq-aa},
	\pref{eq-sa-left}, \pref{eq-sa-right} and \pref{eq-lla} are satisfied.
	If~\pref{eq-lrs} or~\pref{eq-lrs'}
	is satisfied then the structure is indeed a braided pseudomonoid.
	
	It follows that a braided pseudomonoid may be defined using just
	the 2-cells $\aa$, $\ll$, and $\ss$, subject to equations
	\pref{eq-aa}, \pref{eq-sa-left}, \pref{eq-sa-right} and~\pref{eq-lla}.
	The 2-cell $\rr$ can be defined, if necessary, using equation~\pref{eq-lrs}
	or~\pref{eq-lrs'}.
\end{propn}
\begin{proof}
	We shall assume equations~\pref{eq-sa-left}, \pref{eq-lla} and~\pref{eq-lrs},
	and derive~\pref{eq-lra}.
	Consider the 2-cell
	
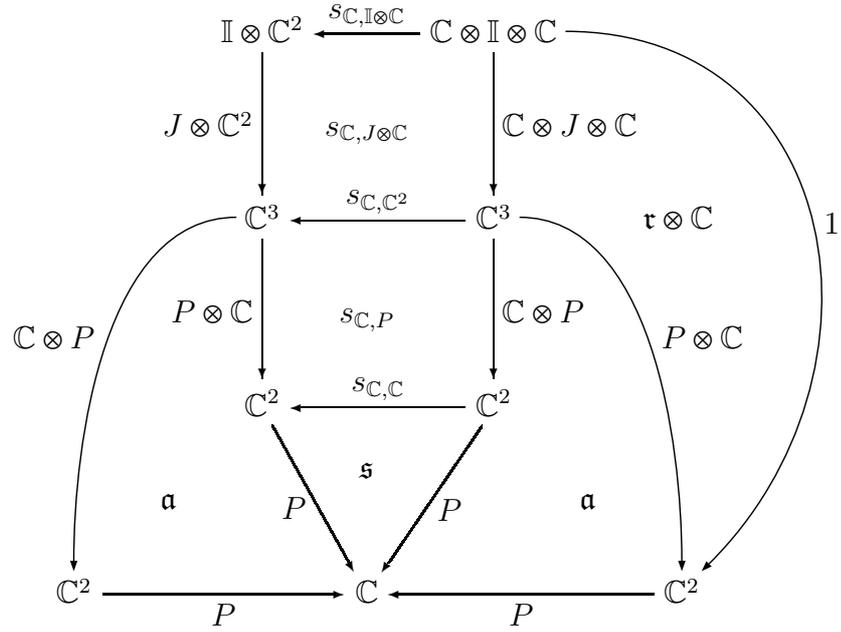
\begin{diagram}
		&&\I\tn \C^{2} & \lTo^{s_{\C,\I\tn \C}} & \rnode{CIC}{\C\tn \I\tn \C} \\
		&&\dTo<{J\tn \C^{2}} & s_{\C,J\tn\C} & \dTo>{\C\tn J\tn \C} \\
		&&\rnode{tl}{\C^{3}} & \lTo^{s_{\C,\C^{2}}} & \rnode{tr}{\C^{3}}
			&& \hbox to 2em{\hss$\rr\tn\C$} \\
		&&\dTo<{P\tn \C} & s_{\C,P} & \dTo>{\C\tn P} \\
		&&\C^{2} & \lTo^{{s_{\C,\C}}} & \C^{2} \\
		&\aa && \raise 1em\hbox{$\ss$} \rdTo[nohug](1,2)_{P} \ldTo[nohug](1,2)_{P}
			&& \aa\\
		\rnode{bl}{\C^{2}} && \rTo_{P} & \rnode{C}{\C} & \lTo_{P} && \rnode{br}{\C^{2}} \\
		\nccurve[angleA=180,angleB=90]{->}{tl}{bl}\Bput{\C\tn P}
		\nccurve[angleA=0,angleB=90]{->}{tr}{br}\Aput{P\tn \C}
		\nccurve[angleA=0,angleB=45,ncurv=1]{->}{CIC}{br}\Aput{1}
	\end{diagram}
	By equation~\pref{eq-sa-left}, this is equal to
	
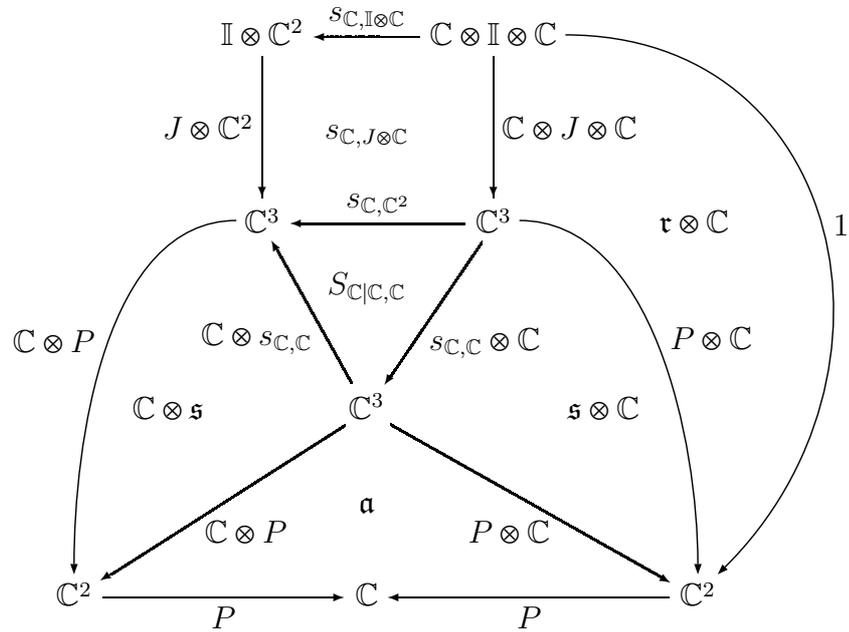
\begin{diagram}
		&&\I\tn \C^{2} & \lTo^{s_{\C,\I\tn \C}} & \rnode{CIC}{\C\tn \I\tn \C} \\
		&&\dTo<{J\tn \C^{2}} & s_{\C,J\tn\C} & \dTo>{\C\tn J\tn \C} \\
		&&\rnode{tl}{\C^{3}} & \lTo^{s_{\C,\C^{2}}} & \rnode{tr}{\C^{3}}
			&& \hbox to 2em{\hss$\rr\tn\C$} \\
		&&& \raise 1em\hbox{$S_{\C|\C,\C}$}
			\luTo[nohug](1,2)_{\C\tn s_{\C,\C}}
			\ldTo[nohug](1,2)_{s_{\C,\C}\tn\C} \\
		&\C\tn\ss && \C^{3} && \ss\tn\C \\
		&&\ldTo(3,2)_{\C\tn P} & \aa & \rdTo(3,2)_{P\tn \C} \\
		\rnode{bl}{\C^{2}} && \rTo_{P} & \rnode{C}{\C} & \lTo_{P}
			&& \rnode{br}{\C^{2}} \\
		\nccurve[angleA=180,angleB=90]{->}{tl}{bl}\Bput{\C\tn P}
		\nccurve[angleA=0,angleB=90]{->}{tr}{br}\Aput{P\tn \C}
		\nccurve[angleA=0,angleB=45,ncurv=1]{->}{CIC}{br}\Aput{1}
	\end{diagram}
	which, since $S$ is a modification, is equal to
	
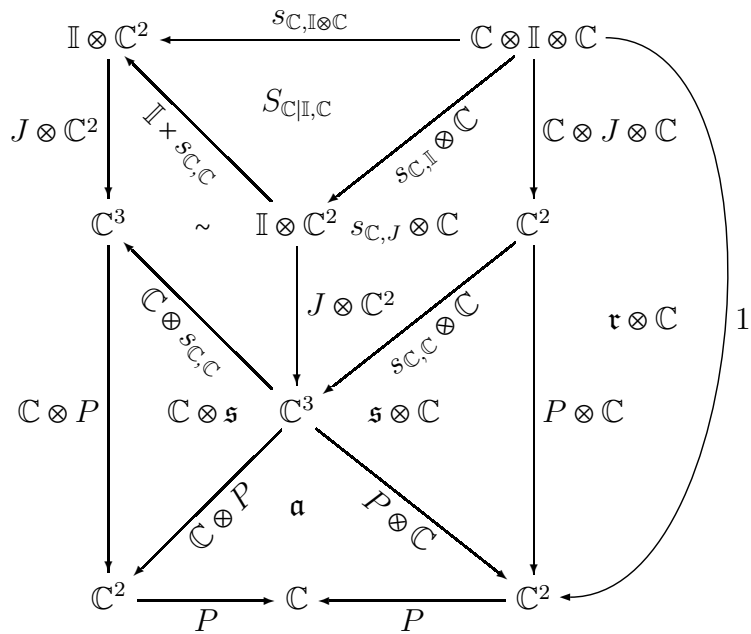
\begin{diagram}[hug]
		\I\tn \C^{2} && \lTo^{s_{\C,\I\tn \C}} && \rnode{CIC}{\C\tn \I\tn \C} \\
		\dTo<{J\tn \C^{2}}&\luTo_{\I\times s_{\C,\C}}
			& \raise 1em\hbox{$S_{\C|\I,\C}$}
			& \ldTo_{s_{\C,\I}\tn\C}
			& \dTo>{\C\tn J\tn \C}\\
		\C^{3} & \sim & \I\tn \C^{2} & s_{\C,J}\tn\C & \C^{2}\\
		&\luTo_{\C\tn s_{\C,\C}} & \dTo[snake=1ex]>{J\tn \C^{2}} & \ldTo_{s_{\C,\C}\tn\C}
			&&\rr\tn\C \\
		\dTo<{\C\tn P} & \C\tn\ss & \C^{3} & \ss\tn\C & \dTo>{P\tn\C} \\
		&\ldTo_{\C\tn P} &\aa& \rdTo_{P\tn \C} \\
		\C^{2} &\rTo_{P}& \C& \lTo_{P}& \rnode{CC}{\C^{2}} \\
		\nccurve[angleA=0,angleB=0]{->}{CIC}{CC} \Aput{1}
	\end{diagram}
	By equation~\pref{eq-lrs}, this is
	
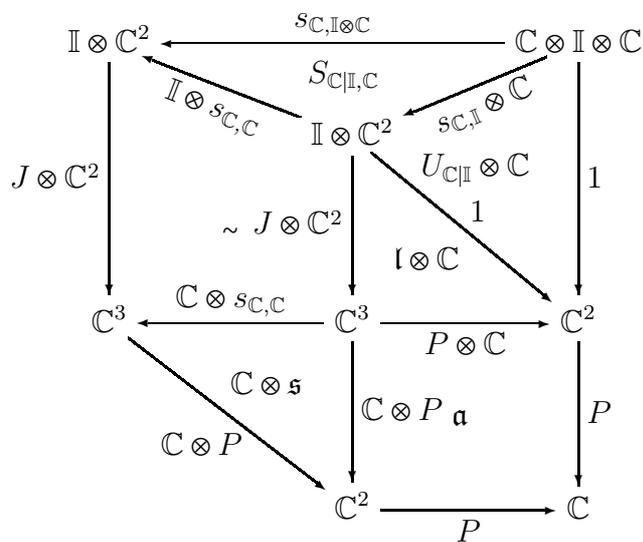
\begin{diagram}[hug,midvshaft]
		\I\tn \C^{2} &&  \raise-1em\hbox to0pt{\hskip5em$S_{\C|\I,\C}$\hss} \lTo^{s_{\C,\I\tn \C}}
			&& \C\tn \I\tn \C \\
		&\luTo(2,1)_{\I\tn s_{\C,\C}} & \I\tn \C^{2} & \ldTo(2,1)_{s_{\C,\I}\tn\C} 
			\raise -1em\hbox to 0pt{\hss\quad$U_{\C|\I}\tn\C$\hss} \\
		\dTo<{J\tn \C^{2}} & \sim & \dTo<{J\tn \C^{2}} & \raise-1em\llap{$\ll\tn\C$}
			\rdTo[nohug]^{1} & \dTo>{1} \\
		\C^{3} & \lTo^{\C\tn s_{\C,\C}} & \C^{3} & \rTo_{P\tn \C} & \C^{2} \\
		& \rdTo[nohug]_{\C\tn P} \raise1em\rlap{$\C\tn\ss$} & \dTo>{\C\tn P} & \aa & \dTo>P \\
		&& \C^{2} & \rTo_{P} & \C
	\end{diagram}
	which, by \pref{eq-lla}, is equal to
	
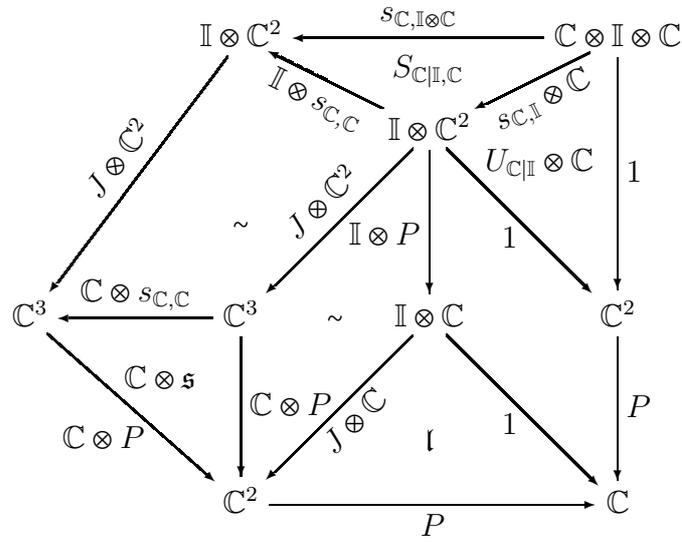
\begin{diagram}[hug,midvshaft]
		&&\I\tn \C^{2} &&  \raise-1em\hbox to0pt{\hskip3.5em$S_{\C|\I,\C}$\hss} \lTo^{s_{\C,\I\tn \C}}
			&& \C\tn \I\tn \C \\
		&\ldTo(2,3)<{J\tn \C^{2}} & &\luTo(2,1)_{\I\tn s_{\C,\C}} & \I\tn \C^{2} & \ldTo(2,1)_{s_{\C,\I}\tn\C} 
			\raise -1em\hbox to 0pt{\hss\quad$U_{\C|\I}\tn\C$\hss} \\
		&& \sim & \ldTo<{J\tn \C^{2}} & \dTo[snake=-1ex]<{\I\tn P}
			& \rdTo[nohug]_{1} & \dTo>{1} \\
		\C^{3} & \lTo^{\C\tn s_{\C,\C}} & \C^{3} & \sim & \I\tn \C & & \C^{2} \\
		& \rdTo[nohug]_{\C\tn P} \raise1em\rlap{$\C\tn\ss$} & \dTo>{\C\tn P}
			& \ldTo_{J\tn \C} & \raise -1em\hbox{$\ll$} & \rdTo[nohug]_{1} & \dTo>P \\
		&& \C^{2} && \rTo_{P} && \C
	\end{diagram}
	which, since the $\sim$ cells are natural, is equal to
	\begin{diagram}[hug,midvshaft]
		&&\rnode{ICC}{\I\tn \C^{2}} &&  \raise-1em\hbox to0pt{\hskip3.5em$S_{\C|\I,\C}$\hss} \lTo^{s_{\C,\I\tn \C}}
			&& \C\tn \I\tn \C \\
		&\ldTo(2,3)<{J\tn \C^{2}} & &\luTo(2,1)_{\I\tn s_{\C,\C}}
			\raise -1.5em\hbox{$\I\tn\ss$}& \I\tn \C^{2} & \ldTo(2,1)_{s_{\C,\I}\tn\C} 
			\raise -1em\hbox to 0pt{\hss\quad$U_{\C|\I}\tn\C$\hss} \\
		& & & & \dTo[snake=-1ex]<{\I\tn P}
			& \rdTo[nohug]_{1} & \dTo>{1} \\
		\C^{3} & & \sim &  & \rnode{IC}{\I\tn \C} & & \C^{2} \\
		& \rdTo[nohug]_{\C\tn P} &
			& \ldTo_{J\tn \C} & \raise -1em\hbox{$\ll$} & \rdTo[nohug]_{1} & \dTo>P \\
		&& \C^{2} && \rTo_{P} && \C
		\ncarc[arcangle=-40,ncurv=.6]{->}{ICC}{IC} \Bput{\I\tn P}
	\end{diagram}
	By~\pref{eq-lla}, this is equal to
	
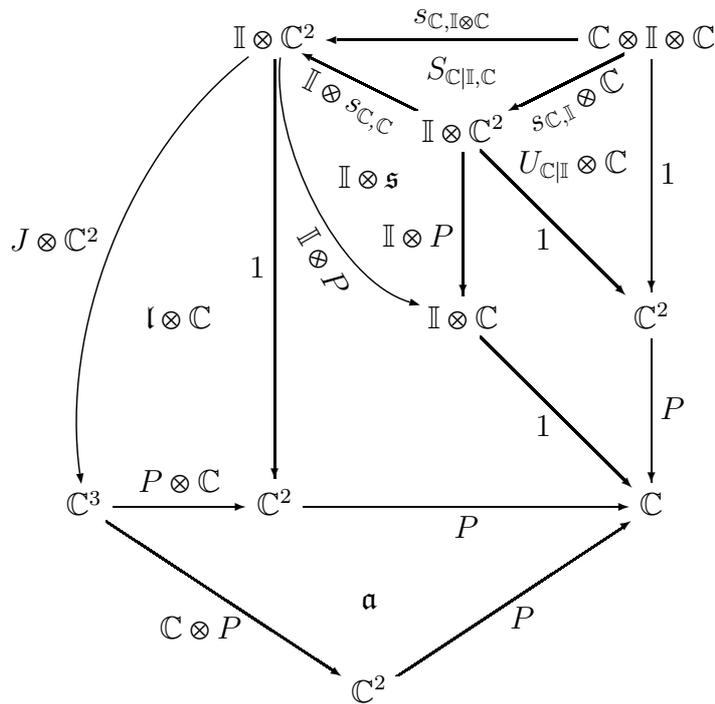
\begin{diagram}[hug,midvshaft]
		&&\rnode{ICC}{\I\tn \C^{2}} &&  \raise-1em\hbox to0pt{\hskip3.5em$S_{\C|\I,\C}$\hss} \lTo^{s_{\C,\I\tn \C}}
			&& \C\tn \I\tn \C \\
		& & &\luTo(2,1)_{\I\tn s_{\C,\C}}
			\raise -1.5em\hbox{$\I\tn\ss$}& \I\tn \C^{2} & \ldTo(2,1)_{s_{\C,\I}\tn\C} 
			\raise -1em\hbox to 0pt{\hss\quad$U_{\C|\I}\tn\C$\hss} \\
		& & & & \dTo[snake=-1ex]<{\I\tn P}
			& \rdTo[nohug]_{1} & \dTo>{1} \\
		& \ll\tn\C & \dTo<1 &  & \rnode{IC}{\I\tn \C} & & \C^{2} \\
		& & & & & \rdTo[nohug]_{1} & \dTo>P \\
		\rnode{CCC}{\C^{3}} & \rTo^{P\tn \C} & \C^{2} && \rTo_{P} && \C \\
		& \rdTo[nohug](3,2)_{\C\tn P} && \aa && \ruTo[nohug](3,2)_{P} \\
		&&& \C^{2}
		\ncarc[arcangle=-40,ncurv=.6,offsetA=3pt]{->}{ICC}{IC} \bput{-60}(.65){\I\tn P}
		\ncarc[arcangle=-30]{->}{ICC}{CCC} \Bput{J\tn \C^{2}}
	\end{diagram}
	which equals
	
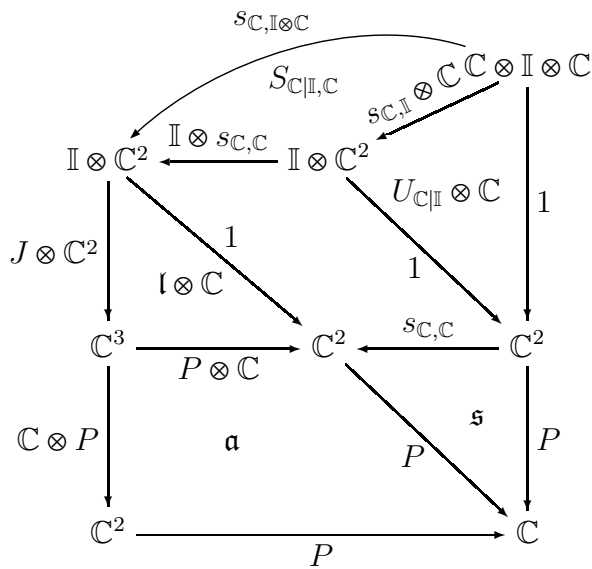
\begin{diagram}
		&&& \raise -1em\hbox{$$} &\rnode{CIC}{\C\tn \I\tn \C} \\
		\rnode{ICC}{\I\tn \C^{2}} & \lTo^{\I\tn s_{\C,\C}} & \I\tn \C^{2}
			& \ldTo[hug,snake=-6pt](2,1)^{s_{\C,\I}\tn\C}
			\raise -1em\hbox to2em{$U_{\C|\I}\tn\C$\hss} & \dTo[midvshaft]>1 \\
		\dTo<{J\tn \C^{2}} & \raise-1em\llap{$\ll\tn\C$} \rdTo^{1} && \rdTo_{1} \\
		\C^{3} & \rTo_{P\tn \C} & \C^{2} & \lTo^{s_{\C,\C}} & \C^{2} \\
		\dTo<{\C\tn P} & \rlap{$\aa$} && \rdTo_{P} \raise.8em\rlap{\hskip1.5em$\ss$}
			& \dTo>P \\
		\C^{2} && \rTo_{P} && \C
		\ncarc[arcangle=-30]{->}{CIC}{ICC} \Bput{s_{\C,\I\tn \C}}
		 	\Aput{\raise-4pt\hbox{$S_{\C|\I,\C}$}}
	\end{diagram}
	By one of the unit axioms in the definition of braiding for
	a monoidal bicategory, this is
	
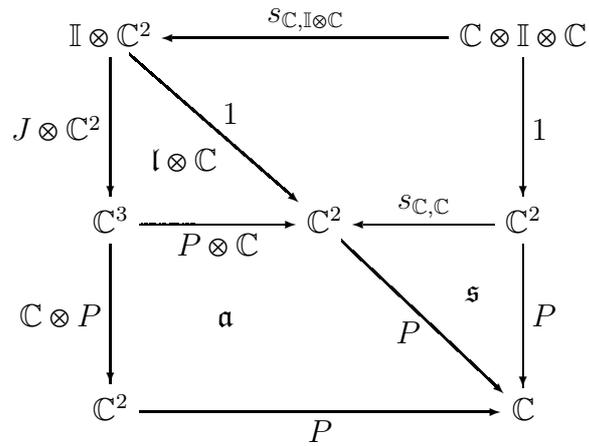
\begin{diagram}
		\rnode{ICC}{\I\tn \C^{2}} & & \lTo^{s_{\C,\I\tn\C}}
			& & \C\tn \I\tn \C \\
		\dTo<{J\tn \C^{2}} & \raise-1em\llap{$\ll\tn\C$} \rdTo^{1} &&& \dTo>1 \\
		\C^{3} & \rTo_{P\tn \C} & \C^{2} & \lTo^{s_{\C,\C}} & \C^{2} \\
		\dTo<{\C\tn P} & \rlap{$\aa$} && \rdTo_{P} \raise.8em\rlap{\hskip1.5em$\ss$}
			& \dTo>P \\
		\C^{2} && \rTo_{P} && \C
	\end{diagram}
	which, since $s$ is pseudo-natural, is equal to
	
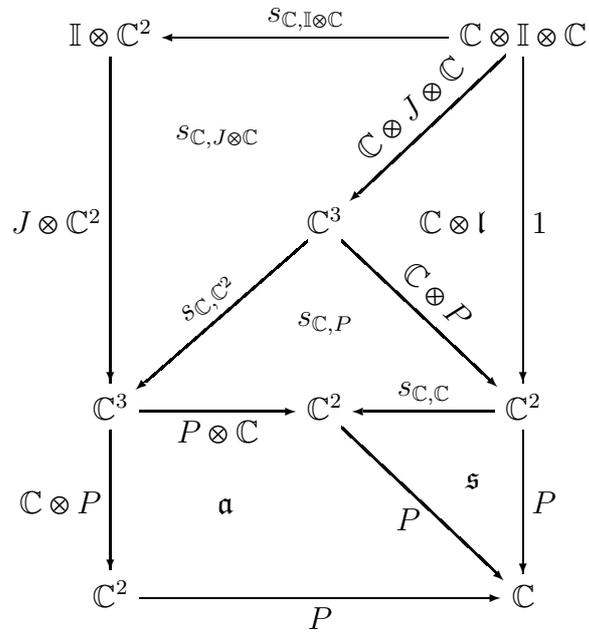
\begin{diagram}
		\I\tn \C^{2} &&\lTo^{s_{\C,\I\tn \C}} && \C\tn \I\tn \C \\
		& s_{\C,J\tn \C} && \ldTo[hug]^{\C\tn J\tn \C} \\
		\dTo<{J\tn \C^{2}} && \C^{3} & \rlap{$\C\tn\ll$} & \dTo>1 \\
		&\ldTo[hug]^{s_{\C,\C^{2}}} & s_{\C,P} & \rdTo[hug]^{\C\tn P} \\
		\C^{3} & \rTo_{P\tn \C} & \C^{2} & \lTo^{s_{\C,\C}} & \C^{2} \\
		\dTo<{\C\tn P} & \rlap{$\aa$} && \rdTo_{P} \raise.8em\rlap{\hskip1.5em$\ss$}
			& \dTo>P \\
		\C^{2} && \rTo_{P} && \C
	\end{diagram}
	If we now compare this with the 2-cell we began with, and cancel the common
	invertible 2-cell
	\begin{diagram}
		&&\I\tn \C^{2} & \lTo^{s_{\C,\I\tn \C}} & \rnode{CIC}{\C\tn \I\tn \C} \\
		&&\dTo<{J\tn \C^{2}} & s_{\C,J\tn\C} & \dTo>{\C\tn J\tn \C} \\
		&&\rnode{tl}{\C^{3}} & \lTo^{s_{\C,\C^{2}}} & \rnode{tr}{\C^{3}}\\
		&&\dTo<{P\tn \C} & s_{\C,P} & \dTo>{\C\tn P} \\
		&&\C^{2} & \lTo^{{s_{\C,\C}}} & \C^{2} \\
		&\aa && \raise 1em\hbox{$\ss$} \rdTo[nohug](1,2)_{P} \ldTo[nohug](1,2)_{P} \\
		\rnode{bl}{\C^{2}} && \rTo_{P} & \rnode{C}{\C} \\
		\nccurve[angleA=180,angleB=90]{->}{tl}{bl}\Bput{\C\tn P}
	\end{diagram}
	we obtain equation~\pref{eq-lra}, as claimed.
\end{proof}

\end{thesischapter}
    
\documentclass{robinthesis}

\begin{thesischapter}{Cayley}{Cayley's Theorem for Pseudomonoids}
In its traditional form, Cayley's theorem says that every finite group
is (isomorphic to) a subgroup of one of the symmetric groups. What the
proof actually shows is that every monoid $M$ is a submonoid of the monoid
of endofunctions $|M|\to|M|$, via the function that maps the element $x\in M$
to the function $x\o-$.

A `Cayley Theorem for monoidal categories' has been used by
\citet[Proposition~1.3]{BTC} to show that every monoidal category
is monoidally equivalent to a \emph{strict} monoidal category.
It turns out that we need the corresponding theorem for
promonoidal categories; and rather than just prove another special
case, it seems wise to generalise. So the purpose of this
chapter is to state and prove a Cayley Theorem for pseudomonoids.

As motivation, we begin by reviewing the one-dimensional case,
of monoids in a monoidal category.

\section{Cayley's theorem for monoids}

I'm not aware of any literature that specifically addresses the
question of giving an analogue of Cayley's theorem for monoids
in a monoidal category, but it is a special case of a standard
construction. A monoid in the monoidal category $\V$ can be
regarded as a one-object $\V$-enriched category.\footnote{This
operation -- reimagining a monoid as a one-object category -- is often
called \emph{suspension}.} Then we can apply the enriched Yoneda
Lemmas of \citet[sections~1.9 and~2.4]{KellyEnriched} to this
$\V$-category. Since it requires some effort to extract the
specifics of the one-object case from Kelly's constructions,
we give an overview here.

Let $\V$ be a monoidal category, and let $M$ be a monoid in $\V$,
with unit $e: I\to M$ and multiplication $m: M\tn M\to M$.
%

The Cayley-Yoneda theorems are best stated in terms of modules:
\begin{definition}\label{def-module}
	A \emph{right $M$-module} consists of an object $X\in\V$ and a map
	\[
		x: X\tn M\to X
	\]
	such that the diagrams
	\[
	\begin{diagram}
		X\tn M\tn M & \rTo^{x\tn M} & X\tn M \\
		\dTo<{X\tn m} && \dTo>{x} \\
		X\tn M & \rTo_{x} & X
	\end{diagram}
	\quad\mbox{and}\quad
	\begin{diagram}
		X &\rTo^{\cong} & X\tn I \\
		\dTo<1 && \dTo>{X\tn e} \\
		X & \lTo_{x} & X\tn M
	\end{diagram}
	\]
	commute.
\end{definition}
\begin{remark}
	Note that $(M,m)$ is a right $M$-module.
\end{remark}
\begin{definition}\label{def-module-map}
	Let $(X,x)$ and $(Y,y)$ be right M-modules.
	A \emph{morphism of modules} from $X$ to $Y$ is
	a map $f: X\to Y$ for which the diagram
	\begin{diagram}
		X\tn M & \rTo^{x} & X \\
		\dTo<{f\tn M} && \dTo>f \\
		Y\tn M & \rTo_{y} & Y
	\end{diagram}
	commutes. The set of such maps will be denoted
	$\Mod_{M}(X,Y)$.
\end{definition}
Kelly proves two versions of his Yoneda Lemma, which he refers
to as `weak' and `strong'. We shall call our corresponding theorems
the `external' and `internal' Cayley Theorem, respectively. The
internal theorem requires some additional properties of $\V$, but
the external one is perfectly general.

\subsection{The external Cayley Theorem}
\begin{thm}[External Cayley]\label{thm-1d-external}
	For every right $M$-module $X$, the natural transformation
	\[
		\phi_{X}: \Mod_{M}(M, X) \to \V(I, X),
	\]
	defined as $\phi_{X}(f) = f\cdot e$, is invertible with
	$\phi_{X}^{-1}(z)$ equal to the composite
	\[
		M \rTo^{\cong} I\tn M \rTo^{z\tn M} X\tn M \rTo^{x} X.
	\]
\end{thm}
\begin{proof}
	Let $f: M\to X$ be a map of modules. We'll first show that
	$\phi_{X}^{-1}(\phi_{X}(f)) = f$: consider the diagram
	\begin{diagram}
		M & \rTo^{\cong} & I\tn M & \rTo^{e\tn M} & M\tn M & \rTo^{f\tn M} & X\tn M \\
		&&& \rdTo_{\cong} & \dTo>m && \dTo>x \\
		&&&& M & \rTo_{f} & X
	\end{diagram}
	The triangle commutes by the left-unit law for the monoid $M$,
	and the square because $f$ is a map of modules. The upper edge
	is $\phi_{X}^{-1}(\phi_{X}(f))$ by definition, and the lower edge is
	equal to $f$.
	
	Now let $z: I\to X$, and consider the diagram
	\begin{diagram}
		&&X \\
		&\ruTo^{z} && \rdTo^{\cong} \\
		I & \rTo_{\cong} & I\tn I & \rTo_{z\tn I}& X\tn I \\
		\dTo<e && \dTo<{I\tn e} && \dTo>{X\tn e} \\
		M & \rTo_{\cong} & I\tn M & \rTo_{z\tn M} & X\tn M & \rTo_{x} & X
	\end{diagram}
	The cells commute by naturality and functoriality of tensor,
	and by the unit condition in the definition of module the upper
	edge is equal to $z$. The lower edge is $\phi_{X}(\phi_{X}^{-1}(z))$
	by definition. Thus $\phi_{X}$ and $\phi_{X}^{-1}$ are indeed
	mutually inverse.
	
	We must also show that $\phi_{X}^{-1}(z)$ is a map of modules,
	so let $z:I\to X$ and consider the diagram
	\begin{diagram}
		M\tn M & \rTo^{\cong} & I\tn M\tn M & \rTo^{z\tn M\tn M} & X\tn M\tn M
			& \rTo^{x\tn M} & X\tn M \\
		\dTo<m &&\dTo>{I\tn m} && \dTo>{X\tn m} && \dTo>m \\
		M & \rTo_{\cong} & I\tn M & \rTo_{z\tn M} & X\tn M & \rTo_{x} & X
	\end{diagram}
	The squares commute, from left to right, by the naturality of the
	left-unit isomorphism, the functoriality of tensor, and the fact
	that $X$ is a module. Thus the outer edge commutes, showing that
	$\phi_{X}^{-1}(z)$ is a map of modules.
\end{proof}
\begin{remark}
	More can be said about $\phi_{X}$, as follows. Each of the sets
	$\Mod_{M}(M, X)$ and $\V(I,X)$ is a right $\V(I,M)$-module
	in a natural way, using the $M$-module structure of $X$.
	Then $\phi_{X}$ is a map of modules with respect to these
	module structures. We omit\footnote{An earlier version of
	this chapter included these details: I removed them, as an
	unneeded distraction.} the details of this, but they are
	quite routine.
\end{remark}
\begin{remark} 
	The set $\V(I, M)$ inherits the monoid structure from $M$:
	its unit is $u:I\to M$, and the product of $a$, $b:I\to M$
	is the composite
	\[
		I \rTo^{\cong} I\tn I \rTo^{a\tn b} M\tn M \rTo^{m} M.
	\]
	This is the \emph{underlying ordinary monoid}
	of the abstract monoid $M$.
\end{remark}

To relate Theorem~\ref{thm-1d-external} to the ordinary
Cayley Theorem, take the module $X$ to be $M$ itself.
Then we have that $\V(I, M)$, the underlying ordinary
monoid of $M$, is isomorphic to $\Mod_{M}(M,M)$, which
is a submonoid of $\V(M,M)$.
We should like to be able to say that this $\phi_{M}$
is an isomorphism of monoids, i.e.\ that it preserves
the monoid structure.
\begin{propn}
	The isomorphism
	\[
		\phi_{M}: \Mod_{M}(M,M) \to \V(I, M)
	\]
	is a map of monoids.
\end{propn}
\begin{proof}
	For the unit, $\phi_{M}(1_{M}) = 1_{M}\cdot e = e$.
	For the multiplication, let $f$, $g: M\to M$ be right
	module morphisms: we must show that $\phi_{M}(f\cdot g)$:
	\[
		I \rTo^{e} M \rTo^{g} M \rTo^{f} M
	\]
	is equal to $\phi_{M}(f)\tn\phi_{M}(g)$:
	\[
		I \rTo^{\cong} I\tn I \rTo^{(f\cdot e)\tn(g\cdot e)} M\tn M \rTo^{m} M.
	\]
	Consider the diagram
	\begin{diagram}
		I\tn I & \rTo^{I\tn (g\cdot e)} & I\tn M  & \rTo^{e\tn M} & M\tn M
			 & \rTo^{f\tn M} & M\tn M \\
		\uTo<\cong && \uTo<\cong &\rdTo^{\cong} & \dTo>m && \dTo>m \\
		I  & \rTo_{g\cdot e} & M & \rTo_{1} & M & \rTo_{f} & M
	\end{diagram}
	in which the right-hand square commutes because $f$ is a map of right $M$-modules,
	and the upper triangle by the right-unit law for $M$.
	By the functoriality of tensor, the top edge is equal to $(f\cdot e)\tn(g\cdot e)$,
	so the upper outer edge is $\phi_{M}(f)\tn\phi_{M}(g)$ and the lower
	edge is $\phi_{M}(f\cdot g)$, which are therefore equal as required.
\end{proof}

%
\subsection{The internal Cayley Theorem}
The theorem above is external in the sense that it essentially
concerns the set $\V(I, X)$ rather than the module $X$ itself.
If the monoidal category $\V$ is closed, then we can state an
internal version, purely in terms of arrows in $\V$ itself.
So let $\V$ be biclosed, i.e.\ suppose we are given
functors
\[\begin{array}{l}
	\lolli: \V\op\times\V\to\V \\
	\illol: \V\times\V\op\to\V,
\end{array}\]
and natural isomorphisms with components
\[\begin{array}{l}
	\curry_{A,B,C}:  \V(A\tn B, C) \cong \V(A, B\lolli C), \\
	\curry'_{A,B,C}: \V(A\tn B, C) \cong \V(B, C\illol A).
\end{array}\]
Then the theorem is:
\begin{thm}[Internal Cayley]\label{thm-1d-internal}
	Let $(X,x)$ be a right $M$-module. Then the diagram
	\begin{diagram}
		X & \rTo^{\curry_{X,M,X}(x)} & M\lolli X & \pile{\rTo^{h} \\ \rTo_{k}}
		 & (M\tn M)\lolli X
	\end{diagram}
	is a coequaliser diagram, where the maps $h$ and $k$ are defined as
	follows. The map $h$ is obtained by currying
	\[
		(M\lolli X)\tn M\tn M \rTo^{\ev^{M}_{X}\tn M} X\tn M \rTo^{x} X,
	\]
	and $k$ is obtained by currying
	\[
		(M\lolli X)\tn M\tn M \rTo^{(M\lolli X)\tn m} (M\lolli X)\tn M \rTo^{\ev^{M}_{X}} X,
	\]
	where the map $\ev^{M}_{X}$ is the result of uncurrying the identity
	on $M\lolli X$.
\end{thm}
Although the statement of the theorem uses only the right-closed
structure $\lolli$, the left-closed structure $\illol$ plays a
crucial role in the proof. We will not give a detailed proof here%
\footnote{Again, I did type out a detailed proof. So this is not
	just an excuse for laziness!},
since this is something of a digression, but explain the outline.
The preliminary observations are that:
\begin{enumerate}
	\item For every object $X\in\C$, the object $X\illol X$
	is a monoid in a canonical way, via the definable `internal unit'
	and `internal composition' maps:
	\[\begin{array}{l}
		u: I \to X\illol X, \\
		m: (X\illol X)\tn(X\illol X) \to X \illol X.
	\end{array}\]
	\item A right $M$-module is
	precisely a map of monoids $M\to X\illol X$ for some $X$.
	\item It follows
	from this that a right $M$-module structure on $X$ induces a canonical
	right $M$-module structure on $A\tn X$ for each object $A\in \V$, because
	there is an `internal tensor' map
	\[
		X\illol X \rTo (A\tn X)\illol(A\tn X),
	\]
	which is compatible with the internal unit and compositions.
	\item In a similar
	way, a right $M$-module structure on $X$ induces a canonical
	right $M$-module structure on $X\illol A$, for each object $A$.
\end{enumerate}
Our second observation is trivial to prove: simply curry the diagrams
in Definition~\ref{def-module}. The others are proved in \citet[Section~1.6]{KellyEnriched}.
Now for the proof itself: suppose we are given an object $A\in\V$ and
a map $g: A\to M\lolli X$ such that $hg=kg$. We can curry $g$ to get a map
\[
	g': A\tn M \to X,
\]
and it is not hard to check that, since $hg=kg$, this $g'$ is a
map of modules from $A\tn M$ to $X$; where the module structure on
$A\tn M$ is obtained from that of $M$ by point (3) above. Then we
curry $g'$ with respect to the left-closure, to obtain a map
\[
	g'': M \to X\illol A,
\]
which is a map of modules with respect to the module
structure on $X\illol A$ obtained from (4). By the
external Cayley Theorem, $g''$ is thus equal to
\[
	M \rTo^{\cong} I\tn M \rTo^{f\tn M} (X\illol A)\tn M
		\rTo X\illol A
\]
for some unique $f: I \to X\illol A$, where the unlabelled
arrow above comes from the $M$-module structure on $X\illol A$.
From this it can be deduced that $g'$ is equal to
\[
	A\tn M \rTo^{f'\tn M} X\tn M \rTo^{x} X
\]
for some unique $f': A\to X$ (obtained by uncurrying $f$).
Thus $g$ is equal to
\[
	A \rTo^{f'} X \rTo^{\curry_{X,M,X}(x)} M\lolli X
\]
for this unique $f'$, which is precisely the universal property of
the equaliser diagram.
\begin{remark}
	Above we have deduced the internal theorem from the external one.
	Conversely, the external theorem can be obtained as a corollary of the internal
	one, under an additional assumption on $\V$. The additional assumption is
	that the functor $\V(I,-): \V\to\Set$ should preserve equalisers. In
	particular, this is always so if $\V$ has small coproducts, because in that
	case $\V$ is \emph{tensored} \citep[in the sense of][Section~2.7]{KellyEnriched}
	as a $\Set$-category, and so the functor $\V(I,-)$ has a left adjoint.
	Then we can obtain the external theorem simply by applying the functor
	$\V(I,-)$ to the equaliser diagram of the internal theorem.
\end{remark}

\section{Cayley's Theorem for Pseudomonoids}
Both these theorems, internal and external, admit generalisation to
the higher-dimensional setting. But since it is sufficient for our
applications, we shall confine ourselves to the external version.

\subsection{Modules for pseudomonoids}
First we must give a suitable definition of $\C$-module, where $\C$ is
a pseudomonoid. In fact, we want to define a bicategory $\Mod_{\C}$ of
right $\C$-modules.
\begin{definition}\label{def-psmod} 
	A \emph{right $\C$-module} $(X,x)$ for the pseudomonoid $\C$
	consists of a 1-cell $x: X\tn\C\to X$ together with invertible
	2-cells
	\begin{diagram}[h=2em]
		&&X\tn\C\\
		&\ruTo^{X\tn J}\\
		X\tn\I & \Arr\Nearrow{\chi^{J}_{X}} & \dTo>x \\
		&\rdTo_{r_{X}} \\
		&&X
	\end{diagram}
	and
	\begin{diagram}
		X\tn(\C\tn\C) & \rTo^{a_{X,\C,\C}} & (X\tn\C)\tn\C
			& \rTo^{x\tn\C} & X\tn\C \\
		\dTo<{X\tn P} && \Nearrow{\chi^{P}_{X}} && \dTo>x\\
		X\tn\C && \rTo_{x} && X
	\end{diagram}
	such that the following two equations hold:
	\[
		\begin{diagram}[tight,vtrianglewidth=1.7em,labelstyle=\scriptstyle]
			&&X\tn\C^{3} && \rTo^{x\tn\C^{2}} && X\tn\C^{2} \\
			&\ldTo^{X\tn\C\tn P} && \rdTo^{X\tn P\tn\C}
			 	&&\Arr\Nearrow{\scriptstyle\chi^{P}_{X}\tn\C} && \rdTo^{x\tn\C} \\
			X\tn\C^{2} && \Right_{\displaystyle X\tn\aa} && X\tn\C^{2}
				&& \rTo^{x\tn\C} && X\tn\C \\
			&\rdTo_{X\tn P} && \ldTo_{X\tn P}
				&& \Arr\Nearrow{\chi^{P}_{X}} && \ldTo_{x} \\
			&&X\tn\C && \rTo_{x} && X
		\end{diagram}
		=
		\begin{diagram}[tight,vtrianglewidth=1.7em,labelstyle=\scriptstyle]
			&&X\tn\C^{3} && \rTo^{x\tn\C^{2}} && X\tn\C^{2} \\
			& \ldTo^{X\tn\C\tn P} &&\sim&&\ldTo^{X\tn P}
				&& \rdTo^{x\tn\C} \\
			X\tn\C^{2} && \rTo^{x\tn\C} && X\tn\C
				&& \Right_{\displaystyle\chi^{P}_{X}} && X\tn\C \\
			&\rdTo_{X\tn P} && \Arr\Nearrow{\chi^{P}_{X}}
				&& \rdTo_{x} && \ldTo_{x} \\
			&&X\tn\C && \rTo_{x} && X
		\end{diagram}
	\]
	and
	\[
		\begin{diagram}[w=4em]
			X\tn \I\tn \C & \rTo^{X\tn J\tn \C} & X\tn \C^{2}
				& \rTo^{x\tn \C} & X\tn \C \\
			&\rdTo_{1}\raise 1em\hbox to1em{$\Arr\Nearrow{X\tn\ll}$\hss}
				& \dTo>{X\tn P} & \Arr\Nearrow{\chi^{P}_{X}} & \dTo>x \\
			&&X\tn \C & \rTo_{x} & X
		\end{diagram}
		=
		\begin{diagram}[w=4em]
			X\tn \I\tn \C & \rTo^{X\tn J\tn \C} & X\tn \C^{2} \\
			&\rdTo_{1} \raise 1em\hbox to 1em{$\Arr\Nearrow{\chi^{J}_{X}\tn\C}$\hss}
				& \dTo>{x\tn \C} \\
			&&X\tn \C\\
			&&\dTo>x \\
			&&X
		\end{diagram}
	\]
\end{definition}
\begin{definition}\label{def-psmod-map} 
	Given right $\C$-modules $(X,x)$ and $(Y,y)$, a
	\emph{map of $\C$-modules} from $X$ to $Y$ consists
	of a 1-cell $f: X\to Y$ together with an invertible 2-cell
	\begin{diagram}
		X\tn \C & \rTo^{f\tn \C} & Y\tn \C \\
		\dTo<x & \Arr\Nearrow{\phi^{f}} & \dTo>y \\
		X & \rTo_{f} & Y
	\end{diagram}
	such that
	\[
		\begin{diagram}[tight,vtrianglewidth=1.7em,labelstyle=\scriptstyle]
			&&X\tn\C^{2} && \rTo^{f\tn\C^{2}} && Y\tn\C^{2} \\
			&\ldTo^{X\tn P} && \rdTo^{X\tn\C}
			 	&&\Arr\Nearrow{\scriptstyle\phi^{f}\tn\C} && \rdTo^{y\tn\C} \\
			X\tn\C && \Right_{\displaystyle \chi^{P}_{X}} && X\tn\C
				&& \rTo^{f\tn\C} && Y\tn\C \\
			&\rdTo_{x} && \ldTo_{x}
				&& \Arr\Nearrow{\phi^{f}} && \ldTo_{y} \\
			&&X && \rTo_{f} && X
		\end{diagram}
		=
		\begin{diagram}[tight,vtrianglewidth=1.7em,labelstyle=\scriptstyle]
			&&X\tn\C^{2} && \rTo^{f\tn\C^{2}} && Y\tn\C^{2} \\
			& \ldTo^{X\tn P} &&\sim&&\ldTo^{Y\tn P}
				&& \rdTo^{y\tn\C} \\
			X\tn\C && \rTo^{f\tn\C} && Y\tn\C
				&& \Right_{\displaystyle\chi^{P}_{Y}} && X\tn\C \\
			&\rdTo_{x} && \Arr\Nearrow{\phi^{f}}
				&& \rdTo_{y} && \ldTo_{y} \\
			&&X && \rTo_{f} && Y
		\end{diagram}
	\]
	and
	\[
		\begin{diagram}[h=2em]
			&\rnode{XI}{X\tn \I} & \rTo^{f\tn\I} & Y\tn \I \\
			&\dTo<{X\tn J} & \sim & \dTo>{Y\tn J} \\
			\Right_{\chi^{J}_{X}} & X\tn \C & \rTo^{f\tn \C} & Y\tn \C \\
			&\dTo<x & \Arr\Nearrow{\phi^{f}} & \dTo>y \\
			&\rnode{X}{X} & \rTo_{f} & Y
			\ncarc[arcangle=-90,ncurv=1.3]{->}{XI}{X}\Bput{1}
		\end{diagram}
		=
		\begin{diagram}[h=2em]
			X\tn \I & \rTo^{f\tn \I} & Y\tn \I \\
			&&&\rdTo^{Y\tn J} \\
			\dTo<1 & & \dTo<1 & \Right_{\chi^{J}_{Y}} & Y\tn \C \\
			&&&\ldTo_{y} \\
			X &\rTo_{f} & Y
		\end{diagram}
	\]
\end{definition}
\begin{definition}\label{def-psmod-bicat} 
	The bicategory $\Mod_{\C}$ of right $\C$-modules is defined
	as follows. An object is a right $\C$-module, a 1-cell is a
	map of right $\C$-modules, and a 2-cell $\gamma: f\To g: X\to Y$
	is a 2-cell $f\To g$ in $\B$ such that
	\[
		\begin{diagram}[w=4em]
			& \raise -2em\hbox to 0pt{\hss$\Arr\Uparrow{\gamma\tn\C}$} \\
			\rnode{XC}{X\tn \C} & \rTo_{f\tn\C} & \rnode{YC}{Y\tn \C} \\
			\dTo<x & \Arr\Nearrow{\phi^{f}} & \dTo>y \\
			X & \rTo_{f} & Y \\
			\ncarc[arcangle=50]{->}{XC}{YC}\Aput{g\tn\C}
		\end{diagram}
		\qquad=\qquad
		\begin{diagram}[w=4em]
			\\
			\rnode{XC}{X\tn \C} & \rTo^{g\tn\C} & \rnode{YC}{Y\tn \C} \\
			\dTo<x & \Arr\Nearrow{\phi^{g}} & \dTo>y \\
			\rnode{X}{X} & \rTo^{g} & \rnode{Y}{Y} \\
			& \raise 2em\hbox{$\Arr\Uparrow{\gamma}$}
			\ncarc[arcangle=-60]{->}{X}{Y}\Bput{f\tn\C}
		\end{diagram}
	\]
	Composition is defined as in $\B$, and given $f: X\to Y$
	and $g: Y\to Z$, the 2-cell $\phi^{gf}$ is the pasting
	\begin{diagram}
		X\tn \C & \rTo^{f\tn \C} & Y\tn \C
			& \rTo^{g\tn \C} & Z\tn \C\\
		\dTo<x & \Arr\Nearrow{\phi^{f}} & \dTo>y 
			& \Arr\Nearrow{\phi^{g}} & \dTo>z \\
		X & \rTo_{f} & Y & \rTo_{g} & Z
	\end{diagram}
	which is easily verified to satisfy the necessary equations.
\end{definition}

\subsection{Modules via Components}
The calculus of components is very useful here. We shall consider
what the definitions mean in terms of components, which in practice
is essentially the same thing as considering what they imply in the
case $\B = \Cat$.

A right $\C$-module consists of an object $\X$ together with a functor
\[
	x: \X\tn\C\to\X;
\]
we shall write this functor in infix form as $\bullet$. There are also
invertible 2-cells with components
\[
	\alpha^{\bullet}_{X,A,B}: X\bullet(A\tn B) \to (X\bullet A)\bullet B
\]
and
\[
	\rho^{\bullet}_{X}: X\bullet I \to X
\]
where we take $X$ to be a (generic) element of $\X$,
and $A$, $B$ to be elements of $\C$. These are just the
2-cells that were denoted $\chi^{P}$ and $\chi^{J}$ above.
The coherence conditions say that the diagrams
\begin{diagram}
	X\bullet(A\tn(B\tn C))
		& \rTo^{\alpha^{\bullet}} & (X\bullet A)\bullet(B\tn C)
		& \rTo^{\alpha^{\bullet}} & ((X\bullet A)\bullet B)\bullet C \\
	\dTo<{X\bullet\alpha} &&&& \uTo>{\alpha^{\bullet}\bullet C} \\
	X\bullet((A\tn B)\tn C) && \rTo_{\alpha^{\bullet}}
		&& (X\bullet(A\tn B))\bullet C
\end{diagram}
and
\begin{diagram}[vtriangleheight=3em]
	X\bullet(I\tn A) && \rTo^{\alpha^{\bullet}} && (X\bullet I)\bullet A \\
	& \rdTo[snake=-1ex]_{X\bullet\lambda_{A}}
		&& \ldTo[snake=1ex]_{\rho^{\bullet}_{X}\bullet A} \\
	&& X\bullet A
\end{diagram}
must commute, for all $X\in\X$ and $A$, $B$, $C\in\C$.
It is clear that, in particular, $\C$ itself is a right
$\C$-module, when equipped in the obvious way.

Now, if we have two right $\C$-modules $\X$ and $\Y$, a
map of modules $F: \X\to\Y$ is a 1-cell $F$ equipped with
an invertible 2-cell with components
\[
	F^{\bullet}_{X,A}: F(X\bullet A)\to FX\bullet A
\]
such that the diagrams
\begin{diagram}
	F(X\bullet(A\tn B)) && \rTo^{F^{\bullet}_{X,A\tn B}}
		&& FX\bullet(A\tn B) \\
	\dTo<{F(\alpha^{\bullet}_{X,A,B})}
		&&&& \dTo>{\alpha^{\bullet}_{FX,A,B}} \\
	F((X\bullet A)\bullet B) & \rTo_{F^{\bullet}_{X\bullet A,B}}
		& F(X\bullet A)\bullet B & \rTo_{F^{\bullet}_{X,A}\bullet B}
		& (FX\bullet A)\bullet B
\end{diagram}
and
\begin{diagram}[vtriangleheight=2.5em]
	F(X\bullet I) && \rTo^{F^{\bullet}_{X,I}} && FX\bullet I\\
	&\rdTo[snake=-1ex]_{F(\rho^{\bullet}_{X})}
		&& \ldTo[snake=1ex]_{\rho^{\bullet}_{FX}} \\
	&&FX
\end{diagram}
commute.

Finally, given maps of modules $F$ and $G: \X\to\Y$, a module
2-cell is given by a 2-cell $\gamma: F\to G$ such that the diagram
of components
\begin{diagram}
	F(X\bullet A) & \rTo^{F^{\bullet}_{X,A}} & FX\bullet A \\
	\dTo<{\gamma_{X\bullet A}} && \dTo>{\gamma_{X}\bullet A} \\
	G(X\bullet A) & \rTo_{G^{\bullet}_{X,A}} & GX\bullet A
\end{diagram}
commutes for all $X\in X$ and $A\in \C$.

\subsection{External Cayley for Pseudomonoids}
We shall use the component presentation to prove the External Cayley
theorem.
\begin{propn}[External Cayley for Pseudomonoids]\label{prop-psmoncayley}
	For every right $\C$-module $\X$, the functor
	\[
		\phi_{\X}: \Mod_{\C}(\C,\X)\to \B(\I,\X)
	\]
	defined by $\phi_{\X}(F) = F(I)$ (and in the obvious
	way on 2-cells) is an equivalence of categories, with equivalence-inverse
	$\phi'_{\X}: \B(\I,\X) \to \Mod_{\C}(\C,\X)$ defined as
	\[
		\phi'_{\X}(X()) = X()\bullet-
	\]
	and made into a map of modules
	by the invertible 2-cell $\alpha^{\bullet}_{X(),-,-}$.
\end{propn}
\begin{proof}
	Let $F: \C\to\X$ be a map of $\C$-modules; then
	$\phi'_{\X}(\phi_{\X}(F)) = F(I)\bullet-$
	which, by the natural isomorphism $F^{\bullet}_{I,-}$, is
	naturally isomorphic to $F(I\tn-)$ which, by the right-unit
	isomorphism of $\C$, is naturally isomorphic to $F$.
	
	Let $X()$ (in component notation) be a 1-cell $\I\to\X$.
	Then
	\[
		\phi_{\X}(\phi'_{\X}(X())) = X()\bullet I,
	\]
	which is
	naturally isomorphic to $X()$ by the natural isomorphism
	$\rho^{\bullet}_{X}$.
\end{proof}

\begin{propn}
	The functor $\phi_{\C}: \Mod_{\C}(\C,\C)\to \B(\I,\C)$
	can be equipped with the structure of a strong monoidal
	functor, with respect to the natural monoidal structures
	on the categories $\Mod_{\C}(\C,\C)$ and $\B(\I,\C)$.
\end{propn}
\begin{proof}
	The unit for the monoidal structure of $\Mod_{\C}(\C,\C)$ is
	of course the identity map, and $\phi_{\C}(1) = 1(I) \cong I$.
	For the `tensor' part of the monoidal structure of $\phi_{\C}$,
	we take the composite
	\[
		G(F(I)) \rTo_{\cong}^{G(\lambda_{FI}^{-1})} G(I\tn FI)
			\rTo_{\cong}^{G^{\bullet}_{I,FI}} GI\tn FI.
	\]
	The unit condition holds by the triangle in the definition of map
	of modules, so it remains to show the associativity condition.
	Let $F$, $G$, and $H$ be module maps $\C\to\C$, and consider
	the diagram
	\begin{diagram}
		HGFI & \rTo^{H\lambda_{GFI}} & H(I\tn GFI)
			& \rTo^{H^{\bullet}_{I,GFI}} & HI\tn GFI \\
		\dTo<{HG\lambda_{FI}} && \dTo>{H(I\tn G\lambda_{FI})}
			&& \dTo>{HI\tn G\lambda_{FI}} \\
		HG(I\tn FI) & \rTo_{H\lambda_{G(I\tn FI)}} & H(I\tn G(I\tn FI))
			& \rTo_{H^{\bullet}_{I,G(I\tn FI)}} & HI\tn G(I\tn FI) \\
		\dTo<{H(G^{\bullet}_{I,FI})} && \dTo>{H(I\tn G^{\bullet}_{I,FI})}
			&& \dTo>{HI\tn G^{\bullet}_{I,FI}} \\
		H(GI\tn FI) & \rTo^{H\lambda_{GI\tn FI}} & H(I\tn(GI\tn FI))
			& \rTo_{H^{\bullet}_{I,GI\tn FI}} & HI\tn(GI\tn FI) \\
		\dTo<{H^{\bullet}_{GI,FI}} & \rdTo_{H(\lambda_{GI}\tn FI)}
			& \dTo>{H(\alpha_{I,GI,FI})} \\
		HGI\tn FI && H((I\tn GI)\tn FI) && \dTo>{\alpha_{HI,GI,FI}} \\
		& \rdTo_{H(\lambda_{GI})\tn FI} & \dTo>{H^{\bullet}_{I\tn GI,FI}} \\
		&& H(I\tn GI)\tn FI & \rTo_{H^{\bullet}_{I,GI}\tn FI} & (HI\tn GI)\tn FI
	\end{diagram}
	where the quadrilaterals commute by naturality, the triangle by
	definition of pseudomonoid, and the pentagonal region by the definition
	of morphism of modules.
\end{proof}

So we have shown that, for any pseudomonoid $\C\in\B$, the monoidal
category $\B(\I,\C)$ is monoidally equivalent to $\Mod_{\C}(\C,\C)$.
In the case $\B=\Cat$, this shows that every monoidal category is
monoidally equivalent to a \emph{strict} monoidal category. However
we are more interested in the case $\B=\Prof$, and some of the
implications in this case are explored in the next chapter.

%
%
\end{thesischapter}
\documentclass{robinthesis}

\begin{thesischapter}{Promon}{Linear Logic without Units}
In this, the title chapter, we at last address the question
that we set out to answer at the outset. We consider the notion
of model described in the introduction, and bring to bear all the
machinery of earlier chapters to find a simple formulation of it.

\section{Promonoidal categories}
As we have foreshadowed, we are mainly interested in the monoidal
bicategory $\Prof$ of categories and profunctors. Its objects are
categories, and a 1-cell $\C\rPro\D$ is a profunctor from $\C$ to $\D$;
which is to say, a functor $\C\op\times\D\to\Set$. (Note that some
authors use the converse direction, so what we call a profunctor
from $\C$ to $\D$ they would call a profunctor from $\D$ to $\C$.
There are good arguments in favour of both choices, and we have
simply chosen the one that is easiest for our purposes.)

Profunctors are composed by a convolution operation, where the
composite
\[
	\CB \rPro^{F} \C \rPro^{G} \D
\]
is defined by the coend formula
\[
	GF(B,D) = \int^{C} F(B,C) \times G(C,D).
\]
The 2-cells of $\Prof$ are just ordinary natural transformations.
It is easy enough to check that $\Prof$ is a bicategory. The
monoidal structure is more subtle, and although it has long been
assumed that $\Prof$ is a symmetric monoidal bicategory, the first
rigorous published proof was given very recently by \citet{CartesianBicatsII}.
The monoidal structure is given by the ordinary cartesian product
of categories.

A \emph{promonoidal category} is a pseudomonoid in $\Prof$. More
concretely, a promonoidal category consists of a category $\C$
equipped with profunctors
\[
	P: \C\times\C\rPro\C
\]
and
\[
	J: 1\rPro\C
\]
together with natural isomorphisms with components
\[
	\aa_{A,B,C,D}: \int^{X} P(A,X;D)\times P(B,C;X)
		\to \int^{X} P(A,B;X)\times P(X,C;D),
\]
\[
	\ll_{A,B}: \int^{X} J(X)\times P(X,A;B) \to \C(A,B),
\]
and
\[
	\rr_{A,B}: \int^{X} J(X)\times P(A,X;B) \to \C(A,B),
\]
satisfying the pseudomonoid axioms.

A functor $F: \C\to \D$ induces profunctors $\C(-,F=): \C\rPro \D$
and $\C(F-,=): \D\rPro\C$. These extend in an obvious way to
bijective-on-objects embeddings
\[
	\Cat \to\Prof
\]
and \[
	\Cat\op\to\Prof,
\]
which (by Yoneda) are locally full and faithful. In particular,
the covariant embedding gives every monoidal category the
structure of a promonoidal category in a natural way. Concretely,
one takes $P(A,B;C) = \C(A\tn B,C)$ and $J(A) = \C(I,A)$.

\subsection{Braiding and symmetry}
Clearly a braided promonoidal category must be a braided pseudomonoid
in $\Prof$: that is, a promonoidal category $\C$ additionally equipped
with a natural isomorphism with components
\[
	\sigma_{A,B,C}: P(A,B;C) \to P(B,A;C)
\]
satisfying the braiding axioms.

We have not given a general definition of symmetric pseudomonoid,
because to do so would require venturing into the deep waters of
\emph{sylleptic} monoidal bicategories. However we can easily say
what it means for a braided promonoidal category to be symmetric;
we simply demand that\[
	\sigma_{A,B,C} = \sigma_{B,A,C}^{-1}.
\]
It is clear that, in the symmetric case, each of the two braiding
axioms implies the other.

\section{Modelling linear logic without units}
As explained in the introduction, the appropriate structures for
modelling the unitless fragment of multiplicative intuitionistic
linear logic are promonoidal categories where the multiplication
$P$ is represented by a functor, but the unit $J$ generally is not.
\begin{definition}
	A semi symmetric monoidal category (semi SMC) consists
	of a category $\C$ equipped with functors
	\[
		\tn: \C\times\C\to \C
	\]
	and
	\[
		J:\C\to\Set
	\]
	and natural isomorphisms with components
	\[
		\alpha_{A,B,C}: A\tn(B\tn C),
	\]
	\[
		\sigma_{A,B}: A\tn B\to B\tn A,
	\]
	and
	\[
		\lambda_{A,B}: \int^{X} J(X)\times\C(X\tn A,B) \to \C(A,B)
	\]
	satisfying the axioms for a symmetric promonoidal category.
	The associativity (pentagon) and symmetry (hexagon) axioms do not involve the unit,
	so are the same as for an ordinary symmetric monoidal category.
	In detail, the diagrams
	\begin{diagram}
	 A\tensor \bigl(B\tensor (C\tensor D)\bigr)
	 &\rTo^\alpha&(A\tensor B)\tensor (C\tensor D)
	 &\rTo^\alpha & \bigl((A\tensor B)\tensor C\bigl)\tensor D
	 \\
	 \dTo<{A\tn\alpha} &&&& \uTo>{\alpha\tn D}
	 \\
	 A\tensor\big((B\tensor C)\tensor D\big)
	 && \rTo_\alpha && \bigl(A\tensor(B\tensor C)\bigr)\tensor D
	\end{diagram}
	and
	\begin{diagram}
	 A\tn(B\tn C) & \rTo^{\alpha} & (A\tn B)\tn C & \rTo^{\sigma} & C\tn(A\tn B) \\
	 \dTo<{A\tn\sigma} &&&& \dTo>{\alpha} \\
	 A\tn(C\tn B) & \rTo_{\alpha} & (A\tn C)\tn B  & \rTo_{\sigma\tn B} & (C\tn A)\tn B
	\end{diagram}
	must commute. The unit axiom says that
	the diagram
	\begin{diagram}
        \int^X J(X)\times\C((X\tn B)\tn C, Z) & \rTo^{\int^X J(X)\times\C(\alfa XBC, Z)}
                & \int^X J(X)\times\C(X\tn (B\tn C), Z)
        \\
        \dTo<{\cong}
        \\
        \int^{X,Y} J(X)\times\C(X\tn B, Y)\times \C(Y\tn C, Z)&
                &\dTo>{\lambda_{B\tn C,Z}}
        \\
        \dTo<{\int^Y\lambda_{B,Y}\times\C(Y\tn C,Z)}
        \\
        \int^Y\C(B,Y)\times\C(Y\tn C,Z) &\rTo_\cong&\C(B\tn C,Z)
	\end{diagram}
	must commute for all $B$,$C$,$Z\in\C$.
\end{definition}
\begin{definition}
	A semi symmetric monoidal closed category (semi SMCC) is a semi SMC
	$\C$ together with a functor
	\[
		\lolli: \C\op\times\C\to\C
	\]
	and a natural isomorphism with components
	\[
		\curry_{A,B,C}: \C(A\tn B,C)\to \C(A,B\lolli C).
	\]
\end{definition}

The purpose of this chapter is to find a simpler formulation of
the notion of semi SMCC. We should like to formulate the notion
in a way that does not involve coends, for one thing. In fact we
give two such formulations, which are superficially very different.
Before delving into the details, we give a summary of our two
presentations.

\subsection{The presentation via linear elements}\label{s-linel}
There is an obvious notion of \emph{unitless SMCC}, essentially
obtained  taking the ordinary definition of SMCC
and erasing those parts that mention the unit object.
In the introduction, we saw that unitless SMCCs are not
adequate to model the unitless fragment of IMLL, because of
the need to represent sequents whose left-hand side is empty.
However, it turns out that the gap between unitless SMCCs and
semi SMCCs is smaller than it seems: surprisingly, no additional
structure is actually required! In fact a semi SMCC can be described
as a unitless SMCC satisfying a certain \emph{property}. Here we
briefly describe the property in question.

Given a unitless SMCC $\C$, and an object $A\in\C$, let us define
a \emph{linear element} of $A$ to be a natural transformation with
components
\[
	\gamma_{X}: X \to A\tn X
\]
such that, for all $X$ and $Y\in\C$, the diagram
\begin{diagram}
	& X\tn Y \\
	\ldTo(1,2)<{\gamma_{X\tn Y}} && \rdTo(1,2)>{\gamma_{X}\tn Y} \\
	A\tn(X\tn Y) & \rTo_{\alpha_{A,X,Y}} & (A\tn X)\tn Y
\end{diagram}
commutes. It's easy to see that, in an ordinary monoidal category,
the linear elements of $A$ are in bijective correspondence with
morphisms $I\to A$. Now, there is an obvious functor
\[
	\Lin: \C\to \Set
\]
that takes each object $A\in\C$ to the set of linear elements on
$\C$. (In an ordinary monoidal category, it is isomorphic to the
hom functor $\C(I,-)$.) Furthermore, there is a canonical natural
transformation with components
\[
	\ell_{A,B}: \Lin(A\lolli B) \to \C(A,B);
\]
if $\gamma$ is a linear element of $A\lolli B$, then
$\ell_{A,B}(\gamma)$ is just the composite
\[
	A \rTo^{\gamma_{A}} (A\lolli B)\tn A \rTo^{\ev^{A}_{B}} B
\]
where $\ev^{A}$ is the counit of the adjunction $-\tn A \dashv A\lolli-$.

Our additional condition, then, is as follows: a unitless SMCC
$\C$ can be regarded as a semi SMCC just when the natural transformation
$\ell$ is invertible, so that every arrow $A\to B$ comes from a
unique linear element of $A\lolli B$. It is easy to see that this
condition is satisfied when $\C$ has a unit object, and in
Section~\ref{s-promon-unit} we shall prove that in fact it
is satisfied if and only if $\C$ has a promonoidal unit.

\subsection{The `$\psi$' presentation}\label{s-psi-presentation}
Our second presentation is given in terms of structure rather
than properties. We show that to give a semi SMCC is to give
a category $\C$ equipped with an associative, symmetric\footnote{
      \emph{I.e.}, equipped with the standard associativity and symmetry 
      isomorphisms and coherence laws of a symmetric monoidal category.}
functor
$\tn:\C\times\C\to\C$, a functor $\lolli:\C\op\times\C\to\C$
through which hom factors up to isomorphism,
\begin{diagram}
       \C\op\times\C &\rTo^\lolli& \C\\
       &\rdTo_{\mathrm{hom}}\raise1em\hbox{\qquad$\cong$}&\dTo.>{\exists J}\\
       &&\Set
\end{diagram}
and a natural isomorphism
\[
       \psi_{A,B,C}\;\;:\;\; (A\tn B)\lolli C \;\;\to\;\;
       A\lolli(B\lolli C)
\]
such that the diagram
\begin{diagram}[h=1.5em] \bigl(A\tn(B\tn C)\bigr)\lolli D
&\rTo^{\alpha_{A,B,C}\lolli D} &\bigl((A\tn B)\tn
C\bigr)\lolli D\\ &&\dTo>{\psi_{A\tn B,C,D}}\\
\dTo<{\psi_{A,B\tn C,D}}&&(A\tn B)\lolli(C\lolli D)\\
&&\dTo>{\psi_{A,B,C\lolli D}}\\ A\lolli\bigl((B\tn C)\lolli
D\bigr)&\rTo_{A\lolli\psi_{B,C,D}}
&A\lolli\bigl(B\lolli(C\lolli D)\bigr) \end{diagram} commutes.


\section{The promonoidal unit}\label{s-promon-unit}
Although we are primarily interested in semi SMCCs, the basic argument
here works in any braided promonoidal category, so we shall work at
that level of generality and specialise to our intended application
at the end.

The essential tool here is the Cayley Theorem for pseudomonoids:
in the case $\B=\mathbf{Cat}$, it shows that every monoidal category is
monoidally equivalent to a \emph{strict} monoidal category. In the case $\B=\mathbf{Prof}$,
which is the case we're interested in here, it allows us to construct a canonical
representation for the promonoidal unit. (This turns out to be particularly powerful in the
braided, or symmetric, case.)

In summary, the story goes as follows.
It is a familiar fact that, in a semigroup, all units must be equal: if $i$ and $j$ are
both units, then $i=ij=j$. The existence or otherwise of a unit is really a \emph{property}
of a semigroup; there is no choice of how to define the unit. By a similar argument, in a
pseudomonoid (or `pseudosemigroup', a pseudomonoid without the unit structure) all units
are isomorphic. Being good category theorists, we don't care about the difference
between isomorphic structures; so we have no real choice in how to define the unit of
a pseudomonoid: given the tensor structure ($P$ and $\aa$), either it is possible to
define a unit, or it isn't. We might say that the unit is `essentially property-like'
\citep[cf.][]{proplike}. This point may seem rather irrelevant, since the easiest way
to demonstrate the existence of a unit is often to exhibit one. However, it turns out
that in the case of promonoidal categories that is not necessarily true. There is a
canonical representative for the unit -- the functor here denoted $\Lin$ -- defined
solely in terms of the tensor structure ($P$ and $\alpha$) that is isomorphic to the
unit whenever there is a unit to be isomorphic to. Furthermore, in the case of braided
or symmetric promonoidal categories, we can define a simple test of whether or not
there is a unit. So we may \emph{define} a braided promonoidal category purely in
terms of the tensor, associator and braiding, subject to a condition that determines
the existence of a unit. Should an actual unit be required, we are free to use the
canonical unit $\Lin$.

Now for the details. Let $\C$ be a promonoidal category. The monoidal category $\mathbf{Prof}(1, \C)$
is $[\C, \Set]$, equipped with Day's convolution tensor. The monoidal category
$\Mod_{\C}(\C,\C)$ does not really have a simpler description than the general one given above:
an object is a profunctor $\C\rPro\C$ together with a natural transformation
\begin{diagram}
	\C\times\C & \rPro^P & \C\\
	\dPro<{F\times\C}&\Arr\Swarrow{\phi^F}&\dTo>F\\
	\C\times\C&\rPro_P & \C
\end{diagram}
satisfying the conditions of Definition~\chref{Cayley}{def-module-map}.
So $\phi^F$ is a natural isomorphism with components
\[
	\phi^F_{A,B,C}: \int^X P(A,B;X)\times F(X;C) \to \int^X F(A;X)\times P(X,B;C).
\]
Now let us write $\E$ as an abbreviation for the monoidal category $\Mod_{\C}(\C,\C)$,
and let $A\in\C$. We have a sequence of natural isomorphisms
\[\begin{array}{rclp{5cm}}
	JA &\cong& [\C,\Set](\C(A,-), J) & {by Yoneda} \\
	&\cong& \E(\phi'_\C(\C(A,-)), \phi'_\C(J)) & {since $i_\C$ is full and faithful} \\
	&\cong& \E(\phi'_\C(\C(A,-)), I) & {applying $m_I$}
\end{array}\]
where $\phi'_{\C}$ is the monoidal equivalence $[\C,\Set]\to\E$
described by Proposition~\chref{Cayley}{prop-psmoncayley}.
If we define the functor $\Lin: \C\to\Set$ as $\Lin(A) := \E(\phi'_\C(\C(A,-)), I)$,
then we have exhibited a natural isomorphism $\theta: J \To \Lin$.
An element of the set $\Lin(A)$ is a natural transformation $\gamma$ with components
\[
	\gamma_{X,Y}: P(A,X;Y) \to \C(X,Y)
\]
such that the diagram
\begin{equation}\label{diag-linel}
\begin{diagram}[h=2em]
	\int^X P(L,M;X)\times P(A,X;N) & \rTo^{\int^X P(L,M;X)\times\gamma_{X,N}}
		& \int^X P(L,M;X)\times\C(X,N) \\
	&&\dTo>\cong \\
	\dTo<{\alpha_{A,L,M}} && P(L,M;N) \\
	&&\uTo>\cong \\
	\int^X P(A,L;X)\times P(X,M;N) & \rTo_{\int^X \gamma_{L,X}\times P(X,M;N)}
		& \int^X\C(L,X)\times P(X,M;N)
\end{diagram}
\hskip-2em 
\end{equation}
commutes for all $L$, $M$ and $N$ in $\C$. We shall refer to the elements
of this set as the \defns{linear element} of $A$.

For the purposes of this thesis, of course, we're particularly interested
in the case where the profunctor $P$ is represented by a functor $\tn: \C\times\C\to\C$.
In this case, a linear element is a natural transformation with components
\[
	\gamma_{X,Y}: \C(A\tn X, Y) \to \C(X,Y)
\]
which, by Yoneda, can be represented by a natural transformation with components
\[
	\gamma_X: X \to A\tn X.
\]
With this representation, the condition boils down to the requirement that
the diagram
\begin{diagram}
	& X\tn Y \\
	\ldTo(1,2)<{\gamma_{X\tn Y}} && \rdTo(1,2)>{\gamma_{X}\tn Y} \\
	A\tn(X\tn Y) & \rTo_{\alpha_{A,X,Y}} & (A\tn X)\tn Y
\end{diagram}
should commute, for all $X$, $Y\in\C$.

Since an element of $\Lin(A)$ is a natural transformation 
\[
	\gamma_{X,Y}: P(A,X;Y) \to \C(X,Y),
\]
for every $A$ there is an obvious natural transformation $\lambda'^A$
with components
\[
	\lambda'^A_{X,Y}: \Lin(A)\times P(A,X;Y) \to \C(X,Y),
\]
dinatural in $A$.

\begin{propn}
	The diagram
	\begin{diagram}[h=2em]
		JA\times P(A,X;Y) \\
		&\rdTo^{\lambda^A_{X,Y}}\\
		\dTo<{\theta_A\times P(A,X;Y)} && \C(X,Y) \\
		&\ruTo_{\lambda'^A_{X,Y}} \\
		\Lin A \times P(A,X;Y)
	\end{diagram}
	commutes, for all $A$, $X$ and $Y\in\C$.
\end{propn}
\begin{proof}
	This is a direct consequence of the definition of $\theta$: any
	apparent complexity here is notational rather than mathematical.
	First we shall calculate the effect of $\theta_A: JA\to\Lin A$
	on an element $e\in JA$. We'll consider in sequence the three
	isomorphisms that define $\theta$. The first takes $e$ to the
	natural transformation with $X$-component
	\[
		(f:A\to X) \mapsto J(f)(e);
	\]
	this must then be mapped by the second to a natural transformation
	\[
		P(A,X;Y) \to \int^Z JZ\times P(Z,X;Y)
	\]
	natural in $X$ and $Y$.
	The elements of $\int^Z JZ\times P(Z,X;Y)$ are equivalence classes
	of pairs $\langle j, p\rangle$, with $j\in JZ$ and $p\in P(Z,X;Y)$
	for some $Z\in\C$. Our element is mapped to the $A$-indexed natural
	family of functions
	\[
		(p\in P(A,X;Y)) \mapsto [\langle e,p\rangle].
	\]
	where $[\langle e,p\rangle]$ denotes the equivalence class containing
	$\langle e,p\rangle$. The final natural isomorphism takes this element
	to the $A$-indexed natural family of functions
	\[
		(p\in P(A,X;Y)) \mapsto \lambda_{X,Y}([\langle e,p\rangle]).
	\]
	Now it's easy to show that the diagram commutes: let $\langle e,p\rangle$
	be an element of $JA\times P(A,X;Y)$. The vertical arrow maps it to
	the pair $\langle f, p\rangle$, where $f$ is the linear element
	displayed above. $\lambda'$ then maps this to $\lambda_{X,Y}^A(e,p)$,
	as required.
\end{proof}
There is an apparent asymmetry here: although it was easy to define
$\lambda'$, there is no obvious way to define a corresponding $\rho'$
-- unless of course our promonoidal category is braided, of which more
below. This asymmetry derives from the fact that $\E$ is defined using
\emph{right} $\C$-modules,
and of course it would be possible to define a dual version using left
modules, which would also be monoidally isomorphic to $\Prof(1,\C)$,
hence to $\E$. Using this, we could define a `co-linear elements' functor
$\Lin': \C\to\Set$, with a canonical natural isomorphism
\[
	\rho'^A_{X,Y}: \Lin'A\times P(X,A;Y) \to \C(X,Y).
\]
However, in general there is no canonical natural isomorphism between
$\Lin$ and $\Lin'$. In the braided case, there is. So in that case we
may simply take $\Lin$ (or equivalently $\Lin'$) to be the unit, which
role it is able to fulfil if and only if $\lambda'$ (equivalently $\rho'$)
is invertible. More formally we have:
\begin{propn}\label{prop-promonunit}
	Let $\C$ be a category, and $P: \C\times\C\rPro\C$ a profunctor.
	Let $\alpha$ be an associator satisfying the pentagon
	condition, and let $\sigma$ be a braiding satisfying the
	hexagon conditions. There exists a unit $J:1\rPro\C$ (with
	coherent unit isomorphisms $\lambda$ and $\rho$) if and only
	if the natural transformation
	\[
		\lambda': \int^A \Lin A\times P(A,X;Y) \to \C(X,Y),
	\]
	defined above, is invertible.
\end{propn}
\begin{proof}
	We have already established the `only if' direction, so let
	$\C$, $P$, $\alpha$ and $\sigma$ be given, and define $\Lin: \C\to\Set$
	and $\lambda'$. Suppose that $\lambda'$ is invertible.
	
	By the presentation of braided pseudomonoids given in Section~\chref{Psmon}{s-braided-facts},
	it suffices to show that
	\[
		\begin{diagram}
			1\times\C^2 & \rPro^{\Lin\times\C^2} & \C^3 \\
			\dPro<{1\times P} & \sim & \dPro>{\C\times P} \\
			\rnode{IC}{1\times\C} & \rPro^{\Lin\times\C} & \C^2 \\
			&\lambda' & \dPro>P \\
			&&\rnode{C}{\C}
			\nccurve[angleA=270,angleB=180,ncurv=1]{->}{IC}{C}\Bput1
		\end{diagram}
		=
		\begin{diagram}
			\rnode{ICC}{1\times\C^2} & \rPro^{\Lin\times\C^2} & \C^3 & \rPro^{\C\times P} & \C^2 \\
			&\lambda'\times\C &\dPro[snake=1em]>{P\times\C} & \alpha & \dPro>P \\
			&&\rnode{CC}{\C^2} & \rPro_P & \C
			\nccurve[angleA=270,angleB=180,ncurv=1]{->}{ICC}{CC}\Bput1
		\end{diagram}
	\]
	Concretely, this amounts to showing that the diagram
	\begin{diagram}[h=2em,labelstyle=\scriptstyle]
		\int^{A,X} P(A,X;N)\times\Lin A\times P(L,M;X)
		& \rTo^{\int^X \lambda'_{X,N}\times P(L,M;X)}
		& \int^X \C(X,N)\times P(L,M;X) \\
		&&\dTo>\cong \\
		\dTo<{\int^A\Lin A\times\alpha_{A,L,M,N}}
		&& P(L,M;N) \\
		&&\uTo>\cong \\
		\int^{A,X}\Lin A\times P(A,L;X)\times P(X,M;N)
		& \rTo_{\int^X\lambda'_{L,X}\times P(X,M;N)}
		& \int^X\C(L,X)\times P(X,M;N)
	\end{diagram}
	commutes, which is an immediate consequence of (\ref{diag-linel}).
\end{proof}

\subsection{Application to semi SMCCs}
In the case where $\C$ is a semi SMCC, we have the isomorphism
\[
	\int^A \Lin A\times P(A,X;Y) \cong \int^{A} \Lin(A)\times\C(A,X\lolli Y)
	\cong \Lin(X\lolli Y),
\]
and the natural transformation $\lambda'_{X,Y}: \Lin(X\lolli Y)\to\C(X,Y)$
is precisely the natural transformation that we called $\ell$ in Section~\ref{s-linel}.
Therefore Proposition~\ref{prop-promonunit} does indeed justify the presentation
described in Section~\ref{s-linel}.

\section{The $\psi$ presentation}
In the closed case we have a functor $\lolli$ such that
\[
	P(A,B;C) \cong \C(A, B\lolli C),
\]
The left-unit isomorphism of a promonoidal category has components
\[
	\lambda_{A,B}: \int^{X} J(X)\times P(X,A;B) \to \C(A,B),
\]
the left-hand side of which is isomorphic to
\[
	\int^{X} J(X)\times \C(X, A\lolli B)
\]
which in turn is isomorphic to $J(A\lolli B)$. So we can take
$\lambda$ to be an isomorphism
\[
	\lambda_{A,B}: J(A\lolli B) \to \C(A,B),
\]
thereby eliminating the need to mention coends. The problem
(if we want a fully coherent presentation) is to reconcile the
fact that the associativity and symmetry are defined using $\tn$,
whereas the unit is defined using $\lolli$. Abstractly, we may
consider that we have two isomorphic multiplication profunctors,
say $P$ and $Q$ where
\[
	P(A,B;C) = \C(A\tn B,C)
\]
and
\[
	Q(A,B;C) = \C(A, B\lolli C),
\]
with the associativity and symmetry isomorphisms defined on $P$,
and the unit isomorphism defined on $Q$. Of course a unit isomorphism
may be defined on $P$ by using the isomorphism with $Q$, and the
coherence condition for the unit expressed in terms of this composite.
Here it simplifies matters to back up
and address the question at the level of structure internal to
a general monoidal bicategory. (The symmetry does not play an
essential role in this part of the argument, so there is no need
to assume a braiding here.) We shall use the language of
components, and to make the notation more familiar we shall
write $A\tn B$ to mean $P(A,B)$ and $A\odot B$ to mean $Q(A,B)$.
Also we'll write $I$ to mean $J()$. So (symmetry aside)
we have invertible 2-cells with components
\[
	\alpha_{A,B,C}: A\tn(B\tn C) \to (A\tn B)\tn C,
\]
\[
	\lambda_{A}: I\odot A\to A,
\]
and
\[
	\chi_{A,B}: A\tn B \to A\odot B,
\]
this last corresponding to the currying isomorphism.
We assume that $\alpha$ satisfies the pentagon condition, and that the
diagram of components
\begin{diagram}\dlabel[$\boldsymbol{\lambda\alpha\chi}$]{diag-lac}
	I\tn(A\tn B) && \rTo^{\alpha_{I,A,B}} && (I\tn A)\tn B \\
	\dTo<{\chi_{I,A\tn B}} && \dnum && \dTo>{\chi_{I,A}\tn B} \\
	I\odot(A\tn B) & \rTo_{\lambda_{A\tn B}} & A\tn B
		& \lTo_{\lambda_{A}\tn B} & (I\odot A)\tn B
\end{diagram}
commutes. Now, define $\psi$ to be the unique invertible 2-cell
with components
\[
	\psi_{A,B,C}: A\odot(B\tn C)\to (A\odot B)\odot C,
\]
such that the diagram
\begin{diagram}\dlabel[$\boldsymbol{\alpha\chi\psi}$]{diag-acp}
	A\tn(B\tn C) && \rTo^{\alpha_{A,B,C}} && (A\tn B)\tn C \\
	\dTo<{\chi_{A,B\tn C}} &&\dnum&& \dTo>{\chi_{A,B}\tn C} \\
	A\odot(B\tn C) & \rTo_{\psi_{A,B,C}} & (A\odot B)\odot C
		& \lTo_{\chi_{A\odot B, C}} & (A\odot B)\tn C
\end{diagram}
commutes. In the abstract this seems a rather odd thing to construct,
but in our concrete example it corresponds (via Yoneda) to a natural
isomorphism $(X\tn Y)\lolli Z \cong X\lolli(Y\lolli Z)$, an internal
analogue of currying. We shall consider the relationship between
$\chi$ and $\psi$, with the aim of replacing the former by the latter.
\begin{lemma}\label{lemma-chipsiunit}
	If diagram \pref{diag-acp} commutes, then diagram
	\pref{diag-lac} commutes if and only if the following does.
	\begin{diagram}\dlabel[$\boldsymbol{\lambda\chi\psi}$]{diag-lcp}
		I\odot(A\tn B) & \rTo^{\psi_{I,A,B}} & (I\odot A)\odot B \\
		\dTo<{\lambda_{A\tn B}} &\dnum& \dTo>{\lambda_{A}\odot B} \\
		A\tn B & \rTo_{\chi_{A,B}} & A\odot B
	\end{diagram}
\end{lemma}
\begin{proof}
	Consider the diagram
	\begin{diagram}[h=1.5em]
		I\tn(A\tn B) && \rTo^{\alpha_{I,A,B}} && (I\tn A)\tn B \\
		\\
		\dTo<{\chi_{I,A\tn B}} && (I\odot A)\odot B && \dTo>{\chi_{I,A}\tn B} \\
		&\ruTo^{\psi_{I,A,B}} & \dTo>{\scriptstyle\lambda_{A}\odot B} & \luTo^{\chi_{I\odot A,B}}\\
		I\odot(A\tn B) && A\odot B && (I\odot A)\tn B \\
		& \rdTo_{\lambda_{A\tn B}} & \uTo>{\scriptstyle\chi_{A,B}} & \ldTo_{\lambda_{A}\tn B} \\
		&&A\tn B
	\end{diagram}
	The upper region is an instance of \pref{diag-acp}, and the quadrilateral
	at lower-right commutes by naturality. Since all the components are invertible,
	it follows that the outside \pref{diag-lac} commutes if and only if the
	left-hand quadrilateral \pref{diag-lcp} does.
\end{proof}
\begin{lemma}\label{lemma-pentagons}
	If diagram \pref{diag-acp} commutes, then so does
	\begin{diagram}\dlabel[$\boldsymbol{\alpha\psi}$]{diag-ap}
	 A\odot \bigl(B\tensor (C\tensor D)\bigr)
	 &\rTo^\psi&(A\odot B)\odot (C\tensor D)
	 &\rTo^\psi & \bigl((A\odot B)\odot C\bigl)\odot D
	 \\
	 \dTo<{A\tn\alpha} &&\dnum&& \uTo>{\psi\odot D}
	 \\
	 A\odot\big((B\tensor C)\tensor D\big)
	 && \rTo_\psi && \bigl(A\odot(B\tensor C)\bigr)\odot D
	\end{diagram}
\end{lemma}
\begin{proof}
	We can use \pref{diag-acp} to show that \pref{diag-ap} is equivalent
	to the pentagon condition.
	\begin{sidewaysfigure}
	\begin{diagram}
	 A\odot \bigl(B\tensor (C\tensor D)\bigr)
	 &&\rTo^\psi&&(A\odot B)\odot (C\tensor D)
	 &&\rTo^\psi && \bigl((A\odot B)\odot C\bigl)\odot D
	 \\
	 &\luTo &&& \uTo &&& \ruTo
	 \\
	 &&A\tensor \bigl(B\tensor (C\tensor D)\bigr)
	 &\rTo^\alpha&(A\tensor B)\tensor (C\tensor D)
	 &\rTo^\alpha & \bigl((A\tensor B)\tensor C\bigl)\tensor D
	 \\
	 \dTo<{A\odot\alpha} &&\dTo<{A\tn\alpha} &&&& \uTo>{\alpha\tn D} && \uTo>{\psi\odot D}
	 \\
	 && A\tensor\big((B\tensor C)\tensor D\big)
	 && \rTo_\alpha && \bigl(A\tensor(B\tensor C)\bigr)\tensor D
	 \\
	 &\ldTo &&&&&& \rdTo
	 \\
	 A\odot\big((B\tensor C)\tensor D\big)
	 &&&& \rTo_\psi &&&& \bigl(A\odot(B\tensor C)\bigr)\odot D
	\end{diagram}
	\caption{Diagram used in the proof of Lemma~\ref{lemma-pentagons}}
	\label{fig-pentagons}
	\end{sidewaysfigure}
	Consider the diagram shown in Figure~\ref{fig-pentagons}. The
	unlabelled arrows are constructed using repeated instances of $\chi$:
	for example the vertical arrow
	\[
		(A\tn B)\tn(C\tn D) \to (A\odot B)\odot(C\tn D)
	\]
	is equal to the diagonal of the commutative square
	\begin{diagram}
		(A\tn B)\tn(C\tn D) & \rTo^{\chi_{A,B}\tn(C\tn D)} & (A\odot B)\tn(C\tn D) \\
		\dTo<{\chi_{A\tn B,C\tn D}} && \dTo>{\chi_{A\odot B,C\tn D}} \\
		(A\tn B)\odot(C\tn D) & \rTo_{\chi_{A,B}\odot(C\tn D)} & (A\odot B)\odot(C\tn D)
	\end{diagram}
	The cells that involve these arrows thus commute by a combination of
	naturality and condition~\pref{diag-acp}. So the central pentagon
	commutes if and only if the outside does.
\end{proof}
\begin{lemma}\label{lemma-acp}
	If \pref{diag-lcp} and \pref{diag-ap} commute, then so does \pref{diag-acp}.
\end{lemma}
\begin{proof}
	Consider the diagram shown in Figure~\ref{fig-acp}.
	\begin{sidewaysfigure}
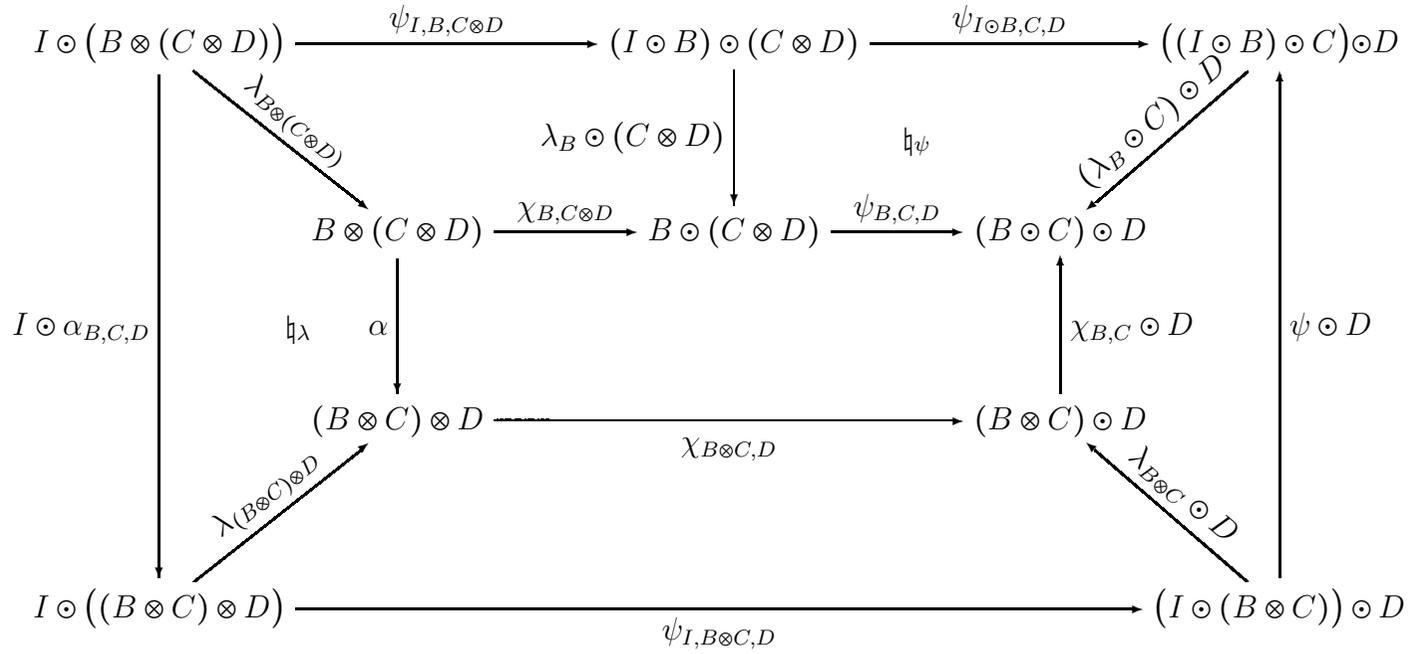

	\begin{diagram}[hug]
	 I\odot \bigl(B\tensor (C\tensor D)\bigr)
	 &&\rTo^{\psi_{I,B,C\tn D}}&&(I\odot B)\odot (C\tensor D)
	 &&\rTo^{\psi_{I\odot B,C,D}} && \bigl((I\odot B)\odot C\bigl)\odot D
	 \\
	 &\rdTo^{\lambda_{B\tn(C\tn D)}} &&& \dTo<{\lambda_{B}\odot(C\tn D)}
		&\natural_{\psi}&& \ldTo^{(\lambda_{B}\odot C)\odot D}
	 \\
	 &&B\tensor (C\tensor D)
	 &\rTo^{\chi_{B,C\tn D}}&B\odot (C\tensor D)
	 &\rTo^{\psi_{B,C,D}} & (B\odot C)\odot D
	 \\
	 \dTo<{I\odot\alpha_{B,C,D}} &\natural_{\lambda}&\dTo<{\alpha} &&&& \uTo>{\chi_{B,C}\odot D}
		&& \uTo>{\psi\odot D}
	 \\
	 && (B\tensor C)\tensor D
	 && \rTo_{\chi_{B\tn C,D}} && (B\tensor C)\odot D
	 \\
	 &\ruTo^{\lambda_{(B\tn C)\tn D}} &&&&&& \luTo^{\lambda_{B\tn C}\odot D}
	 \\
	 I\odot\big((B\tensor C)\tensor D\big)
	 &&&& \rTo_{\psi_{I,B\tn C,D}} &&&& \bigl(I\odot(B\tensor C)\bigr)\odot D
	\end{diagram}
	\caption{Diagram used in the proof of Lemma~\ref{lemma-acp}}
	\label{fig-acp}
	\end{sidewaysfigure}
	The regions marked with the symbol $\natural$ commute by
	naturality, the other three quadrilaterals commute by~\pref{diag-lcp},
	and the outside is an instance of~\pref{diag-ap}. Thus the inner
	pentagonal region commutes, which is precisely~\pref{diag-acp}.
\end{proof}
So $\chi$ and $\psi$ are interdefinable: given $\chi$, we can define
$\psi$ using diagram~\pref{diag-acp}, and alternatively given $\psi$
we can define $\chi$ using diagram~\pref{diag-lcp}. If we take $\psi$
rather than $\chi$ to be part of our defining data, then it suffices
to take~\pref{diag-ap} as an additional axiom (in addition to the
pentagon condition). Diagram \pref{diag-lcp} commutes by construction
so, by Lemma~\ref{lemma-acp}, condition~\pref{diag-acp} holds;
then by Lemma~\ref{lemma-chipsiunit} condition~\pref{diag-lac}
holds too.

In the concrete case we're considering, this justifies the
presentation of Section~\ref{s-psi-presentation}.

\section{The star-autonomous Case}
Finally, we consider full (non-intuitionistic) multiplicative
linear logic. The appropriate notion of model (for the unitless
fragment) is easy to find:
\begin{definition}
	A \emph{semi star-autonomous category} is a semi SMC $\C$
	equipped with an equivalence $-^{*}: \C\to\C\op$ and a natural
	isomorphism with components $\C(A\tn B, C)\cong\C(A, (B\tn\C^{*})^{*})$.
\end{definition}
We shall write $B\lolli C$ as an abbreviation for $(B\tn C^{*})^{*}$.
The $\psi$ presentation becomes remarkably simple in the
star-autonomous case:
\begin{propn}
	To give a semi star-autonomous category is to give a
	category $\C$ equipped with an associative, symmetric
	functor $\tn: \C\times\C\to\C$, an equivalence
	$-^{*}: \C\to\C\op$, and a functor $J: \C\to\Set$
	with a natural isomorphism
	\[
		J((A\tn B^{*})^{*}) \cong \C(A,B).
	\]
\end{propn}
\begin{proof}
	Define $\psi_{A,B,C}$ to be the composite
	\[
		\bigl((A\tn B)\tn C^{*}\bigr)^{*}
		\rTo^{(\alpha_{A,B,C^{*}})^{*}}
		\bigl(A\tn (B\tn C^{*})\bigr)^{*}
		\rTo^{(A\tn n_{B\tn C^{*}})^{*}}
		\Bigl( A\tn \bigl( (B\tn C^{*})^{*} \bigr)^{*} \Bigr)^{*}.
	\]
	With this definition, the diagram~\pref{diag-ap} commutes
	as a consequence of the pentagon condition.
\end{proof}

\end{thesischapter}
	
	{\let\origdiag = \diag
	\def\diag#1{\origdiag{semicc/#1}}
	\let\origcdiag = \cdiag
	\def\cdiag#1{\origcdiag{semicc/#1}}
\documentclass{robinthesis}

\begin{thesischapter}{SemiCC}{Compact Closed Categories without Units}
It is, of course, a routine matter to specialise
this definition of semi star-aut\-on\-om\-ous category to the compact closed case.
This chapter gives an elementary axiomatisation of semi compact closure, and shows its
equivalence to the `abstract' notion. The present definition is completely algebraic,
and is perhaps easier to understand and use. The definition itself is not
really new: \citet[][\S3.5]{HinesSS} has a similar-looking definition, which
seems to be strictly weaker than the present one, and \citet{ProofNetCats}
give a more general version.\footnote{
	The Do{\v s}en-Petri{\'c} axioms are intended to define a
	semi star-autonomous category, therefore assume two tensors $\tn$ and $\parr$,
	related by a linear distributivity. If one takes the two tensors to be equal, and the
	linear distributivity to be the ordinary associativity, then one recovers the present
	definition with some redundancy.}

\begin{definition}
A \emph{category with tensor} $\C$ is a category equipped with a tensor product
\[
	\tn: \C\times\C\to\C,
\]
together with natural isomorphisms having components
\[\begin{array}{l}
	\alpha_{A,B,C}: A\tn(B\tn C) \to (A\tn B)\tn C,\\
	\sigma_{A,B}: A\tn B \to B\tn A
\end{array}\]
such that $\sigma_{B,A}^{-1} = \sigma_{A,B}$, and subject to the pentagon and hexagon conditions
found in the usual definition of symmetric monoidal category.
\end{definition}
Although the development is stated in terms of a symmetric tensor, it is perfectly
possible -- with only a little more work -- to carry it through when the tensor is
merely braided. The string diagrams, in particular, should make it clear which
direction of braiding is required in any particular definition. Note that an additional
axiom is needed in the braided case, specifically the braid dual of the second
cancellation condition, and braid-dual versions of the lemmas need to be proved.
We also introduce an abbreviation that will be useful in the next definition:
let $\theta$ denote the unique canonical natural isomorphism with components
\[
	\theta_{A,B,C}: A\tn(B\tn C) \to (C\tn A)\tn B.
\]
(This may be defined as either
$\alpha_{C,A,B}.\sigma_{A\tn B,C}.\alpha_{A,B,C}$
or
$(\sigma_{A,C}\tn B)\alpha_{A,C,B}(A\tn\sigma_{B,C})$;
the hexagon condition says precisely that these must be equal.)
\begin{definition}
A \emph{semi compact closed category} is a category $\C$ with tensor, equip\-ped
with:
for every object $A\in\C$, a \emph{dual object} $A^*$, and
natural transformations $\eta^A$ and $\e^A$ with components
\[\begin{array}l
	\eta^A_X: X \to X\tn(A^*\tn A)\\
	\e^A_X: (A\tn A^*)\tn X \to A
\end{array}\]
These natural transformations are called the \emph{unit} and \emph{counit}
of $A$, and are required to satisfy the four axioms shown in Fig.~\ref{fig-ax}.
\end{definition}
\begin{figure}
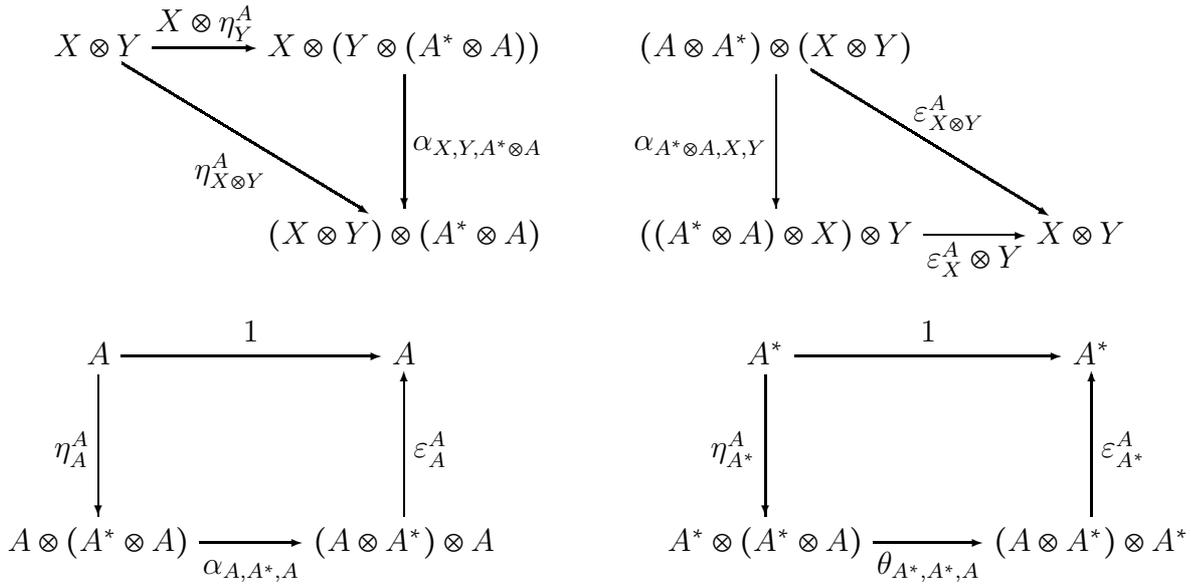

	\[\begin{array}{cc}
	\hskip-3em
	\begin{diagram}
		X\tn Y &\rTo^{X\tn \eta^A_Y} & X\tn(Y\tn(A^*\tn A))\\
		&\rdTo_{\eta^A_{X\tn Y}}&\dTo>{\alpha_{X,Y,A^*\tn A}}\\
		&&(X\tn Y)\tn(A^*\tn A)
	\end{diagram}
	&
	\hskip 2em
	\begin{diagram}
		(A\tn A^*)\tn(X\tn Y)\\
		\dTo<{\alpha_{A^*\tn A,X,Y}}&\rdTo^{\e^A_{X\tn Y}}\\
		((A^*\tn A)\tn X)\tn Y &\rTo_{\e^A_X\tn Y}&X\tn Y
	\end{diagram}
	\\[5em]
	\hskip-3em
	\begin{diagram}
		A &\rTo^1& A\\
		\dTo<{\eta^A_A}&&\uTo>{\e^A_A}\\
		A\tn(A^*\tn A) &\rTo_{\alpha_{A,A^*,A}}&(A\tn A^*)\tn A
	\end{diagram}
	&
	\hskip 2em
	\begin{diagram}
		A^* &\rTo^1& A^*\\
		\dTo<{\eta^A_{A^*}}&&\uTo>{\e^A_{A^*}}\\
		A^*\tn(A^*\tn A) & \rTo_{\theta_{A^*,A^*,A}}&(A\tn A^*)\tn A^*
	\end{diagram}
	\end{array}\]
	\caption{Coherence conditions for a semi compact closed category}
	\label{fig-ax}
\end{figure}%
The plan for the rest of this chapter is as follows. In \S\ref{s-direct} we develop the
theory of semi compact closed categories directly from the axioms, since it is instructive
to see how readily this may be done, and how similar it is to the ordinary theory of
compact closure. (But see later for an alternative, indirect, approach.) \S\ref{s-hhs}
then shows that every semi compact closed category is (degenerately) semi star-%
autonomous in the sense of Chapter~\refchapter{Promon}.

\S\ref{s-embed} is independent of the previous sections, and shows
how an arbitrary semi compact closed category may be fully and faithfully embedded in an
ordinary compact closed category (which has one additional object playing the role of
the unit). This embedding preserves the tensor and duality on the nose, which makes it
possible to transfer most of our knowledge about compact closed categories to the
unitless situation, and in particular to deduce the main results of \S\ref{s-direct}.

\section{Direct development}\label{s-direct}
We shall use string diagrams \citep{GTC}, to make the calculations easier
to follow. Our diagrams are to be read from left to right, and we notate $\eta$
and $\e$ as in Fig.~\ref{fig-diagdef}.
\begin{figure}
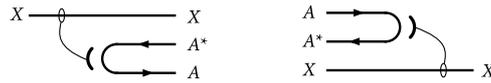

	\hbox to\columnwidth\bgroup\hss
		\begin{tabular}{c@{\hskip 3em}c}
		\diag{d-eta} & \diag{d-eps}
		\end{tabular}
	\hss\egroup
	\caption{Diagrammatic notation for $\eta$ and $\e$}\label{fig-diagdef}
\end{figure}
Diagrammatic forms of the axioms are shown in Figs.~\ref{fig-diagnat}--\ref{fig-diagcanc}.
	\begin{figure}
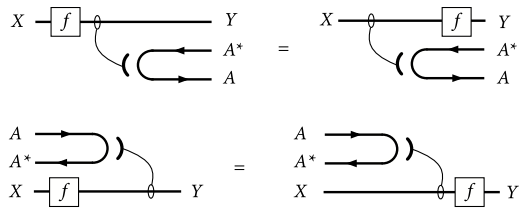

	\hbox to\columnwidth\bgroup\hss
		\begin{tabular}{c}
		\diag{d-eta-nat}
		\\[1em]
		\diag{d-eps-nat}
		\end{tabular}
	\hss\egroup
	\caption{Diagrammatic form of the naturality conditions}\label{fig-diagnat}
\end{figure}
\begin{figure}
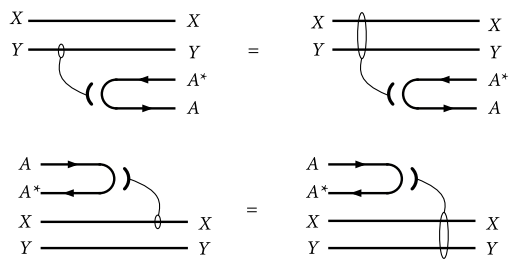

	\hbox to\columnwidth\bgroup\hss
		\begin{tabular}{c}
		\diag{d-eta-alph}
		\\[1em]
		\diag{d-eps-alph}
		\end{tabular}
	\hss\egroup
	\caption{Diagrammatic form of the associativity conditions}\label{fig-diagass}
\end{figure}
\begin{figure}
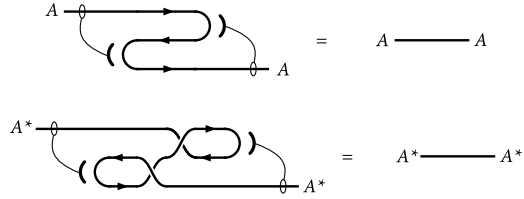

	\hbox to\columnwidth\bgroup\hss
		\begin{tabular}{c}
		\diag{d-ax1}
		\\[1em]
		\diag{d-ax2}
		\end{tabular}
	\hss\egroup
	\caption{Diagrammatic form of the cancellation conditions}\label{fig-diagcanc}
\end{figure}
The first task is to show how the duality operation can be extended to a contravariant
functor, in such a way that $\eta$ and $\e$ are both dinatural in $A$. Given an arrow
$f: A\to B$, we define $f^*: B^*\to A^*$ as shown in Fig.~\ref{fig-fstar}.
\begin{figure}
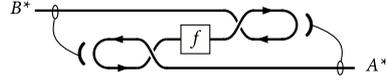

	\hbox to\columnwidth{\hss\diag{d-fstar}\hss}
	\caption{Given $f: A\to B$, we define $f^*: B^*\to A^*$ using this diagram}\label{fig-fstar}
\end{figure}
Note that, directly from the second cancellation axiom, we have $1_A^* = 1_{A^*}$
for all $A\in\C$, thus our putative functor preserves identities (which is a good start).
It is surprisingly complicated to prove directly that it also preserves composition, but
it will be easy once we have the right lemmas.
\begin{lemma}\label{lemma}
	For all $X$, $A$ and $Y\in\C$, the following diagrams commute.
	\begin{diagram}
		(X\tn(A\tn A^*))\tn Y &\rTo^{\alpha_{X,A\tn A^*, Y}^{-1}}& X\tn((A\tn A^*)\tn Y)\\
		\dTo<{\sigma_{A\tn A^*,X}\tn Y}&&\dTo>{X\tn\e^A_Y}\\
		((A\tn A^*)\tn X)\tn Y &\rTo_{\e^A_X\tn Y}&X\tn Y
	\end{diagram}
	\begin{diagram}
		X\tn Y &\rTo^{X\tn\eta^A_Y}&X\tn(Y\tn(A^*\tn A))\\
		\dTo<{\eta^A_X\tn Y}&&\dTo>{X\tn\sigma_{A^*\tn A, Y}}\\
		(X\tn(A^*\tn A))\tn Y &\rTo_{\alpha_{X,A^*\tn A,Y}^{-1}}&X\tn((A^*\tn A)\tn Y)
	\end{diagram}
\end{lemma}
\begin{proof}
	The proof of the first diagram, by string diagram manipulation, is shown in Fig.~\ref{fig-lemma}.
	(Perhaps the least obvious step is the penultimate one, which uses the
	naturality of $\sigma$.) The second is proved by a symmetrical argument:
	Fig.~\ref{fig-lemma'} shows the diagrammatic form of its statement.
\end{proof}
\begin{figure}
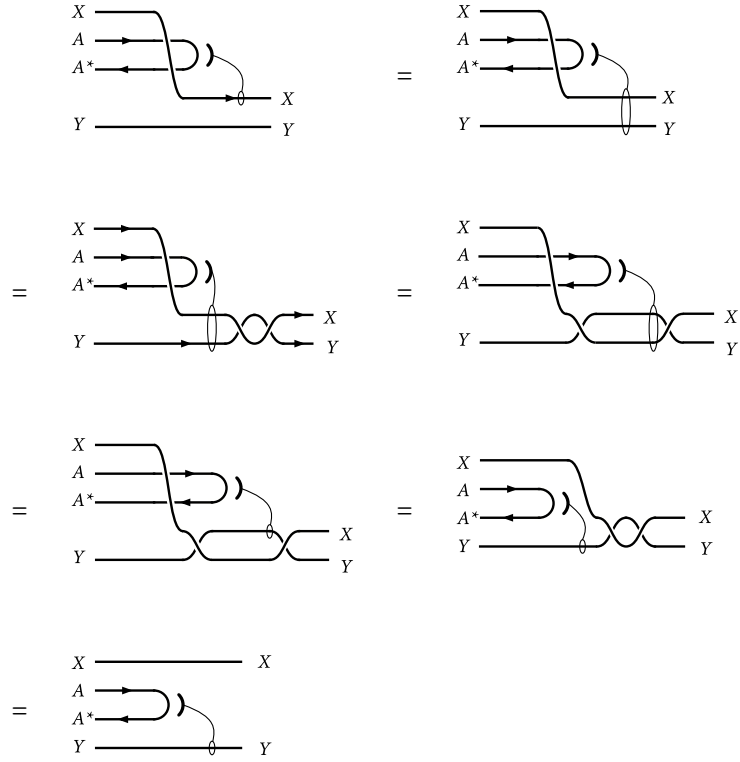

\[\begin{array}{rlcl}
	& \cdiag{d-l1} &=&  \cdiag{d-l2}\\[4em]
	=& \cdiag{d-l3} &=&  \cdiag{d-l4}\\[4em]
	=& \cdiag{d-l5} &=&  \cdiag{d-l6}\\[4em]
	=& \cdiag{d-l7}
\end{array}\]
\caption{A diagrammatic proof of Lemma~\ref{lemma}}\label{fig-lemma}
\end{figure}
\begin{figure}
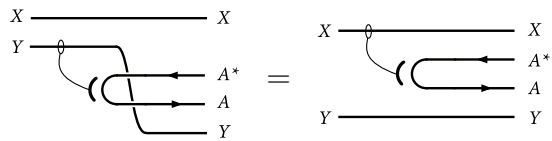

	\hbox to\columnwidth\bgroup\hss
		\cdiag{d-l-eta-lhs} = \cdiag{d-l-eta-rhs}
	\hss\egroup
	\caption{The second part of Lemma~\ref{lemma}}\label{fig-lemma'}
\end{figure}
\begin{lemma}\label{sec}
For any objects $X$,$A$,$B$,$Y$, and arrow $f:A\to B$, the following diagram
commutes. (The associativities have been suppressed to make it more comprehensible.)
\begin{diagram}
	X\tn B^*\tn Y & \rTo^{X\tn f^*\tn Y} & X\tn A^*\tn Y\\
	\dTo<{\eta^A_X\tn B^*\tn Y} && \uTo>{X\tn A^*\tn\e^B_Y}\\
	X\tn A^*\tn A \tn B^*\tn Y &\rTo_{X\tn A^*\tn f \tn B^*\tn Y}& X\tn A^*\tn B\tn B^*\tn Y
\end{diagram}
In string diagram terms, this says
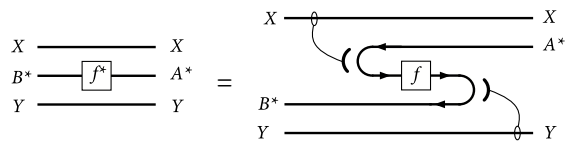
\[
	\cdiag{d-sec-rhs} = \cdiag{d-sec-lhs}
\]
\end{lemma}
\begin{proof}
	The proof is again by string diagram manipulation, shown in Fig.~\ref{fig-sec}.
	Both parts of Lemma~\ref{lemma} are used.
\end{proof}
\begin{figure}
\hbox to \columnwidth{\hss$\begin{array}{rlcl}
	& \cdiag{d-sec-rhs} &=&  \cdiag{d-sec1}\\[4em]
	=& \cdiag{d-sec2} &=&  \cdiag{d-sec3}\\[4em]
	=& \cdiag{d-sec4} &=&  \cdiag{d-sec5}\\[4em]
	=& \cdiag{d-sec-lhs}
\end{array}$\hss}
\caption{A diagrammatic proof of Lemma~\ref{sec}}\label{fig-sec}
\end{figure}%
\begin{lemma}\label{third}
	For all $X$, $A$, $Y\in\C$, the following diagram commutes.
	\begin{diagram}
	X\tn A\tn Y & \rTo^{1} & X\tn A\tn Y\\
	\dTo<{\eta^A_X\tn A\tn Y} && \uTo>{X\tn A\tn\e^A_Y}\\
	X\tn A^*\tn A \tn A\tn Y & \rTo_{X\tn\sigma_{A^*,A\tn A, Y}} & X\tn A\tn A\tn A^*\tn Y 
	\end{diagram}
	(The associativities have again been suppressed.)
\end{lemma}
\begin{proof}
s	See Fig.~\ref{fig-third}. The first step uses both parts of Lemma~\ref{lemma}.%
	\footnote{In the braided case, it uses the braid-dual analogue of that lemma.}
\end{proof}
\begin{figure}
\hbox to \columnwidth{\hss$\begin{array}{rlcl}
	& \cdiag{d-third1} &=&  \cdiag{d-third2}\\[4em]
	=& \cdiag{d-third3} &=&  \cdiag{d-third4}\\[4em]
\end{array}$\hss}
\caption{Proof of Lemma~\ref{third}}\label{fig-third}
\end{figure}%
All the hard work was in the lemmas: everything else is comparatively straightforward.
\begin{propn}\label{prop-dinat}
	The natural transformations $\eta$ and $\e$ are also dinatural in
	the superscript variable.
\end{propn}
\begin{proof}
	See Fig.~\ref{fig-dinat} for a proof that $\e$ is dinatural.
	The proof for $\eta$ may be obtained by turning the string diagrams upside down. 
\end{proof}
\begin{figure}
\hbox to \columnwidth{\hss$\begin{array}{rlcl}
	& \cdiag{d-dinat-rhs} &=&  \cdiag{d-dinat1}\\[4em]
	=& \cdiag{d-dinat2} &=&  \cdiag{d-dinat3}\\[4em]
	=& \cdiag{d-dinat-lhs}
\end{array}$\hss}
\caption{Proof that $\e$ is dinatural (Prop.~\ref{prop-dinat})}\label{fig-dinat}
\end{figure}%
\begin{propn}\label{prop-comp}
	The duality preserves composition, i.e.\ given $f:A\to B$ and $g: B\to C$,
	we have $(gf)^* = f^*g^*$.
\end{propn}
\begin{proof}
	See Fig.~\ref{fig-comp}. The third equality uses the dinaturality of $\eta$.
\end{proof}
\begin{figure}
\hbox to \columnwidth{\hss$\begin{array}{rlcl}
	& \cdiag{d-comp1} &=&  \cdiag{d-comp2}\\[4em]
	=& \cdiag{d-comp3} &=&  \cdiag{d-comp4}\\[4em]
	=& \cdiag{d-comp5}
\end{array}$\hss}
\caption{Proof of Prop.~\ref{prop-comp}}\label{fig-comp}
\end{figure}%
\begin{propn}\label{prop-starstar}
	There is a natural isomorphism $A\cong A^{**}$.
\end{propn}
\begin{proof}
Fig.~\ref{fig-starstar} shows how to construct a natural isomorphism $A\cong A^{**}$.
\begin{figure}
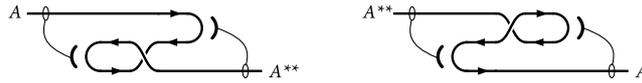

	\hbox to \columnwidth\bgroup\hss
	\begin{tabular}{c@{\qquad}c}
	\diag{d-starstar1} & \diag{d-starstar2}
	\end{tabular}
	\hss\egroup
	\caption{How to construct a natural isomorphism $A\cong A^{**}$}\label{fig-starstar}
\end{figure}%
Naturality is immediate from the naturality and dinaturality of $\eta$ and $\e$,
and the naturality of $\sigma$. Figs.~\ref{fig-starstar-inva} and~\ref{fig-starstar-invb}
show that these maps are indeed mutually inverse, hence determine an isomorphism.
Notice that the second step in Fig.~\ref{fig-starstar-inva} uses Lemma~\ref{third}.
\begin{figure}
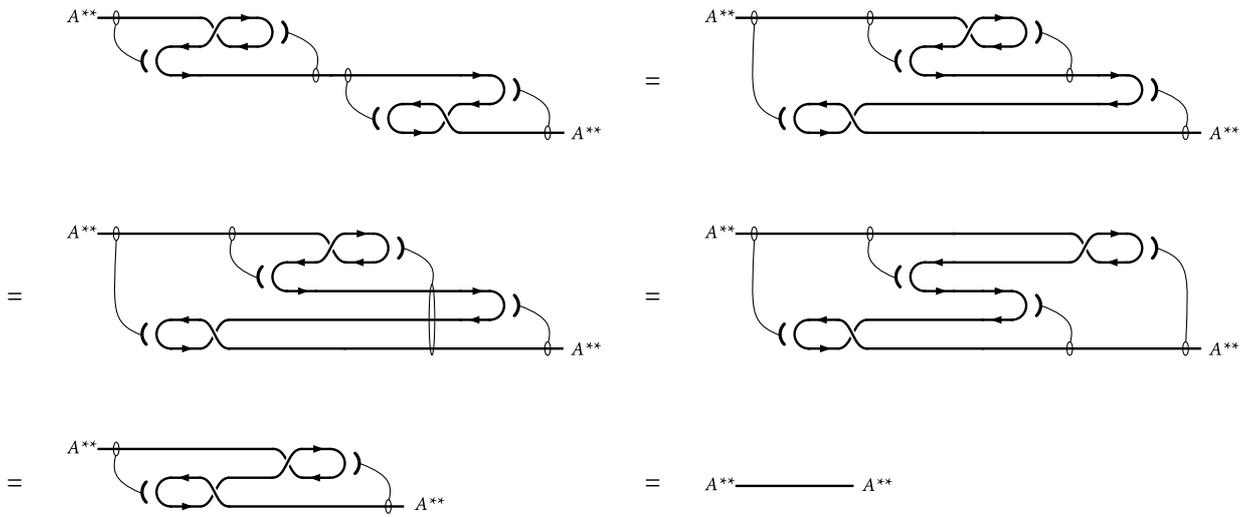

	\hbox to \columnwidth{\hss$\begin{array}{rlcl}
		& \cdiag{d-starinv-a1} &=&  \cdiag{d-starinv-a2}\\[4em]
		=& \cdiag{d-starinv-a3} &=&  \cdiag{d-starinv-a4}
	\end{array}$\hss}
	\caption{The maps from Fig.~\ref{fig-starstar} compose to give the identity on $A$}
	\label{fig-starstar-inva}
\end{figure}%
\begin{figure}
	\hbox to \columnwidth{\hss$\begin{array}{rlcl}
		& \cdiag{d-starinv-b1} &=&  \cdiag{d-starinv-b2}\\[4em]
		=& \cdiag{d-starinv-b3} &=&  \cdiag{d-starinv-b4}\\[4em]
		=& \cdiag{d-starinv-b5} &=&  \cdiag{d-starinv-b6}
	\end{array}$\hss}
	\caption{The maps from Fig.~\ref{fig-starstar} compose to give the identity on $A^{**}$}
	\label{fig-starstar-invb}
\end{figure}
\end{proof}
\begin{propn}\label{prop-adj}
	For each object $A$, there is an adjunction $A\tn- \dashv A^*\tn-$, which
	determines a natural isomorphism
	\[
		\C(B\tn A, C)\cong\C(A, B^*\tn C)
	\]
\end{propn}
\begin{proof}
There are obvious natural transformations
\[\begin{array}l
	\gamma_{A,B,C}: \C(A\tn B^*, C) \to \C(A, C\tn B),\\
	\delta_{A,B,C}: \C(A, B^*\tn C) \to \C(B\tn A, C)
\end{array}\]
illustrated in Figs.~\ref{fig-gamma}--\ref{fig-delta}.
\begin{figure}
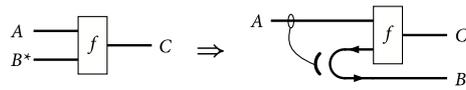

\[
	\cdiag{d-f} \To \cdiag{d-gamma-f}
\]
\caption{The natural transformation $\gamma$}\label{fig-gamma}
\end{figure}%
\begin{figure}
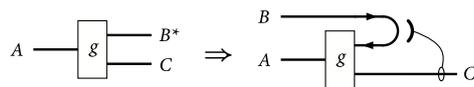

\[
	\cdiag{d-g} \To \cdiag{d-delta-g}
\]
\caption{The natural transformation $\delta$}\label{fig-delta}
\end{figure}%
We need to show that one of these natural transformations is invertible, which
we shall do by showing that they are in some sense mutually inverse. We begin
by showing that the composite
\vskip\abovedisplayskip\vbox{%
	\hbox to \columnwidth{$
		\C(A\tn B^*, C) \rTo^\gamma \C(A, C\tn B)
			\rTo^\cong \C(A, C\tn B^{**})$\hss}%
	\vskip\baselineskip
	\hbox to \columnwidth{\hss$
		\rTo^\cong \C(A, B^{**}\tn C)
			\rTo^\delta \C(B^*\tn A, C) \rTo^\cong \C(A\tn B^*, C)
$}}\vskip\belowdisplayskip\noindent
is the identity (where the unlabelled isomorphisms are symmetry or involution maps).
Consider some $f: A\tn B^*\to C$: the result of applying this composite to $f$ is
shown in Fig.~\ref{fig-dg}.
\begin{figure}
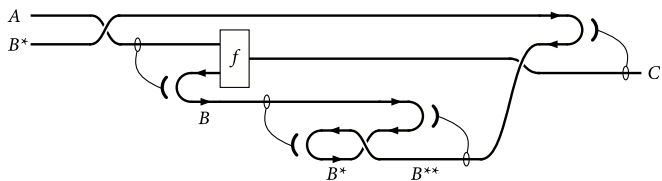

\hbox to \columnwidth{\hss\diag{d-delta-gamma-f}\hss}
\caption{The result of applying $\gamma$ and then $\delta$ to some $f: A\tn B^*\to C$}
\label{fig-dg}
\end{figure}%
Fig.~\ref{fig-dg-proof} shows that this is equal to $f$. (The first step combines
several uses of naturality and associativity conditions.)
\begin{figure}
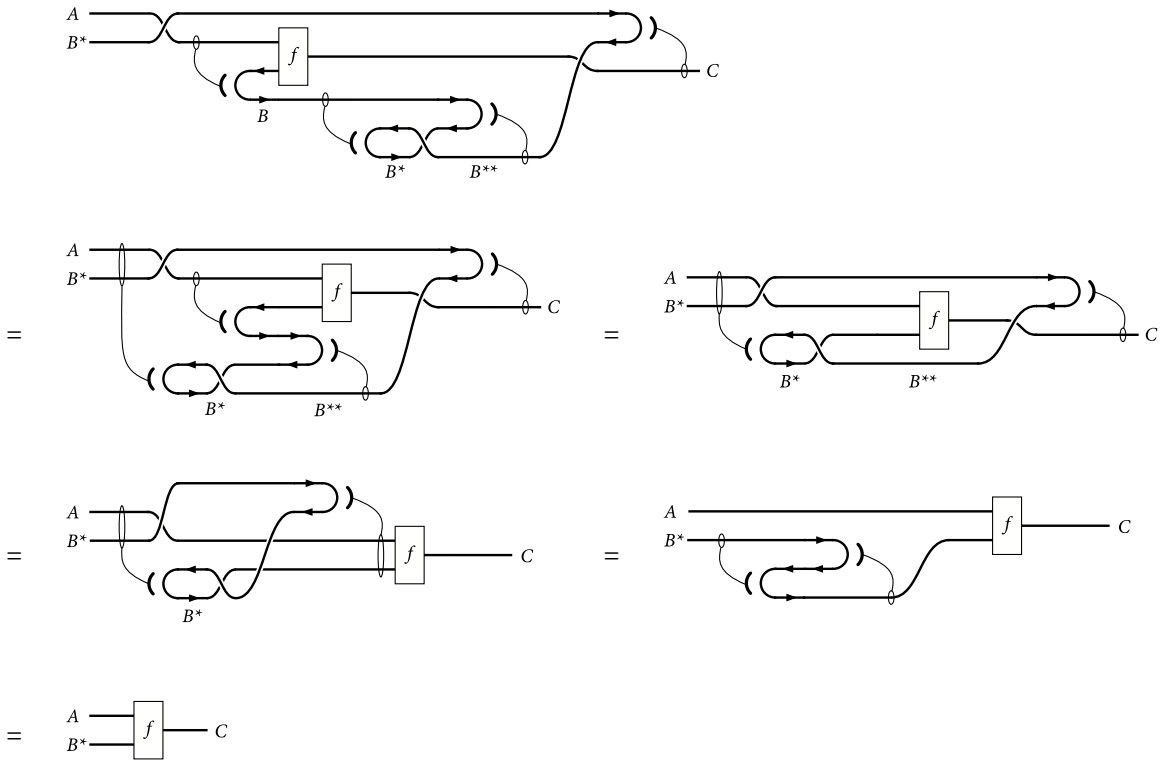

	\hbox to \columnwidth{\hss$\begin{array}{rlcl}
		& \multicolumn 3l{\cdiag{d-delta-gamma-f}}\\[4em]
		=& \cdiag{d-dg-1} &=&  \cdiag{d-dg-2}\\[4em]
		=& \cdiag{d-dg-3} &=&  \cdiag{d-dg-4}\\[4em]
		=&\cdiag{d-f}
	\end{array}$\hss}
	\caption{The map shown in Fig.~\ref{fig-dg} is equal to $f$}
	\label{fig-dg-proof}
\end{figure}%
Therefore $\gamma$ has a post-inverse and $\delta$ a pre-inverse.

Similarly one may take a map $g: A\to B^*\tn C$, and apply to it the composite
\vskip\abovedisplayskip\vbox{%
	\hbox to \columnwidth{$
	\C(A, B^*\tn C) \rTo^\delta \C(B\tn A, C) \rTo^\cong \C(B^{**}\tn A, C)
	$\hss}%
	\vskip\baselineskip
	\hbox to \columnwidth{\hss$
	\rTo^\cong \C(A\tn B^{**}, C)
		\rTo^\gamma \C(A, C\tn B^*) \rTo^\cong \C(A, B^*\tn C)
$\hss}}\vskip\belowdisplayskip\noindent
as shown in Fig.~\ref{fig-gd}.
\begin{figure}
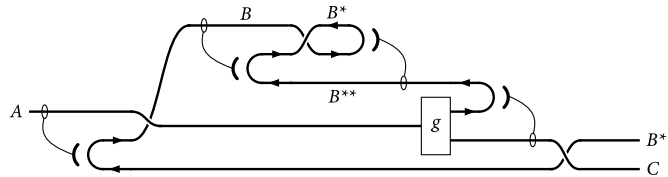

	\hbox to \columnwidth{\hss\diag{d-gamma-delta-g}\hss}
	\caption{The result of applying $\delta$ and then $\gamma$ to some $g: A\to B^*\tn C$,}
	\label{fig-gd}
\end{figure}%
This is equal to $g$ -- the proof is obtained by turning all the diagrams in Fig.~\ref{fig-dg-proof}
upside down -- hence $\gamma$ also has a pre-inverse and $\delta$ a post-inverse.
Therefore both are invertible, as claimed.
\end{proof}

\section{The promonoidal structure}\label{s-hhs}
This section shows that a semi compact closed category is semi star-autonomous
in the sense of Chapter~\refchapter{Promon}. The proof relies on the characterisation
of semi SMC categories via linear elements, as given in Section~\chref{Promon}{s-linel}.
With this machinery available it is easy to prove the main result of this section:
\begin{propn}\label{prop-ssa}
	A semi compact closed category is semi star-autonomous.
\end{propn} 
\begin{proof}
	Let $\C$ be a semi compact closed category. By assumption it
	is equipped with a symmetric tensor, and if we define
	\[
		A\lolli B := A^*\tn B
	\]
	then Prop.\ref{prop-adj} shows that we have a natural isomorphism
	\[
		\delta_{A,B,C}:\C(B\tn A,C) \lTo^\cong \C(A, B\lolli C).
	\]
	It remains only to construct an inverse to the function
	$\ell_{A,B}: \Lin(A^*\tn B) \to \C(A,B)$.
	
	If we represent a linear element $x\in\Lin(A^*\tn B)$ as shown in Fig.~\ref{fig-x},
	\begin{figure}
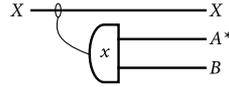

		\hbox to \columnwidth{\hss\cdiag{d-lin4}\hss}
		\caption{The diagrammatic representation of a linear element $x\in\Lin(A^*\tn B)$}
		\label{fig-x}
	\end{figure}%
	note that, by the definition of $\delta$, the arrow $\ell_{A,B}(x)$ is as shown
	in Fig.~\ref{fig-lx}.
	\begin{figure}
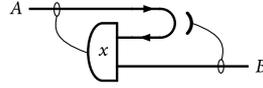

		\hbox to \columnwidth{\hss\cdiag{d-lx}\hss}
		\caption{The diagrammatic representation of $\ell_{A,B}(x)$}
		\label{fig-lx}
	\end{figure}%
		
	Given a map $f: A\to B$,
	define $\ell_{A,B}^{-1}(f)$ to be the natural transformation whose
	component at $X$ is
	\[
		X \rTo^{\eta^A_X} X\tn(A^*\tn A) \rTo^{X\tn(A^*\tn f)} X\tn(A^*\tn B).
	\]
	For any $f:A\to B$, the arrow $\ell_{A,B}(l_{A,B}^{-1}(f))$ is the composite
	\[
		A \rTo^{\eta^A_A} A\tn(A^*\tn A) \rTo^{A\tn(A^*\tn f)} A\tn(A^*\tn B)
			\rTo^{\alpha_{A,A^*,B}} (A\tn A^*)\tn B \rTo^{\e^A_B} B.
	\]
	By the naturality of $\alpha$ and $\epsilon$, and the first cancellation
	condition, this is indeed equal to $f$. Conversely suppose we have a
	linear element $x\in\Lin(A^*\tn B)$. Since it will be convenient to use
	a string diagram calculation here, we introduce a diagrammatic notation
	for this linear element, shown in Fig.~\ref{fig-x}.
	Now Fig.~\ref{fig-lin} shows a proof that $\ell_{A,B}^{-1}(l_{A,B}(x))$ is equal to~$x$.
	\begin{figure}
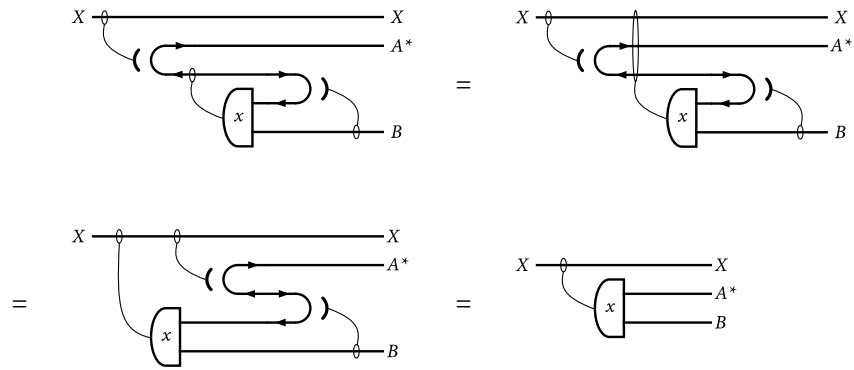

	\hbox to \columnwidth{\hss$\begin{array}{rlcl}
		& \cdiag{d-lin1} &=&  \cdiag{d-lin2}\\[4em]
		=& \cdiag{d-lin3} &=&  \cdiag{d-lin4}
	\end{array}$\hss}
		\caption{Proof that $\ell_{A,B}^{-1}(\ell_{A,B}(x)) = x$. (See Prop.~\ref{prop-ssa})}
		\label{fig-lin}
	\end{figure}%
\end{proof}

\section{Embedding theorem}\label{s-embed}
If we wanted to add a unit object $I$ to a semi star-autonomous category $\C$, we would
also have to add an infinite family of other objects such as $I^*$, $I^*\tn A$ for $A\in\C$,
and so on. In the compact closed case, there is no such obstacle, since $I^*$ is always
isomorphic to $I$, and we may take $I^* = I$ without essential loss of generality. This
raises the hope that it may always be possible to fully embed any semi compact closed
category $\C$ into a compact closed category $\C'$, in such a way the objects of $\C'$
are essentially just the objects of $\C$ plus a unit object.
It turns out that such an embedding is possible, as this section shows.

Recall the `$\mathbf e$' construction of \cite{BTC}, there used to prove
the case $\B=\Cat$ of the Cayley Theorem (our Proposition~\chref{Cayley}{prop-psmoncayley}).
Given a monoidal category $\C$,
the category $\mathbf{e}(\C)$ is defined as follows. An object of $\mathbf{e}(\C)$ is
a pair $(F, \gamma^F)$ of a functor $F:\C\to\C$ and a natural isomorphism with
components
\[
	\gamma^F_{A,B}: F(A\tn B) \to F(A)\tn B.
\]
A morphism $\delta: (F, \gamma^F) \to(G,\gamma^G)$ is a natural transformation
$F\To G$ such that the diagram
\begin{diagram}
	F(A\tn B) & \rTo^{\gamma^F_{A,B}} & F(A)\tn B
	\\
	\dTo<{\delta_{A\tn B}} && \dTo>{\delta_A\tn B}
	\\
	G(A\tn B) & \rTo^{\gamma^G_{A,B}} & G(A)\tn B
\end{diagram}
commutes for all $A$ and $B\in \C$.

The tensor product $(F, \gamma^F)\tn (G, \gamma^G)$ is defined to be $(FG, \gamma^{FG})$,
where $\gamma^{FG}_{A,B}$ is the composite
\[
	FG(A\tn B) \rTo^{F\gamma^G_{A,B}} F(GA \tn B) \rTo^{\gamma^F_{GA,B}} FGA\tn B.
\]
The tensor product of two arrows is their horizontal composite as natural transformations.
The tensor unit $I$ is simply the identity functor, with the identity natural transformation.

There is an functor $L: \C\to\mathbf{e}(\C)$, where $L(A) := (A\tn-, \alpha_{A,-,-})$
for objects $A\in \C$, and $L(f) := f\tn-$ for arrows $f$. Note that the unit object of $\C$ plays no part in the construction of the category $\mathbf{e}(\C)$ or the functor $L$, so that everything so far makes sense for a semi compact closed category. Furthermore:

\begin{propn}\label{prop-L-ff}
	When $\C$ is a semi compact closed category, thr functor $L$ is full and faithful.
\end{propn}
\begin{proof}
	\citeauthor{BTC}'s proof of this claim (for $\C$ a monoidal category) uses the tensor unit in an essential way, so we
	need to find a new proof that uses semi compact closure instead. The functor $L$ induces,
	for every $X$ and $Y\in\C$, a function $\C(X,Y) \to \mathbf{e}(\C)(LX, LY)$. We'll
	describe an inverse to this function, showing that it is invertible and hence that $L$ is
	full and faithful.
	
	Let $\delta$ be a natural transformation $LX\To LY$.
	Thus $\delta$ consists of components $\delta_A: X\tn A \to Y\tn A$, natural in $A$
	and such that the diagram
	\begin{diagram}
		X\tn (A\tn B) & \rTo^{\alpha_{X,A,B}} & (X\tn A)\tn B\\
		\dTo<{\delta_{A\tn B}} && \dTo>{\delta_A\tn B}\\
		Y\tn (A\tn B) & \rTo^{\alpha_{Y,A,B}} & (Y\tn A)\tn B
	\end{diagram}
	commutes for all $A$, $B\in\C$.
	
	It will be convenient to use string diagrams in the proof: we'll picture
	$\delta_A$ as
	\[
		\cdiag{d-L-delta}.
	\]
	In string diagram terms, the commutative square above is a rewiring condition
	of the sort we have seen above:
	\[\begin{array}[c]{lcr}
		\cdiag{d-L-delta-rewiring-lhs} &=& \cdiag{d-L-delta-rewiring-rhs}.
	\end{array}\]
	The naturality of $\delta$ means that functions can pass through the loop:
	\[\begin{array}[c]{lcr}
		\cdiag{d-L-delta-nat-lhs} &=& \cdiag{d-L-delta-nat-rhs}.
	\end{array}\]
	Now we can define our inverse to the action of L, to take $\delta$ to
	the following arrow $f: X\to Y$:
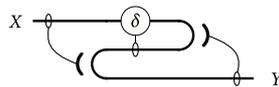
	\[
		\cdiag{d-L-f}
	\]
	We must show that $f\tn A = \delta_A$, for any $A\in \C$. The proof is
	a routine string diagram manipulation, shown in Fig.~\ref{fig-ftnA}.
	\begin{figure}
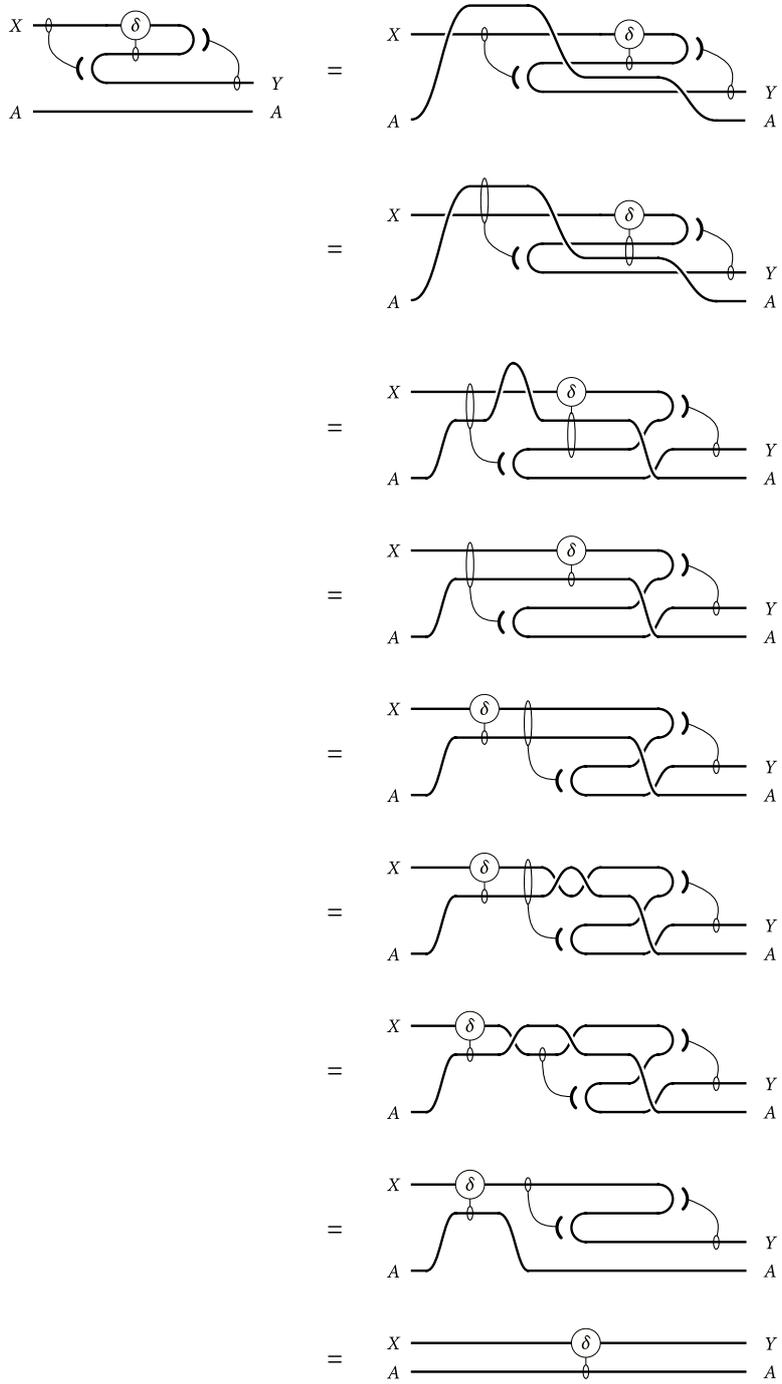

		\[\begin{array}[c]{rcl}
			\cdiag{d-ftnA-1}	
			&=&\cdiag{d-ftnA-2}	\\ \strut\\
			&=&\cdiag{d-ftnA-3}	\\ \strut\\
			&=&\cdiag{d-ftnA-4}	\\ \strut\\
			&=&\cdiag{d-ftnA-5}	\\ \strut\\
			&=&\cdiag{d-ftnA-6}	\\ \strut\\
			&=&\cdiag{d-ftnA-7}	\\ \strut\\
			&=&\cdiag{d-ftnA-8}	\\ \strut\\
			&=&\cdiag{d-ftnA-9}	\\ \strut\\
			&=&\cdiag{d-ftnA-10}
		\end{array}\]
		\caption{Proof that $f\tn A = \delta_A$, used in Prop.~\ref{prop-L-ff}.}
		\label{fig-ftnA}
	\end{figure}
	This shows that the passage  $\mathbf{e}(\C)(LX, LY)\to \C(X,Y)\to\mathbf{e}(\C)(LX, LY)$
	is the identity. For the other direction, we need to show that
	\[\begin{array}[c]{c}
		\cdiag{d-L-other-lhs} = \cdiag{d-L-other-rhs}
	\end{array}\]
	which is immediate.
\end{proof}
Now define $\E$ to be the full subcategory of $\mathbf{e}(\C)$ determined by the
objects that have adjoints. This is clearly a compact closed category.
\begin{propn}
	The image of $L$ is contained in $\E$. Specifically, for every $X\in\C$, the object
	$L(X)$ is adjoint to $L(X^*)$ in $\mathbf{e}(\C)$.
\end{propn}
\begin{proof}
	Define $\eta^{LX}: I \to L(X^*)L(X)$ to have components
	\[
		\cdiag{d-eta-LX}
	\]
	and $\e^{LX}: L(X)L(X^*)\to I$ to have components
	\[
		\cdiag{d-eps-LX}
	\]
	These are clearly natural in $A$, and it's easy to verify that they satisfy the
	condition making them maps of $\mathbf{e}(\C)$.
	To show that they really do define an adjunction between $L(X)$ and $L(X^*)$,
	we need to show that
	\[
		\cdiag{d-Ladj-X}
		\qquad \mbox{and} \qquad
		\cdiag{d-Ladj-Xstar}
	\]
	are both equal to the identity. This is an easy exercise in manipulations of the
	sort that are by now routine.
\end{proof}
Finally, let $\C'$ be the full subcategory of $\E$ determined by those objects that are
either isomorphic to $L(X)$ for some $X$, or isomorphic to $I$. This subcategory is
closed under the tensor and adjoint operations, so it's compact closed. The image
of the (full and faithful) functor $L$ is contained in $\C'$ by definition. Thus $\C$ is
embedded, in a structure-preserving fashion, in a compact closed category that
has essentially only one extra object, the unit object. (If in fact $\C$ had a unit object
all along, this functor will be an equivalence.)






\end{thesischapter}}
\end{thesis}